\documentclass[english,12pt,a4paper,twoside,openright]{book}
\pdfoutput=1
\usepackage[T1]{fontenc}
\usepackage[latin1]{inputenc}
\usepackage[letterpaper]{geometry}
\usepackage{color}
\usepackage{amsmath}
\usepackage{graphicx}
\usepackage{amssymb}
\usepackage{babel}
\usepackage{float}
\usepackage{mathtools}
\usepackage{amsmath}
\usepackage{chngcntr}
\newcommand{\dd}{\mathrm{d}}

\begin{document}

\begin{titlepage}
\thispagestyle{empty}
\begin{center}
\null
\vspace{4.4cm}
 \noindent\rule{16cm}{1.5pt}
  \\
  \vspace{0.3cm}
 {\huge \textbf{Composite dark matter and direct-search experiments}}
  \\
    \noindent\rule{16cm}{1.5pt}
\end{center}
\end{titlepage}

\newpage
\thispagestyle{empty}
\null
\newpage

\begin{titlepage}
\begin{center}
\includegraphics[scale=0.4]{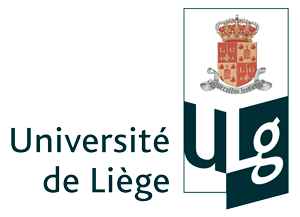}
\vspace{0.5cm}
\\
{\Large Ph.D. Thesis}
\vspace{1.cm}
\\
 \noindent\rule{16cm}{1.5pt}
  \\
  \vspace{0.3cm}
 {\huge \textbf{Composite dark matter and direct-search experiments}}
  \\
    \noindent\rule{16cm}{1.5pt}
    \vspace{1.cm}
    \\
    {\LARGE Quentin Wallemacq}
    \vspace{0.6cm}
  \\
  {\large Interactions Fondamentales en Physique et en Astrophysique}
  \vspace{0.4cm}
  \\
  {\large D\'epartement d'Astrophysique, G\'eophysique et Oc\'eanographie}
    \vspace{0.4cm}
  \\
  {\large Facult\'e des Sciences}
  \vspace{1cm}
  \\
  {\large \it Members of the jury}
  \vspace{0.5cm}
  \\
 {\large \it Prof. Jean-Ren\'e Cudell} \hfill {\large \it Advisor}
 \\
 {\large \it Prof. Pierre Magain} \hfill {\large \it President}
 \\
 {\large \it Dr. ~~Catarina Sim\~oes} \hfill {\large \it Secretary}
 \\
 {\large \it Prof. Rita Bernabei} \hfill {\large \it Examiner}
 \\
 {\large \it Prof. Maxim Yu. Khlopov} \hfill {\large \it Examiner}
 \\
 {\large \it Prof. Joseph Cugnon} \hfill {\large \it Examiner}
  \vfill
   {\large A dissertation submitted in fulfillment of the requirements} 
   \\
   {\large for the degree of Ph.D. in Sciences}
  \vspace{1cm}
  \\
  {\large September 2015}
\end{center}
\end{titlepage}

\newpage
\thispagestyle{empty}
\null
\newpage

\pagenumbering{roman}

\newenvironment{abstract}{\cleardoublepage\null\vfill\begin{center}\LARGE Abstract\end{center}}{\vfill\null}
\begin{abstract}
The main objective of this work was to reinterpret the results of the direct searches for dark matter in a new framework, in order to solve the contradictions between these experiments when they are interpreted in the scenario of the Weakly Interacting Massive Particles (WIMPs). For this purpose, we chose the direction of composite dark matter, i.e. dark matter particles that form neutral bound states, generically called ``dark atoms'', either with ordinary particles, or with other dark matter particles.

Three different scenarios have been investigated: the O-helium scenario, milli-interacting dark matter and dark anti-atoms. In each of them, dark matter interacts sufficiently strongly with terrestrial matter to be stopped in it before reaching underground detectors, which are typically located at a depth of 1 km. As they drift towards the center of the earth because of gravity, these thermal dark atoms are radiatively captured by the atoms of the active medium of underground detectors, which causes the emission of photons that produce the signals through their interactions with the electrons of the medium. This provides a way of reinterpreting the results in terms of electron recoils instead of nuclear recoils as in the WIMP scenario.

Although O-helium gives interesting explanations to some indirect observations, we studied in detail the interactions of O-helium with ordinary matter and found that it was not an acceptable candidate for dark matter because of the absence of a repulsion mechanism preventing it from falling into the deep nuclear wells of nuclei.

The two other models involve milli-charges and use some ingredients of the O-helium scenario together with novel characteristics. They are able to reconcile the most contradictory experiments and we determined, for each model, the regions in the parameter space that reproduce the experiments with positive results in full consistency with the constraints of the experiments with negative results. We also paid attention to the experimental and observational constraints on milli-charges and discussed some typical signatures of the models that could be used to test them.

\end{abstract}

\newenvironment{resume}{\cleardoublepage\null\vfill\begin{center}\LARGE R\'esum\'e\end{center}}{\vfill\null}
\begin{resume}
L'objectif principal de ce travail \'etait de r\'einterpr\'eter les r\'esultats des exp\'eriences de recherche directe de mati\`ere noire dans un nouveau contexte, afin de r\'esoudre les contractions entre celles-ci lorsqu'elles sont interpr\'et\'ees dans le cadre du sc\'enario des Weakly Interacting Massive Particles (WIMPs). Pour ce faire, nous avons envisag\'e le cas de la mati\`ere noire composite, c'est-\`a-dire des particules de mati\`ere noire qui forment des \'etats li\'es, appel\'es g\'en\'eriquement << atomes noirs >>, soit avec des particules ordinaires, soit avec d'autres particules de mati\`ere noire.

Trois sc\'enarios diff\'erents ont \'et\'e \'etudi\'es : l'O-h\'elium, la mati\`ere noire milli-charg\'ee et les anti-atomes noirs. Dans chaque mod\`ele, la mati\`ere noire interagit suffisamment avec la mati\`ere terrestre pour y \^etre stopp\'ee avant d'atteindre les d\'etecteurs souterrains, lesquels sont typiquement situ\'es \`a 1 km de profondeur. Alors qu'ils d\'erivent vers le centre de la Terre \`a cause de la gravit\'e, ces atomes noirs thermiques sont captur\'es de mani\`ere radiative par les atomes constituant le milieu actif des d\'etecteurs, ce qui provoque l'\'emission de photons qui produisent les signaux en interagissant avec les \'electrons du milieu. Cela permet de r\'einterpr\'eter les r\'esultats en termes de reculs \'electroniques au lieu de reculs nucl\'eaires tels que dans le sc\'enario des WIMPs.

Bien que l'O-h\'elium fournisse des explications int\'eressantes \`a certaines observations indirectes, l'\'etude d\'etaill\'ee de ses interactions avec la mati\`ere ordinaire nous a montr\'e qu'il n'\'etait pas un candidat acceptable en raison de l'absence de m\'ecanisme de r\'epulsion l'emp\^echant de tomber dans les profonds puits nucl\'eaires des noyaux.

Les deux autres mod\`eles impliquent des milli-charges et ont certains ingr\'edients en commun avec l'O-h\'elium, tout en comportant des caract\'eristiques nouvelles. Ils sont \`a m\^eme de r\'econcilier les exp\'eriences les plus contradictoires et nous avons d\'etermin\'e, pour chaque mod\`ele, les r\'egions dans l'espace des param\`etres qui reproduisent les exp\'eriences \`a r\'esultats positifs tout en ne contredisant pas celles \`a r\'esultats n\'egatifs. Nous avons \'egalement pr\^et\'e attention aux contraintes exp\'erimentales et observationnelles sur les milli-charges et avons discut\'e de certaines signatures propres aux mod\`eles qui pourraient \^etre utilis\'ees pour les tester.
\end{resume}

\newenvironment{acknowledgements}{\cleardoublepage\null\vfill\begin{center}\LARGE Acknowledgements\end{center}}{\vfill\null}
\begin{acknowledgements}
I would like to thank Jean-Ren\'e Cudell for his great availability during these four years of research under his supervision. Besides having fulfilled his role of advisor through all his comments, ideas and advice during our uncountable discussions, he could also create a good atmosphere thanks to his sense of humor and his perpetual good mood. This has for sure made of this doctorate a pleasant time.

My thanks then go to Maxim Yu. Khlopov, who inspired part of this thesis, and whose wide knowledge of physics has been beneficial to me all along our collaboration.

I am grateful to all the other members of the jury, who accepted to devote time to read this thesis.

I thank the members of the IFPA group whose passage in Li\`ege coincided with some period of my doctorate. They all contributed to give a social dimension to my stay here during lunches, coffee breaks or drinks outside.

I express my gratitude to the Fonds National de la Recherche Scientifique for having provided me the financial support for this project. 

Thank you to Perrine for her support and for being part of my life for more than ten years now.

I thank my parents for having given me the opportunity to realize my dreams and to be who I am today.

To my grandparents, who are so proud of me. I feel lucky that they are still there to witness this moment of my life.

My last thoughts go to my family and long-time friends, for who dark matter may look obscure, but who could in their own way support me and contribute to my personal growth.
\end{acknowledgements}

\tableofcontents

\newpage
\pagenumbering{arabic}

\chapter*{Introduction}
\markboth{\MakeUppercase{Introduction}}{}
\addcontentsline{toc}{chapter}{Introduction}

Dark matter is one of the most intriguing aspects of cosmology and astrophysics. Since Fritz Zwicky noticed in 1933 that the total mass of stars and gas in the Coma cluster was far from sufficient to explain the high velocities of its constituent galaxies, the evidence of a ``missing mass'' in the universe has been accumulating at all scales, from the inner kiloparsecs of galaxies out to the size of galaxy clusters and to cosmological scales. This invisible mass, betraying its presence through its gravitational effects only is today called ``dark matter'' and is known to represent about 80$\%$ of the matter of the universe even if there is still no consensus about a direct observation.

Some standard explanations have been proposed, such as hot gas within clusters that would have been unseen in the visible spectrum. Early studies have also suggested the presence in the galactic halos of faint heavy objects such as black holes, neutron stars or white dwarfs that may have escaped observations until now and offering therefore a solution to the missing mass in terms of baryonic matter. These objects, called MACHOs (Massive Compact Halo Objects), have been looked for in the early 1990s in the framework of the MACHO \cite{Alcock:2000ph} and EROS \cite{Moniez:2008zza} projects, which observed the Magellanic clouds. The idea was to detect potential MACHOs in the halo of our Galaxy by microlensing, i.e. via the amplification of the light of background stars caused by a passing object between the source and the observer. But the results were in contradiction and not conclusive, and the MACHO hypothesis has been abandoned since then.

Today, the nature of dark matter remains widely unknown, except the rather well accepted fact that it is non-baryonic and hence not made of ordinary matter that would have remained undetected until now. As soon as this became relatively clear, the problem was moved to the composition of dark matter, which should be made of new particles that are not part of the Standard Model. Hence, dark matter became not only a problem, but also an evidence that the Standard Model had to be extended. In this context, astrophysicists and cosmologists started to list the main characteristics of these particles to explain the observations: they have to be massive, stable on cosmological timescales, neutral, non-relativistic at decoupling and give the right relic density. They therefore ended up with the general idea of a collisionless Cold Dark Matter (CDM).

This type of dark matter explains remarkably well many different kinds of observations. This is the only model that is able to reproduce at the same time the flatness of the rotation curves of spiral galaxies, the formation of a galaxy from an initial halo of CDM, the Large Scale Structure (LSS) of the universe and the anisotropies of the Cosmic Microwave Background (CMB). Concerning the latter, the recent results of Planck \cite{Ade:2013zuv} provide a flagrant example of the success of CDM in astrophysics and cosmology, with the reproduction of the angular power spectrum of the CMB to a degree of accuracy never reached before and allowing for a very precise determination of the density of CDM today in the universe.

Up to this point, the presence of dark matter has been suggested by gravitational effects only, and one could argue that the problem does not come from a missing mass but directly from our theory of gravitation. Although modifications or alternatives to general relativity exist, none of these theoretical frameworks have been able to reproduce as many observations as the CDM does. In this thesis, we therefore adopt the point of view that dark matter exists, and that it is made of new particles.

In order to have a precise idea of the nature of the particles constituting dark matter, fundamental theories are necessary. To satisfy the requirements of CDM, many candidates have been proposed: WIMPs, sterile neutrinos, hidden sectors (e.g. mirror matter), etc., but they can today all be put under the same generic denomination ``WIMP'', for Weakly Interacting Massive Particle. Here, ``Weakly'' has to be taken in the general sense and is not restricted only to the case of weak interactions: any interaction mechanism between dark and standard particles which is sufficiently weak is concerned, even the absence of any interaction other than gravitation. WIMPs can play the role of CDM and benefit from its successes, but they could also be detected indirectly or directly. Indirect detection would correspond for example to annihilation processes from which we would detect the products. Depending on the complexity of the signal, special attention to the details of each specific model could then help discriminate between them and retain only a few of them or even a single one. The direct detection of WIMPs would consist in observing in underground detectors the nuclear recoils induced by the collisions between WIMPs and nuclei, and could be used to some extent with the same aim as indirect detection, even if the generated signals would a priori be less powerful in discriminating between WIMP models. In any case, direct or indirect detection would indicate to a certain level of confidence that WIMPs do exist, and that so does dark matter. In this thesis, I will focus on the direct detection of dark matter.

From the idea of proving the existence of WIMPs and hence the reality of CDM, earth-based experiments started to be performed in the 1990s, in order to detect directly the recoils caused by WIMPs as they collide with nuclei in underground detectors isolated from cosmic rays. Of course, this implicitly assumes the presence of a non-gravitational interaction between WIMPs and ordinary matter, but it is the only hope we have to detect them. As the earth moves in the dark matter halo of our galaxy, it feels a ``wind'' of WIMPs that hits its surface. If their interaction is very weak, then most of them pass from one side of the earth to the other without being affected, but sometimes a particle can interact in the earth. If this happens just in the active volume of an underground detector, then a nuclear recoil can be detected. For a dark particle of incident velocity $v$ in the frame of the detector, the maximum recoil kinetic energy $E_R^{max}$ that can be transmitted to a nucleus initially at rest is:

\begin{equation}
E_R^{max}=\frac{2\mu^2 v^2}{m_N},\label{Emax}
\end{equation}
which corresponds to a head-on collision. Here, $\mu=\frac{m_N m}{m_N+m}$ is the reduced mass, $m$ is the WIMP mass and $m_N$ is the mass of the nucleus. Typically, $m_N\simeq 100$ GeV and $v\simeq 10^{-3}c$, so that for WIMP masses between 10 GeV and 1 TeV, $E_R^{max}$ can go roughly from 1 keV to 100 keV. To fix the ideas, a WIMP of the halo interacting with a nucleus will deposit an energy of ten to several tens of keV inside a detector, which thus requires a rather low experimental threshold.

In the generic WIMP scenario, a short-range interaction is induced by the exchange of some heavy mediator with the quarks of the nucleus. In the non-relativistic limit, which is the case here with velocities $v\simeq 10^{-3}c$, a fermionic WIMP interacts in an effective way with the nucleons and one can consider two interaction terms of the form:

\begin{eqnarray}
\mathcal{L}_{s}&=&f_{p,n}\bar{\chi}\chi \bar{\psi}_{p,n}\psi_{p,n}, \\
\mathcal{L}_{v}&=&a_{p,n}\bar{\chi}\gamma^{\mu}\gamma_{5}\chi \bar{\psi}_{p,n}\gamma_{\mu}\gamma_{5}\psi_{p,n},
\end{eqnarray}
where $\mathcal{L}_{s}$ and $\mathcal{L}_{v}$ are respectively scalar and axial-vector terms and give spin-independent and spin-dependent interactions. $\chi$ is the WIMP field, $\psi_{p,n}$ are the proton and neutron fields, $f_{p,n}$ are the scalar couplings to the proton and the neutron respectively and $a_{p,n}$ are the axial-vector couplings to the proton and the neutron. In the particular case of the scalar coupling, the scattering amplitudes with the nucleons add coherently over the whole size of the nucleus, so that the spin-independent WIMP-nucleus elastic cross section is given by:

\begin{equation}
\frac{\mathrm{d}\sigma_{\chi N}^{SI}}{\mathrm{d}q^2}=\frac{1}{\pi v^2}\left[Zf_p+(A-Z)f_n\right]^2F^2(q),
\end{equation}
where $q=\sqrt{2m_N E_R}$ is the transferred momentum, $E_R$ is the recoil energy of the nucleus, $Z$ is the atomic number of the nucleus and $A$ its mass number. $F(q)$ is the nuclear form factor, normalized to $F(0)=1$. In many WIMP models, the couplings to the proton and to the neutron are equal, which gives:

\begin{equation}
\frac{\mathrm{d}\sigma_{\chi N}^{SI}}{\mathrm{d}q^2} \propto A^2,\label{A2enhancement}
\end{equation}
where typically $A^2\simeq 5000-15000$. Because of this $A^2$ factor, spin-independent interactions dominate the event rates with respect to spin-dependent interactions in detectors with heavy nuclei. To study the latter, specific detectors with lighter nuclei are necessary, which exist, but I will concentrate here on spin-independent interactions.

Defining the spin-independent WIMP-proton elastic cross section at zero momentum $\sigma_{\chi p}^{SI}=\frac{4\mu_p^2}{\pi}f_p^2$, where $\mu_p$ is the reduced mass of the WIMP-proton system, the WIMP-nucleus cross section can be re-expressed as:

\begin{equation}
\frac{\mathrm{d}\sigma_{\chi N}^{SI}}{\mathrm{d}q^2}=\frac{\sigma_{\chi p}^{SI}}{4\mu_p^2 v^2}A^2F^2(q).
\end{equation}
This allows to express the cross section with any nucleus in terms of a common quantity $\sigma_{\chi p}^{SI}$ and hence to compare different experiments with each other. The differential rate of nuclear recoils in an underground detector in cpd/kg/keV (counts per day, per kg and per keV) is then given by:

\begin{align}
\frac{\mathrm{d}R}{\mathrm{d}E_R}&=n_N \frac{\rho_{\odot}}{m}\int_{v>v_{min}}\frac{\mathrm{d}\sigma_{\chi N}^{SI}}{\mathrm{d}E_R}f(\vec{v},t)v\mathrm{d}\vec{v}\label{dRdE1} \\
&=n_N\frac{m_N\rho_{\odot}}{2m\mu_p^2}\sigma_{\chi p}^{SI}A^2F^2(q)\int_{v>v_{min}}\frac{f(\vec{v},t)}{v}\mathrm{d}\vec{v},\label{dRdE}
\end{align}
where $n_N$ is the number of nuclei per kg of detector, $\rho_{\odot}=0.3$ GeV/cm$^3$ is the density of dark matter at the location of the solar system (local dark matter density), the function $f(\vec{v},t)$ is the velocity distribution of WIMPs, depending on the time $t$, in the frame of the detector and $v_{min}=\sqrt{\frac{m_N E_R}{2\mu^2}}$ is the minimum velocity needed to create a recoil of energy $E_R$ and is obtained from \eqref{Emax}. The velocity distribution of WIMPs in the detector frame is obtained from the velocity distribution in the frame of the dark halo, taken to be Maxwellian with a velocity dispersion of the order of the rotation velocity of the sun around the galactic center ($10^{-3}c$), to which we apply boosts to take into account the movement of the sun around the galactic center and the rotation of the earth around the sun. Provided one adopts a common geometry and the same velocity distribution, the whole problem is reduced to the two parameters $m$ and $\sigma_{\chi p}^{SI}$ (the latter being rather a WIMP-nucleon cross section in the case where $f_p\ne f_n$), and each experiment can reproduce its observed event rate with \eqref{dRdE} and obtain a region in the $(m,\sigma_{\chi p}^{SI})$ plane which is directly comparable to the others.

\begin{figure}
\begin{center}
\includegraphics[scale=0.45]{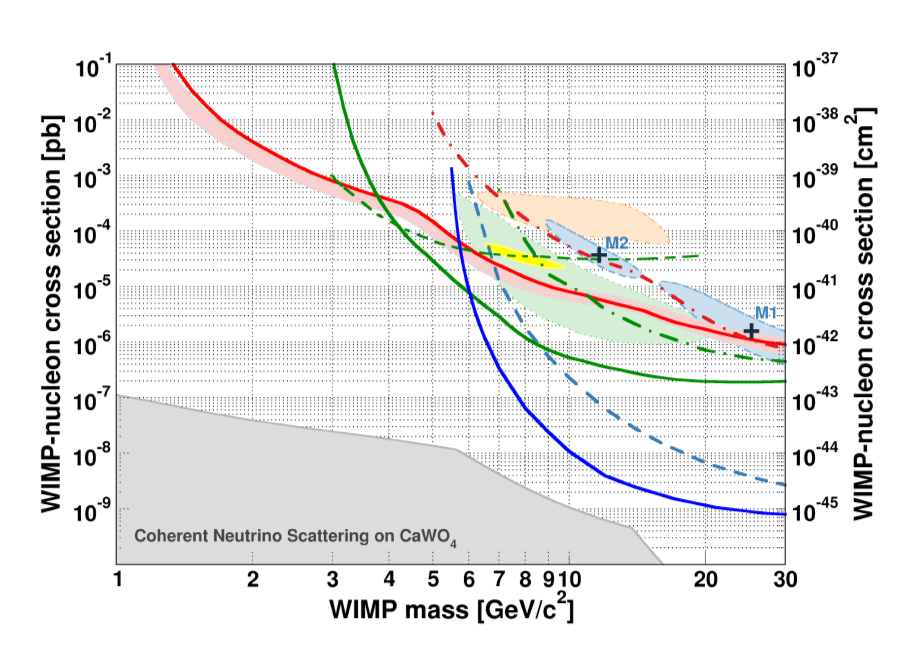}
\end{center}
\caption{WIMP parameter space in the case of spin-independent WIMP-nucleon scattering. The 90$\%$ C.L. upper limits from LUX \cite{Akerib:2013tjd} and XENON100 \cite{Aprile:2012nq} are respectively shown in solid blue and dashed blue. The same limit for superCDMS \cite{Agnese:2014aze} is in solid green and the one from CRESST-II (phase 2) \cite{Angloher:2014myn} is plotted in solid red. The regions favored by DAMA at $3\sigma$ C.L. \cite{Savage:2008er}, by CoGeNT at 99$\%$ C.L.  \cite{Aalseth:2012if} and by CDMS-II/Si at 90$\%$ C.L. \cite{Agnese:2013rvf} are respectively shown in light red, yellow and light green. Also shown in light blue is the previous $2\sigma$ C.L. region of CRESST-II (phase 1) \cite{Angloher:2011uu}, and the 90$\%$ C.L. upper limits from the re-analyzed commissioning run data in dash-dotted red \cite{Brown:2011dp}. Figure extracted from \cite{Angloher:2014myn}.}\label{FigWIMPnucleon}
\end{figure}

There have been many experiments since the late 1990s, and Figure \ref{FigWIMPnucleon} summarizes the current situation. As we can see, instead of crossing each other and defining a common closed region in the parameter space where we would expect the WIMP parameters to lie, the constraints are more complicated. Some experiments, i.e. DAMA, CoGeNT, CDMS-II/Si, observe a signal but their regions do not overlap while some others, i.e. LUX, XENON100, superCDMS, CDMS-II/Ge, CRESST-II, do not see any signal and are only able to put an upper limit on the WIMP-nucleon cross section. And on top of that, the preferred regions of the experiments with positive results lie in the forbidden regions of the ones with negative results, which give very stringent constraints, in particular the LUX experiment. The same kind of analysis in the case of the spin-dependent WIMP-nucleon interaction gives similar tensions and contradictions \cite{Akerib:2005za,Aprile:2013doa}. It should be noted, however, that all these tensions go together with many uncertainties, e.g. on the actual experimental thresholds or on the knowledge of nuclear form factors, and a better understanding of these aspects could help to improve the status of WIMPs in explaining the results of the direct-search experiments.

At this point, and in view of these clear inconsistencies, several attitudes can be adopted: 1) relax the hypotheses about the velocity distribution of WIMPs in the halo and play on the uncertainties on the astrophysical quantities of the problem to try to reduce the tensions; 2) assume that one or several experiments are wrong and put them aside in order to get a more coherent picture; or 3) consider that the experiments are correct but that the problem comes from the interpretation we make of them, which would mean that the contradictions of Figure \ref{FigWIMPnucleon} are only apparent and should vanish in another scenario. The first attitude, which is the most natural, has been tested and it turns out that it is very difficult to get rid of the contradictions while remaining reasonable about the assumptions made in the study. Concerning the second point, the major problem comes from the fact that two experiments with very high confidence levels contradict each other: DAMA, which observes a signal at a confidence level (C.L.) of $9.3\sigma$ \cite{Bernabei:2013xsa}, and LUX, the 90$\%$ C.L. \cite{Akerib:2013tjd} exclusion region of which is very far below the DAMA region. To select only some experiments in order to avoid contradictions would necessarily imply to reject either an experiment with a very high confidence level of detection, or another one with very stringent limits, which does not seem reasonable. Therefore, the third way will be followed here and will be the general thinking throughout this work.

In response to the problems caused by the WIMP interpretation of the direct-search experiments, alternative scenarios appeared in order to reinterpret their results \cite{Panci:2014gga,Cline:2012is,Foot:2013msa,McCullough:2013jma,Khlopov:2010pq,Khlopov:2010ik,Kaplan:2011yj,resDM}. Keeping in mind that CDM seems to be a very powerful solution to the problems of modern cosmology and astrophysics, it could be that dark matter is made of a more complex kind of particles, which would behave like CDM on large scales but would interact in a less naive way in underground detectors. To go further with this idea, one could even imagine that dark particles are not elementary but consist of structures made of exotic particles, that we could generically call ``composite dark matter''. This composite dark matter could either saturate the dark matter density, or be reduced to a fraction of it, the one which produces the signals of direct detection, while the rest would be made of inert particles that act like CDM without producing any recoil, i.e. a particular case of WIMPs that do not interact with standard particles except gravitationally. Of course, we can expect that, if such composite structures exist, they interact with each other and form a kind of self-interacting dark matter which evolves in a complex dark sector with a rich phenomenology and which has common characteristics with ours. Actually, self-interacting dark matter can provide a solution to some small-scale problems of CDM, such as the core/cusp \cite{Flores:1994gz,Moore:1994yx} and the missing-satellite problems. It is a viable candidate if it represents only a subdominant part of dark matter, but its self-interactions should be reduced in the case where it is dominant in order to be consistent with the stringent constraints on self-interacting dark matter \cite{2002ApJ...564...60M,Buote:2002wd,Markevitch:2003at,Randall:2007ph,Harvey:2015hha}. The models that will be presented in this thesis present all these features and all aim at solving the discrepancies between direct-search experiments when they are interpreted in terms of WIMPs.

Among the inspiring alternative models, one counts milli-charged atomic dark matter \cite{Cline:2012is}, mirror matter \cite{Foot:2013msa} and exothermic double-disk dark matter \cite{McCullough:2013jma}, which we shall now respectively describe. In the milli-charged atomic dark matter scenario, a remaining unbroken dark $U(1)_d$ gauge symmetry is added to the standard model and provides a dark massless photon $\dot{\gamma}$, which is kinetically mixed with the standard photon $\gamma$ through the mixing Lagrangian:

\begin{equation}
\mathcal{L}_{mix}=\frac{\epsilon}{2}F_{\mu \nu}\dot{F}^{\mu \nu},\label{Lmix}
\end{equation}
where $\epsilon$ is the the dimensionless mixing parameter, $F$ is the electromagnetic field strength of the standard photon and $\dot{F}$ that of the dark one. This kind of mixing has been originally proposed by Holdom \cite{Holdom:1985ag} and gives rise, after diagonalisation of the Lagrangian and renaming of the fields, to the situation where the gauge field $A_{\mu}$ of $\gamma$ couples both to the standard and to the dark currents $J^{\mu}$ and $\dot{J}^{\mu}$ while the gauge field $\dot{A}_{\mu}$ of $\dot{\gamma}$ couples only to the dark current:

\begin{equation}
\mathcal{L}_{int}=A_{\mu}(eJ^{\mu}+\epsilon e\dot{J}^{\mu})+\dot{A}_{\mu}g\dot{J}^{\mu},
\end{equation}
where $e$ is the standard $U(1)$ coupling constant and $g$ is the dark $U(1)_d$ coupling constant. Introducing a dark proton $\dot{p}$ and a dark electron $\dot{e}$, these behave therefore like electric milli-charges of values $+\epsilon e$ and $-\epsilon e$. In this model, $\dot{p}$ and $\dot{e}$ bind to each other through the dark $U(1)_d$ and form hydrogen-like dark atoms $\mathrm{\dot{H}}$ that realize the full dark matter density. Two cases are considered, depending on the masses $m_{\dot{p}}$ and $m_{\dot{e}}$ of $\dot{p}$ and $\dot{e}$: the case $m_{\dot{e}}\ll m_{\dot{p}}$ and the particular case $m_{\dot{e}}=m_{\dot{p}}$.

In the first case, the screening of the central charge of $\dot{p}$ by $\dot{e}$ produces, in the low momentum transfer approximation, a contact interaction between a dark atom $\mathrm{\dot{H}}$ and the standard proton, with the elastic scattering cross section:

\begin{equation}
\sigma_p = 4\pi \alpha^2\epsilon^2\mu^2 \dot{a}_{0}^{4},
\end{equation}
where $\mu$ is the reduced mass of the proton-$\mathrm{\dot{H}}$ system, $\alpha=\frac{e^2}{4\pi}$ is the fine structure constant, $\dot{a}_{0}=\frac{1}{m_{\dot{e}}\dot{\alpha}}$ is the Bohr radius of $\mathrm{\dot{H}}$ and $\dot{\alpha}=\frac{g^2}{4\pi}$. So, these dark atoms act like WIMPs since they penetrate the earth and collide on nuclei in underground detectors. But there, $\mathrm{\dot{H}}$ interacts only with the protons of the nuclei and not with all nucleons. Compared with WIMPs, there is thus a weakening factor $\left(\frac{Z}{A}\right)^2$ on $\sigma_p$ and we can define an effective cross section $\sigma_{p,eff}=\left(\frac{Z}{A}\right)^2\sigma_p$ which can directly replace the WIMP-nucleon cross section in Figure \ref{FigWIMPnucleon}, while the WIMP mass has to be renamed to the mass $m_{\dot{H}}$ of $\mathrm{\dot{H}}$ ($m_{\dot{H}}\simeq m_{\dot{p}}$). In terms of $\sigma_p$, Figure \ref{FigWIMPnucleon} has therefore to be shifted vertically by a factor $\left(\frac{A}{Z}\right)^2$ which depends on the experiment (e.g. $\left(\frac{A}{Z}\right)^2\simeq 6$ for LUX and XENON100 and 4.4 for DAMA if it is interpreted in terms of collisions with its sodium component), but this is not sufficient to solve the contradictions due to the several orders of magnitude that separate the cross sections excluded or favored by these experiments.

The particular case $m_{\dot{e}}=m_{\dot{p}}$ is interesting since it proposes an explanation to the excess of CoGeNT \cite{Aalseth:2012if}. In that situation, the matrix element for elastic scattering vanishes in the Born approximation, which leaves the possibility that inelastic processes are dominant. The hyperfine splitting $E_{hf}$ of the fundamental state of a dark atom, which is equal to 5.9$\times 10^{-6}$ eV in the case of standard hydrogen (21-cm line), can be brought up to the keV range if the masses of the two particles are identical and in the GeV range:

\begin{align}
E_{hf}&=\frac{2}{3}g_{\dot{e}}g_{\dot{p}}\dot{\alpha}^{4}\frac{m_{\dot{e}}^2m_{\dot{p}}^2}{(m_{\dot{e}}+m_{\dot{p}})^3}\nonumber \\
&=\frac{1}{6}\dot{\alpha}^{4}m_{\dot{H}},
\end{align}
where $g_{\dot{e}}=g_{\dot{p}}=2$ are the gyromagnetic ratios of $\dot{e}$ and $\dot{p}$ and $m_{\dot{H}}\simeq 2m_{\dot{p}}=2m_{\dot{e}}$. The collisions between dark atoms and the germanium nuclei of CoGeNT, accompanied by the excitation of $\mathrm{\dot{H}}$ from its spin singlet to its triplet state, can reproduce the nuclear-recoil-event rate of CoGeNT with the parameters:

\begin{eqnarray}
\epsilon & \simeq & 10^{-2}\nonumber \\
m_{\dot{H}} & = & 6~\mathrm{GeV} \\
E_{hf} & = & 15~\mathrm{keV} \Rightarrow \dot{\alpha}=0.062. \nonumber
\end{eqnarray}
So, even if this model can provide a reinterpretation of the direct-search experiments and a solution to one of them in a particular case, together with being consistent with all the constraints related to milli-charges (accelerators, CMB, self-interaction, etc.), it either keeps all the contradictions or it explains one experiment at the expense of the others, which is not satisfactory in the framework of this thesis.

The mirror matter scenario was first imagined by Lee and Yang \cite{PhysRev.104.254} with the goal to introduce a global symmetry of parity in nature. The Lagrangian of the Standard Model $\mathcal{L}_{SM}$ is not invariant under the parity symmetry since by a parity transformation all its fields acquire the opposite chirality and hence become right-handed instead of left-handed. The idea is therefore to add to $\mathcal{L}_{SM}$ the same Lagrangian, with the same coupling constants, but with the right-handed fields. By doing so, we introduce an exact ``mirror'' sector, made of ``mirror particles'' that have no gauge interactions with the standard ones, so that the mirror sector interacts with ours only through gravitation, which makes mirror matter a potential dark matter candidate. It has been shown \cite{Ciarcelluti:2012zz,Ciarcelluti:2014scd} that mirror matter can actually reproduce the anisotropies of the CMB and the LSS as well as CDM and is therefore a serious candidate to consider from the cosmological point of view. But in its simplest version, this scenario cannot provide any solution to direct detections. To do so, one possibility is to add the same kinetic mixing as in \eqref{Lmix} between the standard and the mirror photons. This creates a class of milli-charged particles of known masses that can penetrate through the earth straight to underground detectors and produce nuclear recoils.

More precisely, the galactic dark matter halo is supposed to be made of a pressure-supported-multi-component plasma containing mirror nuclei such as $\mathrm{\dot{H}}$, $\mathrm{\dot{He}}$, $\mathrm{\dot{O}}$, $\mathrm{\dot{Fe}}$, etc. Each component has a Maxwellian velocity distribution of temperature $T$ roughly given by:

\begin{equation}
T\simeq \frac{1}{2}\bar{m}v^2_{rot},
\end{equation}
where $\bar{m}\simeq$ 1.1 GeV is the mean mass of mirror particles in the halo and $v_{rot}$ is the galactic rotational velocity. The rate of nuclear recoils in each experiment is assumed to be dominated by a single component, which is not $\mathrm{\dot{H}}$ nor $\mathrm{\dot{He}}$ since these elements are too light to produce a significant signal, but rather a heavier mirror ``metal''. The elastic scattering cross section between a mirror nucleus and an ordinary one is given by the Rutherford formula, multiplied by the nuclear form factors $F_{\dot{N}}$ and $F_N$ of the mirror and ordinary nuclei to take their finite size into account:

\begin{equation}
\frac{\mathrm{d}\sigma}{\mathrm{d}E_R}=\frac{2\pi \epsilon^2 Z^2 \dot{Z}^{2}\alpha^2 F_N^2 F^2_{\dot{N}}}{m_N E_R^{2} v^2},
\end{equation}
where $\dot{Z}$ is the atomic number of the mirror nucleus. The rate of nuclear recoils can therefore be calculated in a similar way as \eqref{dRdE1} and depends on three free parameters: the mass $m_{\dot{N}}$ of the mirror nucleus, $\epsilon \sqrt{\xi_{\dot{N}}}$, where $\xi_{\dot{N}}$ is the halo mass fraction of species $\mathrm{\dot{N}}$, and $v_{rot}$. The positive results of DAMA, CoGeNT, CDMS-II/Si and CRESST-II (when the latter was still reporting a signal) can be reproduced with:

\begin{eqnarray}
v_{rot} & \simeq & 200~\mathrm{km/s}\nonumber \\
\frac{m_{\dot{N}}}{m_p} & \simeq & 55.8 \Rightarrow \mathrm{\dot{N}}=\mathrm{\dot{Fe}} \\
\epsilon \sqrt{\xi} & \simeq & 2\times 10^{-10}\nonumber
\end{eqnarray}
It is also noted in \cite{Foot:2013msa} that a preferred value of $\epsilon$, which is consistent with astrophysics and cosmology, can be obtained independently from the requirement that ordinary supernovae supply the necessary energy to the mirror plasma in order to balance the energy dissipated mainly through thermal bremsstrahlung. This gives $\epsilon \simeq 10^{-9}$, and hence $\xi\simeq 0.04$. So, this model of mirror matter, enriched by a kinetic mixing between the standard and the mirror photons, can bring an original explanation to the experiments with positive results in terms of collisions with $\mathrm{\dot{Fe}}$ nuclei which are subdominantly present in the halo, but one expects that experiments with higher thresholds such as XENON100 or LUX should be able to detect the tail of the velocity distribution of $\mathrm{\dot{Fe}}$. XENON100, for example, should have seen several dozen of events, which is clearly not the case. It seems therefore that the favored regions of the parameter space have significant tensions with the experiments with negative results and this is the reason why we must continue our non-exhaustive review of alternative models.

In the exothermic double-disk dark matter scenario \cite{McCullough:2013jma}, it is postulated that a small fraction of the dark matter has arbitrarily strong self-interactions and that there exist an additional unbroken $U(1)$ symmetry, and hence a massless gauge boson (typically the same as in the previous models), that allows for dissipative dynamics in the dark sector, while the rest of dark matter is made of conventional collisionless CDM. In the presence of a cooling mechanism, Ref. \cite{Fan:2013yva} argues that the self-interacting component should condensate and form a dark disk, just as baryonic matter does. The existing constraints on self-interacting dark matter from halo shapes \cite{2002ApJ...564...60M,Buote:2002wd} or from colliding galaxy clusters \cite{Markevitch:2003at,Randall:2007ph,Harvey:2015hha} can all be avoided if one requires that it reduces to less than 10$\%$ of the total amount of the halo dark matter. Below this value, the effects on the dynamics of the galactic halos are too small for them to be probed by the gravitational lensing observations. But in the case of a dark disk, there is a more stringent constraint from the kinematics of nearby stars which leads to a limit on the mass of the self-interacting component of 5$\%$ of the total dark mass of the Milky Way halo.

If one endows this sector with an interaction with nucleons, it could in principle be possible to detect the dark particles directly. But for current experiments, which have thresholds of a few keV in nuclear recoil, equation \eqref{Emax} indicates that the velocity of a dark particle in the frame of a detector should be of the order of $10^{-4}c$ in order to produce a detectable signal. For WIMPs, which have typical velocities of $10^{-3}c$, this is not a problem. But in this scenario, even in the most optimistic case where the solar system is located in the dark disk, both the baryonic and the dark disks should be in approximately the same circular orbit, so that their large rotation velocities should be equal and hence the relative velocities between the earth and the dark particles are expected to be suppressed. There are several sources of relative velocities, such as the motion of the earth around the sun or the peculiar velocity of the sun in its spiral arm, but they are not sufficient to produce a detectable signal, unless one could lower the current experimental thresholds. In order to make this candidate detectable in existing experiments, the model is enhanced with a mass splitting $\delta$ of the fundamental dark matter state:

\begin{equation}
M_{+}-M_{-}=\delta,
\end{equation}
where $M_{-}$ and $M_{+}$ are respectively the actual ground and excited states. This gives the so-called exothermic double-disk dark matter model \cite{McCullough:2013jma}, in which a dark particle enters the detector in the excited state, collides with a nucleus and de-excites to the lower state. This additional energy deposit is transmitted to the nucleus and contributes to its recoil energy, which allows to reduce the minimum velocity $v_{min}$ needed to produce a recoil $E_R$ from the one given by \eqref{Emax} to:

\begin{equation}
v_{min}=\frac{1}{\sqrt{2m_N E_R}}\left|\frac{m_N E_R}{\mu}-\delta \right|,
\end{equation}
where $\mu$ is the dark matter-nucleus reduced mass. So, for an experimental threshold of value $E_{th}$, the minimum velocity that can be probed is $v_{min}(E_{th})$ and if $\delta$ is sufficiently large, it can reach the low velocities that are characteristic of this model. In the case where $\delta>\frac{m_N E_R}{\mu}$, $v_{min}(E_{th})=0$ and the whole velocity distribution in the dark disk can be accessed by the experiment.

There are several distinctive features of this model, due to the particular phase-space distribution in the dark disk and to the different collisional kinematics with respect to standard dark matter halo models: 1) as the reduction of $v_{min}(E_{th})$ is greater for smaller $m_N$, detectors with light target nuclei are more sensitive; 2) due to the low velocity dispersion in the dark disk, of the order of 25 km/s to be compared to the 300 km/s of typical WIMPs, the recoil spectra are expected to be very narrow; 3) in the best situation for direct detection where the two disks are aligned and have the same circular orbit, there should be some relative velocity between them which should be close to zero if the collapse and cooling mechanisms are similar. In this case, the period of the year when the flux of dark matter particles falling on earth reaches its maximum is not the date at which the rotation vector of the earth around the sun is aligned with the velocity vector of the sun with respect to a fixed halo, but the moment at which the former is aligned with the peculiar velocity of the sun in its spiral arm, which shifts the date of maximum flux by approximately 100 days with respect to the standard halo models.

Because of the first two points, the model can reproduce the events seen by CDMS-II/Si, since it is made of light silicon nuclei and since the observed recoils lie in a narrow energy window. Moreover, the stringent constraints from the XENON100 experiment are relaxed with respect to the WIMP scenario since the experimental thresholds are higher and the constituent xenon nuclei are much heavier, which makes these experiments less sensitive to probing the low velocities in the dark disk. So much so that the region of the parameter space of the exothermic double-disk dark matter model that is favored by CDMS-II/Si, which lies around:

\begin{eqnarray}
m & = & 5.5~\mathrm{GeV}\nonumber \\
\delta & = & 50~\mathrm{keV} \\
\sigma_n & = & 10^{-43}~\mathrm{cm}^2,\nonumber
\end{eqnarray}
where $m$ is the mass of a dark particle and $\sigma_n$ is the dark matter-nucleon elastic scattering cross section at zero-momentum, is almost fully consistent with the null results of the other experiments. However, a consistent interpretation of CDMS-II/Si and DAMA seems impossible, since the large shift of the date of maximum flux that results from the low relative velocity between the visible baryonic and the dark disks is incompatible with the phase measured by DAMA. And even if the relative velocity was increased in order to make them compatible, the cross section needed for CDMS-II/Si would be several orders of magnitude too low to reproduce the rate of DAMA. Again, a coherent reinterpretation of the full set of experiments seems out of reach.

Throughout this thesis, I will present three dark matter models that are directly inspired by the ones discussed here. However, although they include some ingredients coming from these, they feature very specific and novel characteristics. In particular, the interactions in underground detectors, instead of consisting in elastic or inelastic collisions with fast-moving particles, rely on a mechanism of bound-state formation between thermalized composite dark matter particles and the atoms of the active medium. As these bound states are formed by radiative capture, the emitted photons produce electron recoils instead of nuclear recoils. This enriches somewhat the phenomenology and allows to reinterpret in a completely different way the results of the experiments, with the aim of reconciling all of them. The O-helium model, which was the first to consider this new origin of the signal, presented in Chapter \ref{OHe}, is shown to be unable to do so, while milli-interacting dark matter, discussed in Chapter \ref{Millicharges}, realizes this task quite well, although the events seen by CDMS-II/Si remain difficult to explain. Dark anti-atoms, in Chapter \ref{AntiH}, can be seen as a simplification of the milli-interacting model with fewer parameters and reproduce also well the results of the direct-search experiments, up to some personal bias on the results of CoGeNT and CDMS-II/Si. But before discussing these models, I review non-exhaustively the experimental setups and their raw results (i.e. without any interpretation) in Chapter \ref{DirectSearches}.

\chapter{The direct-search experiments}\label{DirectSearches}

To directly detect the interactions of dark matter particles from the galactic halo with standard particles, several earth-based experiments have been designed, which are all located deeply underground, at a typical depth of 1 km, in order to isolate best a possible signal from the natural background at the surface, which is mostly due to cosmic rays. Because of the combined motions of the earth and the sun in the halo, the earth is subjected to a continuous flux of dark matter particles that may, in a way depending on the considered scenario, reach underground detectors and produce nuclear or electron recoils. All these detectors use different techniques to convert the energy released during an interaction into a visible signal, as well as different event-selection criteria, background modelization and suppression, or shieldings against environmental radioactivity and muons. In this chapter, I review some of these experiments and present their basic characteristics and results.

\section{The DAMA experiment}\label{secDAMA}

\subsection{The set-up}

The DAMA experiment is located at the Gran Sasso National Laboratory in Italy. It is one of the first to have started to collect results as it is running since 1996. The experiment has been performed in two phases, DAMA/NaI during the period 1996-2002, and DAMA/LIBRA (Large sodium Iodide Bulk for RAre processes) from 2003 to 2013\footnote{An update of the second phase is planned in the coming years.}.

\begin{figure}
\begin{center}
\includegraphics[scale=0.25]{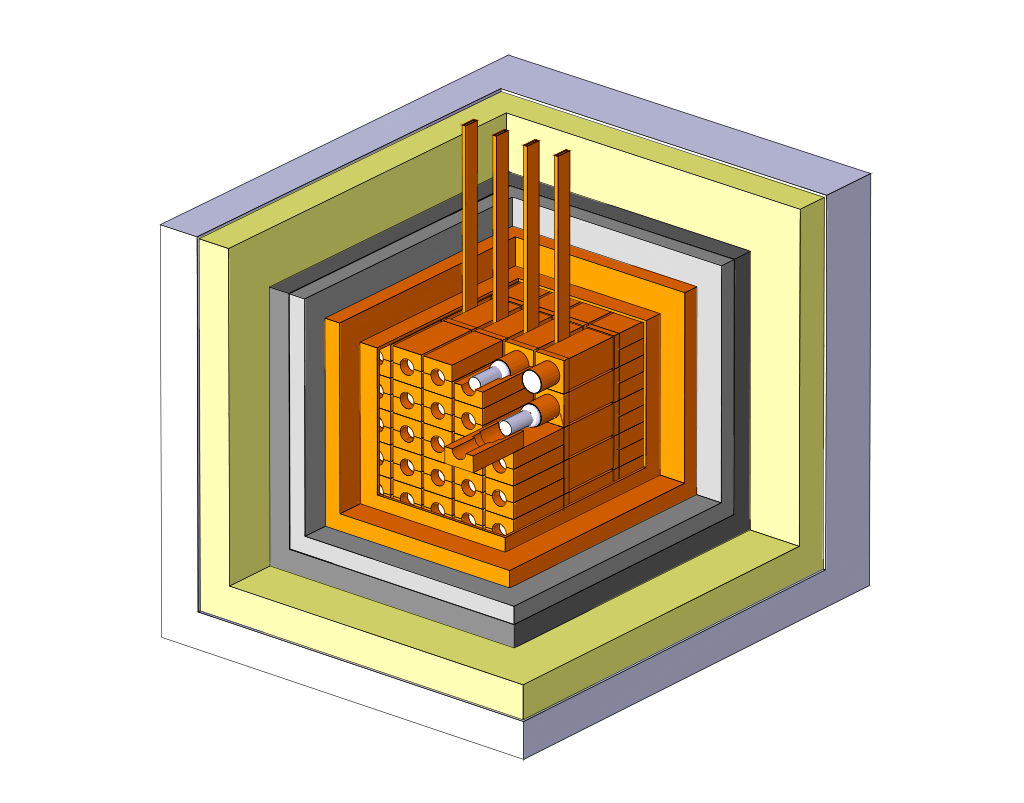}
\end{center}
\caption{DAMA/LIBRA apparatus. From inside to outside: $5\times 5$ matrix of the scintillators together with their PMTs, internal copper box, and passive shield made of 10 cm of copper, 15 cm of lead, 1.5 mm of cadmium and an external layer of polyethylene/paraffin. Figure extracted from \cite{Bernabei:2013xsa}.}\label{FigDAMAsetup}
\end{figure}

In its second version, the sensitive part of the detector consists of 25 highly radiopure sodium-iodide scintillators doped with thallium (NaI(Tl)), organized in a 5-row-by-5-column matrix, each of them having a mass of 9.7 kg and a size of $10.2\times 10.2\times 25.4$ cm$^3$ \cite{Bernabei:2008yh}. This gives a total mass of about 250 kg for a volume of 0.0625 m$^3$. Each detector, enclosed in a copper brick, is coupled to two photomultiplier tubes (PMTs) at its opposite faces, and the whole matrix (detectors+PMTs) is sealed in a low-radioactive copper box. In addition to the surrounding Gran Sasso rock, the apparatus is passively protected from the environmental radioactivity by a succession of 10 cm of copper, 15 cm of lead, 1.5 mm of cadmium and by an external layer of polyethylene/paraffin. The whole system operates at room temperature (300 K). Figure \ref{FigDAMAsetup} is a schematic view of the set-up.

\subsection{Detection method}

One particularity of the DAMA experiment is that it exploits the model-independent annual modulation signature. In many scenarios, as long as one can consider that the earth moves, throughout its revolution around the sun, in a locally uniform environment of dark matter particles with an isotropic velocity distribution, one expects that the flux of dark matter particles is maximum around June 2, when the velocities of the earth around the sun and of the sun around the galactic center are aligned, and that it is minimum around December 2 when they are anti-aligned. Hence, the event rate in an underground detector should be modulated with a period of one year. Since no other source is known to have such a behavior, this is considered as a typical signature of dark matter.

The DAMA/LIBRA detector is an inorganic sodium iodide scintillator activated with thallium. The presence of these impurities at the $10^{-3}$ level in the NaI crystal has the effect of creating allowed energy levels for the electrons in the band gap at about 3 eV from the valence band. If an interacting particle (X/$\gamma$-ray photon, charged particle, dark matter particle, etc.) passes through the crystal, it can excite electrons from the valence band to these excited intermediate states, which then fall back to the valence band by emitting visible photons with wavelengths centered on $\lambda(\mathrm{nm})=\frac{2\pi \times 197}{3~\mathrm{eV}}\simeq 415~\mathrm{nm}$, i.e. with a maximum emission in violet light.

A scintillation photon is then collected by a PMT, where it produces one or a few photoelectrons at the photocathode. These are multiplied by successive secondary electron emissions from one dynode to the next in the PMT, which gives a measurable electric pulse as output. After calibration of the set-up with X/$\gamma$-ray photons of known energies coming from nuclear and internal-electron transitions, one can access, for a measured pulse, the energy that was deposited by the incident particle through interactions with the electrons of the crystal. This energy, which is the only one that is really measured by the detector, is usually expressed in electron equivalent energy $E_{ee}$ with the units eV$_{ee}$. If the initial interaction occurred with a nucleus (e.g. the incident particle was a neutron), only a part of the recoil energy $E_R$ in eV$_{nr}$ was transferred to the electrons, the rest being dissipated in the ion lattice. The ratio between the detected energy in eV$_{ee}$ and the kinetic energy of the recoiling nucleus in eV$_{nr}$ is the quenching factor $q$, i.e. $q=\frac{E_{ee}}{E_{R}}$. The latter depends on the nucleus and on the recoil energy. DAMA reports values of $q_{Na}=0.3$ and $q_I=0.09$, respectively for sodium and iodine, over the energy range of interest. These values are obtained from a $^{252}$Cf neutron source, by fitting the low-energy recoil spectrum produced by the elastic scattering of neutrons on Na and I nuclei \cite{Bernabei:1996vj}. Note that in the next chapters, we will not use eV$_{ee}$ and eV$_{nr}$ but we will only refer to eV, the context being always sufficiently clear to make a distinction.

One should note that, in view of the detection technique discussed above, electron recoils cannot be distinguished from nuclear recoils in the case of the DAMA experiment. This will be an important point when interpreting the results in the framework of the models presented in Chapters \ref{OHe}, \ref{Millicharges} and \ref{AntiH}. The detector is calibrated from the MeV region down to the keV range, with a detection threshold at 2 keV$_{ee}$. In order to identify the events due to dark matter, only those which occurred in a single scintillator over the 25 available are retained, as dark matter particles are supposed to interact rarely with standard matter.

\subsection{Results}

\begin{figure}
\begin{center}
\includegraphics[scale=0.3]{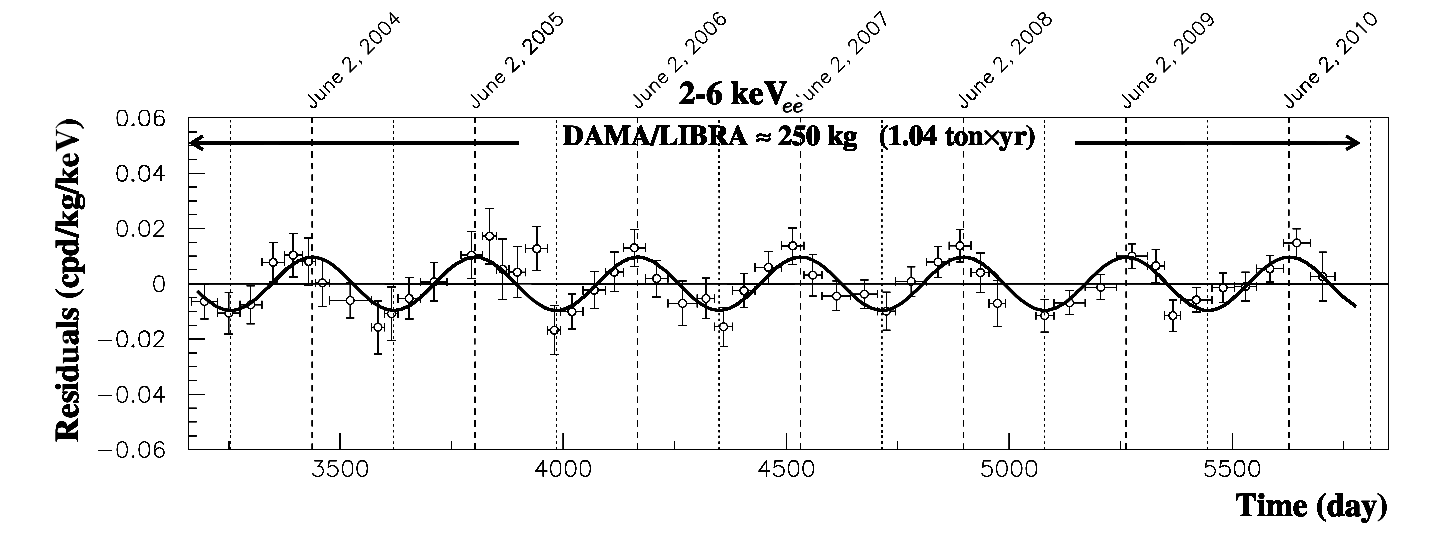}
\caption{Residual rate of the single-hit events measured during the phase DAMA/LIBRA, in the energy interval $\left(2-6\right)$ keV$_{ee}$, as a function of time. The experimental points are shown with circles and errorbars while the superimposed solid curve is the function $A\cos\left(\frac{2\pi}{T}\left(t-t_0\right)\right)$ with $T=1$ yr, $t_0=152.5$ days $=$ June 2 and the central value of $A$ obtained by fitting the data points of the phase DAMA/LIBRA. Figure extracted from \cite{Bernabei:2013xsa}.}
\label{FigDAMAresiduals}
\end{center}
\end{figure}

\begin{figure}
\begin{center}
\includegraphics[scale=0.3]{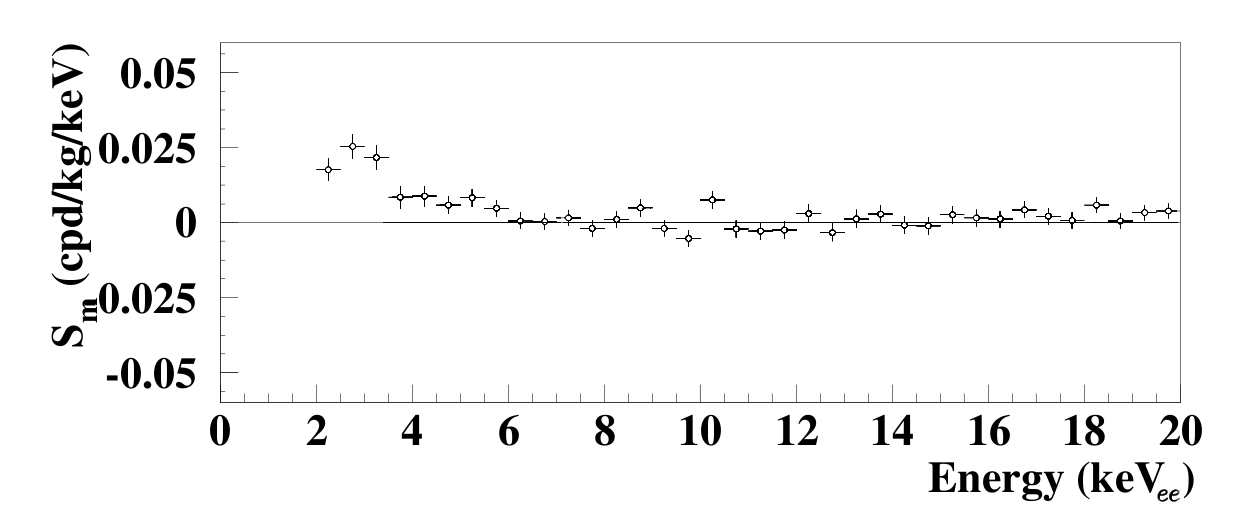}
\caption{Energy distribution of the modulation amplitude $S_{m,k}$ for an energy bin of 0.5 eV$_{ee}$, obtained by fitting the total exposure of DAMA/NaI and DAMA/LIBRA with fixed $T=1$ yr and $t_0=152.5$ days $=$ June 2. Figure extracted from \cite{Bernabei:2013xsa}.}
\label{FigDAMAspectrum}
\end{center}
\end{figure}

When taking the results of both DAMA/NaI and DAMA/LIBRA, which corresponds to a total exposure of 1.33 ton$\times$yr, a modulation of the rate of single-hit events in the $\left(2-6\right)$ keV$_{ee}$ energy range is found at a confidence level of $9.3\sigma$ \cite{Bernabei:2013xsa}. This is obtained by fitting the data with the formula $A\cos\left(\frac{2\pi}{T}\left(t-t_0\right)\right)$ and by letting the amplitude $A$, the period $T$ and the phase $t_0$ free. The latter is defined as the moment of maximum rate, with an origin of time on January 1. In that case, the measured amplitude is $A=\left(0.0112\pm 0.0012\right)$ cpd/kg/$\mathrm{keV}_{ee}$ (cpd: counts per day) while the period and phase are respectively $T=\left(0.998\pm 0.002\right)$ yr and $t_0=\left(144\pm 7\right)$ days, in agreement with the annual modulation scenario. Figure \ref{FigDAMAresiduals} shows the residual rate of the single-hit scintillation events between 2 and 6 keV$_{ee}$ for the phase DAMA/LIBRA, where we can see a clear modulation. Figure \ref{FigDAMAspectrum} shows the energy distribution of the modulation amplitude, obtained by fitting the whole data in the $k$-th energy interval with the form $S_k=S_{0,k}+S_{m,k}\cos\left(\frac{2\pi}{T}\left(t-t_0\right)\right)$, with fixed $T=1$ yr and $t_0=152.5$ days $=$ June 2 and an energy bin of 0.5 eV$_{ee}$. It confirms the presence of a modulation in the same energy window while at energies above 6 keV$_{ee}$ the amplitudes are compatible with zero.

\section{The CoGeNT experiment}\label{secCoGeNT}

\subsection{The set-up}

The CoGeNT (Coherent Germanium Neutrino Technology) experiment is based at the Soudan Underground Laboratory, in the Soudan iron mine in Minnesota. It has collected data over a period of 3.4 years between December 2009 and April 2013, resulting in a 1129 live-days data set\footnote{An expansion of the experiment, involving a larger active mass as well as lower threshold and background, is planned in the coming years.}.  

\begin{figure}
\begin{center}
\includegraphics[scale=0.3]{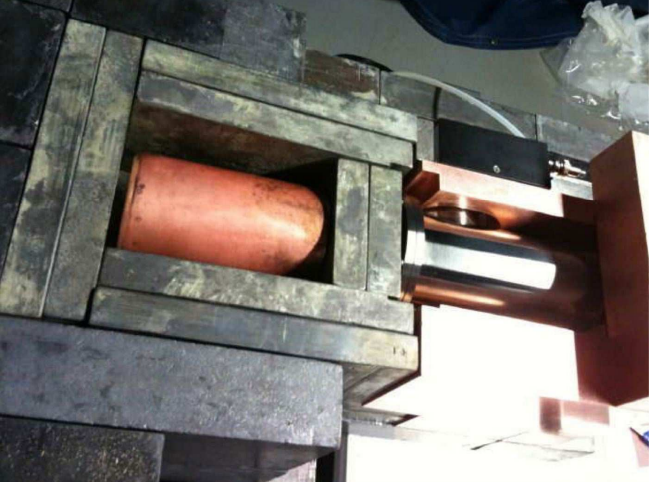}
\end{center}
\caption{CoGeNT detector in its cylindrical copper can, surrounded by the innermost 5 cm-thick layer of ancient lead with very low radioactivity. Figure extracted from \cite{Aalseth:2012if}.}
\label{FigCoGeNTsetup}
\end{figure}

The detector is a single 440-gram $p$-type Point Contact (PPC) germanium crystal of high purity, cooled down at the temperature of liquid nitrogen (77 K under atmospheric pressure). It is a cylinder of diameter 60.5 mm and length 31 mm, for a fiducial mass of 330 grams, contained in a cylindrical copper can which is connected to the cryostat through a copper cold finger. The innermost passive shield is made of three different layers of lead from inside to outside: 5 cm of ancient lead with a very low radioactivity and two layers, of 10 cm each, of contemporary lead, for a total width of 25 cm of lead surrounding the detector in all directions. Figure \ref{FigCoGeNTsetup} is a picture of the CoGeNT detector and its lead shield.

\subsection{Detection method}

The CoGeNT apparatus is a germanium semiconductor detector measuring the ionization produced by an incident particle (X/$\gamma$-ray photon, charged particle, dark matter particle, etc.). Due to its semiconductor nature, a germanium crystal has a small gap of 0.66 eV between the valence and the conduction bands. At room temperature, some electrons can therefore  have sufficient energies to jump to the conduction band, which has the effect of blurring any interesting signal in the electronic noise, justifying the cooling at cryogenic temperatures as well as the use of high-purity crystals, in order to avoid as much as possible the presence of free-charge carriers. In addition, highly doped materials of $p$ and $n$ types, denoted by $p^+$ and $n^+$ to underline their strong doping, are joined at the two opposite faces of the germanium semiconductor, which is of type $p$ despite its intrinsic high purity. This gives rise to a diffusion of the holes to the region $n^+$ and of the electrons to the region $p$, creating an area around the $p-n^+$ junction that is empty of charge carriers. A high voltage is then applied between the two opposite faces of the germanium crystal in reverse bias (negative electrode at the $p^+$-side and positive electrode at the $n^+$-side), which produces the desertion of the holes and of the electrons respectively to the extreme sides of the $p^+$ and $n^+$ regions, increasing the size of the area empty of free charge carriers to the whole germanium crystal.

In these conditions, if an interacting particle passes through the detector, it creates electron/hole pairs (i.e. it ionizes the crystal by placing electrons in the valence band). These primary charges are attracted, according to their nature, by the positive or by the negative electrode and create avalanches of secondary charges, which produce an electric pulse that is measured. The number of secondary charges is proportional to the number of primary charges, which reflects the energy deposited in the crystal through ionization. After calibration, it is therefore possible to make the correspondence between the measured pulse and the ionization energy, knowing that the reported experimental threshold is 0.5 keV$_{ee}$. In the case of a nuclear recoil, a well accepted quenching factor for CoGeNT is $q_{Ge}=0.2$ \cite{Aalseth:2012if,Barbeau:2007qi}. However, similarly to DAMA, we see from the detection mechanism that no discrimination between nuclear and electron recoils is possible for such a detector.

\subsection{Results}

\begin{figure}
\begin{center}
\includegraphics[width=9cm,height=13cm]{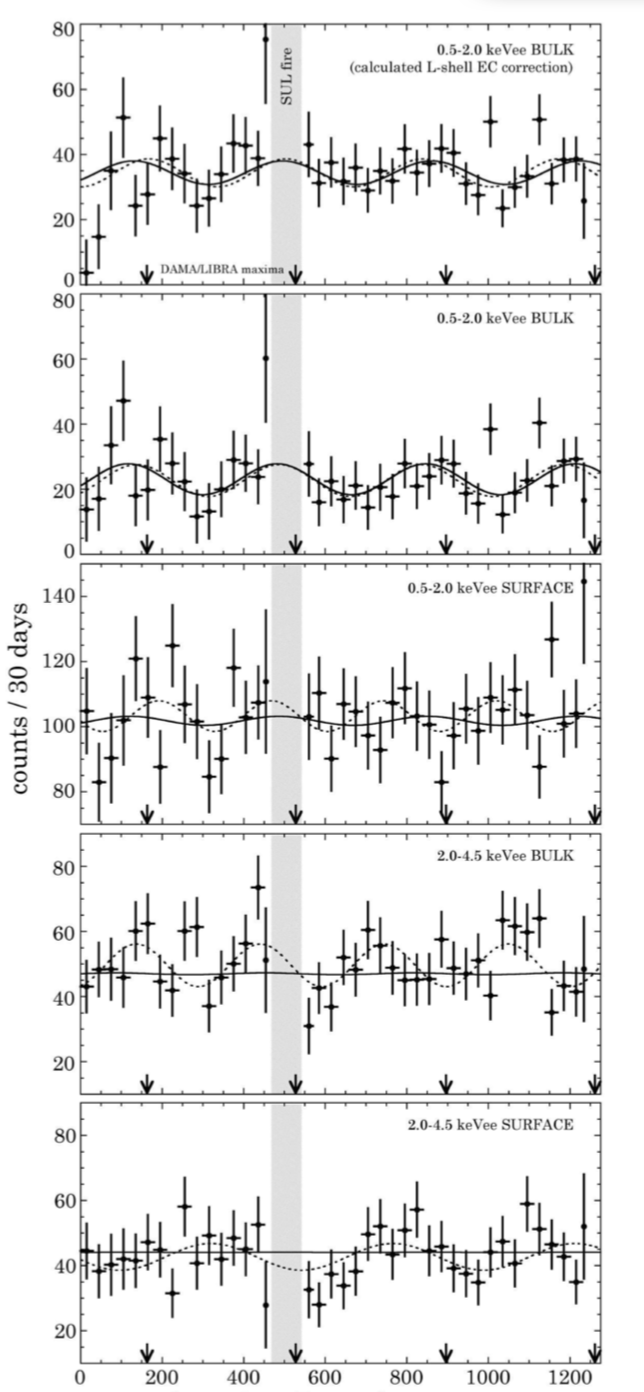}
%\begin{minipage}{0.498\linewidth}
%\centerline{\includegraphics[width=1\linewidth]{CoGeNT_modulation1}}
%\end{minipage}
%\begin{minipage}{0.498\linewidth}
%\centerline{\includegraphics[width=1\linewidth]{CoGeNT_modulation2}}
%\end{minipage}
%\begin{minipage}{0.498\linewidth}
%\centerline{\includegraphics[width=1\linewidth]{CoGeNT_modulation3}}
%\end{minipage}
%\begin{minipage}{0.498\linewidth}
%\centerline{\includegraphics[width=1\linewidth]{CoGeNT_modulation4}}
%\end{minipage}
%\begin{center}
%\includegraphics[scale=0.7]{CoGeNT_modulation5}
\end{center}
\caption{Time-dependence (days) of the rate of ionization events (counts/30 days) observed by CoGeNT for the different groups of events: 0.5-2 keV$_{ee}$ bulk events after subtraction of the calculated L-shell electron capture by cosmogenically activated $^{68}$Ge (top); 0.5-2 keV$_{ee}$ bulk events after subtraction of a decaying background with a free half-time (second from top); 0.5-2 keV$_{ee}$ surface events after subtraction of a decaying background with a free half-time (third from top); 2-4.5 keV$_{ee}$ bulk events after subtraction of a decaying background with a free half-time (fourth from top); 2-4.5 keV$_{ee}$ surface events after subtraction of a decaying background component with a free half-time (bottom). The origin of time is on 3$^\mathrm{rd}$ of December 2009. The date points are represented by circles with errorbars while the dotted lines are for the fits with free constant part, modulation amplitude $A$, period $T$ and phase $t_0$ and the solid lines are fits of the data points with fixed $T=1$ yr. The arrows indicate the times of maximum rate of DAMA \cite{Bernabei:2013xsa}. The vertical grey bands correspond to the period during which the data acquisition had to be stopped because of a fire in the Soudan Underground Laboratory. Figure extracted from \cite{Aalseth:2014eft}.}
\label{FiGCoGeNTmodulation}
\end{figure}

In 2011, CoGeNT reported for the first time the observation of an irreducible excess of low-energy events below 3 keV$_{ee}$ \cite{Aalseth:2010vx}, after which they performed a temporal analysis and found an annual modulation of the events \cite{Aalseth:2011wp} with a period of one year and a phase compatible with the one observed by DAMA. The statistical significance was of $2.8\sigma$ at that time, revised at $2.2\sigma$ with the whole data set in 2014 \cite{Aalseth:2014eft}.

Part of the analysis consists in identifying and separating the events occurring near the surface (outermost 1 mm-thick layer of the active part of the detector), called surface events, and the ones happening in the bulk, called bulk events. Indeed, a dark matter particle is more likely to interact in the larger volume of the bulk than just at the edge. For that reason, the surface events are considered as background events and rejected. The identification is based on the measure of the rise-time $t_{10-90}$ of the ionization events, defined as the time needed for an electric pulse to pass from 10$\%$ of its maximum height to 90$\%$. An event occurring in the bulk of the detector will generate more secondary charges during the drift towards the electrodes than a surface event and will therefore, on average, have a faster growth, and hence a smaller rise-time. According to simulations performed by the CoGeNT collaboration \cite{Aalseth:2012if}, the rise-times of these two populations follow log-normal distributions and one can, by fitting the number of events as a function of the rise-time with a sum of two log-normal functions, separate the bulk events, peaked on smaller rise-times, from the surface events. The latter are expected to be more present at low energy, as the efficiency of the conversion of the ionization energy into an electric pulse is less than 1 near the surface. This is verified by comparing the two considered energy windows 0.5-2 keV$_{ee}$, where an excess was found, and 2-4.5 keV$_{ee}$. Indeed, the two distributions are better separated and the number of events with large rise-times drops in the second one. The selection criterion is fixed at the point where the two distributions have equal values: the events with $t_{10-90}<0.7$ $\mu$s correspond to bulk events in the 0.5-2 keV$_{ee}$ energy window, with an expected contamination of 13.6$\%$ by surface backgrounds, while the separation is at 0.6 $\mu$s in the 2-4.5 keV$_{ee}$ range, with a contamination of 4.4$\%$.

Figure \ref{FiGCoGeNTmodulation} shows the temporal evolution of the number of ionization events in each group, after subtraction of the different decaying background components. Two kinds of fits are performed: one with free constant part for the rate, modulation amplitude $A$, period $T$ and phase $t_0$ (dotted lines), the other with a fixed period at $T=1$ yr (solid lines). A modulation can be found only for the low-energy bulk events (first two panels), with a confidence level of $2.2\sigma$, confirming the previously observed excess. In the case of the calculated L-shell electron capture, the preferred values for the period and the amplitude are $T=\left(336\pm 24\right)$ days and $A=\left(12.4\pm 5\right)\%$ while they are $T=\left(350\pm 20\right)$ and $A=\left(21.7\pm 15\right)\%$ for the second group, which is consistent with the annual modulation expected from a dark matter component. When the period is fixed to $T=1$ yr, the preferred phase, defined as the moment of maximum rate and with an origin of time on January 1, is $t_0=\left(102\pm 47\right)$ days, which is compatible with DAMA, as seen by its times of maximum rate represented by vertical arrows in Figure \ref{FiGCoGeNTmodulation}. Fits to the other groups give random values for the modulation parameters while for fixed $T=1$ yr, no modulation is favored.

\section{The XENON100 and LUX experiments}

\subsection{The set-up}

\begin{figure}
%\begin{minipage}{0.498\linewidth}
%\centerline{\includegraphics[width=0.7\linewidth]{XENON100_setup}}
%\end{minipage}
%\begin{minipage}{0.498\linewidth}
%\centerline{\includegraphics[width=1\linewidth]{LUX_setup}}
%\end{minipage}
\begin{center}
\includegraphics[scale=0.25]{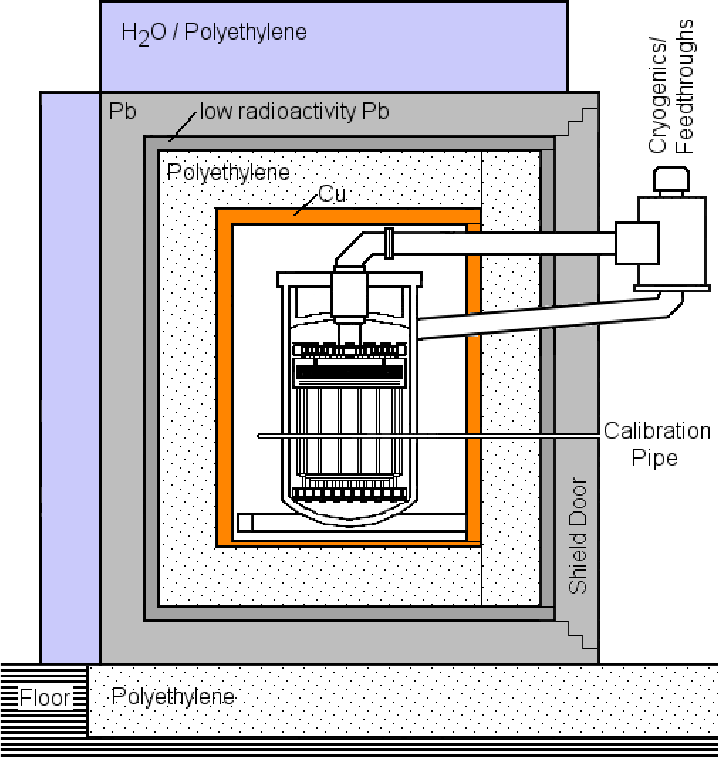}\hspace{1cm}\includegraphics[scale=0.3]{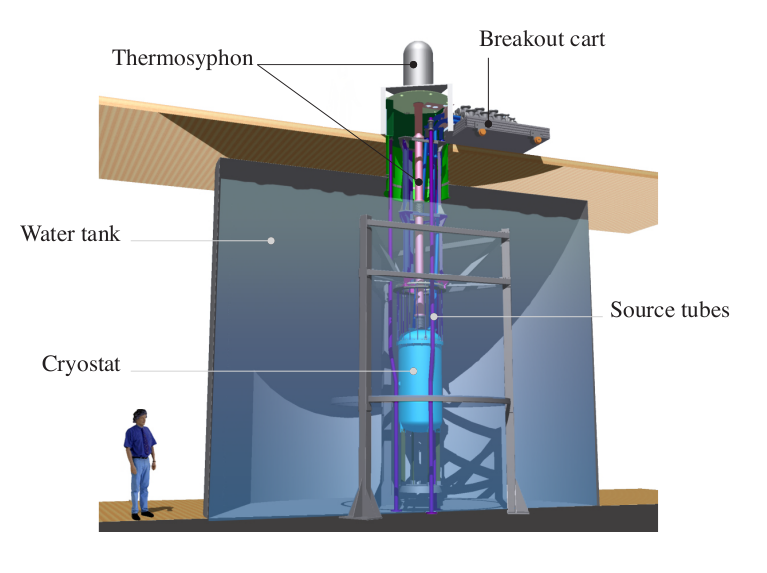}
\end{center}
\caption{Left: schematic view of the XENON100 apparatus, with the cylindrical two-phase TPC at the center together with the LXe veto, surrounded by a passive shield of 5 cm of copper, 20 cm of polyethylene and 20 cm of lead from inside to outside. Additional 20-cm thick layers of H$_2$O or polyethylene are placed on some sides and the entire system rests on a polyethylene slab of 25 cm. The cryogenic system is held outside of the shielding to avoid background from radioactive materials. Right: LUX detector (cryostat) inside its water tank used as a passive shield and as an active muon veto. The cryostat contains the TPC and the upper and lower copper layers. The cryogenic system is also visible (thermosyphon), as well as the calibration facility (source tubes). Figures extracted from \cite{Aprile:2011dd} and \cite{Akerib:2012ys} (left and right respectively).}
\label{FigXENON100LUXsetups}
\end{figure}

The XENON100 and LUX (Large Underground Xenon) experiments, respectively located at the Gran Sasso National Laboratory in Italy and at the Sanford
Underground Research Facility in South Dakota, present similar detectors. The former has acquired data during the period 2011-2012 for a total of 225 live days\footnote{The construction of a larger detector, of a mass of 1 t, started in 2012. The first results of this upgraded experiment, called XENON1T, are expected in 2017.} while the latter is currently still collecting data and has already released results from a first 85 live-day run between April and August 2013.

The major difference between both experiments relies on their active masses:\break XENON100 features 62 kg of target liquid xenon (LXe), for a fiducial mass of 34 kg, while LUX contains 250 kg of LXe, for a final fiducial mass of 118 kg, which gives the latter a greater sensitivity. Also, the LUX experiment has a slightly smaller electron-recoil background of $3.1\times 10^{-3}$ cpd/kg/$\mathrm{keV}_{ee}$ while it is of $5.3\times 10^{-3}$ cpd/kg/$\mathrm{keV}_{ee}$ for XENON100.

The detectors are cylindrical two-phase (liquid-gaseous xenon) time projection chambers (TPCs) of 30.5 cm height and 30.6 cm diameter for XENON100, and of 48 cm height and 47 cm diameter in the case of LUX, in which the gaseous phase occupies the top part and the active liquid phase fills the majority of the lower volume. Xenon acts as a scintillator, observed by 178 and 122 PMTs respectively for XENON100 and LUX (one array in the gaseous phase and one immersed at the bottom of the liquid phase). The LXe is maintained at the condensation temperature of xenon, i.e. $182~\mathrm{K}$ for XENON100 and $175~\mathrm{K}$ for LUX, depending on the pressure conditions of the gaseous phase.

The TPC of the XENON100 detector is surrounded on all sides by a 4 cm-thick layer of LXe, of a total mass of 99 kg, acting as an active veto observed by 64 PMTs. This allows to reduce the background in the target xenon of the TPC, in complement to a passive shield made of 5 cm of copper, 20 cm of polyethylene and 20 cm of lead from the inside to the outside.

The passive shield of the LUX apparatus consists of a 5 cm-thick copper disk at the top of the TPC and of a 15 cm-thick one at its bottom as well as a water tank with a diameter of 7.6 m and a height of 6.1 m in which the whole detector is immersed. This water tank also serves as an active muon veto, since it is observed by 20 PMTs and used as a Cherenkov detector: if a nuclear recoil is detected in the fiducial volume of the LUX detector in coincidence with the Cherenkov light of a muon, it is considered as being due to a muon-induced neutron \cite{Araujo:2004rv} and the event is rejected. Figure \ref{FigXENON100LUXsetups} shows schematic views of the XENON100 (left) and LUX (right) detectors, together with their passive shields.

%The gaseous phase at the top of the chamber is at room temperature while the liquid phase is maintained at the condensation temperature of xenon under the present pressure conditions. In the case of XENON100 for example, the gas pressure is of more than 2 atm, in which case the boiling point is at 182 K: the chamber is connected to the cryogenic system through an insulated pipe in which the gas can reach the cold finger at 170 K and condense. The drops are then collected by a funnel and flow back into the detector by gravity inside a pipe of smaller diameter passing at the center of the insulated one, refilling continuously the liquid phase and keeping it at a constant temperature of 182 K.

\subsection{Detection method}

The XENON100 and LUX detection principles are identical. An incoming particle (X/$\gamma$-ray photon, charged particle, dark matter particle, etc.) can interact with a xenon atom in the liquid volume of the TPC. This produces the excitation of atomic electrons, that then de-excite by emitting a direct scintillation light of wavelength $178~\mathrm{nm}$ (vacuum ultraviolet), and ionization electrons. The scintillation photons are detected by the PMTs and produce photoelectrons at the photocathodes, giving a direct scintillation signal S1 expressed in number of photoelectrons (PE). An electric field is applied vertically between the upper and lower arrays of PMTs in the TPC, so that the ionization electrons produced during the interaction drift towards the liquid-gas interface. There, a stronger electric field extracts them from the liquid to the gas, where they are accelerated and produce a scintillation light proportional to their total charge by collisions with the xenon atoms of the gaseous phase, giving rise to a second scintillation signal S2.

The energy $E_{ee}$ deposited in the detector during the interaction with the electrons (via the recoiling nucleus in case of a nuclear recoil) can be measured, after calibration, using S1 and S2, but the knowledge of S1 and S2 together allows, contrarily to DAMA and CoGeNT, to distinguish between electron and nuclear recoils and to eventually convert $E_{ee}$ into the kinetic recoil energy $E_R$ in the latter case, with the appropriate measured quenching factor. Indeed, for the same direct scintillation signal S1, a nuclear recoil will produce fewer ionization electrons and hence a lower second scintillation S2 than an electron recoil. Nuclear recoils feature therefore smaller S2/S1 ratios than electron recoils and in an (S1,S2/S1) plane they form two different bands, which can be used for discrimination.

Moreover, the XENON100 and LUX detectors allow to locate spatially the events occurring in their cylindrical TPC. As the drift velocity of the ionization electrons is constant and fixed by the intensity of the electric field, one can access the $z$-coordinate of the event by measuring the time difference between the S1 and S2 signals. The $(x,y)$ location is deduced from the hit pattern of the second scintillation light in the top PMT array. Using this 3-dimensional localization, it is possible to identify multiple-scattering events, happening at different places, and to reject them, since dark matter particles are expected to interact only once. For example, double-scattering events are separable when their $z$-coordinates differ by more than 3 mm in the case of XENON100. It is also possible to recognize the events that occur near the edges of the active region, which are mostly due to the background, and to exclude therefore a surface volume, in order to keep an internal volume (fiducial volume) where dark matter particles are assumed to interact most of the time. This external region of the active medium therefore acts as self-shielding that complements the passive shields and vetos and is the reason why the active mass goes from 62 kg to a fiducial mass of 34 kg for XENON100, and from 250 kg to 118 kg for LUX.

Note that the detection thresholds on the S1 signals are at 3 and 2 PE respectively for XENON100 and LUX, corresponding to minimal deposited energies of 1.3 and 0.9 keV$_{ee}$.

\subsection{Results}

\begin{figure}
%\begin{minipage}{0.498\linewidth}
%\centerline{\includegraphics[width=1\linewidth]{XENON100_results}}
%\end{minipage}
%\begin{minipage}{0.498\linewidth}
%\centerline{\includegraphics[width=1\linewidth]{LUX_results}}
%\end{minipage}
\begin{center}
\includegraphics[width=7.7cm,height=6cm]{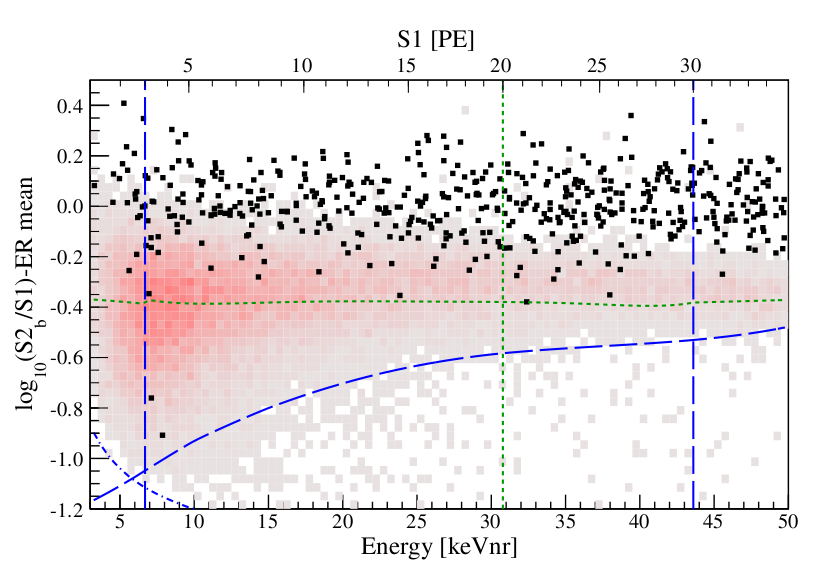}\includegraphics[width=7.7cm,height=5.6cm]{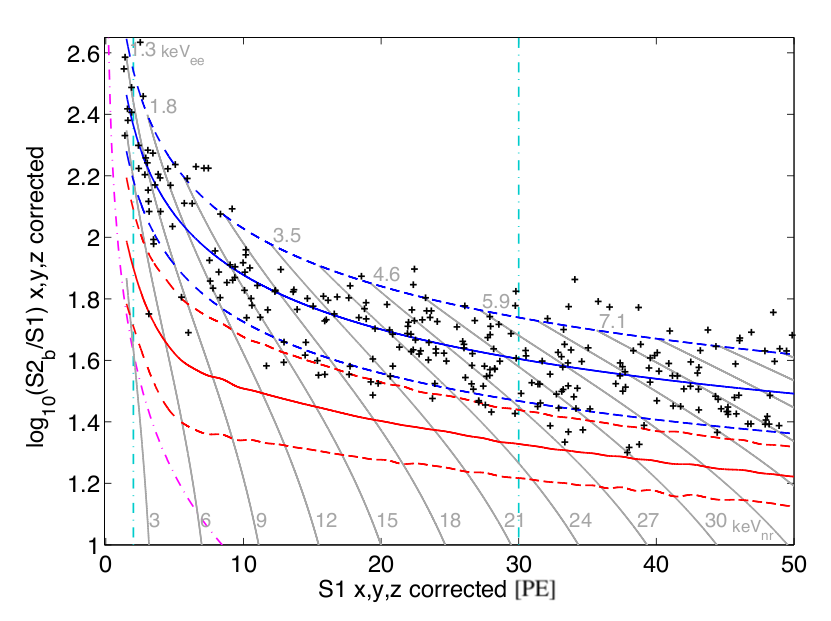}
\end{center}
\caption{Left: event distribution in the discrimination plane (S1,$\log_{10}(\mathrm{S2/S1})$) after all selection cuts, for the final 225 live days $\times$ 34 kg exposure of XENON100. The lower and upper thresholds (resp. 3 PE $\equiv$ 6.6 keV$_{nr}$ and 20 PE $\equiv$ 30.5 keV$_{nr}$) are represented by the vertical dashed blue and dotted green lines. The lower dashed blue curve is the 97$\%$-nuclear-recoil-acceptance line while the upper dotted green one is the 99.75$\%$-electron-recoil-rejection line. Right: event distribution of LUX after the first 85 live days $\times$ 118 kg exposure. The electron-recoil and nuclear-recoil bands are respectively delimited by the dashed blue and dashed red lines, which are at $\pm 1.28\sigma$ from the mean values (solid lines), from Gaussian fits to slices in S1. The lower and upper thresholds (2-30 PE) are represented by the vertical dash-dotted light blue lines. Figures extracted from \cite{Aprile:2012nq} and \cite{Akerib:2013tjd} (left and right respectively).\label{FigXENON100LUXresults}}
\end{figure}

A series of cuts are applied to the data taken in the TPCs before they are represented in the discrimination plane (S1,S2/S1). In the case of XENON100, quality tests are first performed in order to reject, for example, events with an excessive level of energy due to a high voltage discharge or very high energy events. Cuts to identify the single-hit events are then realized by counting the number of S1 and S2 peaks and using the information from the active veto, and consistency checks allow to clean the data further, e.g. by ensuring that the width of the S2 pulse, which is due to the diffusion of the ionization electrons in the LXe, is consistent with the $z$-coordinate of the event. The fiducial volume is then defined and events that occur outside of it are rejected. This gives the remaining events of Figure \ref{FigXENON100LUXresults} (left) in the (S1,$\log_{10}(\mathrm{S2/S1})$) plane, obtained for the final 225 live days $\times$ 34 kg exposure \cite{Aprile:2012nq}. As XENON100 is searching for WIMPs in particular, producing nuclear recoils, a region is defined in this discrimination plane where nuclear recoils are expected to be located: lower and upper thresholds (vertical dashed blue and dotted green lines) on the energy (3-20 PE for S1 or equivalently 6.6-30.5 keV$_{nr}$ for the recoil energy $E_R$), an upper line on $\log_{10}(\mathrm{S2/S1})$ (dotted green), below which the electron-recoil hypothesis can be rejected at more than 99.75$\%$, and a lower bound (dashed blue), above which the nuclear-recoil acceptance from neutron calibration data is larger than 97$\%$. This gives the lower region of Figure \ref{FigXENON100LUXresults} on the left, where two events present all the features of nuclear recoils induced by WIMPs. But both nuclear- and electron-recoil-background events can contaminate this region. The former background is determined by Monte Carlo simulations to reproduce the emission of neutrons by the radioactive components of the detector as well as the flux of muon-induced neutrons. This gives an expected nuclear-recoil background of $(0.17^{+0.12}_{-0.07})$ events for the given exposure in the WIMP region. The latter is due to the radioactivity of all components ($\beta$ and $\gamma$), which can produce electron recoils that have the characteristics of nuclear recoils. This is estimated with calibration data from $^{60}$Co and $^{232}$Th and it turns out that ($0.79\pm 0.16$) electron-recoil events can leak anomalously into the nuclear-recoil band. In total, this gives a total expected background of (1.0 $\pm$ 0.2) events in the WIMP region. But the probability that the background fluctuates to two events is of 26$\%$, and that is why the XENON100 collaboration reports null results concerning the search for WIMPs. Note that the consistency between the electron recoils in the upper region and the expected background is also checked.

LUX, following its own analysis method, obtains the same kind of results, shown in Figure \ref{FigXENON100LUXresults} (right), for an exposure of 85 live days $\times$ 118 kg \cite{Akerib:2013tjd}. The electron-recoil (dashed blue lines) and nuclear-recoil bands (dashed red lines) are obtained by calibration from $\beta^{-}$ and neutron emitting sources. The solid lines correspond to the mean values, from Gaussian fits to slices in S1, while the dashed lines are at $\pm1.28\sigma$ from them. The lower and upper thresholds (2-30 PE) are represented by the vertical dash-dotted light blue lines. All the events are consistent with the background-only hypothesis. In particular, the observed and the expected electron-recoil backgrounds in the energy region of interest 0.9-5.3 keV$_{ee}$ are consistent, since they are respectively of $(3.1\pm 0.2_{stat})$ and $(2.6\pm 0.2_{stat}\pm 0.4_{sys})$ in units of 10$^{-3}$ cpd/kg/$\mathrm{keV}_{ee}$.

\section{Other experiments}

The experiments that we have just discussed are very representative of the direct-search experiments since they incorporate the two experiments with positive results (DAMA and CoGeNT) as well as experiments that put limits on WIMP parameters which are among the strongest ones (XENON100 and LUX, see Figure \ref{FigWIMPnucleon} of the Introduction). Moreover, these examples allowed us to see that there are two kinds of experiments: the ones that do not make any distinction between electron and nuclear recoils, and the ones that have a discrimination power. The latter is due to the use of a more complex detection technique, based on two different signals per event (S1 and S2 signals for XENON100 and LUX), which allow them to identify nuclear recoils and hence to focus their research on WIMPs. There are other experiments that exploit several signals to do so, and they use varied detection techniques. We shall briefly describe some of them here, which will be considered later in this thesis, but we should keep in mind that the list will still be non-exhaustive.

\subsection{The CDMS-II experiment}

The Cryogenic Dark Matter Search (CDMS) experiment, located at the Soudan Underground Laboratory in Minnesota, has, in its version II, acquired data during the period 2003-2008. The final results were released in 2009 \cite{Ahmed:2008eu}, but they were based on a recoil-energy threshold of 10 keV$_{nr}$. In order to be sensitive to lower WIMP masses, the threshold has been decreased to 2 keV$_{nr}$ in a low-energy analysis in 2011 \cite{Ahmed:2010wy}.

The CDMS-II detector consists of an array of 19 germanium (230 g each) and 11 silicon (105 g each) disks arranged in five towers of six detectors each and cooled down to cryogenic temperatures ($<50$ mK). At such low temperatures, a particle that interacts in the detector through a nuclear or an electron recoil causes a measurable increase in temperature, which can be detected by a superconducting thin film (phonon sensor) at the surface of each Ge or Si crystal: as the superconducting material is maintained at its critical temperature, its electrical resistance strongly depends on the temperature, and a small variation of the latter due to a particle interaction produces a large change of the former, which is constantly monitored. The thermal sensors are called ``phonon sensors'' because thermal agitation in a crystal can be seen as waves that propagate through it, which can be decomposed in vibration modes made of quasi-particles called ``phonons''. Therefore, a particle interacting in the detector causes the emission of phonons of total energy $E_{ph}$ that are absorbed by the phonon sensors, which constitutes the first signal and is used to precisely determine the energy that has been deposited. So, in case of a nuclear recoil, the nuclear recoil energy is directly accessible and there is no need to define a quenching factor in such a detector (or, equivalently, the quenching factor is equal to 1).

The second signal of the event comes from the ionization electrons produced by the particle during the interaction in the semiconductor. These are extracted with a small electric field applied across the crystal which produces a pulse allowing to measure the energy that has been transferred to the electrons, denoted by $E_{ee}$. The ratio between $E_{ee}$ and $E_{ph}$, called ionization yield, is used as a primary discrimination parameter to distinguish between electron and nuclear recoils, as a nuclear recoil of a certain energy $E_{ph}$ will produce fewer ionization electrons than an electron recoil of the same $E_{ph}$.

Using the data from the Ge detectors only, the CDMS collaboration did not detect any nuclear recoil that could not be explained by the background, and hence reported null results about the search for WIMPs in \cite{Ahmed:2008eu,Ahmed:2010wy}. In the following, we will refer to this Ge-only analysis as CDMS-II/Ge.

In contrast, the analysis of the data from the Si detectors only, called CDMS-II/Si in the following, was released in 2013 \cite{Agnese:2013rvf} and revealed that three nuclear recoils, at energies 8.2, 9.5 and 12.3 keV$_{nr}$, passed all the selection and quality cuts. The interpretation of these events in terms of WIMPs gave the 90$\%$ C.L. region from the CDMS-II/Si experiment in Figure \ref{FigWIMPnucleon} of the Introduction, with a highest likelihood for a WIMP mass of 8.6 GeV and a spin-independent WIMP-nucleon cross section of $1.9\times 10^{-41}$ cm$^2$. However, it is worth mentioning that simulations of the known background showed that there is a probability of 5.4$\%$ that three background events leak into the signal region.

\subsection{The superCDMS experiment}

SuperCDMS, currently operating at the Soudan Underground Laboratory, is an upgrade of the former CDMS-II experiment, with an increased mass in order to improve the sensitivity: it counts now 15 cylindrical germanium crystals of 600 g each\footnote{A next phase of superCDMS is planned at SNOLAB in the coming years, which will reach a target mass of 200 kg, allowing to increase the sensitivity by another order of magnitude.}, for a total target mass of 9 kg, arranged in five towers of 3 detectors. It looks for low-mass WIMPs by analyzing nuclear recoils in the range $(1.6-10)$ keV$_{nr}$.

The detection principle is basically the same as CDMS-II, except that now it is possible to separate surface events, due to background radiation, from bulk events, where dark matter is supposed to mainly interact, and hence to reject the former in the analysis. While the phonon sensors were, in the CDMS-II detector, placed only on one face of the crystals with the ionization sensors on the other one, both types of sensors are now interleaved on the two faces of the germanium cylinders. This allows a discrimination in the $z$-coordinate inside each cylinder since for an event occurring in the bulk, the ionization electrons and holes will be collected symmetrically on both faces, while they will be collected only on the closest face for a surface event. Moreover, the separation of the sensors in outer (along the periphery of the disks) and inner (filling the central part of the disks) sensors allows to make a radial discrimination of the events, as an event taking place near the side of a cylinder will leave more energy in the outer sensors. With this identification of the peripheral regions, one can therefore define a fiducial volume and reject surface events, since these were often suffering from a reduced ionization signal in the CDMS-II detector, leading to possible misinterpretations of surface events as nuclear recoils, and thus polluting the WIMP-search region.

SuperCDMS started to operate in 2012 and released its first results in 2014 \cite{Agnese:2014aze}. After all selection cuts, all the observed events are consistent with the background and the resulting upper limits on the spin-independent WIMP-nucleon cross section for low-mass WIMPs are shown in Figure \ref{FigWIMPnucleon} of the Introduction. Due to the low threshold of 1.6 keV$_{nr}$, the experiment can access very-low-mass WIMPs of about 3 GeV.

\subsection{The CRESST-II experiment}

The Cryogenic Rare Event Search with Superconducting Thermometers (CRESST) experiment, taking place at the Gran Sasso National Laboratory in Italy, uses, in its version II, 33 scintillating CaWO$_4$ crystals of cylindrical shape and 330 g each, for a total target mass of about 10 kg. The results from an exposure between 2009 and 2011 were published in 2012 (phase 1) \cite{Angloher:2011uu}, were an irreducible excess of events was reported. An upgrade of the CRESST-II setup (phase 2) acquired its first data from August 2013 to January 2014 and did not confirm the previous excess, reporting null results in early 2015 \cite{Angloher:2014myn}.

Where CDMS-II uses the phonon and the ionization signals coincidently produced during an event to discriminate between electron and nuclear recoils, the CRESST-II detector detects phonons and scintillation light: when an interacting particle scatters off a nucleus or interacts with electrons, it heats the crystal and produces the emission of phonons which are absorbed by phonon sensors equipped on each crystal, with the same kind of technology as CDMS-II and superCDMS, except that the temperature is even lower (10 mK). This phonon signal gives a precise measure of the deposited energy (independently of the kind of recoil) while a small fraction of it is converted into scintillation light, detected by light absorbers also mounted on each crystal. This second signal allows to define the light yield as the ratio of the energy measured by the light detector to the energy measured by the phonon detector. As for the same phonon signal, nuclear recoils are expected to produce less scintillation light than electron recoils, the former can be distinguished from the latter in a light yield-deposited energy plane since they occupy a lower band.

An advantage of the CRESST-II detector is the use of three nuclei of different masses: oxygen (O), calcium (Ca) and tungsten (W). We saw in equation \eqref{A2enhancement} of the Introduction that coherent scattering cross sections were proportional to $A^2$ ($A$ being the mass number of the target nucleus), so that elastic scattering is more efficient with heavy nuclei and hence the event rate in CRESST-II is expected to be dominated by the recoils of W. This is true for WIMPs of rather large mass (above 30 GeV from \cite{Angloher:2011uu}), but we also know from equation \eqref{Emax} of the Introduction that a heavy nucleus tends to receive less recoil energy from lighter WIMPs and in an experiment with a fixed threshold, it can be that scattering of light WIMPs on a too heavy nucleus gives a signal below the threshold. Therefore, the presence of lighter constituents allows to produce recoils above the threshold and makes these become important despite the coherence enhancement. In such a way, the event rate of CRESST-II phase 1 is dominated by Ca recoils for WIMP masses between 12 and 20 GeV (W events are below threshold and O recoils are less efficient), while O recoils completely dominate the event rate below 12 GeV (both Ca and W are then below threshold). This is how CRESST-II phase 1 is able, even with a rather high (with respect to CDMS-II for example) threshold of 12 keV$_{nr}$, to access very low WIMP masses of about 5 GeV. Note that, as we go to lower WIMP masses, the coherent enhancement decreases together with the mass of the relevant constituent and so does the expected event rate, which degrades the sensitivity of the experiment at low-WIMP masses.

In the final results of CRESST-II phase 1 \cite{Angloher:2011uu}, the events observed in the acceptance region that could not be explained by the known background gave rise to two regions in the WIMP parameter space as seen in Figure \ref{FigWIMPnucleon} of the Introduction, each one being centered on a different maximum of the likelihood function. One of the two maxima is only slightly disfavored with respect to the other. The maximum at lower mass corresponds to a WIMP mass of 11.6 GeV and to a spin-independent WIMP-nucleon cross section of $3.7\times 10^{-5}$ pb while the other favors a mass of 25.3 GeV for a cross section of $1.6\times 10^{-6}$ pb. The fact that one has two possibilities is due to the different nuclei in the detector. For the largest mass, the WIMPs are heavy enough to scatter off tungsten while in the other case the dominant contributions come from the oxygen and calcium nuclei. But due to the coherent enhancement, the WIMP-nucleon cross section that is needed to reproduce the observed rate is smaller for the largest WIMP mass.

However, the upgraded phase 2 of CRESST-II did not confirm the previously seen excess in its first results \cite{Angloher:2014myn}, and even rejected the maxima from phase 1, as can be seen in Figure \ref{FigWIMPnucleon}. This upgrade uses modules (CaWO$_4$ crystal plus phonon and light sensors) made of materials with a lower internal radioactivity, resulting in a 2 to 10 times lower background in the region of interest. But the disappearance of the signal could mainly come from a background uncertainty of phase 1 that is better controlled in phase 2: in the metal structure of the previous modules, $^{210}$Po could emit $\alpha$ particles that were absorbed by the metal and remained undetected, while the recoiling $^{206}$Pb nuclei could produce nuclear recoils in the CaWO$_4$ crystal, looking therefore like single-hit events, typical from WIMPs. In phase 2, these metal parts have been replaced by CaWO$_4$ material and hence $\alpha$ particles emit an additional scintillation light correlated to the $^{206}$Pb recoils, which allows to reject the events. With a threshold lowered to 0.6 keV$_{nr}$ and the presence of three nuclei of different masses, CRESST-II phase 2 can probe WIMPs masses below 3 GeV and hence constrain regions of the parameter space that were not covered by the other experiments.

\chapter{The O-helium scenario}\label{OHe}

The O-helium model is a very important composite dark matter model for this thesis, and one of the simplest that one can imagine since it is widely based on known physics. It has been proposed in 2006 to solve the discrepancies between experiments as contradictory as DAMA and XENON100 when they are interpreted in terms of WIMPs. Even if a precise study of the interactions of O-helium with ordinary matter has, during this thesis, shown it to be unable to fulfill its primary role, in particular because of the fact that we found no evidence for a repulsion between the O-helium and a nucleus, this model has very interesting features that were re-used for the models of Chapters \ref{Millicharges} and \ref{AntiH}. Also, it can account for some challenging indirect-detection observations. In this chapter, I present the main features of the O-helium scenario. Then I study in detail the O-helium system and show that it can give interesting explanations to some indirect-detection observations, after which I discuss the different methods that were used to study the interactions of O-helium in underground detectors. In view of the absence of a dipole barrier, I finally consider the consequences on earth and in galactic halos of the inelastic processes involving O-helium in the early universe.

\section{O-helium dark matter}

One alternative to the popular WIMPs as the constituents of dark matter consists in new heavy stable charged particles bound in neutral ``dark atoms''. Cosmological arguments indicate that these charged particles should be of charge $-2$ only. Indeed, the main problem with negatively charged particles as dark matter is the suppression of the abundance of positively charged antiparticles bound to ordinary electrons, the result being anomalous isotopes of hydrogen or helium. This problem is insurmountable if the particles are of charge $-1$: in 2005, Glashow \cite{Glashow:2005jy} proposed a model in which stable tera-quarks U (of mass of the order of 1 TeV) of charge $+2/3$ formed clusters UUU  bound with tera-electrons E of charge $-1$ in neutral UUU-EE tera-helium behaving like WIMPs. The problem is that as soon as primordial helium is formed in Big Bang Nucleosynthesis (BBN), it captures all the free E in positively charged HeE$^+$ ions, preventing any further suppression of the positively charged antiparticles. The acceptable solution is in fact obtained by considering particles of charge $-2$ only. Such particles are actually predicted by several exotic theories \cite{Khlopov:2005ew,Fargion:2005ep,Khlopov:2006uv,Khlopov:2008ty} and they will be generically denoted by O. In all these models, O, of mass $m_{O}= \mathcal{O}(1~\mathrm{TeV})$, behaves either as a lepton or as a heavy quark cluster of fourth or fifth generation with strongly suppressed hadronic interactions. Just after it is formed in BBN, helium screens the O in composite $^4$He$^{++}$O$^{--}$ neutral dark atoms, called O-helium (OHe) atoms \cite{Khlopov:2010pq,Khlopov:2010ik}. Due to the suppressed hadronic interactions of O and the global neutrality of the OHe atom, the interactions of OHe with ordinary matter are dominated by the nuclear interactions of the He nucleus. This makes the OHe scenario very attractive since, assuming that nuclear parameters are sufficiently well known, the only free parameter of the model is the mass of O.

Starting from 2006 \cite{Khlopov:2010pq,Khlopov:2010ik,Khlopov:2005ew,Fargion:2005ep,Khlopov:2006uv,Khlopov:2008ty}, it was proposed that OHe represents the whole dark matter of the universe and is responsible for the signals seen by some direct-search experiments. Provided the existence of an interaction mechanism preventing O and/or He from falling into the deep nuclear potential wells of ordinary nuclei, such as an electric dipole barrier appearing when an approaching nucleus is sufficiently close to invert the polarization of OHe, the interactions of OHe with normal matter are reduced to elastic collisions and the cosmological scenario can be led successfully from its formation just after BBN until its presence in galactic dark matter halos today. If the size of the He nucleus is neglected, an OHe atom can be seen as a first approximation as a hydrogen-like atom of binding energy:

\begin{equation}
E_{o}=-\frac{1}{2}m_{He}\left(Z_{O}Z_{He}\alpha\right)^2\simeq -1.6~\mathrm{MeV},\label{EbindingOHe}
\end{equation}
where $m_{He}$ ($\ll m_{O}$) is the mass of a helium nucleus, and $Z_{O}=Z_{He}=2$ are the electric charges of O and He in absolute value. Note that, as $|E_o|\ll m_{He}\ll m_O$, one has for the mass $m_o$ of OHe:
\begin{equation}
m_o=m_O+m_{He}+E_o\simeq m_O,\label{mO}
\end{equation}
so that we will rather use the mass $m_O$ of O instead of the mass of OHe in the following. The Bohr radius $r_o$ of such a structure is given by:

\begin{equation}
r_o=\frac{1}{m_{He}Z_{O}Z_{He}\alpha}\simeq 2~\mathrm{fm},\label{ro}
\end{equation}
so that OHe atoms, in their self-interactions at low momentum, can be seen as hard spheres of radius $r_o$ and elastic cross section:

\begin{equation}
\sigma_{o} =4\pi r_o^2\simeq 10^{-25}~\mathrm{cm}^2.\label{sigmao}
\end{equation}
As OHe is assumed to saturate the full dark matter density, one has to pay attention to the constraints on self-interacting dark matter. These usually constrain the ratio of the self-interaction cross section to the mass of the dark matter particle, with the strongest upper limit from \cite{2002ApJ...564...60M} of $10^{-25.5}$ cm$^2$/GeV\footnote{Note, however, that recent simulations showed that this constraint may have been overestimated and could be weakened by one or two orders of magnitude \cite{Peter:2012jh}.}. Here, $\sigma_o/m_{O}\simeq 10^{-28}$ cm$^2$/GeV, so that OHe behaves like collisionless dark matter and the shapes of dark matter halos are therefore not perturbed by the self-interactions of OHe. The cross section \eqref{sigmao} can be used to estimate the elastic cross section between OHe and standard matter. In presence of a specific interaction mechanism, the interactions of OHe with ordinary matter are dominantly elastic, and the OHe atoms falling onto earth because of its motion in the galactic halo have purely elastic collisions with terrestrial nuclei. With the value \eqref{sigmao} of the cross section and the equations \eqref{penLength} and \eqref{dEdx} that will be introduced in Chapter \ref{Millicharges}, we calculate that OHe atoms are slowed down and completely stopped in the terrestrial crust after a few tens of meters. This means that the earth, during its orbital motion around the sun, stores all the in-falling flux of OHe atoms, which thermalize just below the surface and eventually fall to its center.

\section{OHe in underground detectors}\label{OHeunderground}

According to the scenario in \cite{Khlopov:2010pq,Khlopov:2010ik}, as soon as they are thermalized, OHe atoms start to drift down towards the center of the earth, driven by gravity, at the drift velocity $V_d$\footnote{Although the expression \eqref{Vdrift} has been used throughout this thesis, we realized through recent Monte Carlo simulations of particles undergoing thermal collisions in a uniform medium submitted to gravity that the correct formula is actually $V_d=\frac{m_O}{\mu_{\oplus}}\frac{g}{n_{\oplus}\sigma_o v_t}$, where $\mu_{\oplus}$ is the reduced mass of the system composed of the diffusing particle and of the particle constituting the medium. Here, $m_O\simeq 1~\mathrm{TeV}$ and the average mass of the nuclei in the terrestrial crust is close to the mass of silicon, i.e. $\mu_{\oplus}\simeq 26$ GeV, which brings a correction factor $\frac{m_O}{\mu_{\oplus}}$ of about 40. According to equation \eqref{fluxes}, this decreases the number density $n$ of thermalized dark particles below the surface by the same factor, which requires capture processes of the form \eqref{capture} to be 40 times more efficient in order to reproduce the event rate \eqref{capturerate}. This will not affect the results of this chapter since we will only focus on the interaction mechanisms between OHe and ordinary matter (interaction potential) and not on capture processes.}:

\begin{equation}
V_d=\frac{g}{n_{\oplus}\sigma_o v_t}= \mathcal{O}(10~\mathrm{cm/s}),\label{Vdrift}
\end{equation}
until they reach underground detectors, typically located at a depth of 1 km, $10^5/10=10^4$ s $\simeq 3$ h later. In \eqref{Vdrift}, $g=980$ cm/s$^2$ is the acceleration of gravity on earth, $n_{\oplus}\simeq 10^{23}$ cm$^{-3}$ is the number density of terrestrial nuclei in the crust and $v_t=\mathcal{O}(10^4~\mathrm{cm/s})$ is the mean relative velocity between thermal OHe atoms and terrestrial nuclei. For a stationary state, an equilibrium is set between the incident in-falling and the down-drifting fluxes:

\begin{equation}
\frac{1}{4}n_{\odot} |\vec{V}_{\odot}+\vec{V}_{\oplus}|=nV_d,\label{fluxes}
\end{equation}
where $n_{\odot}(\mathrm{cm}^{-3})=0.3/m_{O}$(GeV) is the number density of OHe atoms in the solar system, $\vec{V}_{\odot}$ is the velocity vector of the sun with respect to the galactic dark matter halo, $\vec{V}_{\oplus}$ is the orbital velocity vector of the earth around the sun and $n$ is the number density of thermalized OHe atoms. The factor $\frac{1}{4}$ in \eqref{fluxes} takes into account the average of the incident flux over all latitudes (factor $\frac{1}{2}$) and the mean diurnal cycle due to the occultation of the dark matter flux by the earth during its rotation on itself (factor $\frac{1}{2}$). In standard halo models, the Local Standard of Rest (LSR) moves at about 220 km/s with respect to the dark matter halo, and the sun has a small peculiar velocity with respect to the LSR which we can neglect, so that we can say that $|\vec{V}_{\odot}|=V_{\odot}=220$ km/s. For $|\vec{V}_{\oplus}|=V_{\oplus}$, we use a value of 29.5 km/s.

From \eqref{fluxes}, the equilibrium number density $n$ adjusts to the incident flux of OHe atoms, which is annually modulated due to the orbital motion of the earth around the sun. As $V_{\oplus}\ll V_{\odot}$, one can develop the norm $|\vec{V}_{\odot}+\vec{V}_{\oplus}|$ as:

\begin{equation}
|\vec{V}_{\odot}+\vec{V}_{\oplus}|\simeq V_{\odot}+V_{\oplus}\cos i \cos(\omega(t-t_0)),\label{norm}
\end{equation}
where $i\simeq 60^{\circ}$ is the inclination angle of the ecliptic plane on the galactic plane, $\omega=2\pi/T_{orb}$ is the angular frequency of the orbital motion of the earth, with a period $T_{orb}=1$ yr, and $t_0\simeq$ June 2 is the day of the year when the projection of $\vec{V}_{\oplus}$ on the galactic plane is parallel to $\vec{V}_{\odot}$ and in the same direction. Therefore, the number density of thermalized OHe atoms in underground detectors is modulated with a period of one year and a phase $t_0$ at which it reaches a maximum:

\begin{equation}
n(t)=n^0+n^m\cos(\omega(t-t_0)),\label{neqmod}
\end{equation}
with the constant and modulated parts $n^0$ and $n^m$ defined as:

\begin{eqnarray}
n^0&=&\frac{n_{\odot}}{4V_d}V_{\odot}, \\
n^m&=&\frac{n_{\odot}}{4V_d}V_{\oplus}\cos i.
\end{eqnarray}

The thermal energies of the OHe atoms when they arrive in underground detectors do not allow them to produce nuclear recoils above the experimental thresholds. To explain consistently the positive results of DAMA and the negative results of XENON100 or CDMS-II, it had been proposed in \cite{Khlopov:2010pq,Khlopov:2010ik} that the signals may have another origin: if OHe has weakly bound states (in the keV range) with ordinary matter, an event could correspond to the radiative capture of a thermalized OHe atom by a nucleus N of the active medium. The emitted photon $\gamma$, of energy equal to the energy of the bound state in absolute value, would then deposit its energy in the detector by electron recoil and produce a signal. The capture reaction we are interested in is:

\begin{equation}
\mathrm{N}~+~^4\mathrm{He}^{++}\mathrm{O}^{--}\rightarrow \mathrm{N}^4\mathrm{He}^{++}\mathrm{O}^{--}~+~\gamma.\label{capture}
\end{equation}
If $\sigma_{capt}$ is the cross section of the process \eqref{capture}, then the bound-state-formation rate $\Gamma$ is given by:

\begin{equation}
\Gamma(t)=n_N n(t) \left<\sigma_{capt}v\right>=\Gamma^0 +\Gamma^m \cos(\omega(t-t_0)),\label{capturerate}
\end{equation}
where $n_N$ is the number density of nuclei N in the active medium of the detector and $v$ is relative velocity between OHe and N. As the capture occurs at thermal energy, the average of $\sigma_{capt}v$ has to be performed over Maxwellian velocity distributions at the operating temperature for the OHe and N species. So, the event rate $\Gamma$ is modulated because of the annual modulation of the number density $n$ of OHe, with constant and modulated parts:

\begin{eqnarray}
\Gamma^0&=&n_N n^0 \left<\sigma_{capt}v\right>, \\
\Gamma^m&=&n_N n^m \left<\sigma_{capt}v\right>.
\end{eqnarray}
Therefore, the annual modulation seen by experiments such as DAMA and CoGeNT can be explained with the right period and the right phase by the motion of the earth in a dark matter halo made of OHe atoms that thermalize in the terrestrial crust and form bound states with the constituent nuclei of underground detectors.

Note that, as electron recoils are produced here, experiments such as DAMA and CoGeNT, which do not discriminate between electron and nuclear recoils, can be reinterpreted directly. For the others that can distinguish between both types of recoils and look for nuclear recoils, the events fall in the electron-recoil background, which should therefore be studied thoroughly. In the case where such an experiment has negative results (XENON100, LUX, CDMS-II/Ge, superCDMS, CRESST-II), one only has to ensure that the predicted electron-recoil background is consistent with the observed one, which weakens considerably the constraints. In contrast, if an experiment reports the observation of nuclear recoils (CDMS-II/Si), these cannot be explained by this kind of model.

In \cite{Khlopov:2010pq,Khlopov:2010ik}, it was shown that the existence of weakly bound states and the capture rate depend on the detector composition and on the operating temperature. The study was performed via an effective OHe-nucleus interaction potential with spherical symmetry, which was simplified by a square well potential, hence allowing to solve the radial Schr\"odinger equation analytically. It was found that, for some parameters of the potential, energy levels at around $-3$ keV existed with sodium, while no bound states were present with xenon, providing a natural explanation to the events of DAMA in the $\left(2-6\right)$ keV range together with the absence of signal in XENON100 and LUX, without even having to consider their electron-recoil backgrounds. Also, the capture rate was shown to be proportional to the temperature, so that the formation of bound states was strongly suppressed in cryogenic detectors such as CDMS and CRESST, which is consistent with the negative results from the latter and from the CDMS-II/Ge and superCDMS versions of the former. In my undergraduate thesis \cite{Wall2011}, I confirmed these results for a more realistic effective potential, through the semi-classical WKB approximation.

But all these considerations relied on the presence of a dipole barrier in the potential which, in addition to preventing inelastic processes during the early universe and the formation of anomalously heavy isotopes of known light elements, allowed weakly bound states to form and to be stable with respect to the deep nuclear well at short distance, and prevented the deposition of a great amount of energy in the nucleus. This barrier was assumed to appear at a distance of the order of 8 fm when the He nucleus is sufficiently close to the nucleus N, so that the former gets attracted towards the latter by nuclear force, which creates an O-He-N configuration that is repulsive in terms of electrostatic forces. Dipolar repulsion at such distances gives a barrier with a height of the order of 1 MeV, and this was expected to dominate the OHe-N potential until the OHe enters in the nucleus, i.e. at a distance of the order of 4 fm. Before we look for such a necessary barrier through several methods in Section \ref{secRepulsion}, I study more precisely the OHe system in Section \ref{secSpectrum}, taking into account the size of the helium component, in order to improve the hydrogen-like approximation and to determine its full spectrum and its allowed transitions. This shows that OHe can give interesting explanations to some indirect-detection observations and gives rise to typical indirect signatures that can be tested. Section \ref{secSpectrum} is based on papers \cite{Cudell:2014wca} and \cite{Cudell:2014eta}, while Section \ref{secRepulsion} contains the results of paper \cite{Cudell:2012fw}.

\section{The OHe spectrum and transitions}\label{secSpectrum}

\begin{figure}
\begin{center}
\includegraphics[scale=0.7]{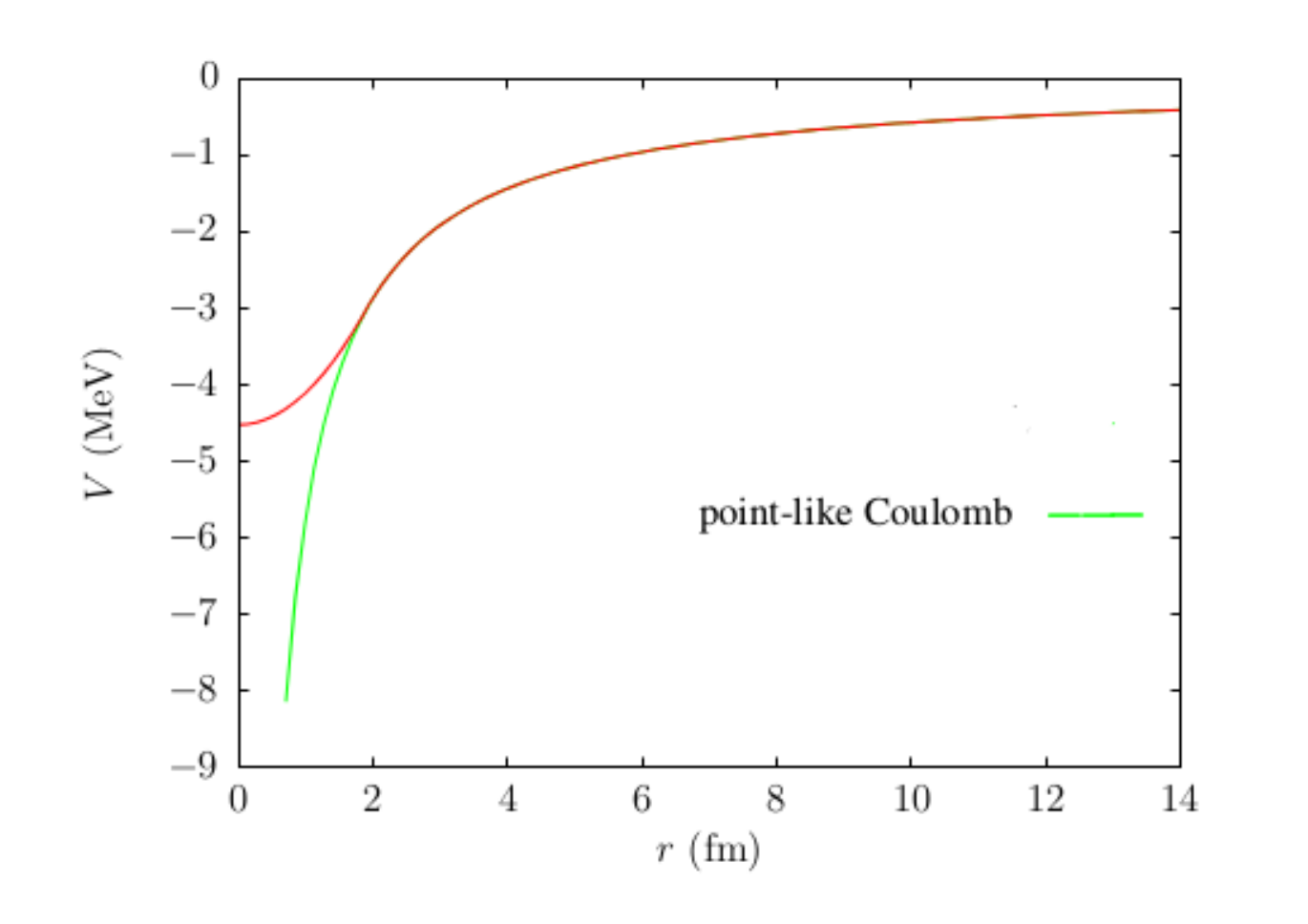}
\end{center}
\caption{The O-helium interaction potential (red) and the elementary Coulomb potential (green) obtained for a point-like helium nucleus, as a function of the distance $r$ between O and the center of the helium nucleus.}
\label{FigOHePot}
\end{figure}

As already mentioned, an OHe atom is made of a heavy O particle and a He nucleus, of respective masses $m_O$ and $m_{He}$ such that $m_O\gg m_{He}$, bound by the
Coulomb interaction in a hydrogen-like structure. While the O particle, of charge $Z_O=2$, is point-like, we approximate the charge distribution
of the helium nucleus as a uniformly charged sphere of radius $R_{He}(\mathrm{fm})=1.2\times 4^{1/3}$. This allows us to take the O-helium
attractive interaction potential as that between a point and a sphere, i.e.:

\begin{align}
V(r) & = -\frac{Z_O Z_{He}\alpha}{r},~\mathrm{for}~r\geq R_{He}\nonumber \\
& =-\frac{Z_O Z_{He}\alpha}{2R_{He}}\left(3-\frac{r^2}{R_{He}^2}\right),~\mathrm{for}~r<R_{He}\label{OHepot}
\end{align}
where $r$ is the distance between O and the center of the helium nucleus. The potential is represented in Figure \ref{FigOHePot} together with the elementary Coulomb potential obtained when the helium nucleus is assumed to be point-like.

To obtain the eigenstates of an OHe atom, we solve the radial time-independent Schr\"odinger equation with the potential \eqref{OHepot}, written in the center-of-mass frame of the system, which corresponds almost exactly to the O particle as $m_O\gg m_{He}$. At zero angular momentum $l$, we use the non-modified WKB approximation as the potential $V$ is regular at the origin ($\lim\limits_{r \to 0}rV(r)=0$). For the states at higher $l$, we use the modified WKB approximation, which consists in replacing $l(l+1)$ by $(l+\frac{1}{2})^2$ in the centrifugal term of the effective potential \cite{froman}.

The eigenvalues $E_{n,l}$ are shown in Table \ref{TabOHelevels} for the first ten values of the principal quantum number $n$ and the corresponding angular momenta $l=0,...,n-1$, together with the pure hydrogen-like solutions $E^H_n=-\frac{1}{2}m_{He}(Z_O Z_{He}\alpha)^2/n^2$. We notice a lift of degeneracy on $l$ because the potential \eqref{OHepot} is no longer proportional to $1/r$, so that the energy levels now depend on both $n$ and $l$. Also, the energy of the ground state $1s$ has been increased significantly with respect to $E^H_1$ (denoted by $E_o$ in \eqref{EbindingOHe} in this particular case), due to the finiteness of the well at $r=0$ when the charge distribution of the helium nucleus is taken into account. In general, at fixed $n$, the pure hydrogen-like energy level constitutes a lower limit to which the levels $E_{n,l}$ tend as $l$ increases, i.e. as the states are excited and thus as O and He lie further apart from each other, making the helium nucleus increasingly point-like. At very high $n$, the differences between the two types of levels are very small and we recover the pure hydrogen-like case when $n\rightarrow \infty$. Note also that $E_{n,l}>E_{n',l'}$ for all $l,l'$ if $n>n'$.  As it is well known that the WKB approximation is less accurate for the deeper bound states, we computed the energy of the ground state by a variational method using up to $11$ hydrogen-like $s$-orbitals and found the result $-1.17715$ MeV, which shows that the error on the values given in Table \ref{TabOHelevels} is less than 1\textperthousand.

\begin{table}
\begin{center}
%\tbl{Energy levels $E_{n,l}$ (MeV) of the OHe atom, for the first ten values of the principal quantum number $n$ and the corresponding angular momenta $l=0,...,n-1$. In the last column are also shown the pure hydrogen-like solutions $E^H_n$ (MeV) obtained when the helium nucleus is assumed to be point-like. The exponents indicate the power of $10$ by which the numbers have to be multiplied to obtain the energy in MeV.}{
\begin{tabular}{c|c|c|c|c|c|c}
\hline
\hline
 $n$ & $l=0$ & $l=1$ & $l=2$ & $l=3$ & $l=4$ & $E^H_n$ \\
\hline
\hline
1 & -1.17597  & - & - & - & - & -1.58790 \\
2 & -0.34463  & -0.39691 & - & - & - & -0.39698 \\
3 & -0.16070  & -0.17642 & -0.17642 & - & - & -0.17643 \\
4 & -9.2538$^{-2}$  & -9.9230$^{-2}$ & -9.9240$^{-2}$ & -9.9239$^{-2}$ & - & -9.9244$^{-2}$ \\
5 & -6.0057$^{-2}$  & -6.3511$^{-2}$ & -6.3511$^{-2}$ & -6.3511$^{-2}$ & -6.3510$^{-2}$ & -6.3516$^{-2}$\\
6 & -4.2097$^{-2}$ & -4.4106$^{-2}$ & -4.4106$^{-2}$ & -4.4106$^{-2}$ & -4.4106$^{-2}$  & -4.4108$^{-2}$  \\
7 & -3.1136$^{-2}$ & -3.2404$^{-2}$ & -3.2404$^{-2}$ & -3.2404$^{-2}$ & -3.2404$^{-2}$ &  -3.2406$^{-2}$\\
8 & -2.3957$^{-2}$ & -2.4808$^{-2}$ & -2.4811$^{-2}$ & -2.4810$^{-2}$ & -2.4810$^{-2}$ & -2.4811$^{-2}$ \\
9 & -1.9002$^{-2}$ & -1.9602$^{-2}$ & -1.9602$^{-2}$ & -1.9602$^{-2}$ & -1.9602$^{-2}$ & -1.9604$^{-2}$ \\
10 & -1.5439$^{-2}$ & -1.5878$^{-2}$ & -1.5878$^{-2}$ & -1.5878$^{-2}$ & -1.5878$^{-2}$ & -1.5879$^{-2}$\\
 \hline
 \hline
  $n$ & $l=5$ & $l=6$ & $l=7$ & $l=8$ & $l=9$ & $E^H_n$ \\
\hline
\hline
 1 & -  & - & - & - & - & -1.58790 \\
2 & - & - & - & - & - & -0.39698 \\
3 & -  & - & - & - & - & -0.17643 \\
4 & -  & - & - & - & - & -9.9244$^{-2}$ \\
5 & -  & - & - & - & - & -6.3516$^{-2}$\\
6 & -4.4105$^{-2}$ & - & - & - & - & -4.4108$^{-2}$  \\
7 & -3.2403$^{-2}$ & -3.2406$^{-2}$ & - & - & - &  -3.2406$^{-2}$\\
8 & -2.4810$^{-2}$ & -2.4810$^{-2}$ & -2.4810$^{-2}$ & - & - & -2.4811$^{-2}$ \\
9 & -1.9604$^{-2}$ & -1.9602$^{-2}$ & -1.9602$^{-2}$ & -1.9602$^{-2}$ & - & -1.9604$^{-2}$ \\
10 & -1.5878$^{-2}$ & -1.5878$^{-2}$ & -1.5878$^{-2}$ & -1.5878$^{-2}$ & -1.5879$^{-2}$ & -1.5879$^{-2}$ \\
  \hline
 \hline
\end{tabular}
\end{center}
\caption{Energy levels $E_{n,l}$ (MeV) of the OHe atom, for the first ten values of the principal quantum number $n$ and the corresponding angular momenta $l=0,...,n-1$. In the last column are also shown the pure hydrogen-like solutions $E^H_n$ (MeV) obtained when the helium nucleus is assumed to be point-like. The exponents indicate the power of $10$ by which the numbers have to be multiplied to obtain the energy in MeV.\label{TabOHelevels}}
%\label{Table1}}
\end{table}

OHe-OHe collisions in the central parts of galaxies can lead to the excitation of OHe. Indeed, for velocities $v=\mathcal{O}(10^{-3})$ in such regions, equation \eqref{Emax} indicates that the energy transfers can be as large as $\Delta E=\mathcal{O}(1~\mathrm{MeV})$ for $m_O=1$ TeV, which is just of the order of the binding energy.

If OHe levels with non-zero angular momentum are excited through OHe-OHe collisions, X and gamma lines should appear at their de-excitations from electric dipole (E1) transitions. Such spontaneous transitions from a state $(n,l)$ to a state $(n',l')$ require $\Delta l=l'-l=\pm 1$ and $E_{n,l}>E_{n',l'}$. In view of Table \ref{TabOHelevels}, this means that if $n=n'$, then $\Delta l=+1$, and if $n>n'$, then $\Delta l=\pm 1$ (no transition with $n<n'$ being energetically allowed). These predictions may be of interest for the analysis of the possible nature of unidentified lines, observed in the center of the galaxy. The lines with energies above 20 keV can be searched for in the INTEGRAL data \cite{Teegarden:2004ct}, while a forest of lines of lower energy may be found in X-ray observations. In particular, high-level transitions can lead to a number of lines around 3.5 keV. Such a line was actually identified by several groups in the XMM-Newton data \cite{Bulbul:2014sua,Boyarsky:2014jta}, which cover the range $(0.1-12)$ keV, as coming from the central regions of a sample of galaxies.

\begin{figure}
\begin{center}
\includegraphics[scale=0.7]{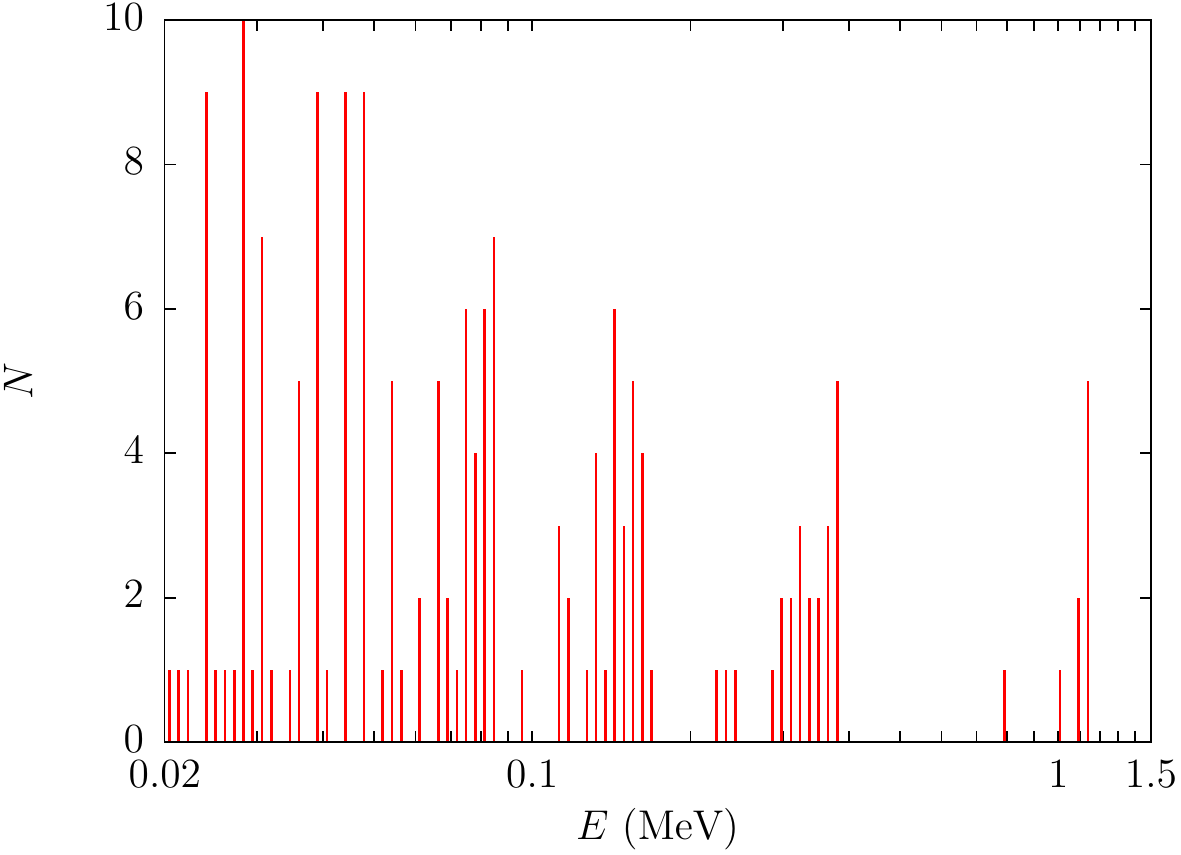}
\end{center}
\caption{Number $N$ of E1 transitions with energies contained within logarithmic bins in energy of width $\Delta \log E=\log(1162~\mathrm{keV}/20~\mathrm{keV})/100$, as a function of the energy $E$ of the transition.}
\label{FigOHeForest}
\end{figure}

By considering all the possible E1 transitions between the states presented in Table \ref{TabOHelevels}, we find several hundreds of allowed lines, with energies from the eV range to the MeV range, even for the limited sample $n\leq 10$. In Figure \ref{FigOHeForest}, we show the number $N$ of E1 transitions as a function of the energy $E$, from $20$ keV to $1.162$ MeV, the latter being the largest energy of the sample and corresponding to the transition from the state $(n=10,l=1)$ to the ground state $(n=1,l=0)$. This was obtained by counting the number of E1 transitions with energies contained within logarithmic bins in energy of width $\Delta \log E=\log (1162~\mathrm{keV}/20~\mathrm{keV})/100$. The lines with energies between $3$ and $4$ keV are listed in Table \ref{Tab34keV} and are in the right energy region to account for the 3.5-keV line seen in the XMM-Newton data. More generally, the comparison of these predictions with the observations may provide an efficient tool to test the OHe composite-dark-matter model.

\begin{table}
\begin{center}
%\tbl{E1 transitions of an OHe atom with energies between $3$ and $4$ keV.}{
\begin{tabular}{c|c|c}
\hline
\hline
Initial state $(n,l)$ & Final state $(n',l')$ & Energy (keV) \\
\hline
\hline
$(5,0)$ & $(5,1)$ & 3.4542 \\
\hline
$(10,1)$ & $(9,0)$ & 3.1236 \\
         & $(9,2)$ & 3.7243 \\
\hline
$(10,2)$ & $(9,1)$ & 3.7243 \\
         & $(9,3)$ & 3.7241 \\
\hline
$(10,3)$ & $(9,2)$ & 3.7245 \\
         & $(9,4)$ & 3.7242 \\
\hline
$(10,4)$ & $(9,3)$ & 3.7243 \\
         & $(9,5)$ & 3.7258 \\
\hline
$(10,5)$ & $(9,4)$ & 3.7246 \\
         & $(9,6)$ & 3.7243 \\
\hline
$(10,6)$ & $(9,5)$ & 3.7260 \\
         & $(9,7)$ & 3.7245 \\
\hline
$(10,7)$ & $(9,6)$ & 3.7240 \\
         & $(9,8)$ & 3.7240 \\
\hline
$(10,8)$ & $(9,7)$ & 3.7245 \\
\hline
$(10,9)$ & $(9,8)$ & 3.7233 \\
\hline
\hline
\end{tabular}
%\label{Table2}
\end{center}
\caption{E1 transitions of an OHe atom with energies between $3$ and $4$ keV.\label{Tab34keV}}
\end{table}

Also, it was first noted in \cite{Khlopov:2008ki} that OHe collisions in the galactic bulge can lead to OHe de-excitation not by photon emission, but by electron-positron-pair production. It is estimated in \cite{Finkbeiner:2007kk} that an electron-positron-production rate of $3\times 10^{42}$ s$^{-1}$ is sufficient to explain the excess in the electron-positron-annihilation line from the galactic bulge measured by INTEGRAL. OHe can provide such a rate: if the $2s$-level is excited in an OHe-OHe collision, pair production dominates over the two-photon channel in the de-excitation to the $1s$-state, so that the electron-positron production is not accompanied by a strong gamma signal. The rate of electron-positron production $\Gamma_{ee}$ strongly depends on the velocity and density profiles of OHe in the center of the galaxy and we found a range of parameters for which this rate is sufficient to explain the INTEGRAL data. These results are shown in Figure \ref{FigeeProduction} and the details of the calculation of $\Gamma_{ee}$ are given in Appendix \ref{AppendixA}\footnote{It should be noted that, even if the energy difference between the $1s$ and the $2s$ states is sufficient to produce an $e^{+} e^{-}$ pair in the hydrogen-like case, it is not in the real one (see Table \ref{TabOHelevels}). The calculation of Appendix \ref{AppendixA} leading to Figure \ref{FigeeProduction} has been made in the hydrogen-like case for the levels $1s$ and $2s$ for illustration, but the same kind of developments for $3s\rightarrow 1s$ transitions, which allow to produce $e^{+} e^{-}$ pairs even in the real case, give similar results with higher required central densities.}.

\begin{figure}
\begin{center}
\includegraphics[scale=0.7]{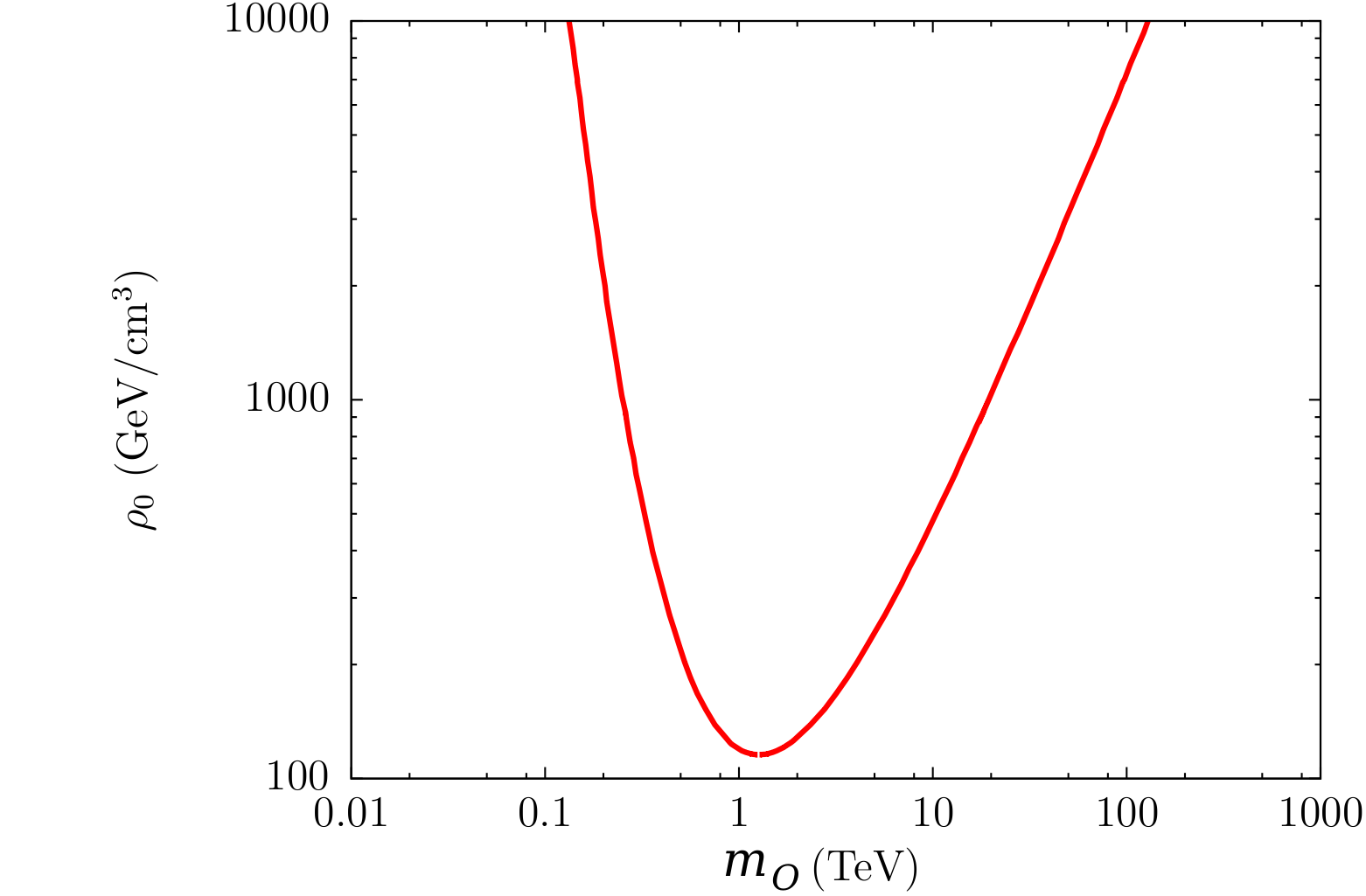}
\end{center}
\caption{Values of the central dark matter density $\rho_{0}$, assuming a Burkert profile \cite{Burkert:1995yz}, and of the OHe mass $m_O$ reproducing the excess of $e^{+} e^{-}$-pair production from the galactic bulge measured by INTEGRAL \cite{Teegarden:2004ct,Finkbeiner:2007kk}. Below the red curve, the predicted rate is too low.}
\label{FigeeProduction}
\end{figure}

\section{Search for a repulsion in the OHe-nucleus interaction}\label{secRepulsion}

\subsection{Classical model in one dimension}\label{subsecClassical}

\begin{figure}
\begin{center}
\includegraphics[scale=0.25]{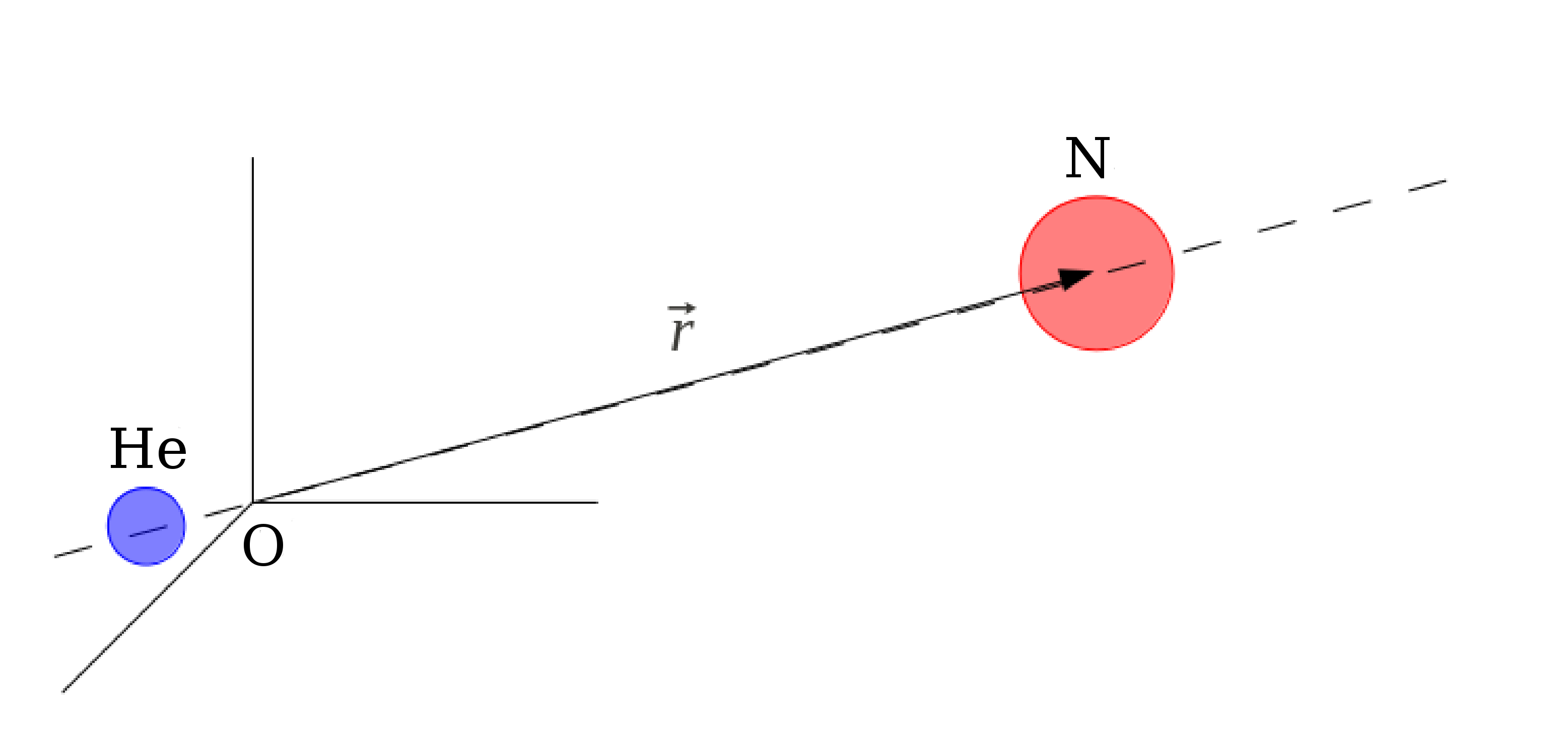}
\includegraphics[scale=0.25]{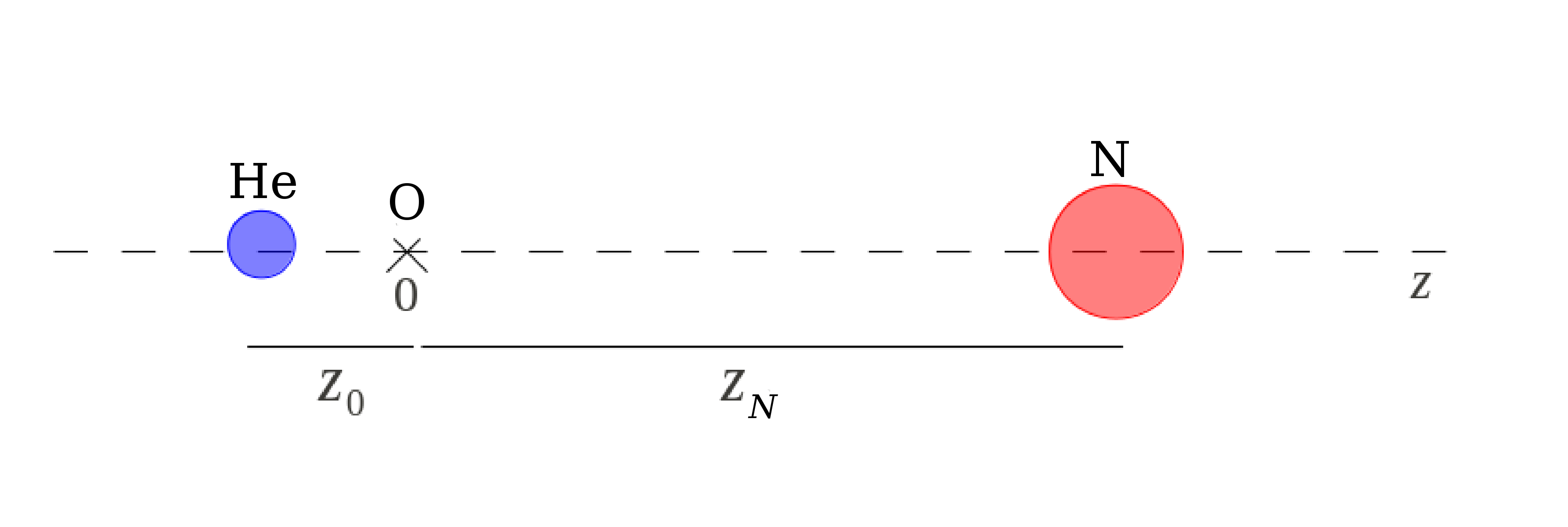}
\end{center}
\caption{One-dimensional OHe atom polarized along the O-N axis, denoted $z$. $z_{0}$ is the distance between O and He and $z_{N}$ is the distance of the N nucleus along the $z$-axis.}
\label{Fig1Dgeometry}
\end{figure}

To study the polarization of the OHe atom under the influence of an approaching nucleus N, we can first treat OHe as a classical structure and neglect the effects of the motions of O and of the nucleus. The polarization of OHe is then fixed by the equilibrium of forces acting on the He nucleus. For every position of the nucleus N, we can work on the O-N axis, in the rest frame of the O particle, as shown in Figure \ref{Fig1Dgeometry}.

We take the He and N nuclei as uniformly charged spheres of charges $Z_{He}=2$ and $Z$. Throughout this section, as we will be looking for a crucial repulsion mechanism in the OHe-nucleus interaction, we will pay a particular attention to radial distances, for which we will be as accurate as possible. For this reason, we will estimate nuclear radii with a specific experimental parametrization from scattering experiments \cite{PhysRev.106.126}. This is of the form $R_{He}(\mathrm{fm})=1.35\times 4^{1/3}$ and $R_{N}(\mathrm{fm})=1.35\times A^{1/3}$, $A$ being the mass number of N, instead of the usual expression $1.2\times A^{1/3}$ used in Section \ref{secSpectrum} and in Chapters \ref{Millicharges} and \ref{AntiH}. Then, as O is assumed to be point-like, we obtain the electrostatic potential for the interactions with O:

\begin{align}
V_{ON}(z_{N}) & = -\frac{ZZ_{O}\alpha}{z_{N}}, {\rm\ for\ } z_{N}>R_N\nonumber\\
 & = -\frac{ZZ_{O}\alpha}{2R_N}\left(3-\frac{z_{N}^{2}}{R_N^{2}}\right),{\rm\ for\ } z_{N}<R_N
 \end{align}
for the interaction between O and the nucleus N, and a similar expression for $V_{OHe}$. The potential between the two nuclei has both electrostatic and nuclear contributions. In the former, we neglect the He size, and for the latter we use the experimental parametrisation \cite{PhysRev.106.126} of the He-nucleus potential:

\begin{align}
V_{HeN}(z_{N}-z_{0}) & = \frac{Z_{He}Z\alpha}{|z_{N}-z_{0}|}+\frac{-V_{0}}{1+e^{\left(|z_{N}-z_{0}|-R_{*}\right)/a}},{\rm\ for\ }|z_{N}-z_{0}|>R_N\\
 & = \frac{Z_{He}Z\alpha}{2R_N}\left(3-\frac{\left(z_{N}-z_{0}\right)^{2}}{R_N^{2}}\right)+\frac{-V_{0}}{1+e^{\left(|z_{N}-z_{0}|-R_{*}\right)/a}},{\rm\ for\ } |z_{N}-z_{0}|<R_N.\nonumber
 \end{align}
The nuclear interaction is represented in a Woods-Saxon form, with parameters $V_{0}=30~\mathrm{MeV}$, $a=0.5$ fm and $R_{*}(\mathrm{fm})=1.35\times A^{1/3}+1.3$. This parametrization allows to take into account the finite size of the He nucleus.

\begin{figure}
\begin{centering}
\includegraphics[scale=0.7]{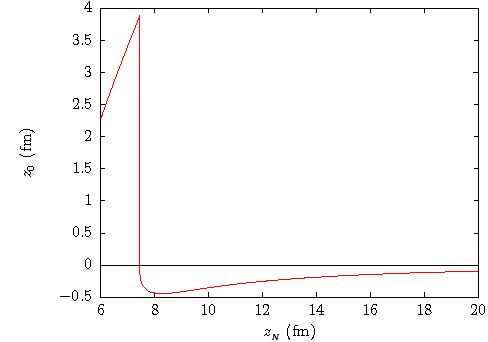}
\par\end{centering}
\caption{Polarization $z_{0}$ of the one-dimensional OHe atom as a function of the distance $z_{N}$ of an approaching sodium nucleus. The break in the curve corresponds to the fact that for a distance of about 9 fm, He falls into the nuclear potential of N. \label{FigPolarization1D}}
\end{figure}

The equilibrium position $z_{0}$ of the He nucleus will be at the minimum of the potential $V_{HeN}+V_{OHe}$ and will depend on $z_{N}$. At that point, the Coulomb forces balance the nuclear force:

\begin{equation}
\vec{F}_{OHe}+\vec{F}_{HeN}=\vec{0}.\label{Fres}
\end{equation}
When the equilibrium position is determined, the OHe-N potential is obtained by adding the dipole potential to the Woods-Saxon one:

\begin{equation}
V_{OHeN}(z_{N})=V_{dip}(z_{N})+V_{WS}(|z_{N}-z_{0}|),
\end{equation}
where

\begin{equation}
V_{dip}(z_{N})=\frac{Z_OZ\alpha z_{0}(z_{N})}{z_{N}\left(z_{N}-z_{0}(z_{N})\right)}
\end{equation}
is the dipole potential of the polarized OHe atom.

\begin{figure}
\begin{centering}
\includegraphics[scale=0.7]{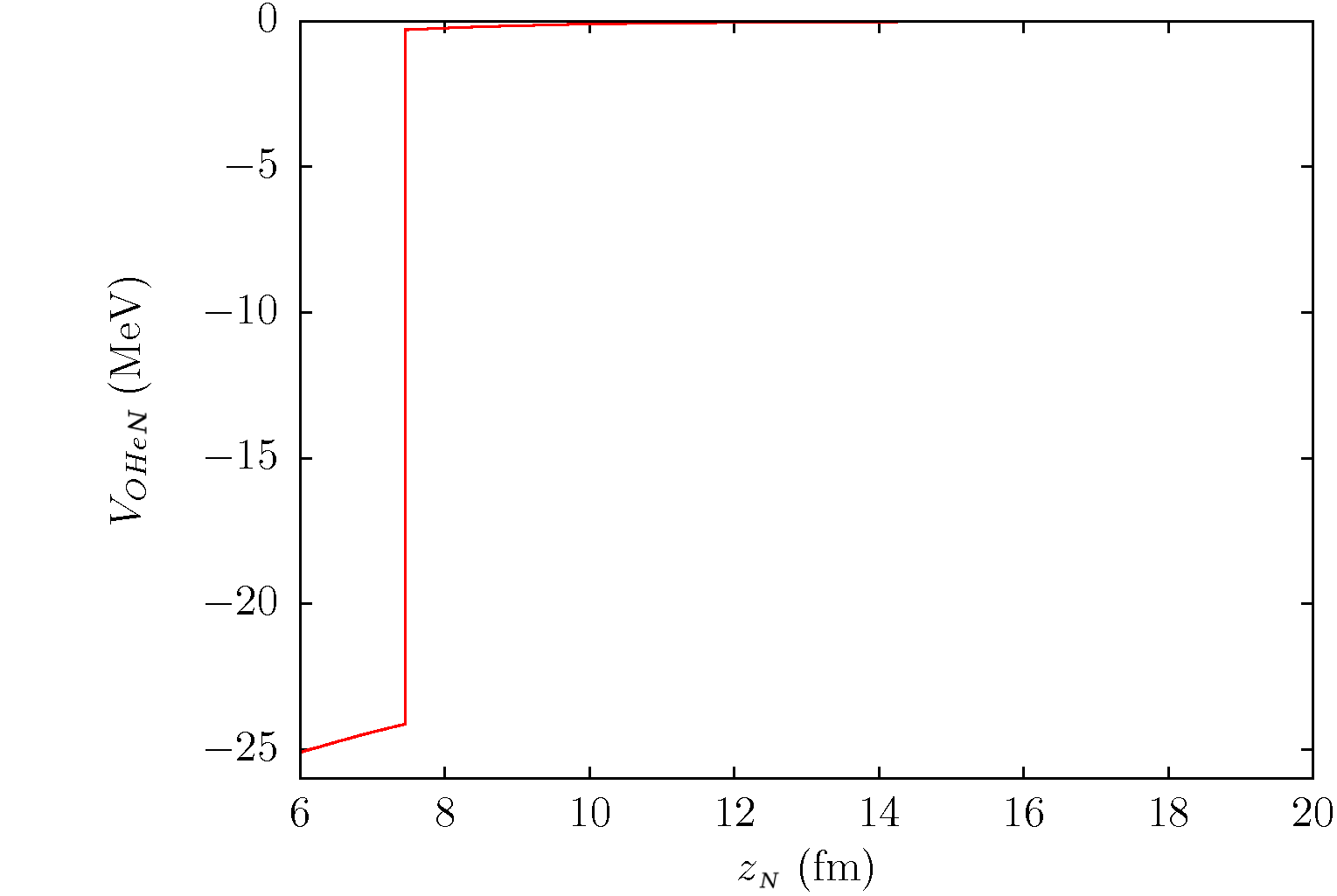}
\par\end{centering}
\caption{Total OHe-N potential for sodium in the one-dimensional approach.\label{FigTotalOHeApot}}
\end{figure}

Figure \ref{FigPolarization1D} shows the polarization $z_{0}$ of the OHe atom as a function of the position $z_{N}$ along the $z$-axis for an approaching sodium nucleus with $Z=11$. We see that it is negative at large distance, giving rise to an attractive dipole potential, and that the nuclear force starts to reverse the dipole when the nucleus gets closer to O. This situation corresponds to a repulsive electromagnetic dipole that acts against the nuclear force, but it can be seen in Figure \ref{FigTotalOHeApot} that this repulsive force is not sufficient to overcome the nuclear force between the two nuclei and to give rise to a repulsive global potential. When $z_{N}\lesssim7.5$ fm, equation \eqref{Fres} projected along the $z$-axis loses its initial solution and another one remains, that is located in the nuclear well, giving rise to a jump in the polarization and therefore in the total potential. Similar results and pictures can be obtained for other nuclei.

Hence we see that, classically, no repulsive potential appears, even if the electrostatic force of OHe repels the nucleus. In fact one can argue that this is a generic classical result which does not depend on the details of the calculation. If the configuration of the three objects is He-O-N, then clearly the force is attractive. If the configuration is O-He-N, then that means that the nuclear force on He, directed to the right, is larger than the electrostatic forces from O and N, directed to the left. Again, a net attraction between OHe and N results. To settle this classical picture of permanent attraction in the OHe-nucleus system, a semi-classical or a full quantum treatment of the problem is therefore needed.

\subsection{Semi-classical model}\label{subsecSemiclass}

The best treatment of the OHe-nucleus interaction would be to solve the quantum three-body O-He-N problem, with the simplifying assumption that one of the three components, O, is infinitely massive ($m_O\gg m_{He},m_N$), so that it could be fixed at the origin of the center-of-mass frame. But the problem involves very different scales: from \eqref{EbindingOHe}, the OHe binding energy is of the order of one MeV, and we are looking for OHe-nucleus bound states of about one keV. To obtain both in the same framework would imply a solution of the three-body problem at better than one per thousand, which is clearly very hard. A further simplification would be to consider that the nucleus N is much more massive than the He nucleus ($m_O,m_N\gg m_{He}$), which is often the case since for most detector compositions $Z>Z_{He}$. In that case, N could also be fixed in the frame and for each position, the He nucleus would rapidly re-arrange around the O-N axis, giving a cylindrical symmetry to the problem. The dynamics of N would therefore reduce to a single parameter, i.e. the distance between O and N, as a function of which we would study the evolution of the wavefunction of He. This approach will be followed in Section \ref{subsecPert}.

For now, as we are interested in the very excited bound states, we can use a simplified method. For these states, the OHe atom will not dissociate, so we can treat that system as a whole, allowing a small polarization along the O-N direction. The general three-body problem reduces therefore to a two-body OHe-N problem. Furthermore, the interaction potential OHe-N can be taken as radial, as the polarization of OHe will always be in the direction of N. Hence we can use spherical coordinates, with the O fixed at the origin of the center-of-mass frame and $\vec{r}$ the position of the center of the nucleus N. We know in this case that the solutions of the Schr\"odinger equation take the form $\psi_{k,l,m}(r,\theta,\varphi)=\frac{u_{k,l}(r)}{r}Y_{l}^{m}(\theta,\varphi)$ where $Y_{l}^{m}(\theta,\varphi)$ are the spherical harmonics and where the radial part $u_{k,l}(r)$ has to satisfy the radial Schr\"odinger equation:

\begin{equation}
\frac{d^{2}u_{k,l}(r)}{dr^{2}}+2m_{N}\left[E_{k,l}-V(r)-\frac{l\left(l+1\right)}
{2m_{N}r^{2}}\right]u_{k,l}(r)=0,\label{radialSchro}
\end{equation}
where $m_N$ is the mass of the nucleus N, $l$ is the relative angular momentum, $E_{k,l}$ is the
total energy in the center-of-mass frame, and $V(r)$ is the sum of the nuclear and of the electrostatic potentials between OHe and N. The index $k=1,2,3,...$ classifies the energies for some $l$ in ascending order and is related to the principal quantum number $n$ via $n=k+l$.

The next simplification comes from the fact that one is looking for weakly bound states, for which the WKB method applies, and considerably simplifies the solution. Finally, we further simplify the problem by approximating the He wavefunction in the OHe bound state by a $1s$ hydrogenoid wavefunction (which corresponds to neglecting the size of He). As in Section \ref{subsecClassical}, the nucleus N is seen as a uniformly charged sphere of charge $Z$ and of radius $R_{N}$(fm)$=1.35\times A^{1/3}$. Its mass $m_{N}$ is corrected by the nuclear binding energy $B$ given by the Bethe-Weizs\"acker formula: $m_{N}=Zm_{p}+(A-Z)m_{n}-B$, where $m_{p}$ and $m_{n}$ are the masses of the proton and of the neutron respectively.

The interactions between the OHe atom and the nucleus take two forms: a nuclear attraction between the helium and the N nucleus at distances $r\lesssim R_{N}$ and an electrostatic interaction due to the electrical charges of the components at distances $r\gtrsim R_{N}$.

Out of the nuclear region, the electrostatic interaction is dominant and can be separated into two contributions : 1) the electrostatic interaction between the spherical charge distribution of the OHe atom in its ground state and the spherical charge distribution of the nucleus; 2) the electrostatic interaction between the polarized OHe atom and the nucleus due to the Stark effect. Therefore, we can write:

\begin{equation}
V_{elec}=V_{coul}+V_{stark},\label{Velec}
\end{equation}
where $V_{Coul}$ corresponds to Coulomb attraction between O screened by the helium charge distribution and N, and $V_{stark}$ represents the interaction term of the charged nucleus and the dipole.

Outside the nucleus, i.e. for $r\geqslant R_{N}$, we find for the Coulomb term:
\begin{eqnarray}
V_{coul}(r) & = & \frac{3}{8}\left(\frac{-Z_{O}Z\alpha}{\rho^3 r}\right)e^{-2r/r_{o}}\left[e^{-2\rho}\left\{ \rho^{2}+\frac{5}{4}+\frac{5}{2}\rho +\left(\frac{1}{2}+\rho\right)\frac{r}{r_o}\right\} \right.\nonumber\\
 & +&\left.e^{2\rho}\left\{ -\rho^{2}-\frac{5}{4}+\frac{5}{2}\rho +\left(-\frac{1}{2}+\rho\right)\frac{r}{r_o}\right\} \right],\label{Vcoul}
\end{eqnarray}
with  the Bohr radius of the OHe atom $r_{o}$ given by \eqref{ro} and $\rho=R_N/r_o$. This expression can be considered as an improvement of the form from \cite{Khlopov:2010pq,Khlopov:2010ik} where the nucleus was assumed to be a point-like particle.

For the Stark potential, we use the formula for the quadratic effect in a constant electric field $\vec{E}$ \cite{Landau}, taken to be the field of the nucleus at the position of O. The average dipole moment of the OHe atom in its perturbed ground state can then be written:

\begin{equation}
<q\vec{R}>=\frac{9}{2}r_{o}^{3}\vec{E},
\end{equation}
so that, for $r\geqslant R_{N}$,

\begin{equation}
V_{stark}(r)=-q\vec{R}.\vec{E}=-\frac{9}{2}r_{o}^{3}E^{2}=-\frac{9}{2}r_{o}^{3}\frac{Z^{2}\alpha}{r^{4}}.\label{Vstark}
\end{equation}
Expressions \eqref{Vcoul} and \eqref{Vstark} are valid when the nuclear effects are negligible.

In the nuclear region, we take a trapezoidal nuclear well, which will simplify the WKB solution:

\begin{equation}
\begin{array}{cccc}
V_{nucl}(r) & = & -V_{0,}&{\rm for\ }r\leq R_{*} \\
 & = & \frac{V_{elec}(R_{*}+2a)+V_{0}}{2a}\left(r-R_{*}\right)-V_{0},&{\rm for\ }R_{*}\leq r\leq R_{*}+2a \\
 & = & 0,&{\rm for\ }r>R_{*}+2a
 \end{array}\label{Vnucl}
 \end{equation}
characterized by its depth $V_{0}$ and its diffuseness parameter $a$ representing the region of $r$ in which it goes linearly from $-V_{0}$ to $V_{elec}(R_{*}+2a)$. From diffusion experiments of $\alpha$ particles on nuclei \cite{PhysRev.106.126}, one gets $V_{0}\simeq 30$ MeV for nuclei with $Z\leq  25$ and $V_{0}=45$ MeV for $Z>25$, as well as $a=0.5$ fm. In the following, we shall not need the transition region $Z\in [21,29]$, which does not contain any nucleus used for direct dark matter detection. For the nuclear radius parameter $R_{*}$, we added $r_o$ to the previous parametrization (Section \ref{subsecClassical}) in order to take the size of the OHe atom into account, which now gives $R_{*}(\mathrm{fm})=R_{N}(\mathrm{fm})+1.3+r_{o}(\mathrm{fm})$.

So, finally, the total OHe-nucleus potential in this semi-classical model is given by:

\begin{figure}
\begin{center}
\includegraphics[scale=0.24]{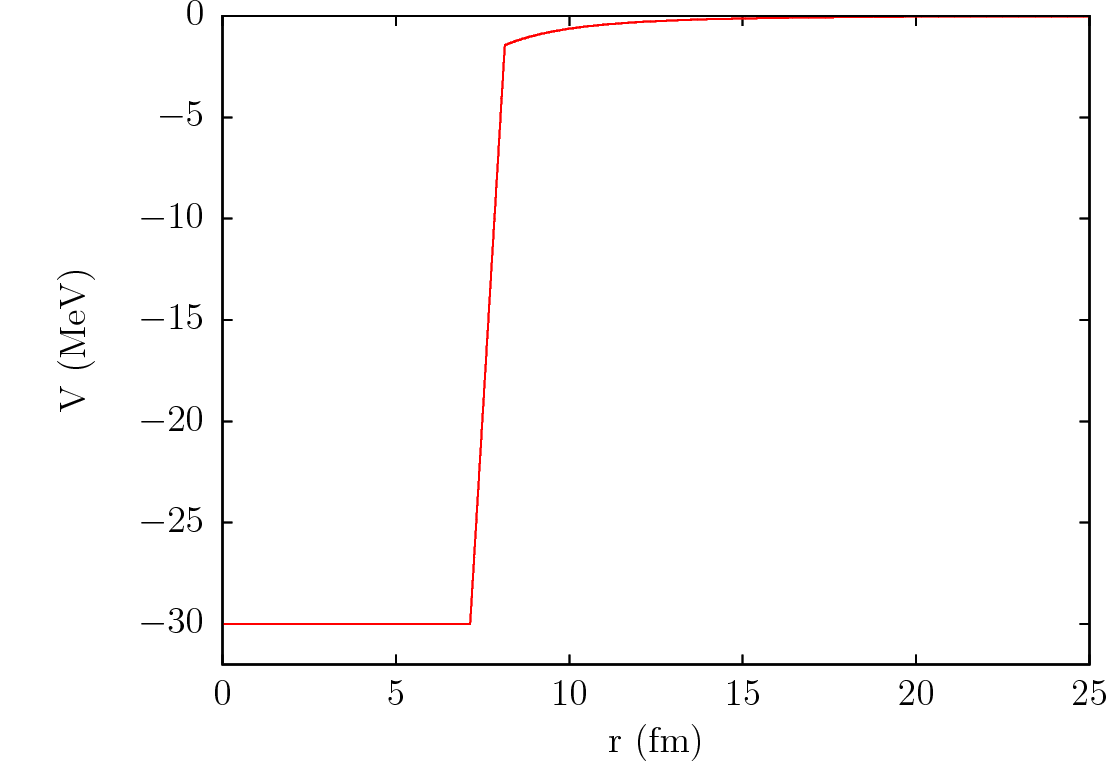}
\end{center}
\caption{Shape of the total interaction potential OHe-sodium, from \eqref{Vtot}.\label{FigVsemiclassical}}
\end{figure}

\begin{align}
V(r)&=V_{nucl}(r),~{\rm for\ }r\leq R_{*}+2a\nonumber \\
&=V_{elec}(r),~{\rm for\ }r> R_{*}+2a\label{Vtot}
\end{align}
and is represented in Figure \ref{FigVsemiclassical} for a sodium nucleus. Between the nuclear interaction at short distance and the dominant Stark effect at large distance, it is not clear from this semi-classical approach using reasonable arguments where a repulsive barrier could appear. Figure \ref{FigVsemiclassical} shows therefore a fully attractive effective potential, the implications of which we are going to study.

To find the spectrum corresponding to the potential of Figure \ref{FigVsemiclassical}, we use the approximate WKB solutions which, once applied to each region, give a quantization condition for the energy. For $l=0$, we obtain:
\begin{equation}
\lambda=\left(k-\frac{1}{4}\right)\pi,\,\,\,\, k=1,2,3,...\label{WKB0}
\end{equation}
where $\lambda=\int_{0}^{b}\sqrt{2m_{N}\left(E-V(r)\right)}\mathrm{d}r$ and $b$ is the turning point such that $E=V(b)$. Equation \eqref{WKB0} has to be solved numerically by browsing the trial energies $E$ until we get an equality between the right- and the left-hand sides. The energy that is solution for $k$ is therefore the eigenvalue $E_{k,0}$ in \eqref{radialSchro}. At $l\neq0,$ we know from \cite{froman} that the behavior of the effective potential $V_{eff}(r)=V(r)+\frac{l\left(l+1\right)}{2m_{N}r^{2}}$ at the origin requires, as in Section \ref{secSpectrum}, to modify the WKB method by applying its solutions to $u(r)$ after having changed $l\left(l+1\right)$ to $\left(l+\frac{1}{2}\right)^{2}$ in the radial equation : $V_{eff}(r)\rightarrow\tilde{V}_{eff}(r)=V(r)+\frac{\left(l+\frac{1}{2}\right)^{2}}{2m_{N}r^{2}}$.
Therefore, the quantization condition becomes:

\begin{equation}
2e^{2\tilde{\tau}}\cos\tilde{\lambda}+\sin\tilde{\lambda}=0, \label{WKBl}
\end{equation}
where $\tilde{\lambda}=\int_{a}^{b}\sqrt{2m_{N}(E-\tilde{V}_{eff}(r))}\mathrm{d}r$, $\tilde{\tau}=\int_{0}^{a}\sqrt{2m_{N}(\tilde{V}_{eff}(r)-E)}\mathrm{d}r$, and $a$ and $b$ are the turning points such that $E=\tilde{V}_{eff}(a)=\tilde{V}_{eff}(b)$. The numerical resolution of \eqref{WKBl} gives the eigenvalues $E_{k,l\ne 0}$. Note that, contrary to the OHe system, there is a finite number of energy levels for each $l$ and there is a value of $l$ from which there is no solution anymore. However, it is expected to be high as the nuclear well is deep, so we did not attempt to determine it and considered a maximum value of $l=5$ in the following.

\subsubsection{Spectra from a screened Coulomb potential}

\begin{figure}
\begin{center}
\includegraphics[scale=0.25]{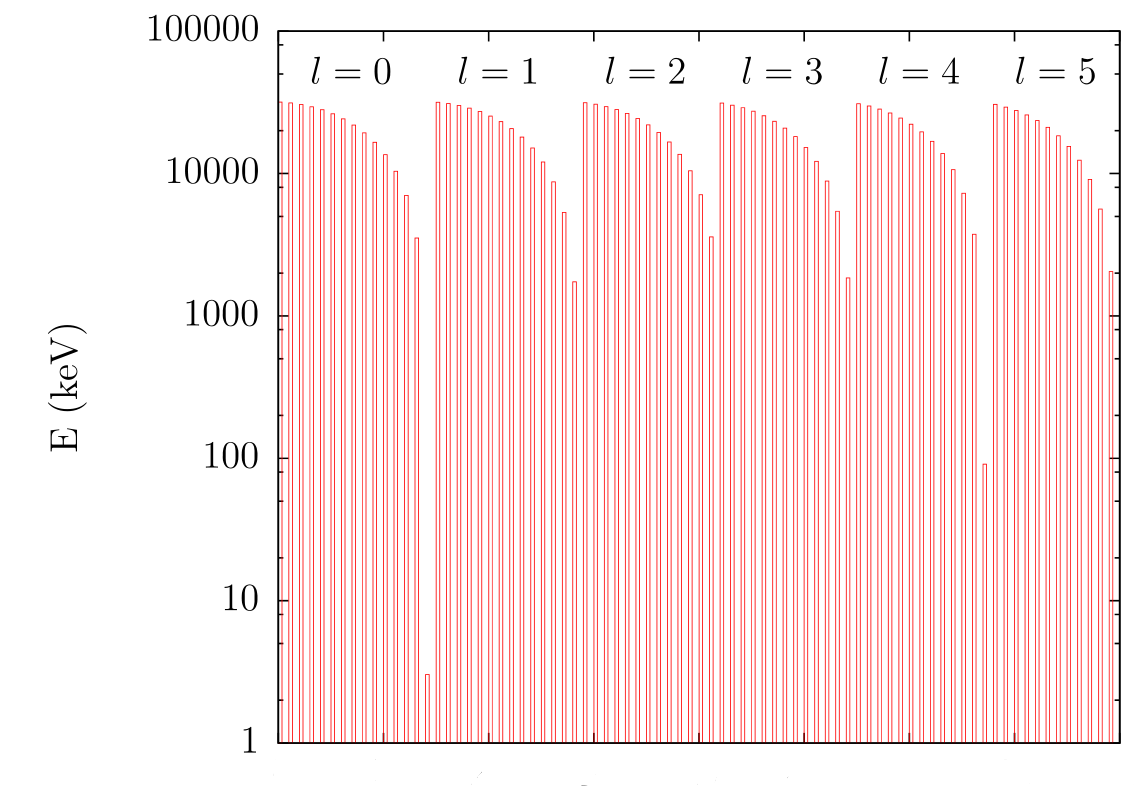}
\end{center}
\caption{Spectrum of the OHe-$^{23}$Na system at different values of $l$, when $V_{elec}=V_{coul}$. The energies are in absolute value. $V_{0}=31.9$ MeV and $a=0.5$ fm.\label{FigEnNaCoul}}
\end{figure}

In \cite{Khlopov:2010pq,Khlopov:2010ik}, the spectrum was considered for a screened Coulomb potential at long distance, as in \eqref{Vcoul}, except that the nucleus was then considered to be point-like. We reanalyze this question here with our WKB formalism. For small nuclei $Z\leq 20$, we first fix the exact value of $V_{0}$ to obtain the highest level at $-3$ keV for $^{23}$Na from DAMA for $l=0$. We obtain $V_{0}=31.9$ MeV, in good agreement with the 30 MeV from \cite{PhysRev.106.126}, and use this value for all nuclei with $Z\leq 20$. The spectrum of the OHe-$^{23}$Na system is shown in Figure \ref{FigEnNaCoul}.

\begin{figure}
\begin{center}
\includegraphics[scale=0.7]{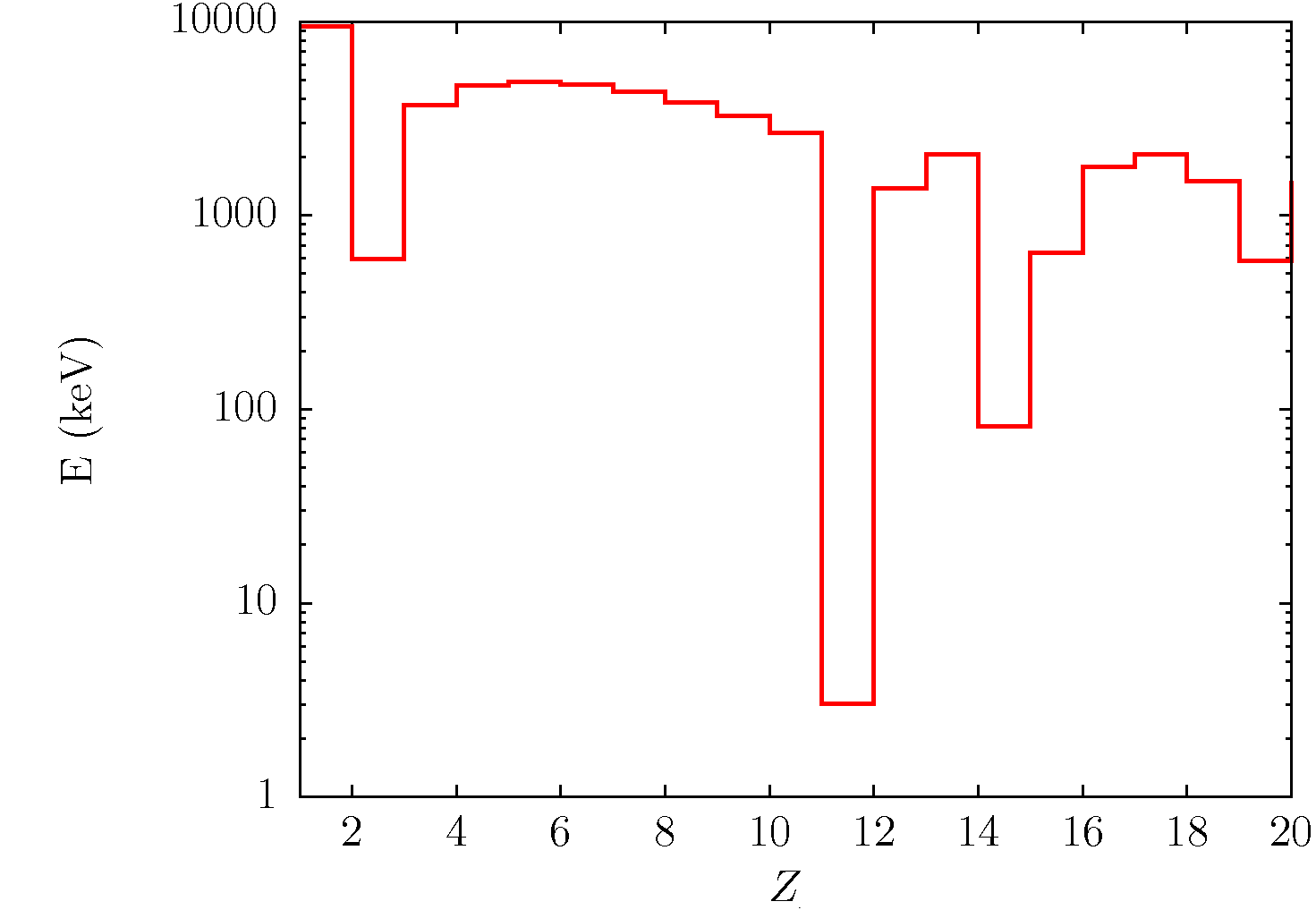}
\end{center}
\caption{Highest-energy level at $l=0$ for stable nuclei from $Z=1$ to $Z=20$, when $V_{elec}=V_{coul}$. The energies are in absolute value. $V_{0}=31.9$ MeV and $a=0.5$ fm.}
\label{FigHighestCoul}
\end{figure}

We see a rich spectrum with many levels in the MeV region, corresponding to nuclear levels. The only level in the keV region is at $l=0$. It can be considered as being due to the presence of the electrostatic potential in the sense that it disappears if one only keeps the nuclear well and sets $V_{elec}$ to zero. It is remarkable that other nuclei such that $Z_{A}\leq 20$ do not have keV bound states. Figure \ref{FigHighestCoul} shows the highest level at $l=0$ for the most stable nuclei for $Z$ going from $1$ to $20$. It turns out that only $^{23}$Na at $Z=11$ has a level in the keV region.

\begin{figure}
\begin{center}
\includegraphics[scale=0.25]{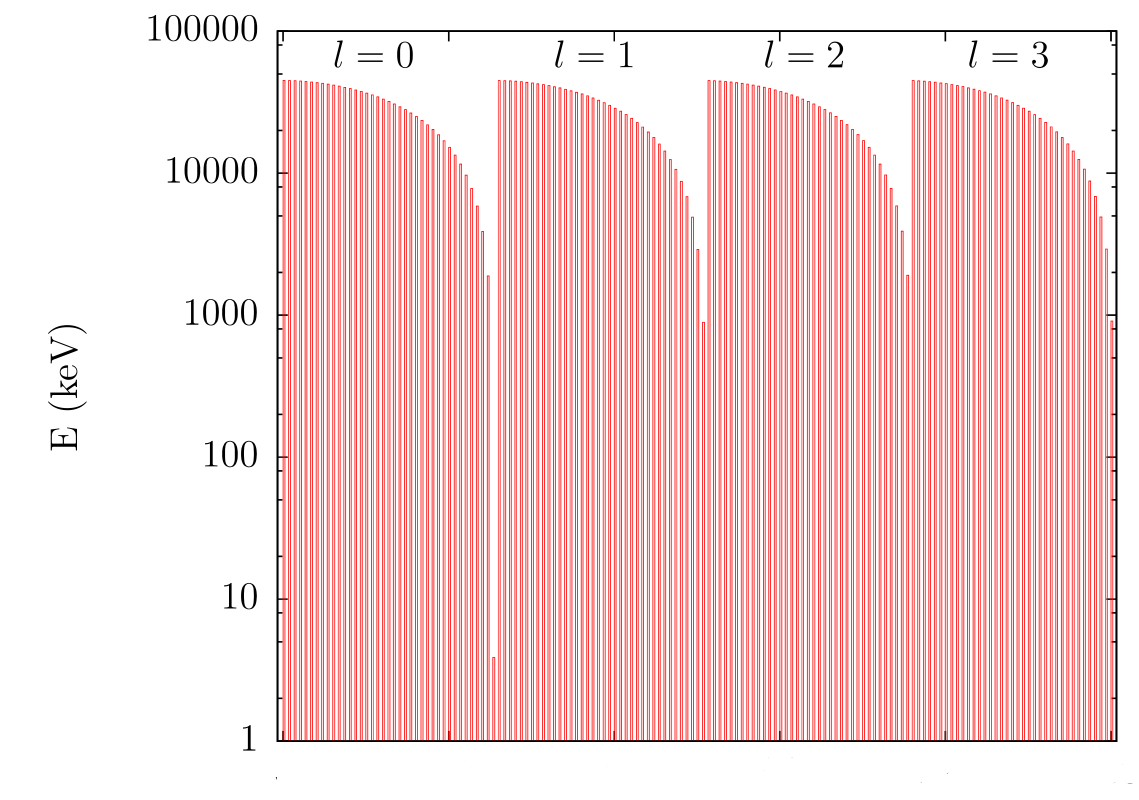}
\end{center}
\caption{Spectrum of the OHe-$^{74}$Ge system at different values of $l$, when $V_{elec}=V_{coul}$. The energies are in absolute value. $V_{0}=45$ MeV and $a=0.5$ fm}
\label{FigEnGeCoul}
\end{figure}

For large nuclei $Z\geq 30$, the data indicate that the nuclear well is deeper. In this case, we take as a reference germanium with $Z=32,\, A=74$ from CoGeNT, for which we find a highest level at $l=0$ in the keV region for $V_{0}=45$ MeV, which is precisely the central value from \cite{PhysRev.106.126} for larger nuclei. This value is used for all nuclei with $Z\geq 30$. Figure \ref{FigEnGeCoul} represents the spectrum of the OHe-$^{74}$Ge system. It is of the same kind as for $^{23}$Na, with only one level in the keV region.

The second column of Table \ref{TabHighestLargeZ} gives the highest-energy level at $l=0$ for the large stable nuclei involved in the experiments of interest. According to this model, iodine and thallium from DAMA each admit one level in the keV region, while xenon from XENON100 and LUX and tungsten from CRESST-II do not.

Therefore, only looking at the energies of the signals, DAMA and CoGeNT could be accounted for by the presence of energy levels in the keV range with their constituent nuclei (sodium, iodine and thallium for DAMA and germanium for CoGeNT) while the negative results of XENON100, LUX and CRESST-II would be due to the absence of weakly bound states with xenon for the first two and with oxygen, calcium and tungsten for the latter. Note that CDMS-II/Ge, which is made of germanium just like CoGeNT but does not see any signal, could be accounted for afterwards firstly by the fact that it is cryogenic, and secondly because it looks for nuclear recoils while here we produce electron recoils that contribute to the background. 

But all this would be possible only if we could find a way to prevent the system from de-exciting (e.g. via electric dipole transitions) to the many deep nuclear levels such as the ones of Figures \ref{FigEnNaCoul} and \ref{FigEnGeCoul}. In the present situation, it seems that the system will either emit gamma rays of several tens of MeV into the detector or bring equivalent amounts of energy to the nuclei, causing for example the ejection of neutrons, which is clearly not observed. We will see below that the situation is not improved when we use the complete expression \eqref{Velec} for $V_{elec}$, i.e. when we include $V_{stark}$.

\begin{table}
\begin{center}
\begin{tabular}{c|c|c}
\hline
\hline
Nuclei  & E(keV) for a screened  & E(keV) for a screened Coulomb\\
  & Coulomb potential  &  potential added to a Stark potential\\
\hline
\hline
$^{74}$Ge      &$3.88$ &$1.16$ \\
$^{127}$I      &$0.500$ &$2.31$ \\
$^{132}$Xe     &$540.$ &$2.33$ \\
$^{184}$W      &$350.$ &$1.86$ \\
$^{201}$Tl     &$15.6$&$52.7$ \\
\hline
\hline
\end{tabular}
\caption{Highest-energy level at $l=0$ for some heavy stable nuclei from the experiments of interest, when $V_{elec}=V_{coul}$ (second column), or $V_{elec}=V_{coul}+V_{stark}$ (third column). The energies are in absolute value. $V_{0}=45$ MeV and $a=0.5$ fm.}\label{TabHighestLargeZ}
\end{center}
\end{table}

\subsubsection{Spectra from a screened Coulomb potential and a Stark potential}

\begin{figure}
\begin{center}
\includegraphics[scale=0.25]{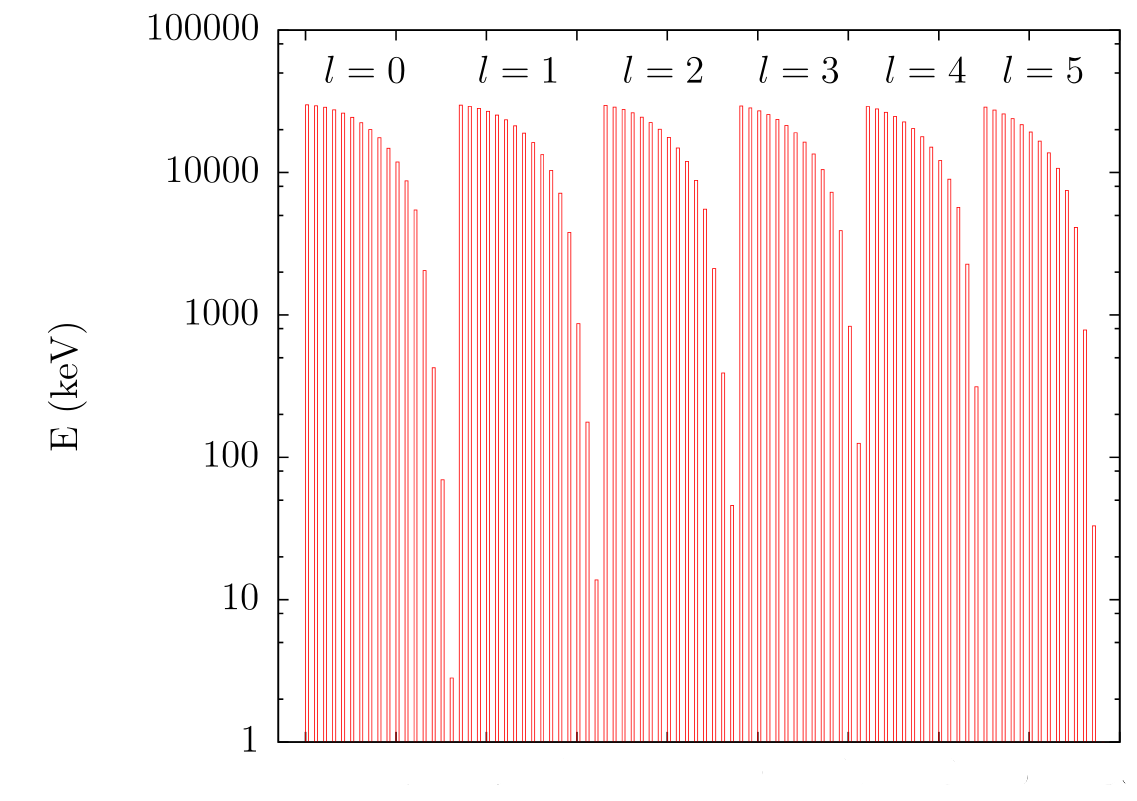}
\end{center}
\caption{Spectrum of the OHe-$^{23}$Na system at different values of $l$, when $V_{elec}=V_{stark}+V_{coul}$. The energies are in absolute value. $V_{0}=30$ MeV and $a=0.5$ fm.\label{FigEnNaCoulStark}}
\end{figure}

The results can de discussed in the same way when $V_{elec}=V_{stark}+V_{coul}$ is used in the calculations, and the values of $V_{0}$ are identical to the central experimental values, i.e. $30$ and $45$ MeV, for small and large nuclei respectively. Figure \ref{FigEnNaCoulStark} illustrates the results in the particular case of the OHe-$^{23}$Na system.

\begin{figure}
\begin{center}
\includegraphics[scale=0.7]{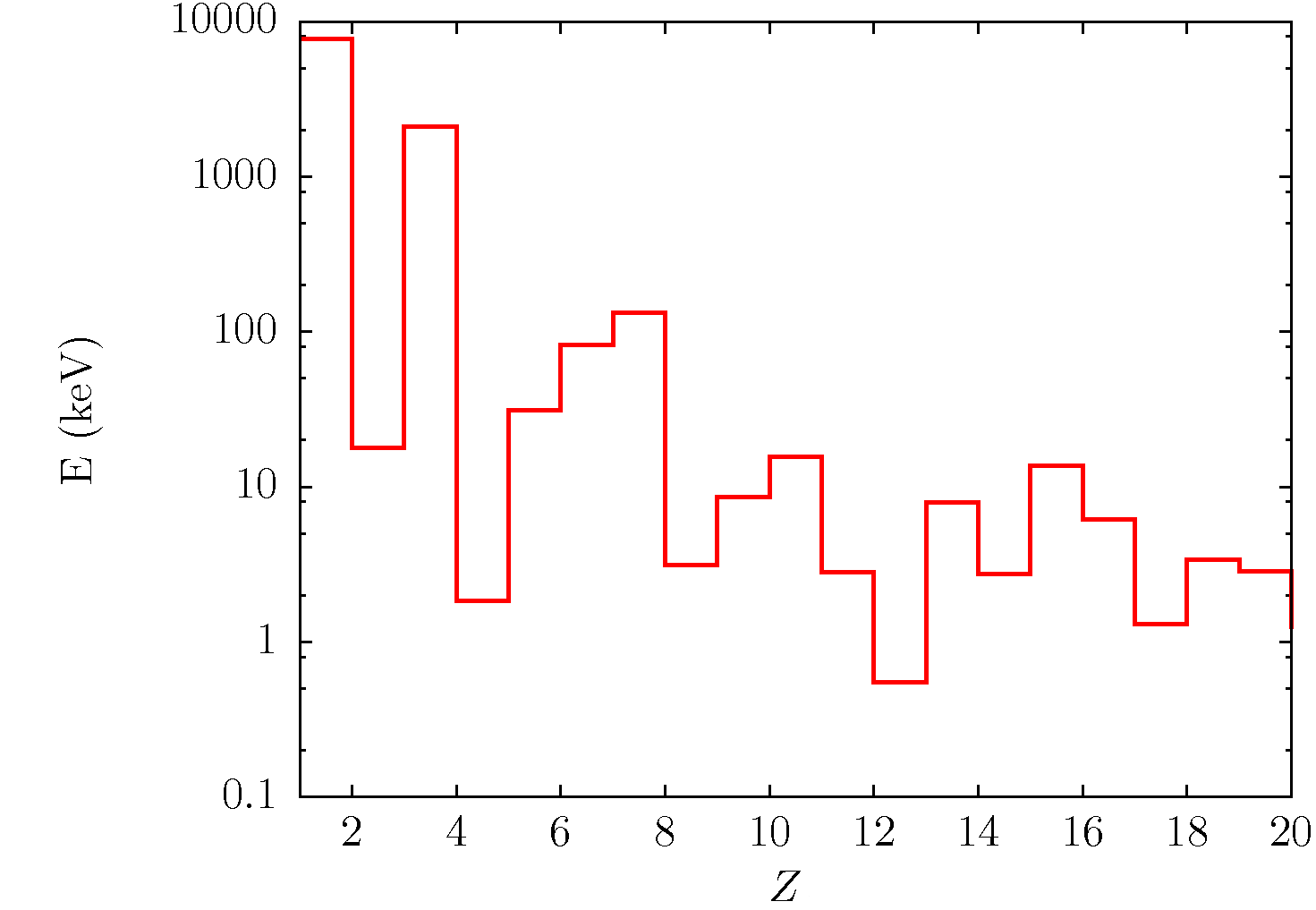}
\end{center}
\caption{Highest level at $l=0$ for stable nuclei from $Z=1$ to $Z=20$, when $V_{elec}=V_{stark}+V_{coul}$. The energies are in absolute value. $V_{0}=30$ MeV and $a=0.5$ fm.}
\label{FigHighestCoulStark}
\end{figure}

The major difference lies in the fact that, in this case, the levels in the keV region are obtained more easily, with sometimes several keV levels for the same nucleus, especially for large nuclei. The reason lies in the shape of $V_{elec}$, that is deeper and less steep when $V_{stark}$ is used. Figure \ref{FigHighestCoulStark}, as well as the third column of Table \ref{TabHighestLargeZ}, show that most nuclei now have keV bound states. Hence the inclusion of the Stark potential seems to destroy the previous interpretation of the data, which relied on Na and Ge to be very special nuclei. It seems therefore crucial to find a repulsive mechanism in the OHe-nucleus interaction, which we shall undertake with a fully quantum calculation in Section \ref{subsecPert}.

\subsection{Perturbation analysis}\label{subsecPert}

As already mentioned at the beginning of Section \ref{subsecSemiclass}, the three-body O-He-nucleus bound-state problem can be simplified by noting that helium is much lighter than most of the nuclei N: $m_{He}\ll m_N$. Given this, one can simplify the total Hamiltonian of the system, written in the reference frame of the O particle, and choosing the $z$-axis in the direction of N to:

\begin{equation}
H\simeq -\frac{1}
{2m_{He}}\triangle_{1}+V_{OHe}\left(r_{1}\right)+V_{ON}\left(R\right)+V_{HeN}\left(r_{12}\right),\label{Ham1}
\end{equation}
in which the kinetic energy term of the nucleus N has been neglected and where $\vec{r}_{1}$ is the position of the He nucleus, $R$ is the distance of N on the positive part of the $z$-axis and $r_{12}$ is the distance between He and N. $V_{IJ}$ stands for the interaction potential between I and J. We are thus left with the one-body problem of He in a total potential depending on the parameter $R$. We shall consider here the contribution of the external nucleus as a perturbation to the OHe atom. The Hamiltonian \eqref{Ham1} can be re-written as the sum of an unperturbed part $H_{0}$ and a perturbation $W$:

\begin{equation}
H=H_{0}+W\label{Ham2},
\end{equation}
where $H_{0}=-\frac{1}{2m_{He}}\triangle_{1}+V_{OHe}\left(r_{1}\right)$ corresponds to the isolated OHe atom and where $W=V_{ON}\left(R\right)+V_{HeN}\left(r_{12}\right)$ is due to the presence of the external nucleus. We use here the following two-body interaction potentials :

\begin{eqnarray}
V_{OHe}(r_{1}) & = & -\frac{Z_{O}Z_{He}\alpha}{r_{1}}\\
&&\nonumber\\
V_{ON}(R) & = & -\frac{Z_{O}Z\alpha}{R},~\mathrm{for}~R>R_{N}\nonumber\\
 & = & -\frac{Z_{O}Z\alpha}{2R_{N}}\left(3-\frac{R^{2}}{R_{N}^{2}}\right),~\mathrm{for}~R<R_{N}\label{VON}\\
 &&\nonumber\\
 V_{HeN}(r_{12}) & = & \frac{Z_{He}Z\alpha}{r_{12}}+\frac{-V_{0}}{1+e^{\left(r_{12}-R_{N}\right)/a}},~\mathrm{for}~r_{12}>R_{N}\nonumber\\
  & = & \frac{Z_{He}Z\alpha}{2R_{N}}\left(3-\frac{r_{12}^{2}}{R_{N}^{2}}\right)+\frac{-V_{0}}{1+e^{\left(r_{12}-R_{N}\right)/a}},~\mathrm{for}~ 
 r_{12}<R_{N}
\end{eqnarray}
where, in order to simplify the calculation, the He nucleus is assumed to be point-like, so that the OHe atom can be treated as a hydrogenoid system, and where the nucleus N is seen as a sphere of radius $R_{N}$(fm)$=1.35\times A^{1/3}$ and charge $Z$ in the definitions of the potentials $V_{ON}$ and $V_{HeN}$, which are therefore point (O or He) -sphere (N) interaction potentials. A Woods-Saxon potential of parameters $V_{0}$ and $a$ has been added in $V_{HeN}$ to take the nuclear interaction of both nuclei into account.

We are studying the perturbed ground state $E_{0}(R)$ of the OHe atom under the influence of the external perturbation $W(R)$. The perturbed energy is therefore an approximation of the energy of the total O-He-N system, described by the Hamiltonian \eqref{Ham2}. If this energy presents
a minimum for some $R_{min}$, then the system will tend to this configuration to minimize its energy, and we will get a stable OHe-nucleus bound
state of length $R_{min}$ and energy $E_{0}(R_{min})$. If there is no minimum, then we will have to conclude that no stable bound state can form at those distances.

In the following, we shall go to 3rd-order perturbation theory. We assume that $H_0$ has a spectrum $|\psi_{p,i}^{0}>$ of eigenfunctions with eigenvalues $E_p^0$, and that the unperturbed ground level $E_0^0$ that we are studying is non-degenerate, $|\psi_0^0>$ being its eigenfunction. The $|\psi_{p,i}^{0}>$ are supposed to form an orthonormal basis of the state space of the system: $<\psi_{p,i}^0|\psi_{p',i'}^0>=\delta_{p,p'}\delta_{i,i'}$ and $\sum_{p,i}|\psi_{p,i}^0><\psi_{p,i}^0|=1$. Following mainly the method presented in \cite{CohenTann}, we find that the perturbed wavefunction $|\psi_0>$ at order 2 and the perturbed energy $E_0$ at order 3 are given by:

\begin{eqnarray}
E_{0} & = & E_{0}^{0}+<\psi_{0}^{0}|W|\psi_{0}^{0}>+\sum_{i,{p\neq 0}}\frac{|<\psi_{p,i}^{0}|W|\psi_{0}^{0}>|^{2}}{E_{0}^{0}-E_{p}^{0}}\nonumber\\
&-& <\psi_{0}^{0}|W|\psi_{0}^{0}>\sum_{i,{p\neq 0}}\frac{|<\psi_{p,i}^{0}|W|\psi_{0}^{0}>|^{2}}{(E_{0}^{0}-E_{p}^{0})^{2}}\nonumber\\
&+& \sum_{i,{p\neq 0}}\sum_{i',{p'\neq 0}}\frac{<\psi_{p',i'}^{0}|W|\psi_{0}^{0}><\psi_{p,i}^{0}|W|\psi_{p',i'}^{0}><\psi_{0}^{0}|W| \psi_{p,i}^{0}>}{\left(E_{0}^{0}-E_{p}^{0}\right)\left(E_{0}^{0}-E_{p^{,}}^{0}\right)},\label{En}\\
&&\nonumber\\
&&\nonumber\\
|\psi_{0}> & = & |\psi_{0}^{0}>+\sum_{i,p\neq 0}\frac{<\psi_{p,i}^{0}|W|\psi_{0}^{0}>}{E_{0}^{0}-E_{p}^{0}}|\psi_{p,i}^{0}> \nonumber\\
&-&<\psi_{0}^{0}|W|\psi_{0}^{0}>\sum_{i,{p\neq 0}}\frac{<\psi_{p,i}^{0}|W|\psi_{0}^{0}>}{(E_{0}^{0}-E_{p}^{0})^{2}}|\psi_{p,i}^{0}>\nonumber\\
&-&\frac{1}{2}\sum_{i,{p\neq 0}}\frac{|<\psi_{p,i}^{0}|W|\psi_{0}^{0}>|^{2}}{(E_{0}^{0}-E_{p}^{0})^{2}}|\psi_{0}^{0}>\nonumber\\
&+& \sum_{i,{p\neq 0}}\sum_{i',{p'\neq 0}}\frac{<\psi_{p',i'}^{0}|W|\psi_{0}^{0}><\psi_{p,i}^{0}|W|\psi_{p',i'}^{0}>}{\left(E_{0}^{0}-E_{p}^{0}\right)\left(E_{0}^{0}-E_{p'}^{0}\right)}|\psi_{p,i}^0>.\label{psin}
\end{eqnarray}

In our case, the non-degenerate unperturbed energy $E_{0}^{0}$ is the ground level of the hydrogen-like OHe atom and was given in \eqref{EbindingOHe}:

\begin{equation}
E_{0}^{0}\equiv E_{o}=-\frac{1}{2}m_{He}\left(Z_{O}Z_{He}\alpha\right)^{2}\simeq -1.6~\mathrm{MeV},
\end{equation}
and the unperturbed eigenvalues $E_p^0$ and eigenfunctions $|\psi^{0}_{p,i}>$ are those of the hydrogen atom: 
\begin{equation}
E_p^0\equiv E^H_n = \frac{E_o}{n^2},
\label{EH}
\end{equation}
\begin{equation}
\psi^{0}_{p,i}(\vec{r}_1)\equiv \psi^H_{n,l,m}(\vec{r}_1)=R^H_{n,l}(r_{1})Y_{l}^{m}(\theta_{1},\varphi_{1}),
\label{psiH}
\end{equation}
where the superscript $H$ refers to the hydrogen-like solutions as in Section \ref{secSpectrum} and where the generic indices $p,i$ have been replaced by $n,l,m$, which are respectively the principal, azimuthal and magnetic quantum numbers. The $Y_l^m$ are the normalized spherical harmonics and the radial part $R^H_{n,l}$ is given by:

\begin{equation}
R^H_{n,l}(r_{1})=C_{n,l}\times r_{1}^{l}\sum_{q=0}^{n-l-1}c_{q}\left(\frac{r_{1}}{r_{o}}\right)^{q}e^{-\frac{r_{1}}{n r_{o}}},\label{RH}
\end{equation}
$C_{n,l}$ being the normalization coefficient of $R^H_{n,l}$ and $r_{o}$ being the Bohr radius of the OHe atom, given in \eqref{ro}. The coefficients $c_{q}$ in \eqref{RH} are recursively given by $\frac{c_{q}}{c_{q-1}}=-\frac{2\left(1-\frac{q+l}{n}\right)}{q\left(q+2l+1\right)}$. Note that the orthonormality requirement for the basis of unperturbed eigenstates is well satisfied since it is known that the wavefunctions of the hydrogen atom form, after a proper normalization, an orthonormal basis of the Hilbert space. The spherical harmonics and the radial parts are generated numerically so that the matrix elements of \eqref{En} and \eqref{psin} can be computed and summed until convergence.

\subsubsection{Correction to the OHe binding energy}

\begin{figure}
\begin{center}
\includegraphics[scale=0.25]{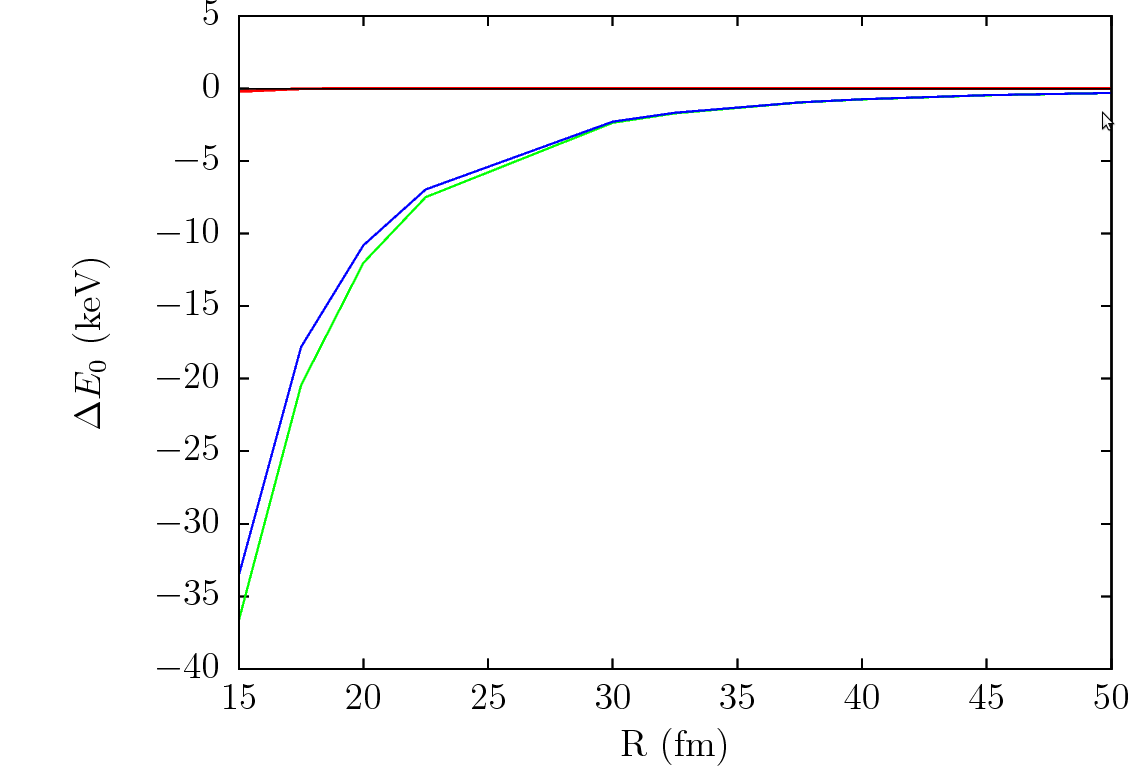}
\end{center}
\caption{$\triangle E_{0}=E_{0}(R)-E_{o}$ up to orders $1$ (red), $2$ (green) and $3$ (blue) for an external sodium nucleus, as a function of its distance $R$. $V_{0}=30$ MeV and $a=0.5$ fm.\label{FigPertEnNa23}}
\end{figure}

First, we consider the effect of an approaching sodium nucleus on the OHe binding energy. Figure \ref{FigPertEnNa23} shows the results for $\triangle E_{0}(R)=E_{0}(R)-E_{o}$ for $V_{0}=30$~MeV and $a=0.5$ fm. As we are looking for a minimum in the curve of $\triangle E_{0}(R)$ with a depth of several keV, we show in Figure \ref{FigPertEnNa23} the first tens of keV, corresponding to $R\geq 15$ fm. We check the validity of the perturbative calculation in this region by calculating the ratio of the difference between the second and the third orders to the binding energy of the unperturbed OHe atom. It remains smaller than 0.025 for $R\geq 15$ fm, which is assumed to be sufficiently small compared with unity for the results to be valid. We see that the first order does not bring a large modification ($\mathcal{O}(10^{-3} ~\mathrm{keV})$ for $R$ between $50$ and $15$ fm). This is due to the sphericity of the unperturbed ground state of the OHe atom:  in case of a uniform electric field $\vec{E}_0$ along the $z$-axis, equation \eqref{En} indicates that the first-order correction to the energy is proportional to the average dipole moment of the unperturbed OHe atom in the $z$-direction ($W=-Z_{He}e\vec{r}_1\cdot \vec{E}_0=-Z_{He}eE_0 z_1$), which is null. Here, at distances $R\geq 15~\mathrm{fm}$ where the nuclear effects are still subdominant, the electric field generated by the nucleus located on the $z$-axis is not exactly but almost uniform, and that is why the first-order correction is not zero but is very small. The second order gives the largest correction ($\mathcal{O}(10~\mathrm{keV})$ for $R$ between $50$ and $15$~fm). This change from first to second order justifies the inclusion of the third order in the calculations, but it turns out that this one does not modify significantly the results from the second order, and that is why the fourth order has not been added. We see in Figure \ref{FigPertEnNa23} that $\triangle E_{0}$ is always decreasing, in other words that there is no minimum in this curve in the region of interest.

\begin{figure}
\begin{center}
\includegraphics[scale=0.25]{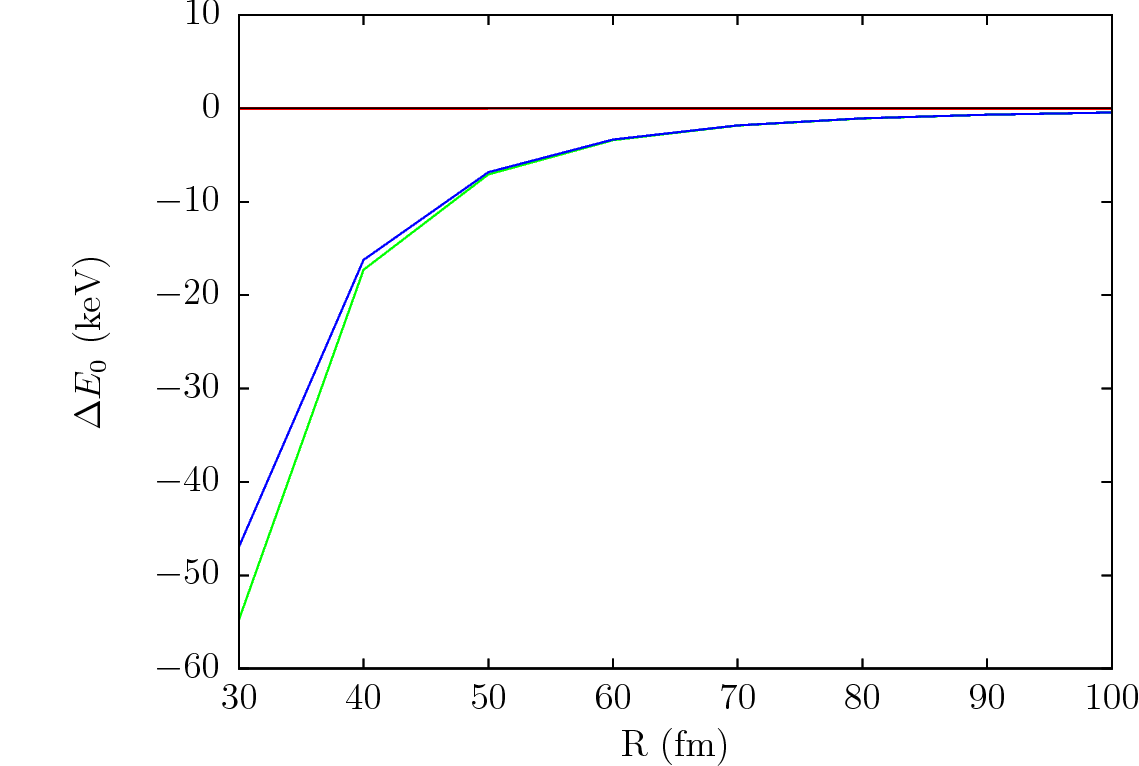}
\end{center}
\caption{$\triangle E_{0}=E_{0}(R)-E_{o}$ up to orders $1$ (red), $2$ (green) and $3$ (blue) for an external iodine nucleus, as a function of its distance $R$. $V_{0}=45$ MeV and $a=0.5$ fm.\label{FigPertEnI127}}
\end{figure}

Similar results hold if one strengthens the nuclear potential, or if one considers different nuclei, as shown in Figure \ref{FigPertEnI127} in the case of iodine.

\subsubsection{Interaction with the incoming nucleus and polarization}

We can now calculate the electrostatic and nuclear interaction energies between the perturbed charge distribution of the helium nucleus and the charge distribution of the nucleus N as a function of its distance $R$, as well as the mean position of He along the $z$-axis, that is the polarization of the OHe under the influence of the external nucleus.

The total electrostatic interaction energy between the external nucleus and the perturbed OHe system is given by:
\begin{equation}
E_{el}=\int\int\frac{\rho_{1}\left(\vec{r}_1\right)\rho_{2}\left(\vec{r}_2\right)}
{4\pi\left|\vec{r}_1-\vec{r}_2\right|}\mathrm{d}\vec{r}_1\mathrm{d}\vec{r}_2+V_{ON},\label{Eel}
\end{equation}
where $\vec{r}_1$ and $\vec{r}_2$ give the positions into the charge distributions $\rho_{1}$ and $\rho_{2}$ of the helium nucleus and of the nucleus N respectively and where $V_{ON}$ defined in \eqref{VON} gives the contribution due to the O-N interaction. In the following, we take the first-order version of $|\psi_0>$ for He (as its matrix elements are responsible for the dominant second-order shift in energy), denoted by $|\psi_{He}>$, so that $\rho_{1}\left(\vec{r}_1\right)=|\psi_{He}\left(\vec{r}_1\right)|^2Z_{He}e$, while the external nucleus is treated as a uniform sphere, i.e. $\rho_2=Z_Ne/(4/3\pi R_N^3)$ inside the sphere and $\rho_2=0$ outside.

\begin{figure}
\begin{center}
\includegraphics[scale=0.8]{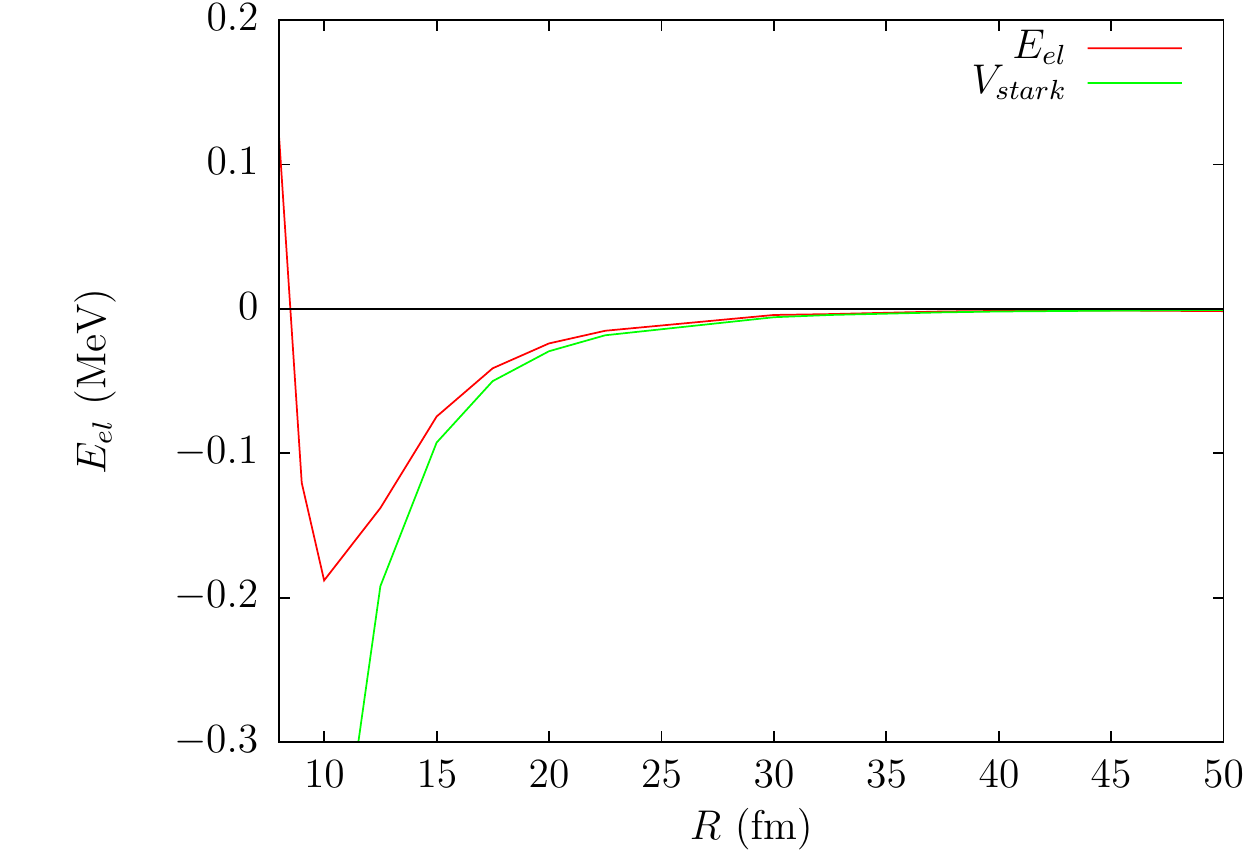}
\end{center}
\caption{Electrostatic interaction energy $E_{el}$ at order $1$ (red) compared to the Stark interaction energy $V_{stark}$ (green) as a function of the distance $R$ of an external sodium nucleus. $V_{0}=30$ MeV and $a=0.5$ fm.\label{FigPertElecNa23}}
\end{figure}

Here, we extend the analysis to smaller values of $R$, since interesting repulsion processes could appear when the external nucleus gets closer. For this purpose, we identify the region where the ratio of the difference between the second and the third order terms of $E_0(R)$ to the binding energy of the isolated OHe atom is smaller than 0.1. This corresponds to $R\geq 8$ fm. The results are compared to the Stark potential \eqref{Vstark} used in Section \ref{subsecSemiclass} in Figure \ref{FigPertElecNa23}. We see that the simplifying assumption of the uniform electric field for the nucleus is reasonable at large distance, while the gap becomes more pronounced around $15$ fm, because the uniform Coulomb field is always stronger than the true one. The fact that $E_{el}$ becomes repulsive at shorter distance is due to the change of the polarization of the OHe atom under the influence of the nuclear force of the sodium nucleus, which makes the helium component turn to positive mean $z_{1}$, that is, towards the external nucleus.

\begin{figure}
\begin{center}
\includegraphics[scale=0.8]{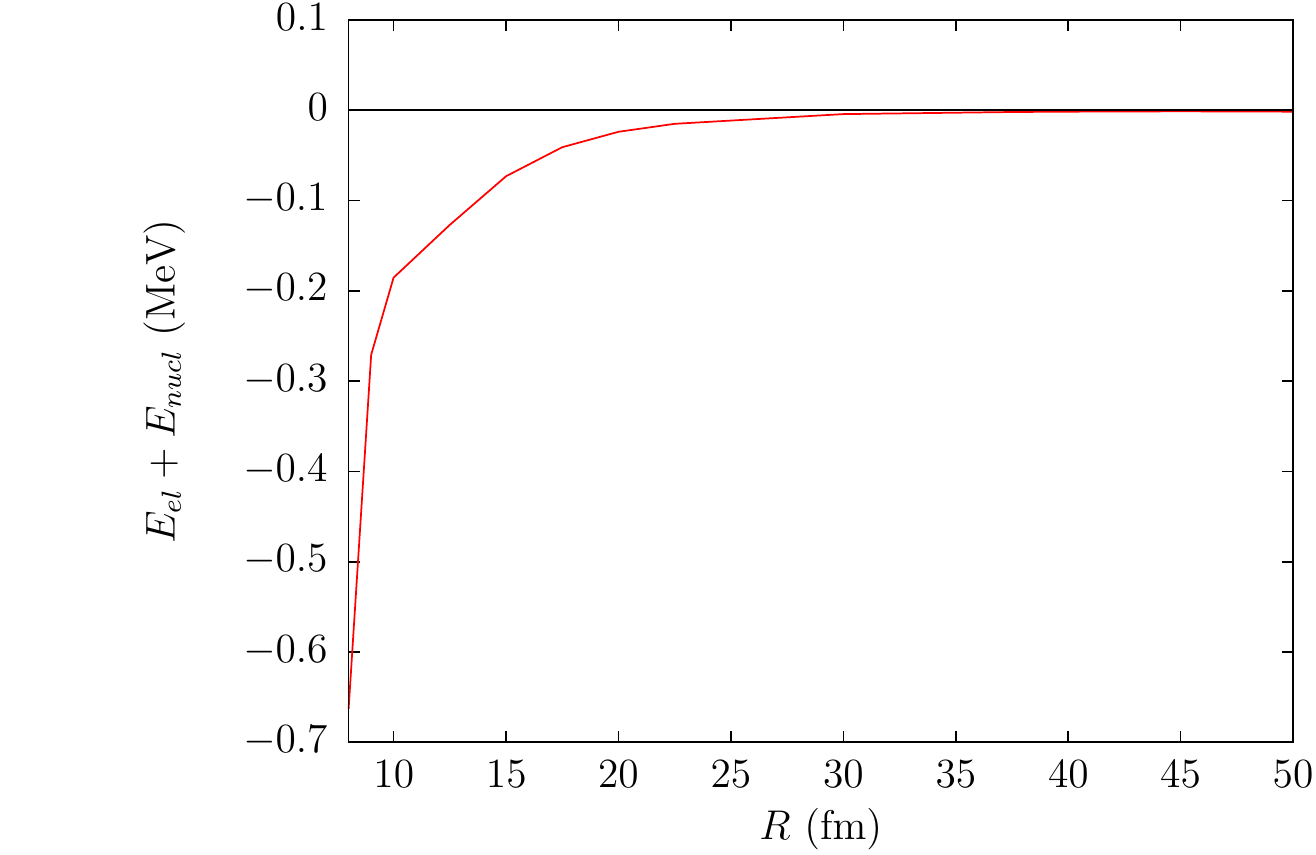}
\end{center}
\caption{Total interaction energy $E_{el}+E_{nucl}$ at order $1$ as a function of the distance $R$ of an external sodium nucleus. $V_{0}=30$ MeV and $a=0.5$ fm.\label{FigEelEnuclNa23}}
\end{figure}

We can also integrate the Woods-Saxon potential $\frac{-V_{0}}{1+e^{\left(r_{12}-R_{N}\right)/a}}$ over the distribution of the helium nucleus to get the total nuclear interaction energy $E_{nucl}$. Adding it to the previous contribution gives us the curve of Figure \ref{FigEelEnuclNa23}, which has no sign of a potential barrier.

\begin{figure}
\begin{center}
\includegraphics[scale=0.8]{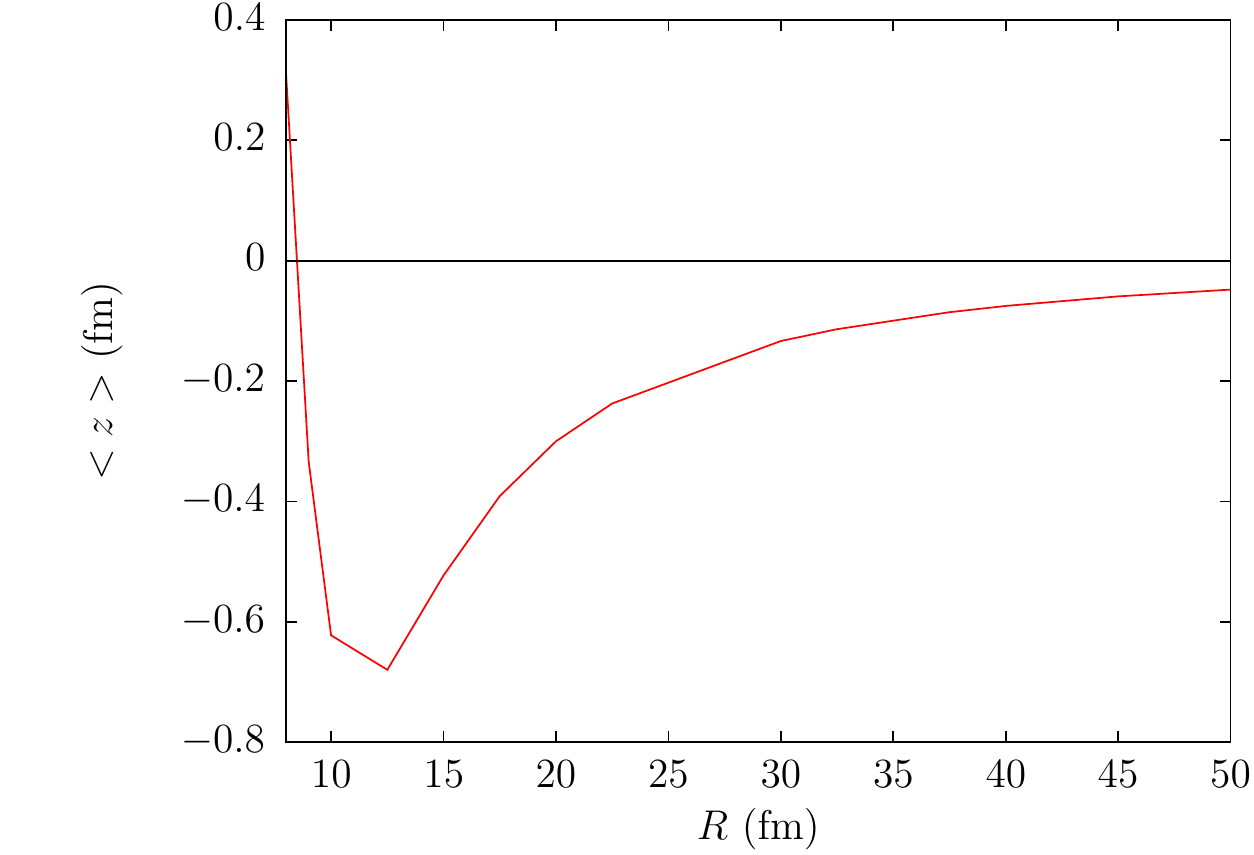}
\end{center}
\caption{Polarization $<z>$ of OHe at order $1$ as a function of the distance $R$ of an external sodium nucleus. $V_{0}=30$ MeV and $a=0.5$ fm.\label{FigPertPol}}
\end{figure}

Finally, we can calculate the  mean value of the position $z_{1}$ of the helium nucleus along the $z$-axis, that is the polarization of the OHe atom, which is simply given by:

\begin{equation}
<z>=<\psi_{He}|z_{1}|\psi_{He}>=\int \mathrm{d}\vec{r}_1\, z_{1}|
\psi_{He}\left(\vec{r}_1\right)|^{2}.\label{zmean}
\end{equation}
Figure \ref{FigPertPol} represents the evolution of this polarization as a function of $R$. It can be seen that, at large distance, the polarization is negative due to Coulomb repulsion between nuclei, as expected. Then, Figure \ref{FigPertPol} shows that the OHe atom gets polarized, for $R\lesssim10$ fm, in a direction that could allow repulsion, provided that the nuclear force is not already too strong at such distance. The addition of the nuclear interaction with $V_{0}=30$ MeV and $a=0.5$~fm in Figure \ref{FigEelEnuclNa23} shows that this condition is in fact not satisfied, giving rise to an attractive force at all distances. A modification of the nuclear parameters $V_{0}$ and $a$ (for example $V_{0}=10,\,100,\,200$ MeV, $a=0.5,\,1.5$ fm) as well as a change of external nucleus do not radically affect the results, modifying only the distance from which the potential falls to nuclear values. Note that, as pointed out in \cite{Wall2011}, the potential barrier that is needed to stabilize the weakly bound states between OHe and nuclei should rise up to a height of the order of 1 MeV or more, so that it is unlikely that we missed it in each of the methods that have been presented.

\section{OHe without a repulsion mechanism}

According to the results of Section \ref{secRepulsion}, it seems rather clear that no dipolar repulsion mechanism takes place in the interaction between an OHe atom and a nucleus N, making the interactions of the former with ordinary matter dominantly inelastic. If OHe could remain under its initial form until today, then there are two possibilities for direct searches: 1) due to the deep nuclear levels resulting from the nuclear interaction between He and N, the weakly bound states are completely unstable and de-excite either by emitting $\gamma$-rays or by ejecting neutrons or $\alpha$-particles from the nuclei in the detectors, which is clearly not observed; 2) as the collisions in the terrestrial crust are mainly inelastic, all the OHe atoms are absorbed by nuclei before reaching underground detectors, and no signal should be observed. Obviously, none of these possibilities are satisfying for direct searches, but the second scenario would not rule out the OHe model.

We will see in this section, based on paper \cite{Cudell:2014dva}, that it is ruled out for another reason: inelastic processes in the early universe between OHe and the elements formed during BBN produce anomalous heavy isotopes of helium that recombine with ordinary electrons, which creates a type of dark matter with atomic cross sections and hence with a strongly suppressed mobility in the terrestrial matter. But this has dramatic consequences, in particular for the constraints on self-interacting dark matter and for the terrestrial searches for anomalous heavy isotopes of helium.

\subsection{Inelastic processes with OHe in the early universe}

As soon as all the OHe is formed in the early universe, inelastic processes between OHe and OHe itself and between OHe and the primordial He 
take place and start consuming the available OHe. The two relevant reactions are:

\begin{eqnarray}
\mathrm{OHe}+\mathrm{OHe}&\rightarrow &\mathrm{O_{2}Be}\label{reac1} \\
\mathrm{OHe}+\mathrm{He}&\rightarrow &\mathrm{OBe}\label{reac2}
\end{eqnarray}
Note that in these reactions  the addition of a He nucleus to the bound OHe system will result in merging the two He nuclei into $^8$Be, since in 
the presence of O, $^8$Be becomes stable: we calculated, as in Section \ref{secSpectrum}, that the energy of OBe is 2.9 MeV smaller than that of OHe+He.

The temperature $T_*$ at which OHe forms depends on its binding energy, which has been accurately evaluated as 1.176 MeV in Section \ref{secSpectrum}, and corresponds approximately to $T_*=50$ keV. As the cosmological time $t$ is related to the temperature through $t(\mathrm{s})\simeq\frac{1}{T^2(\mathrm{MeV})}$, processes \eqref{reac1} and \eqref{reac2} start at a time $t_{*}\simeq\frac{1}{0.05^2}=400$ s after the Big Bang and continue until helium freezes out at $t_{fo}\simeq 10~\mathrm{min} = 600$ s. During these 200 s, the OHe atoms are consumed at a rate:

\begin{equation}
\frac{\mathrm{d}n_{OHe}}{\mathrm{d}t}=-3Hn_{OHe} - n_{OHe}^2\sigma_1 v_1 - n_{OHe}
n_{He}\sigma_2 v_2,
\label{dnOHedt}
\end{equation}
where $n_{OHe}$ and $n_{He}$ are the number densities of OHe and He, $H=\frac{1}{2t}$ is the expansion rate of the universe during the radiation-dominated era, $\sigma_1$ and $\sigma_2$ are the cross sections of processes \eqref{reac1} and \eqref{reac2} respectively and $v_1$ and $v_2$ are the OHe-OHe and OHe-He mean relative velocities. The first term in the right-hand side of equation \eqref{dnOHedt} corresponds to the dilution in an expanding universe. The number of helium nuclei per comoving volume is assumed to be unaffected by reaction \eqref{reac2} since the abundance of helium is more than an order of magnitude higher than that of OHe, so that the only effect on $n_{He}$ is due to the expansion:

\begin{equation}
\frac{\mathrm{d}n_{He}}{\mathrm{d}t}=-3Hn_{He},
\label{dnHedt}
\end{equation}
from which it follows that:

\begin{equation}
n_{He}(t)=n_{He}^{*}\left(\frac{t_{*}}{t}\right)^{3/2},
\label{nHe}
\end{equation}
where $n_{He}^*$ is the number density of He at $t=t_*$.

To take into account the effect of the expansion and calculate the decrease of the fraction of free OHe atoms due to their inelastic reactions, we study the ratio $\zeta$ of the number density of OHe atoms to the number density of He nuclei, $\zeta=\frac{n_{OHe}}{n_{He}}$. From \eqref{dnOHedt} and \eqref{dnHedt}, its evolution is given by:

\begin{equation}
\frac{\mathrm{d}\zeta}{\mathrm{d}t}=-n_{He}\zeta\left(\sigma_1 v_1 \zeta+\sigma_2 v_2\right).
\label{dzetadt}
\end{equation}
The capture cross sections $\sigma_1$ and $\sigma_2$ are of the order of the geometrical ones:

\begin{eqnarray}
\sigma_1&\simeq&4\pi \left(2r_{o}\right)^2,\label{signa1} \\
\sigma_2&\simeq&4\pi \left(r_{o}+R_{He}\right)^2,\label{sigma2}
\end{eqnarray}
where $r_o$  is the Bohr radius of an OHe atom \eqref{ro} and $R_{He}$ is the radius of a He nucleus. As both of them are approximately equal to 2 fm, $\sigma_1 \simeq \sigma_2 \simeq 64\pi \times 10^{-26}$ cm$^2$. Here and if used in the following, geometrical cross section means the elastic cross section of a hard sphere calculated in a quantum framework, i.e. four times the section of that sphere. As the OHe and He species are in thermal equilibrium with the plasma at temperature $T$, the mean relative velocities $v_1$ and $v_2$ are obtained from the Maxwell-Boltzmann velocity distributions of OHe and He and are given by:

\begin{eqnarray}
v_1&=&\sqrt{\frac{8T}{\pi \mu_1}},\label{v1} \\
v_2&=&\sqrt{\frac{8T}{\pi \mu_2}},\label{v2}
\end{eqnarray}
where $\mu_1=m_{O}/2$ and $\mu_2 \simeq m_{He}$ are the reduced masses of the OHe-OHe and OHe-He systems. $m_{O}=1000$ GeV is the mass of an OHe atom \eqref{mO}, and $m_{He}=3.73$ GeV that of a He nucleus. Given the time dependence of the temperature during the radiation-dominated era, $Tt^{1/2}=T_*t_*^{1/2}$, one can use it to express the velocities \eqref{v1} and \eqref{v2} as functions of time and insert the resulting expressions together with \eqref{nHe} in equation \eqref{dzetadt} and get:

\begin{equation}
\frac{\mathrm{d}\zeta}{\mathrm{d}t}=-C\frac{1}{t^{7/4}}\zeta\left(A \zeta+B \right),
\label{dzetadt2}
\end{equation}
with
\begin{eqnarray}
A &=&\frac{\sigma_1}{\sqrt{\mu_1}},\label{A} \\
B &=&\frac{\sigma_2}{\sqrt{\mu_2}},\label{B} \\
C &=&n_{He}^*t_*^{7/4}\sqrt{\frac{8T_*}{\pi}}.\label{C}
\end{eqnarray}
The solution of \eqref{dzetadt2} corresponding to the initial condition $\zeta(t_*)=\zeta_*$ is given by:

\begin{equation}
\zeta(t)=\frac{B \zeta_*}{\exp\left({\frac{4}{3}BC\left({t_*^{-3/4}}-{t^{-3/4}}\right)}\right)\left(A\zeta_*+B\right)-A}.
\label{solzeta}
\end{equation}

The number density of He nuclei at the time of OHe formation, $n_{He}^*$, can be found from its value $n_{He}^{\#}$ today. Helium nuclei represent nowadays approximately 10$\%$ of all baryons, which have an energy density $\rho_B^{\#}$ of about 5$\%$ of the critical density $\rho_c^{\#}$:  $n_{He}^{\#}\simeq 0.1n_{B}^{\#}=0.1\frac{\rho_B^{\#}}{m_p}\simeq 0.1\times 0.05 \frac{\rho_c^{\#}}{m_p}$, where $m_p$  is the mass of the proton. The present critical density is measured to be $\rho_c^{\#}=5.67\times 10^{-6} m_p$/cm$^3$, so that $n_{He}^{\#}\simeq 2.8\times 10^{-8}$ cm$^{-3}$. As it was assumed that the He number density was not affected by reaction \eqref{reac2}, the only effect between $t_*$ and now has been a dilution due to the expansion, and hence $n_{He}\propto \frac{1}{a^3}\propto T^3$, where $a$ is the scale factor. Knowing that the temperature of the CMB today is $T_{\#}=2.7$ K $=2.33\times 10^{-7}$ keV, this gives:

\begin{equation}
n_{He}^*=n_{He}^{\#}\left(\frac{T_*}{T_{\#}}\right)^3\simeq 2.8\times 10^{-8}\left(\frac{50}{2.33\times 10^{-7}}\right)^3\simeq2.8\times 10^{17}~\mathrm{cm}^{-3}.
\end{equation}

At the time of OHe formation, all the O particles were in the form of OHe, i.e. the number density of O at $t=t_*$, $n_{O}^*$,
was equal to that of OHe,  $n_{OHe}^*$. Between $t_*$ and today, O particles may have been bound in different structures, but they have not been created or destroyed, so that their number density has only been diluted by the expansion in the same way as that of He nuclei, so that the ratio of the number density of O particles to the number density of He nuclei remains unchanged: $\frac{n_{O}^*}{n_{He}^*}=\frac{n_{O}^{\#}}{n_{He}^{\#}}$.

Therefore, the initial fraction $\zeta_*$ of OHe atoms can be calculated from the present quantities: $\zeta_*=\frac{n_{OHe}^*}{n_{He}^*}=\frac{n_{O}^*}{n_{He}^*}=\frac{n_{O}^{\#}}{n_{He}^{\#}}$. $n_{O}^{\#}$ is obtained from the fact that O saturates the dark matter energy density, which represents about 25$\%$ of the critical density: $n_{O}^{\#}\simeq 0.25\frac{\rho_c^{\#}}{m_{O}}\simeq 1.3\times 10^{-9}$ cm$^{-3}$. With the previously calculated value of $n_{He}^{\#}$, this gives:

\begin{equation}
\zeta_*\simeq 0.05.
\end{equation}

We can now insert the numerical values into equation \eqref{solzeta} and get the fraction of OHe atoms at the time of helium freeze-out 
 $t_{fo}=600$ s:

\begin{equation}
\zeta(t_{fo})\simeq 5\times 10^{-6133}\ll \zeta_*,
\end{equation}
meaning that no OHe survives reactions \eqref{reac1} and \eqref{reac2}.

More precisely, most of the OHe atoms have captured He nuclei via 
process \eqref{reac2} and are now in the form of OBe. Indeed, the majority of the suppression of $\zeta$ comes from the exponential term present in \eqref{solzeta}, the argument of which, evaluated to be $14127$ in $t_{fo}$, represents the number $N_2$ of reactions \eqref{reac2} that happened between $t_*$ and $t_{fo}$, per OHe atom:

\begin{align}
N_2 & = \int_{t_{*}}^{t_{fo}} n_{He}(t)\sigma_2 v_2(t)\mathrm{d}t\nonumber \\
    & = n_{He}^*t_*^{3/2}\sigma_2 \sqrt{\frac{8T_*t_*^{1/2}}{\pi \mu_2}}\int_{t_{*}}^{t_{fo}}\frac{1}{t^{7/4}}\mathrm{d}t\nonumber \\
    & =  n_{He}^* t_*^{7/4} \sqrt{\frac{8T_*}{\pi}} \frac{\sigma_2}{\sqrt{\mu_2}}\left(-\frac{4}{3}\right)\left(\frac{1}{t_{fo}^{3/4}}-\frac{1}{t_{*}^{3/4}}\right)\nonumber \\
    & = \frac{4}{3}BC \left(\frac{1}{t_{*}^{3/4}}-\frac{1}{t_{fo}^{3/4}}\right),
\end{align}
where we have used \eqref{nHe}, \eqref{v2} and $Tt^{1/2}=T_*t_*^{1/2}$ to pass from the first to the second line and the definitions \eqref{B} and \eqref{C} for the last line.

Therefore, the realization of the scenario of an OHe universe would imply a very strong suppression of reaction \eqref{reac2}, corresponding to $N_2 \ll 1$. Such a suppression needs the development of a strong dipole barrier in the OHe-He interaction, for which we could not find any evidence in Section \ref{secRepulsion}. Therefore, all the O particles seem to be today under the form of OBe, and we will briefly discuss the consequences of this in Section \ref{subsecOBeProblems}.

\subsection{Problems of OBe ``dark'' matter}\label{subsecOBeProblems}

Due to the Coulomb repulsion, further helium capture by OBe is suppressed and one should expect that dark matter is mostly made of doubly charged 
OBe, which recombines with electrons in the period of recombination of helium at the temperature $T=2$ eV, before the beginning of 
matter dominance at $T=1$ eV. It makes anomalous helium the dominant form of dark matter in this scenario.

At momentum values of interest, one finds that elastic cross sections are actually significantly enhanced from their geometrical estimate. In the following, as this will be crucial to determine the viability of OBe, we shall rather use the estimate of \cite{Kaplan:2009de}, based on a compilation of results from general quantum mechanical scattering and from detailed quantum computations of hydrogen scattering, instead of the geometrical approximation:

\begin{equation}
\sigma=4\pi(\kappa a_0)^2,~\kappa=3, ..., 10,\label{sigmaOBe}
\end{equation}
with larger values of $\kappa$ at low momentum. For a size of OBe atoms equal to that of helium $a_0=3\times 10^{-9} $ cm and one obtains an elastic scattering cross section on light elements of the order of $\sigma \simeq 10^{-15} -10^{-14}$ cm$^2$. It makes this ``dark'' matter follow the ordinary baryonic matter in the process of galaxy formation, and makes it collisional on the scale of galaxies. This causes problems with the observations of halo shapes \cite{2002ApJ...564...60M,Buote:2002wd} or of colliding galaxy clusters \cite{Markevitch:2003at,Randall:2007ph,Harvey:2015hha}. The presence of OBe in stars can also influence nuclear processes, in particular helium burning in the red giants. The processes in stars can lead to the capture by OBe of additional nuclei, thus creating anomalous isotopes of elements with higher $Z$. OBe atoms can also be ionized in the galaxy, but in the following we shall assume that neutral OBe atoms are the dominant part of this dark matter on earth, considering also that slowing down anomalous nuclei in the atmosphere leads to 
ionization and their neutralization through electron capture.

OBe atoms falling down on earth are slowed down and, due to the atomic cross section of their collisions, have a very low mobility. After they fall down to the terrestrial surface, the OBe atoms are further slowed down by their elastic collisions with matter. They drift, sinking down towards the center of the earth with a drift velocity similar to \eqref{Vdrift} except that $\sigma_o$ is replaced by $\sigma$:

\begin{equation}
V_d=\frac{m_O}{m_{SiO_2}}\frac{g}{n_{\oplus}\sigma v_t}\leq 4.8\times 10^{-10}~\mathrm{cm/s}=4800~\mathrm{fm/s}.\label{Vdrift2}
\end{equation}
Here, we assimilate the crust of the earth as made of SiO$_2$ of mass $m_{SiO_2}$, and get the number density to be $n_{\oplus}=0.27\times 10^{23}$ molecules/cm$^3$. Using \eqref{sigmaOBe}, and taking the geometrical radius to be that of SiO$_2$, i.e. $a_0\simeq 2$ \AA, we obtain $\sigma\geq 4.5\times 10^{-14}$ cm$^2$, and for the thermal collisions on SiO$_2$ $v_t\simeq 3\times 10^4 $cm/s.

The OBe abundance in the earth is determined by the equilibrium between the in-falling and down-drifting fluxes, as in \eqref{fluxes}, except that for simplicity we do not take into account the annual modulation of the in-falling flux and just consider the constant part. From \eqref{fluxes}, the equilibrium concentration $n_{OBe}$ of OBe, which is established in the terrestrial matter, is given by:

\begin{equation}
    n_{OBe}=\frac{1}{4}n_{\odot}\frac{|\vec{V}_{\odot}+\vec{V}_{\oplus}|}{V_d},
    \label{noE}
\end{equation}
with $n_{\odot}=3\times 10^{-4}$ cm$^{-3}$ for OBe atoms of a mass of 1 TeV and $|\vec{V}_{\odot}+\vec{V}_{\oplus}|=200$ km/s, which, by use of \eqref{Vdrift2}, gives a ratio of anomalous helium isotopes to the total amount of SiO$_2$:

\begin{equation}
\frac{n_{OBe}}{n_{\oplus}}=\frac{1}{4}n_{\odot}\frac{m_{SiO_2}}{m_O}\frac{|\vec{V}_{\odot}+\vec{V}_{\oplus}|\sigma v_t}{g}\geq 1.3\times 10^{-10},
\end{equation}
being independent on the atomic number density of the matter. The upper limits on the anomalous helium abundance are very stringent since they require $\frac{n_{OBe}}{n_{\oplus}}\leq 10^{-19}$ \cite{PhysRevLett.92.022501}, and our rough estimate is nine orders of magnitude too large.

Together with the other problems of an OBe universe stipulated above, this rules out the OBe scenario and, together with it, the OHe scenario, since we could not find in Section \ref{secRepulsion} any mechanism that could prevent all the OHe from forming OBe during the early universe. And even by finding some way to avoid the formation of OBe, the study of the interactions of OHe in underground detectors showed us that great amounts of energy should be deposited due to the transitions to nuclear levels during the captures by nuclei, leading to dramatic effects that are not observed. It seems therefore that there is no way to keep the OHe model in its present form. However, many of its aspects are very interesting, such as the thermalization of dark matter in terrestrial matter, the drift mechanism, the explanation of the signals by radiative captures instead of elastic collisions, and will be used in Chapters \ref{Millicharges} and \ref{AntiH} to build more sophisticated models that will try to deal with the challenges of the direct-search experiments.

\chapter{Milli-interacting dark matter}\label{Millicharges}

In view of all the developments that have been made in Chapter \ref{OHe} about O-helium, we saw that this scenario has to be ruled out, although it is not far from being correct from many points of view. For that reason, I build here a model that has common features with O-helium, but also with the models presented in the Introduction, as well as novel characteristics, needed to reconcile as many direct-search experiments as possible. In this model, dark atoms, interact with standard atoms through a kinetic mixing between photons and dark photons and through a mass mixing of $\sigma$ mesons with dark scalars. These dark atoms diffuse elastically in terrestrial matter where they deposit all their initial energy. After reaching underground detectors through gravity with thermal energies, they form bound states with the atoms of the active medium by radiative capture, which causes the emission of photons that produce the observed signals. The scenario reproduces well the positive results from DAMA and CoGeNT and is consistent with the absence of signal in the XENON100, LUX, CDMS-II/Ge, superCDMS and CRESST-II detectors. The present Chapter in based on papers \cite{Wallemacq:2013hsa,Wallemacq:2014lba,Wallemacq:2014lva}.

\section{Towards a new model}\label{secTowards}

It seems important to keep the idea, which comes from the OHe scenario, to explain the signals of the direct-search experiments in terms of electron recoils instead of nuclear recoils. The former are due to the radiative capture of thermalized dark matter particles by nuclei of the active medium and result in the emission of photons that interact with the electrons. Indeed, as DAMA and CoGeNT make no distinction between both kinds of recoils (see Section \ref{secDAMA} and Section \ref{secCoGeNT} of Chapter \ref{DirectSearches}), their positive results can be reinterpreted directly, while the bound-state-formation events contribute to the electron background of the other experiments and one has therefore only to ensure that the expected and observed electron backgrounds are consistent.

Producing signals in the energy ranges measured by DAMA and CoGeNT requires the existence of weakly bound states between matter and dark matter, i.e. of the order of a few keV. The problem of OHe was that these weakly bound states could not be stabilized because of the strong nuclear attraction between its He component and the nuclei. One simple way to avoid that could be to use a nucleus-dark matter potential that is attractive and fully in the keV range, resulting for example from a kind of ``milli-interaction''. Milli-charges discussed in the Introduction could be an interesting track to follow for that purpose.

An unavoidable consequence of a scenario in which dark matter particles form bound states with ordinary nuclei is the formation of anomalous heavy isotopes of known elements on earth, which could cause a conflict with the results of the terrestrial searches for such species. As these constraints are very strong for light elements, such as helium as seen in Section \ref{subsecOBeProblems}, it would be safer to allow the formation of bound states with heavy elements only. Between the sodium ($Z=11$) and the iodine ($Z=53$) components of the DAMA detector, the latter would therefore be preferred to explain the observations. As a consequence, this would ensure that the collisions with the rather light atoms of the terrestrial crust are purely elastic, preventing the dark matter flux to be lost by absorption during the thermalization process towards the underground detectors.

But on the other hand, XENON100 and LUX, which are made of xenon ($Z=54$), do not see any signal. Yet xenon is almost the same nucleus as iodine from the point of view of its mass, of its size, of its number of nucleons and of its electric charge. If the capture rate depends only on the composition of the detector, then one should expect approximately the same event rates in DAMA and in XENON100 or LUX. One could argue that these events are lost in the electron backgrounds of the latter, but the direct transfer of the event rate of DAMA to the XENON100 and LUX detectors would lead to an inconsistency between their expected and observed electron backgrounds and would therefore be detected as an anomaly.

An additional variable is therefore needed to differentiate two experiments with almost the same nuclear composition, such as the operating temperature $T_{op}$, which was also present in the OHe scenario. We saw in Chapter \ref{DirectSearches} that DAMA operates at room temperature ($T_{op}\simeq 300$ K) while XENON100 and LUX are at the condensation temperature of xenon ($T_{op}\simeq 180$ K). Thermalized dark matter particles have therefore energies of about $1/40$ eV in DAMA and half of that value in XENON100 and LUX. If we could find a way to place, in the nucleus-dark matter potential, a barrier at distances larger than the range of the attractive well and with a height slightly larger than $1/40~\mathrm{eV}$, then the incident dark matter particles would have to tunnel through the barrier before reaching the keV-bound states in the well, but this would be much less efficient in colder detectors, reducing drastically the rates in XENON100 and LUX with respect to DAMA. Also, one could expect that the probability of crossing the barrier is almost zero in cryogenic detectors such as CDMS-II/Ge, superCDMS and CRESST-II, where $T_{op}\simeq 0.001$ K, leading basically to null event rates in these, which would be consistent with their negative results.

In the following, I put all these ideas together and give the ingredients allowing to realize them. 

\section{Dark sector}\label{secDarkSect1}

We postulate that a dark hidden sector exists, in which two fermions, denoted by \newcommand{\pplus}{$\dot{p}^{+}$}\pplus~and \newcommand{\eminus}{$\dot{e}^{-}$}\eminus~and of masses \newcommand{\mpp}{m_{\dot{p}^{+}}}$\mpp$~and \newcommand{\mem}{m_{\dot{e}^{-}}}$\mem$, are coupled to a dark massless photon \newcommand{\gammad}{$\dot{\gamma}$}\gammad~with respective couplings $+\dot{e}$ and $-\dot{e}$ ($\dot{e}>0$). In addition, \pplus~is coupled to a dark neutral scalar \newcommand{\hd}{$\dot{h}^0$}\hd~of mass \newcommand{\mhd}{m_{\dot{h}^0}}$\mhd$ via a Yukawa coupling of constant $\dot{g}_Y$, which leads to the dark interaction Lagrangian:

\begin{equation}
\mathcal{L}_{int}^{dark}=\dot{e}\bar{\psi}_{\dot{p}}\gamma^{\mu}\dot{A}_{\mu}\psi_{\dot{p}}+\dot{e}\bar{\psi}_{\dot{e}}\gamma^{\mu}
\dot{A}_{\mu}\psi_{\dot{e}}-\dot{g}_Y\phi_{\dot{h}^0}\bar{\psi}_{\dot{p}}\psi_{\dot{p}},\label{Lintdark}
\end{equation}
where $\psi_{\dot{p}}$ and $\psi_{\dot{e}}$ are the fermionic fields that contain \pplus~and \eminus, $\dot{A}$ is the vectorial field of \gammad~and $\phi_{\dot{h}^0}$ is the real scalar field of \hd. We assume that \gammad~is massless to avoid several constraints holding for massive para-photons, e.g. from the anomalous magnetic dipole moments of the electron and the muon \cite{Cline:2012is}.

In order to produce non-gravitational interactions between the standard and the dark sector, we postulate that the dark photon \gammad~is kinetically mixed with the standard photon $\gamma$ \cite{Cline:2012is,Foot:2013msa,Holdom:1985ag,Feldman:2007wj}, but also that the dark scalar \hd~is mixed with the neutral scalar meson \newcommand{\sigm}{$\sigma$}\sigm~via a mass term, with the mixing Lagrangian:

\begin{equation}
\mathcal{L}_{mix}=\frac{\tilde{\epsilon}}{2}F^{\mu\nu}\dot{F}_{\mu\nu}+\tilde{\eta}(m_{\sigma}^{2}+\mhd^{2})\phi_{\sigma}\phi_{\dot{h}^0},
\label{Lmix2}
\end{equation}
where $F$ and $\dot{F}$ are the electromagnetic-field-strength tensors of $\gamma$ and \gammad~ respectively, $\phi_{\sigma}$ is the real scalar field of the $\sigma$ meson and \newcommand{\msigm}{m_{\sigma}}$\msigm=600$ MeV \cite{Amsler:2008zzb} is its mass. $\tilde{\epsilon}$ and $\tilde{\eta}$ are the dimensionless parameters of the kinetic $\gamma$-\gammad~and mass \sigm-\hd~mixings and are assumed to be small compared with unity.

In principle, the model contains seven free parameters, $\mpp$, $\mem$, $\mhd$, $\dot{e}$, $\dot{g}_Y$, $\tilde{\epsilon}$ and $\tilde{\eta}$, but these can in fact be reduced to four. Indeed, only the products $\tilde{\epsilon}\dot{e}$ and $\tilde{\eta}\dot{g}_Y$ will be directly constrained
by the direct-search experiments, which suggests to define them in terms of the charge of the proton $e$ and of the Yukawa coupling constant of the nucleon to the $\sigma$ meson $g_Y=14.4$ \cite{Erkol:2005jz}: $\tilde{\epsilon}\dot{e}\equiv\epsilon e$ and $\tilde{\eta}\dot{g}_Y\equiv\eta g_Y$,
where $\epsilon$ and $\eta$ are the redefined dimensionless mixing parameters that will be used in the following. Due to the mixing Lagrangian \eqref{Lmix2}, the species \pplus~and \eminus~interact electromagnetically with the standard charged particles and \pplus~interacts with any standard particle coupled to the $\sigma$ meson, e.g. the proton and the neutron in the framework of an effective Yukawa theory, so that $\pm \epsilon e$ and $\eta g_Y$ represent their electric and ``nuclear'' milli-charges. Note that the first idea that comes to mind to create a milli-nuclear interaction is a mixing with the pion, since its exchange between nucleons is dominant in the standard nuclear interaction. But due to its pseudoscalar nature, it is well known that nucleon-pion couplings are of order $v/c$ in the nonrelativistic limit and hence strongly suppressed at the energies of interest in the direct searches for dark matter. Then, the choice of the meson $\sigma$ instead of any other meson involved in the standard nuclear interaction was motivated by the simplicity of its scalar nature.
 
The \pplus~and \eminus~particles bind to each other through the dark $U(1)_d$ and form hydrogen-like dark atoms. In such an atom, \pplus~plays the role of the nucleus while \eminus~plays that of the electron, i.e. we assume $\mpp\gg\mem$, so that the mass $m_{\dot{p}^{+}\dot{e}^{-}}\simeq \mpp$ and the Bohr radius \newcommand{\ad}{\dot{a}_0}$\ad=1/(\dot{\alpha}\mem)$, where $\dot{\alpha}=\dot{e}^2/4\pi$. The Bohr radius will be fixed in the following to 1~\r{A} so that dark atoms have the same size as standard ones and can thermalize in the earth before reaching the underground detectors, as will be specified in Section \ref{subsecThermalization} . As a result, the parameters of the model can be reexpressed as $\mpp$, $\mhd$, $\epsilon$ and $\eta$.

We will see that \pplus~will bind to nuclei in underground detectors and must therefore be sufficiently massive to form bound states in the keV region. For that reason, we will explore masses of \pplus~between $10$ GeV and $10$ TeV. The mass mixing term in \eqref{Lmix2} induces an attractive interaction between \pplus~and nucleons with a range determined by $\mhd^{-1}$. It cannot be too long ranged but it must allow the existence of nucleus-\pplus~bound states of at least the size of the nucleus, so we will seek masses of \hd~between $100$
keV and $10$ MeV. The model parameters that we will consider are therefore:

\begin{equation}
\begin{array}{c}
10~\mathrm{GeV} \leq \mpp \leq 10~\mathrm{TeV} \\
100~\mathrm{keV} \leq \mhd \leq 10~\mathrm{MeV} \\
\epsilon, \eta \ll 1 \\
\ad = 1~${\r{A}}$
\end{array}\label{params1}
\end{equation}

Note that the galactic dark matter halo could also be populated by dark ions \pplus and \eminus, but Ref. \cite{McDermott:2010pa} ensures that if $\epsilon>5.4\times10^{-13}$ ($m_{\dot{p}^{+},\dot{e}^{-}}/$GeV), they have been evacuated from the disk by supernovae shock waves while galactic magnetic fields prevent them from reentering. This condition will clearly be satisfied by the parameters used to reproduce the results of the direct-dark-matter-search experiments and we can consider their signals to be fully due to dark atoms.

When the dark photon is massless, there are several equivalent definitions for the fields $\gamma$ and \gammad. One possibility is to keep the interaction between both sectors as the exchange of a photon that converts into a dark photon or conversely, each of them being coupled only to
its own sector. Another possibility is to diagonalize the Hamiltonian by defining a photon that couples to the visible current with $e$ and to the dark current with $\epsilon e$, while the dark photon couples only to the dark sector with $\dot{e}$. In any case, the particles charged under the dark $U(1)_d$ appear as electric milli-charges on which laboratory, cosmological and astrophysical constraints exist. One can also derive constraints on nuclear milli-charges from unseen disintegrations of vector mesons and all this will be discussed in Section \ref{subsecConstraints} in view of the values of the parameters required to reproduce the results of the direct-search experiments.

It is important to note that we are clearly in presence of a self-interacting dark matter, which, according to Ref. \cite{Fan:2013yva} and as already mentioned in the Introduction, is likely to form dark disks in galaxies provided the existence of a cooling mechanism. The emission of dark photons by the atoms can contribute to this process and we expect dark atoms, similarly to baryons, to concentrate in a disk, aligned or not with that of visible matter. In that particular case, stellar velocities in and out of the galactic plane give stronger bounds on the amount of self-interacting dark matter and the observation of the kinematics of nearby stars leads to a limit on the mass of dark atoms of 5$\%$ of the total dark mass of the Milky Way halo. We will use the value $\rho_{\odot}=0.3$ GeV/cm$^3$ for the local dark matter density and consider that dark atoms realize 100$\%$ of it. This is possible if the dark and baryonic disks are quasi-aligned, as a thin disk can be much more concentrated than a diffuse halo. The rest of the dark matter of the galactic halo, outside the disk, is therefore made of conventional collisionless CDM that do not produce any nuclear recoil in underground detectors, so that the subdominant part of the dark sector presented here is responsible for all the signals.

Finally, while the velocity distribution of the dark particles in the halo cannot be neglected when the direct searches are interpreted in terms of collisions between nuclei and WIMPs, it is much less important here since all the dark atoms thermalize in terrestrial matter and end up with the same thermal distribution. For that reason, we assume that the dark atoms are at rest in the frame of the dark disk, which itself is at rest with respect to the halo of collisionless particles.

\section{Interaction potentials with standard matter}\label{secInteractionPotentials}

\subsection{Interactions of \pplus~and \eminus~with nucleons and electrons}

In the nonrelativistic limit, the couplings induced by the mixings \eqref{Lmix2} give rise to interaction potentials between the dark and the standard particles. At the elementary level, the kinetic $\gamma$-\gammad~mixing produces a Coulomb interaction potential between the milli-charged dark particles and the proton and the electron:

\begin{equation}
V^{k}=\pm\frac{\epsilon\alpha}{r},\label{Vk}
\end{equation}
where the superscript $k$ refers to the kinetic mixing. The plus and minus signs stand respectively for the pairs proton-\pplus~or electron-\eminus~and proton-\eminus~or electron-\pplus.

Since the mass mixing parameter $\eta$ is small, the attractive interaction between \pplus~and the nucleons is dominated by one $\sigma+$\hd~exchange, which gives:

\begin{equation}
V^{m}=-\frac{\eta\left(m_{\sigma}^{2}+\mhd^{2}\right)\beta}{r}\left(\frac{e^{-m_{\sigma}r}-e^{-\mhd r}}{\mhd^{2}-m_{\sigma}^{2}}\right),\label{Vm}
\end{equation}
where the superscript $m$ stands for the mass mixing and $\beta=\frac{g_Y^{2}}{4\pi}=16.5$. Note that because $\mhd\ll m_{\sigma}$, the potential \eqref{Vm} is essentially a Yukawa potential of range $\mhd^{-1}$:

\begin{equation}
V^{m}\simeq-\frac{\eta\beta}{r}e^{-\mhd r}.\label{Vm2}
\end{equation}

\subsection{Interactions of \pplus~with nuclei}

Because of their interactions with nucleons, the dark fermions \pplus~interact with atomic nuclei. Assuming that a nucleus N of mass number
$A$ and atomic number $Z$ is a uniformly charged sphere of radius $R_N=r_0A^{1/3}$, with $r_0=1.2$ fm, and volume $V_N=\frac{4}{3}\pi R_N^{3}$, the integrations of the elementary potentials \eqref{Vk} and \eqref{Vm} over its electric and nuclear charge distributions $\rho^{k}=\frac{Ze}{V_N}$ and $\rho^{m}=\frac{Ag_Y}{V_N}$ give:

\begin{align}
V_{N\dot{p}^{+}}^k(r) & = \int_{V_N}\left(V^{k}(|\vec{r}-\vec{r'}|)/e\right)\rho^{k}\mathrm{d}\vec{r'}\nonumber \\
& = \frac{\epsilon Z\alpha}{2R_N}\left(3-\frac{r^{2}}{R_N^{2}}\right),~\mathrm{for}~r<R_N\nonumber \\
& =  \frac{\epsilon Z\alpha}{r},~\mathrm{for}~r\geq R_N\label{VkN}
\end{align}
and:
\begin{align}
\label{VmN}
V_{N\dot{p}^{+}}^m(r) & = \int_{V_N}\left(V^m(|\vec{r}-\vec{r'}|)/g_Y\right)\rho^{m}\mathrm{d}\vec{r'}\nonumber \\
& = -\frac{C^m}{r}\left[2r\left(m_{\sigma}^{-2}-\mhd^{-2}\right)+\left(R_N+m_{\sigma}^{-1}\right)m_{\sigma}^{-2}\left(e^{-m_{\sigma}r}-e^{m_{\sigma}r}\right)e^{-m_{\sigma}R_N}\right.\nonumber \\
&~~~\left.-\left(R_N+\mhd^{-1}\right)\mhd^{-2}\left(e^{-\mhd r}-e^{\mhd r}\right)e^{-\mhd R_N}\right],~\mathrm{for}~r<R_N \\
& = -\frac{C^m}{r}\left[m_{\sigma}^{-2}e^{-m_{\sigma}r}\left(e^{m_{\sigma}R_N}\left(R_N-m_{\sigma}^{-1}\right)+e^{-m_{\sigma}R_N}\left(R_N+m_{\sigma}^{-1}\right)\right)\right.\nonumber \\
&~~~\left.-\mhd^{-2}e^{-\mhd r}\left(e^{\mhd R_N}\left(R_N-\mhd^{-1}\right)+e^{-\mhd R_N}\left(R_N+\mhd^{-1}\right)\right)\right],~\mathrm{for}~r\geq R_N\nonumber
\end{align}
where the index $N$ indicates the nucleus N, $\vec{r'}$ is the position vector of a charge element in the nucleus and $C^m=3\eta(m_{\sigma}^{2}+\mhd^{2})\beta/(2r_{0}^{3}(\mhd^{2}-m_{\sigma}^{2}))$.

\begin{figure}
\begin{center}
\includegraphics[scale=0.59]{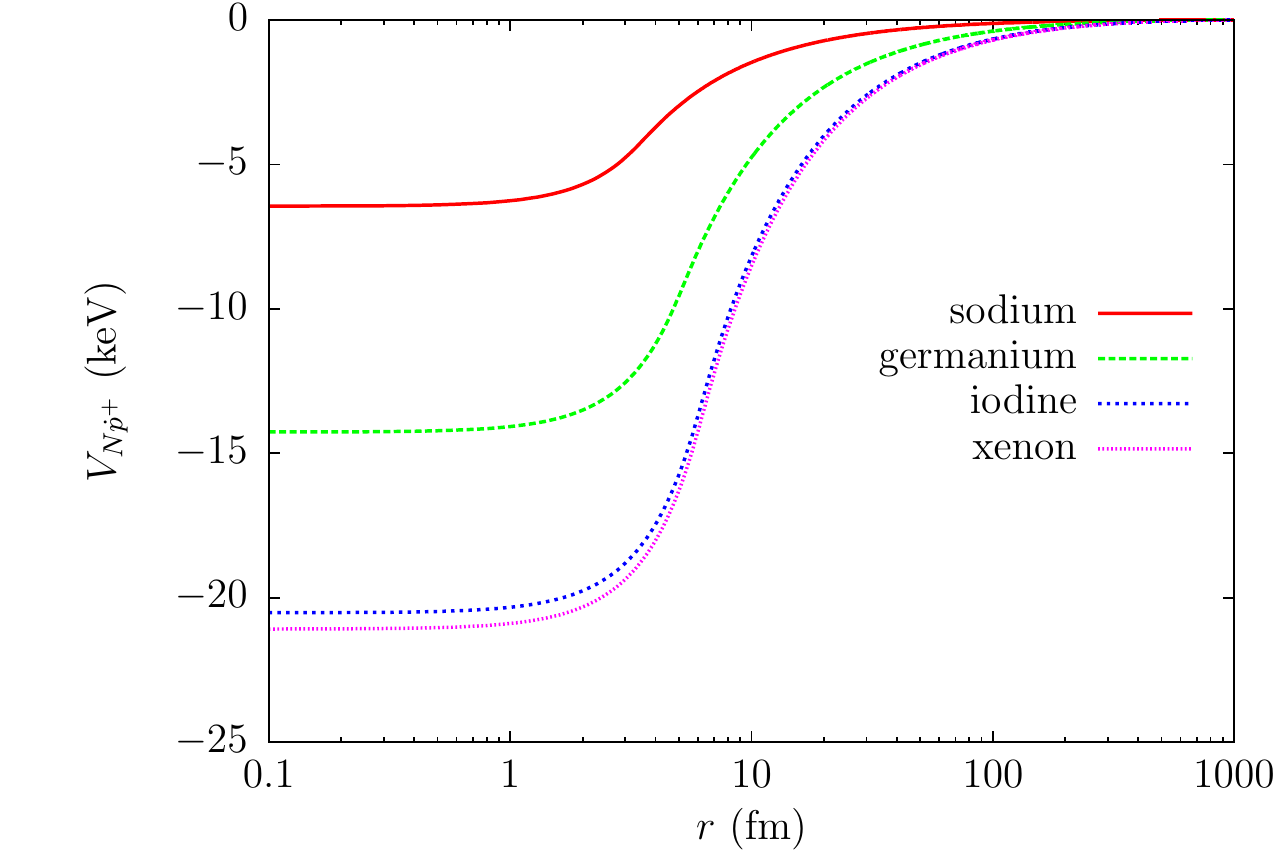}\includegraphics[scale=0.59]{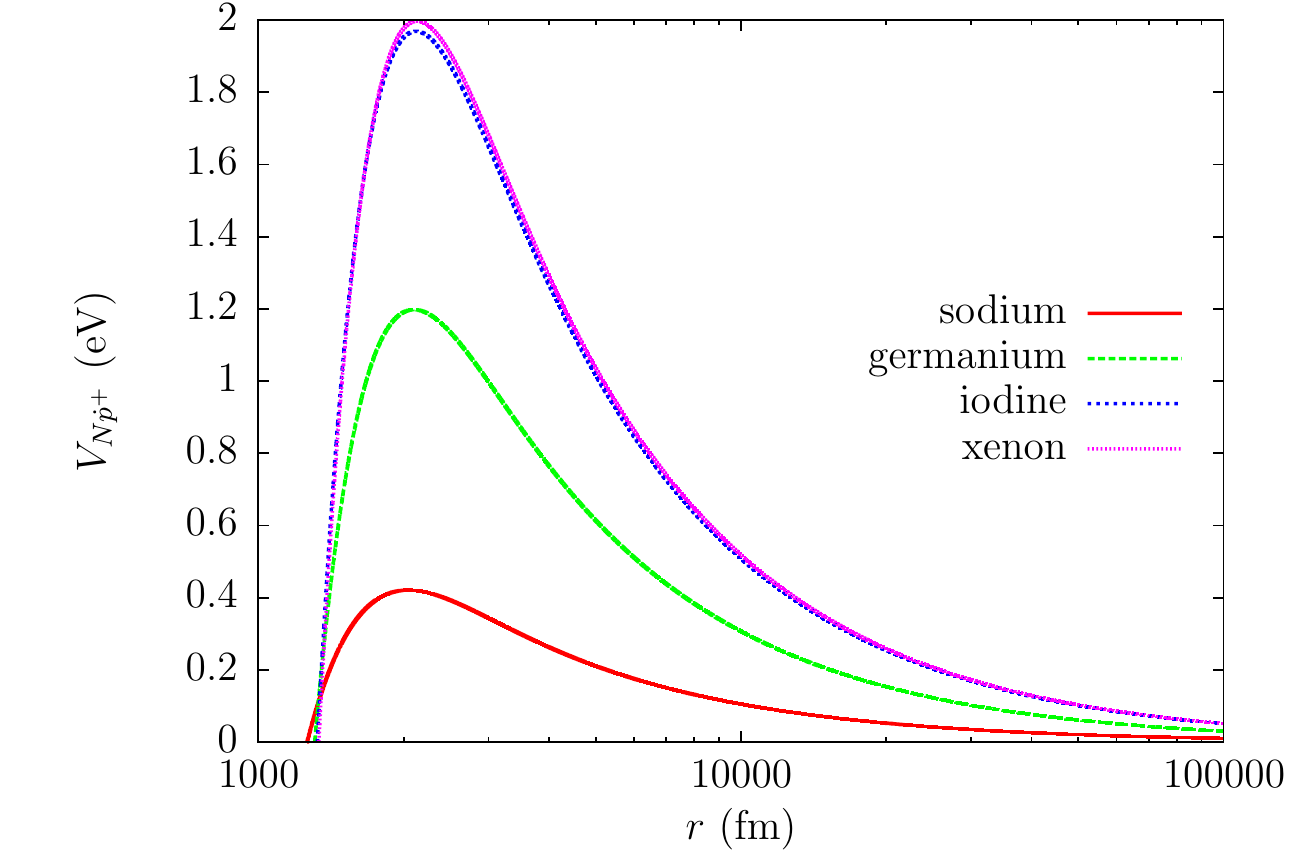}
\end{center}
\caption{Shape of the total nucleus-\pplus~interaction potential \eqref{VN} for light (solid red), intermediate (long dashed green) and heavy (short dashed
blue, dotted magenta) nuclei constituting underground detectors. The attractive part (nuclear well) is on the left (keV) and the repulsive
region (Coulomb barrier) is on the right (eV). The trial parameters $\mpp=650$ GeV, $\mhd=0.426$ MeV, $\epsilon=6.7\times 10^{-5}$ and $\eta=2.2\times 10^{-7}$ have been used.\label{FigPotNucleus}}
\end{figure}

The total nucleus-\pplus~potential $V_{N\dot{p}^{+}}$ is therefore the sum of \eqref{VkN} and \eqref{VmN}:

\begin{equation}
V_{N\dot{p}^{+}}(r)=V_{N\dot{p}^{+}}^k(r)+V_{N\dot{p}^{+}}^m(r).\label{VN}
\end{equation}
$V_{N\dot{p}^{+}}^k$ consists in a repulsive Coulomb potential outside the nucleus and in a harmonic potential, which is concave down, inside. Both parts are continuously connected at $r=R_N$ (as well as their first derivatives), with an inflection point at $r=R_N$ and a maximum reached at $r=0$, where the first derivative is zero. $V_{N\dot{p}^{+}}^m$ corresponds to a finite attractive well, with a size of the order of $\mhd^{-1}$, an inflection point at $r=R_N$ and tending to zero as $\frac{-1}{r}e^{-\mhd r}$ outside the nucleus, when $r\rightarrow \infty$. The total nucleus-\pplus~potential $V_{N\dot{p}^{+}}$ is therefore a negative attractive well at distances $r\lesssim \mhd^{-1}$ that is continuously connected, together with its first derivative, to a positive potential barrier, coming from the repulsive Coulomb potential, at larger distances. As $r\rightarrow\infty$, $V_{N\dot{p}^{+}}^m$ rapidly tends to zero and the total potential is dominated by the Coulomb part. In order to reproduce the direct-search experiments, the depth of the well, mainly determined by the parameter $\eta$, will be of the order of $10$ keV while the barrier, the height of which depends on $\mhd$ and $\epsilon$, will rise up to a few eV. Figure \ref{FigPotNucleus} shows the nuclear wells and the Coulomb barriers for several nuclei involved in direct searches, for trial parameters of the model that will lie in the preferred regions.

\subsection{Interactions of dark atoms with terrestrial atoms}

The galactic dark atoms interact with terrestrial atoms after hitting the surface of the earth because of its motion in the dark matter halo. Although the nucleus-\pplus~potential \eqref{VN} is relevant for the inelastic processes occurring within underground detectors and is the one that will be used in Section \ref{subsecBstateFormation} when these will be discussed, we also have to derive the atom-dark atom interaction potential for the thermalization process in the terrestrial crust. For that purpose, we model an atom and a dark atom as follows: they are seen as uniformly charged and finite spheres of charges $-Ze$ and  $-\epsilon e$ and radii $a_{0}$ and $\ad$ respectively, with $a_0=\ad=1$ \r{A}, and  where $Z$ is the atomic number of the terrestrial atom. This allows us to model the spherical electron distributions, while the nuclei N and \pplus~at their centers are taken as point-like and with opposite charges. Indeed, for the elastic processes in play during thermalization, the size of the nucleus N can be neglected since $qR_N\ll 1$, where $q$ is the transferred momentum, all along the path in the terrestrial crust.

The atom-dark atom interaction potential is therefore the sum of an electrostatic interaction $V_{at}^{k}$ and of a nuclear term $V_{at}^{m}$ due to the $\sigma+$\hd~exchange between the dark nucleus \pplus~and the atomic nucleus N. $V_{at}^{k}$ is obtained by adding the contributions from the four pairs of crossed substructures: nucleus-\pplus~(pointlike-pointlike, pure Coulomb repulsion), nucleus-\eminus~(pointlike-sphere, attractive of form \eqref{VkN}), electron-\pplus~(sphere-pointlike, attractive of form \eqref{VkN}) and electron-\eminus~(sphere-sphere, obtained
by integrating the form \eqref{VkN} over a uniformly charged sphere which center is separated by a distance $r$ from the center of the
first one). This gives:

\begin{align}
V_{at}^k(r) & = \frac{\epsilon Z\alpha}{160\ad^{6}}\left(-r^{5}+30\ad^{2}r^{3}+80\ad^{3}r^{2}-288\ad^{5}+\frac{160\ad^{6}}{r}\right),~\mathrm{for}~r\leq \ad \nonumber \\
& = \frac{\epsilon Z\alpha}{160\ad^{6}}\left(-r^{5}+30\ad^{2}r^{3}-80\ad^{3}r^{2}+192\ad^{5}-\frac{160\ad^{6}}{r}\right),~\mathrm{for}~\ad < r\leq 2\ad \nonumber \\
& = 0,~\mathrm{for}~r> 2\ad \label{Vkat}
\end{align}
where the lower label $at$ stands for atomic. Because the nucleus is here supposed to be point-like, $V_{at}^{m}$ is simply obtained by multiplying \eqref{Vm} by the number $A$ of nucleons in the nucleus:

\begin{equation}
V_{at}^{m}(r)=-\frac{\eta\left(m_{\sigma}^{2}+\mhd^{2}\right)A\beta}{r}\left(\frac{e^{-m_{\sigma}r}-e^{-\mhd r}}{\mhd^{2}-m_{\sigma}^{2}}\right),\label{Vmat}
\end{equation}
and hence the total atom-dark atom interaction potential $V_{at}$ can be expressed as:

\begin{equation}
V_{at}(r)=V_{at}^k(r)+V_{at}^m(r).\label{Vat}
\end{equation}

The atom-dark atom electrostatic potential $V_{at}^{k}$ has three regimes as a function of the distance $r$ between the two centers of the atoms. Each sphere appears neutral from outside because the positive charge at the center compensates exactly the negative charge distributed in the sphere, so that there is no interaction when they are completely separated ($r>2\ad$). As they merge ($\ad <r\leq 2\ad)$, the electrostatic
potential becomes attractive due to the attraction between the nucleus of each sphere and the negatively charged distribution of the other one. When the nuclei enter simultaneously in the approaching spheres ($r=\ad$), an inflection point occurs after which the potential reaches a minimum ($r=0.88$ \AA~for $a_{0}=1$ \AA). As the centers continue to approach each other ($r<0.88$ \AA~for $\ad=1$ \AA), the potential becomes repulsive due to the Coulomb repulsion between nuclei. The potential well appearing when the two atoms merge has a depth of the order of $10^{-4}-10^{-3}$ eV and therefore does not contain any bound state (or if any, not thermally stable), so that it will not contribute in the following to the formation of atom-dark atom bound states.

As $\mhd \ll m_{\sigma},$ $V_{at}^{m}\simeq-\frac{\eta A\beta}{r}e^{-\mhd r}$, which is a purely attractive Yukawa potential. The total atom-dark atom potential $V_{at}$ \eqref{Vat} is therefore essentially equal to its electrostatic part $V_{at}^{k}$ \eqref{Vkat} for $r>\mhd^{-1}$, while the attractive part $V_{at}^{m}$ \eqref{Vmat} dominates at smaller distances.

\section{From space to underground detectors}\label{secFromSpaceTo}

\subsection{Thermalization in the terrestrial crust}\label{subsecThermalization}

Due to its orbital motion around the sun, which turns around the center of the galaxy, the earth receives a wind of dark atoms hitting its surface throughout the year. A dark atom penetrates the surface and starts interacting with terrestrial atoms via the atomic potential \eqref{Vat}. As there is no bound state in the total atomic potential with the relatively light terrestrial atoms, the diffusions are purely elastic. If the elastic scattering cross section is sufficiently large, then the dark atom can deposit all its energy in the terrestrial matter before reaching an underground detector typically located at a depth of $1$ km, just as in the OHe scenario. For the composition of the crust, we use the result of \cite{Sigurdson:2004zp} saying that it is enough to approximate it as being made of pure silicon, with atomic and mass numbers $Z_{Si}=14$ and $A_{Si}=28$.

A differential elastic diffusion cross section $\frac{\mathrm{d}\sigma}{\mathrm{d}\Omega}$ from a two-body-interaction potential $V(\vec{r})$ can be obtained in the framework of the Born approximation via the Fourier transform of the potential:

\begin{equation}
\frac{\mathrm{d}\sigma}{\mathrm{d}\Omega}=\frac{\mu^{2}}{4\pi^{2}}\left|\int \mathrm{d}\vec{r}e^{-i\vec{q}.\vec{r}}V\left(\vec{r}\right)\right|^{2},\label{dsigmaFourier}
\end{equation}
where $\mu$ is the reduced mass of the two-body system and $\vec{q}$ is the transferred momentum. Here, from potentials \eqref{Vkat} and
\eqref{Vmat} and in the center-of-mass frame of the silicon-\pplus~system, we get:

\begin{align}
\left(\frac{\mathrm{d}\sigma}{\mathrm{d}\Omega}\right)_{at} & =\left(\frac{\mathrm{d}\sigma}{\mathrm{d}\Omega}\right)_{at}^{k}+\left(\frac{\mathrm{d}\sigma}{\mathrm{d}\Omega}\right)_{at}^{m}\nonumber \\
&~~~-\frac{4\mu_{Si}^{2}\epsilon\eta Z_{Si}A_{Si}\alpha\beta}{\ad^{6}}\left(\frac{m_{\sigma}^{2}+\mhd^{2}}{\mhd^{2}-m_{\sigma}^{2}}\right)\frac{I}{q^{8}}\left[\frac{1}{m_{\sigma}^{2}+q^{2}}-\frac{1}{\mhd^{2}+q^{2}}\right],\label{dsigmaat}
\end{align}
with:

\begin{eqnarray}
\left(\frac{\mathrm{d}\sigma}{\mathrm{d}\Omega}\right)_{at}^{k} & = & \frac{\mu_{Si}^{2}\epsilon^{2}Z_{Si}^{2}\alpha^{2}}{\ad^{12}}\frac{1}{q^{16}}I^{2},\label{dsigmakat} \\
& &\nonumber \\
I & = & 9\left(q^{2}\ad^{2}+1\right)+9\cos\left(2q\ad\right)\left(q^{2}\ad^{2}-1\right)+12\cos\left(q\ad\right)q^{4}\ad^{4}\nonumber \\
& & -18\sin\left(2q\ad\right)q\ad -12\sin\left(q\ad\right)q^{3}\ad^{3}+2q^{6}\ad^{6},\nonumber
\end{eqnarray}
for the electrostatic interaction and:

\begin{equation}
\left(\frac{\mathrm{d}\sigma}{\mathrm{d}\Omega}\right)_{at}^{m}=4\mu_{Si}^{2}\eta^{2}A_{Si}^{2}\beta^{2}\left(\frac{m_{\sigma}^{2}+\mhd^{2}}{\mhd^{2}-m_{\sigma}^{2}}\right)^{2}\left[\frac{1}{m_{\sigma}^{2}+q^{2}}-\frac{1}{\mhd^{2}+q^{2}}\right]^{2}\label{dsigmamat},
\end{equation}
for the $\sigma+$\hd~exchange. The reduced mass $\mu_{Si}=m_{Si}\mpp/(m_{Si}+\mpp)$, where $m_{Si}$ is the mass of a silicon atom, $q=|\vec{q}|=2k\sin\theta/2$, where $k$ is the initial momentum, and $\theta$ is the deflection angle with respect to the collision axis.

For a dark atom to thermalize between the surface and an underground detector, we have to ensure that its penetration length does not exceed
$1$ km. It is estimated by assuming a linear path of the dark atom through terrestrial matter:

\begin{equation}
x=\int_{E_{f}}^{E_{i}}\frac{\mathrm{d}E}{\left|\mathrm{d}E/\mathrm{d}x\right|}<1~\mathrm{km},\label{penLength}
\end{equation}
where $\frac{\mathrm{d}E}{\mathrm{d}x}$ is the energy loss per unit length in the frame of the earth:

\begin{equation}
\frac{\dd E}{\dd x}=n_{\oplus}\int_{\Omega}\triangle K\left(\frac{\dd\sigma}{\dd\Omega}\right)_{at}\dd\Omega,\label{dEdx}
\end{equation}
obtained by integrating over all diffusion angles. In \eqref{penLength}, the integration is performed from the initial kinetic energy of the dark atom $E_i$ to the thermal energy of the crust $E_f=\frac{3}{2}T_{\oplus}$, where $T_{\oplus}\simeq300$ K. In \eqref{dEdx}, $n_{\oplus}\simeq 5\times 10^{22}$ cm$^{-3}$ is the number density of atoms in the terrestrial crust and $\triangle K=\frac{k^{2}\left(\cos\theta-1\right)}{m_{Si}}$ is the energy lost in the frame of the earth at each collision with a silicon atom at rest in the terrestrial surface. It is clear that the linear path approximation is valid only when $\mpp \gg m_{Si}$, but it gives in the other cases an upper limit on the penetration length of a dark atom through the earth, which is of interest here.

\subsection{Drift down towards underground detectors}

As soon as it has thermalized, a dark atom starts to drift towards the center of the earth by gravity until it reaches an underground detector, just as in the OHe scenario. The drift velocity $V_d$ is given by a similar equation than \eqref{Vdrift} in Section \ref{OHeunderground}\footnote{Here again, as already mentioned in Section \ref{OHeunderground}, there is a correction factor $\frac{\mpp}{m_{Si}}\simeq 0.4-77$ in the drift velocity $V_d$ that was used. This should produce a small displacement of the regions in the parameter space that will be obtained in Section \ref{subsecResultsMilli}, probably not visible at the scale of Figure \ref{FiGDAMACoGeNTResults} due to the high non-linearity of the model because of resonances at the capture, which leads to a strong variation of the results for small changes of the parameters.}:

\begin{equation}
V_d=\frac{g}{n_{\oplus}\left<\sigma_{at}v\right>},\label{Vdrift3}
\end{equation}
where $\sigma_{at}$ is the total atom-dark atom elastic cross section, obtained by integrating numerically \eqref{dsigmaat} over all diffusion angles. $\left<\sigma_{at}v\right>$ is the product of $\sigma_{at}$ and the relative velocity $v$ between thermal atoms and dark atoms, averaged over Maxwellian velocity distributions at temperature $T_{\oplus}=300$ K.

Similarly to Section \ref{OHeunderground}, the equality between the in-falling dark matter flux at the surface and the down-drifting thermalized flux gives the equilibrium number density $n$ of thermalized dark atoms below the surface. As the incident flux is annually modulated, so is $n$. With equations \eqref{fluxes} and \eqref{norm}, we get in the same way the constant and the modulated parts $n^0$ and $n^m$ of $n$:

\begin{eqnarray}
n(t)&=&n^0+n^m\cos(\omega(t-t_0)),\label{neq} \\
n^0&=&\frac{n_{\odot}}{4V_d}V_{\odot},\label{n0} \\
n^m&=&\frac{n_{\odot}}{4V_d}V_{\oplus}\cos i,\label{nm}
\end{eqnarray}
where $n_{\odot}$ is now the number density of \pplus \eminus~dark atoms in the solar system and is defined as $n_{\odot}(\mathrm{cm}^{-3})=0.3/\mpp$(GeV).

Arriving in the detector at room temperature, a dark atom still has to thermalize at the operating temperature. The latter is always lower
than $300$ K, except for the DAMA detector, which operates at room temperature. We will check that this second thermalization at the edge of the detector is realized over a distance much smaller than its typical size and can therefore be considered as instantaneous.

\subsection{Bound-state-formation events}\label{subsecBstateFormation}

In the active medium, the dark atoms undergo thermal collisions with the constituent atoms. Because of the Coulomb barrier due to the electric repulsion between nuclei (potential \eqref{VkN}, Figure \ref{FigPotNucleus}, right), most of these collisions are elastic but sometimes tunneling through the barrier can occur and bring a dark nucleus \pplus~into the region of the potential well present at smaller distance, due to the exchange of $\sigma$ and \hd~between \pplus~and the nuclei of the detector (potential \eqref{VmN}, Figure \ref{FigPotNucleus}, left). There, electric dipole transitions produce the de-excitation of the system to bound states in the keV range by emission of photons that produce electron recoils, causing the signals observed by DAMA or CoGeNT. At this point, only the interaction between nuclei (potential \eqref{VN}) is considered to calculate the capture cross section, since it dominates at small distance ($r\lesssim1$ \AA) and because the long-range part of the atom-dark atom potential \eqref{Vat} is negligible and does not affect the initial diffusion eigenstate.

As the total incident energy $E$ ($>0$) in the center-of-mass frame of the atom-dark atom system is low (thermal), the $s$-wave is the dominant term in the expansion of the incident plane wave into partial waves. Because of this and because of the spin independence of the interaction, magnetic dipole transitions are subdominant and the dominant type of capture is electric dipole (E1) \cite{Segre}. Therefore, the final state of energy $E_p$ ($<0$) has to be a $p$-state due to the selection rules of E1 transitions and the capture causes the emission of a photon of energy $|E-E_p|\simeq |E_p|$.

The transition probability per unit time for an electric multipole radiation of order $a\in \mathbb{N}^{\geq 1}$ is given by \cite{Segre}:

\begin{equation}
\lambda(a,b)=\frac{8\pi(a+1)}{a\left[(2a+1)!!\right]}z^{2a+1}\left|Q_{ab}\right|^{2},\label{lambdaab}
\end{equation}
where $b=-a,...,a$, $z$ is the angular frequency of the emitted radiation and the matrix element:

\begin{equation}
Q_{ab}=\sum_{j=1}^{N}\int e_j r_{j}^{a}Y_{a}^{b*}(\theta_{j},\varphi_{j})\psi_{f}^{*}\psi_{i}\dd\vec{\tau}.\label{Qab}
\end{equation}
The sum in \eqref{Qab} is over the $N$ electric charges $e_{j}$ of the system and the spherical harmonics $Y_{a}^{m}$ are evaluated at the positions $\vec{r}_j$ of each of them. $\psi_{i}$ and $\psi_{f}$ are respectively the initial and final states of the whole system and are, in general, functions of $\vec{r}_1,\vec{r}_2,...,\vec{r}_N$. The volume element $\dd \vec{\tau}$ is therefore defined as $\dd \vec{\tau}=\dd \vec{r}_1 \dd \vec{r}_2 ... \dd \vec{r}_N$.

In the framework of this model, one has for the E1 capture from an $s$-state in the continuum to a bound $p$-state, expressed in the center-of-mass frame of the nucleus-\pplus~system in terms of the relative position $\vec{r}$ between the nucleus and \pplus:

\begin{eqnarray}
\lambda(1,b) & = & \frac{16\pi}{9}z^{3}\left|Q_{1b}\right|^{2},\label{lambda1b}\\
Q_{1b} & = & Ze\left(\frac{\mpp}{m_N+\mpp}\right)\int rY_{1}^{b*}(\theta,\varphi)\psi_{f}^{*}\left(\vec{r}\right)\psi_{i}\left(\vec{r}\right)\dd\vec{r},\label{Q1b}
\end{eqnarray}
where $Z$ is the charge of the nucleus and $m_N$ its mass. In principle, both the nucleus and \pplus~participate in $Q_{1b}$ to the emission of the photon, but the factor $\epsilon$ from the coupling of \pplus~brings a factor $\epsilon^2$ in the transition probability $\lambda(1,b)$, so that the term due to \pplus~has been neglected. The initial and final states are expressed as:

\begin{eqnarray}
\psi_{i}(\vec{r}) & = & \frac{1}{k}R(r),\label{psii}\\
\psi_{f}(\vec{r}) & = & R_{p}(r)Y_{1}^{m}(\theta,\varphi),~m=-1,0,1\label{psif}
\end{eqnarray}
where $R(r)$ is the radial part of the initial diffusion eigenfunction, of energy $E$, and $R_p(r)$ is the radial part of the wave function of the final bound $p$-state. $R(r)$ is obtained by solving numerically the radial Schr\"odinger equation at relative angular momentum $l=0$ for each energy $E$, while $R_p(r)$ and $E_p$ are calculated via the WKB approximation applied to the radial Schr\"odinger equation at $l=1$. $k=\sqrt{2\mu E}$ is the momentum of the incident plane wave, where $\mu=m_N \mpp/(m_N+\mpp)$ is the reduced mass of the nucleus-\pplus~system. The factor $\frac{1}{k}$ comes from the decomposition of a plane wave into partial waves.

We see in \eqref{psif} that there are three possible final states corresponding to the three possible values $m=-1,0,1$ of the magnetic quantum number $m$ when $l=1$, and these have to be summed in $Q_{1b}$ for each value of $b$. From the definition of the spherical harmonics, the integral over $\varphi$ in \eqref{Q1b} takes therefore the form $\sum_{m=-1}^{1}\int_0^{2\pi}e^{-ib\varphi}e^{-im\varphi}\dd\varphi$, each term being equal to zero unless $m=-b$. Since $b=-1,0,1$ as well, there is only one final state \eqref{psif} per value value of $b$ that contributes to $Q_{1b}$.

The link between the transition probability $\lambda(1,b)$ and the capture cross section $\sigma_{capt}(1,b)$ is made via the relation $\lambda(1,b)=\rho\sigma_{capt}(1,b)v$, where $\rho$ is the number density of incident particles and $v$ is the relative velocity between the nucleus and \pplus. The initial state \eqref{psii} is normalized in such a way that there is one incident particle per unit volume ($\rho=1$), by matching the function $R\left(r\right)$ with the asymptotically free amplitude. The final states \eqref{psif} are normalized by using the normalized spherical harmonics and by normalizing numerically the WKB solution in such a way that $\int R_{p}^{2}(r)r^{2}dr=1$. The E1 capture cross section $\sigma_{capt}$ is then obtained by summing the cross sections $\sigma_{capt}(1,b)$ corresponding to the three possible values of $b$ and one finally gets:

\begin{equation}
\sigma_{capt}=\frac{32\pi^{2}Z^{2}\alpha}{3\sqrt{2}}\left(\frac{\mpp}{m_N+\mpp}\right)^{2}\frac{1}{\sqrt{\mu}}\frac{(E-E_{p})^{3}}{E^{3/2}}\left|\mathcal{M}\right|^{2},\label{sigmacapt}
\end{equation}
where $\mathcal{M}=\int_{0}^{\infty}rR_{p}(r)R(r)r^{2}dr$. Note that if there are several levels $E_p$ in the spectrum at $l=1$, the cross section \eqref{sigmacapt} should be summed over all of them in order to get the total capture cross section, but due to the factor $(E-E_{p})^{3}\simeq |E_p|^3$, the lowest $p$-state gives the dominant contribution, as $|E_p|^3$ rapidly tends to zero when $E_p$ approaches zero from the negative side, and we choose to limit the calculation to that one.

After this capture, the system de-excites to an $s$-state of lower energy $E_s$ via a second E1 transition, which causes the emission of a second photon of energy $E_p-E_s$. To avoid the observation of double-hit events in the detector, which are rejected as we saw in Chapter \ref{DirectSearches}, we require that the first emitted photon be not seen, i.e. that its energy be below the threshold $E_{th}$ of the experiment (2 keV for DAMA and 0.5 keV for CoGeNT): $|E-E_p|\simeq |E_p|<E_{th}$. The signal is thus due to the second transition, which in principle can give rise to a spectrum as there are in general several $p$- and $s$-states in the well leading to several transitions of different energies. For simplicity, we shall assume here that the signal is due to the dominant transition, from the lowest $p$-state, of energy $E_p=E_1$, to the ground state, of energy $E_s=E_0$. In view of the energies of the signals of DAMA and CoGeNT, we require:

\begin{eqnarray}
\mathrm{2~keV}\leq & E_1-E_0 & \leq \mathrm{6~keV}~\mathrm{for~DAMA}\label{transitionDAMA} \\
\mathrm{0.5~keV}\leq & E_1-E_0 & \leq \mathrm{2.5~keV}~\mathrm{for~CoGeNT}\label{transitionCoGeNT}
\end{eqnarray}
while the condition related to the double-hit events gives:
\begin{eqnarray}
|E_1| & < & \mathrm{2~keV}~\mathrm{for~DAMA}\label{threshDAMA} \\
|E_1| & < & \mathrm{0.5~keV}~\mathrm{for~CoGeNT}\label{threshCoGeNT}
\end{eqnarray}

Thermal motion in a detector at temperature $T_{op}$ made of nuclei N gives rise to collisions between the N and \pplus~species and hence, as in equation \eqref{capturerate}, to the event counting rate per unit volume:

\begin{equation}
\Gamma(t)=n_{N}n(t)\left<\sigma_{capt}v\right>,\label{capturerate2}
\end{equation}
where $n$ is given by \eqref{neq} and \eqref{n0}, \eqref{nm}, and $n_N$ is the number density of nuclei (or atoms) in the detector. $<\sigma_{capt}v>$ is the thermally averaged capture cross section times the relative velocity $v$. Using Maxwellian velocity distributions at temperature $T_{op}$ in the frame of the detector, passing to center-of-mass and relative velocities $\vec{v}_{CM}$
and $\vec{v}$ and performing the integral over the center-of-mass variables, we get:

\begin{equation}
\Gamma(t)=8\pi n_{N}n(t)\frac{1}{(2\pi T_{op})^{3/2}}\frac{1}{\sqrt{\mu}}\int_{0}^{\infty}\sigma_{capt}(E)Ee^{-E/T_{op}}\dd E.
\end{equation}
Given the modulated form \eqref{neq} of the number density of \pplus, one gets a modulated expression for the event rate:

\begin{equation}
\Gamma(t)=\Gamma^{0}+\Gamma^{m}\cos(\omega(t-t_{0})),
\end{equation}
where the constant and modulated parts $\Gamma^{0}$ and $\Gamma^{m}$, when expressed in counts per day and per kilogram (cpd/kg), are given by: \begin{eqnarray}
\Gamma^{0} & = & Cn^{0}\int_{0}^{\infty}\sigma_{capt}(E)Ee^{-E/T{op}}\dd E,\label{gamma0} \\
\Gamma^{m} & = & Cn^{m}\int_{0}^{\infty}\sigma_{capt}(E)Ee^{-E/T_{op}}\dd E,\label{gammam} \\
C & = & \frac{9.71\times 10^{14}}{M_{mol}\sqrt{\mu}(T_{op})^{3/2}}\nonumber,
\end{eqnarray}
where $M_{mol}$ is the molar mass of the active medium of the detector in g/mol. $n^0$ and $n^m$ given by \eqref{n0} and \eqref{nm} have to be expressed in cm$^{-3}$, $\sigma_{capt}$ in GeV$^{-2}$ and $\mu$, $T_{op}$ and $E$ in GeV in order to get a rate in cpd/kg.

\section{Exploring the parameter space}

\subsection{Reproduction of the results from DAMA and CoGeNT}\label{subsecResultsMilli}

\begin{figure}
\begin{center}
\includegraphics[scale=0.55]{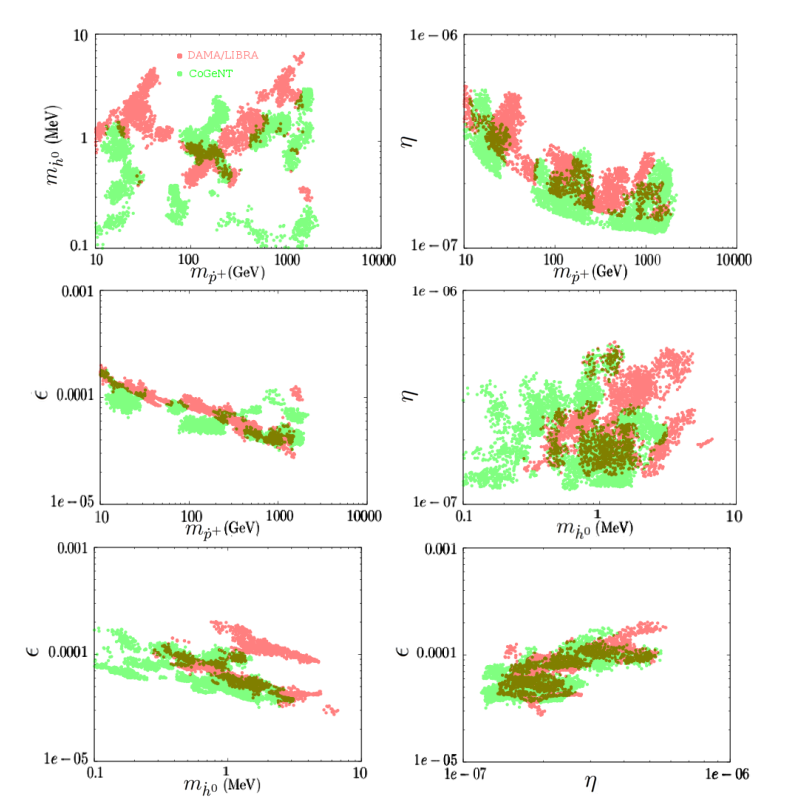}
\end{center}
\caption{Two-dimensional parameter regions reproducing the DAMA (light red) and CoGeNT (light green) results at the $2\sigma$ level. The
overlapping regions stand out in dark green. Top left: $(\mpp,\mhd)$ plane. Top right: $(\mpp,\eta)$ plane. Center left: $(\mpp,\epsilon)$
plane. Center right: $(\mhd,\eta)$ plane. Bottom left: $(\mhd,\epsilon)$ plane. Bottom right: $(\eta,\epsilon)$ plane.\label{FiGDAMACoGeNTResults}}
\end{figure}

The DAMA and CoGeNT experiments observe, as seen in Section \ref{secDAMA} and Section \ref{secCoGeNT} of Chapter \ref{DirectSearches}, integrated modulation amplitudes $\Gamma^m_{DAMA}=\left(0.0464\pm0.0052\right)$ cpd/kg and $\Gamma^m_{CoGeNT}=\left(1.66\pm0.38\right)$ cpd/kg in the energy intervals $(2-6)$ keV and $(0.5-2.5)$ keV respectively.

The four-dimensional parameter space of the model is explored separately for DAMA and CoGeNT in order to reproduce the observed rates and energy intervals at the $2\sigma$ level, which gives corresponding regions for each experiment. We use the isotopes $^{127}$I and $^{74}$Ge respectively for DAMA and CoGeNT, as their detectors are made of NaI and Ge crystals. The choice of the iodine component of the DAMA experiment, rather than $^{23}$Na, as already mentioned in Section \ref{secTowards}, is crucial since it allows to get rid of the formation of bound states with light elements, thus preventing the formation of anomalous heavy isotopes on earth and during Big Bang Nucleosynthesis. A direct consequence is that the collisions in the terrestrial crust are purely elastic.

For each set of parameter and for each experiment, the Schr\"odinger equation independent on time, with potential \eqref{VN} applied to the constituent nucleus, is first solved through the WKB approximation. This gives good approximations for the bound eigenvalues and eigenfunctions of the corresponding nucleus-\pplus~systems, the former allowing us to calculate the energy of the dominant transition (equations \eqref{transitionDAMA} and \eqref{transitionCoGeNT}). The modulated part $n^{m}$ of the number density of \pplus~in the detector is then
computed using \eqref{nm} before finally evaluating the modulated part of the event rate $\Gamma^{m}$ from \eqref{gammam}, at the operating
temperatures $T_{op}=300$ K for DAMA and $T_{op}=77$ K for CoGeNT. To compute the capture cross section $\sigma_{capt}$, given by \eqref{sigmacapt}, at a given energy $E$ in the center-of-mass frame of the nucleus-\pplus~system, one numerically solves the radial Schr\"odinger equation in the continuum to get the radial part $R(r)$ of the initial diffusion eigenstate and calculate the matrix element $\mathcal{M}$.

The regions are projected in two dimensions by combining all the possible pairs of parameters and are given in Figure \ref{FiGDAMACoGeNTResults} . For each model, one has ensured that the first emitted photon has an energy below the threshold of the considered experiment (conditions \eqref{threshDAMA} and \eqref{threshCoGeNT}), that thermalization occurs before $1$ km (equation \eqref{penLength}), that no bound states can form with elements characterized by $Z\leq14$ ($Z=14$ being silicon), and that thermalization at the edge of the CoGeNT detector requires a penetration length much shorter than the size of the detector. For the latter point, we have used \eqref{penLength} and \eqref{dEdx} with $E_{i}=\frac{3}{2}T_{room}$, where $T_{room}=300$ K is the initial room temperature, and $E_{f}=\frac{3}{2}T_{op}$, where $T_{op}=77$ K is the final temperature.

From the overlapping regions in the projected parameter spaces implying $\mpp$, we see that possible values for that parameter are between
$10$ GeV and $2$ TeV. The upper limit comes from the requirement that the penetration length must be less than $1$ km. Analyzing the regions where the parameter $\mhd$ is involved indicates that the values reproducing both the DAMA and CoGeNT experiments at the $2\sigma$ level must lie within $(0.4-3)$ MeV. In the same way, we find that $\eta$ ranges from $1.3\times 10^{-7}$ to $5\times 10^{-7}$ while $\epsilon$ goes from $3\times 10^{-5}$ to $2\times 10^{-4}$.

Note that, as CoGeNT is much colder than DAMA, one would expect that for the models reproducing the latter, the event rate in the former is well below the observed value, due to the Coulomb barrier through which the dark atoms have to tunnel. Two effects allow to compensate that: 1) germanium has a smaller $Z$ than iodine, so its Coulomb barrier is lower and tunneling is more efficient; 2) there are resonances in the capture cross section \eqref{sigmacapt} when the incident energy $E$ is close to the energy of a quasi-bound state of positive energy, so that the radial part of the diffusion eigenstate $R(r)$ is much less damped in the region of the well than out of a resonance, and thanks to this the rate of CoGeNT can be reproduced with parameters common to DAMA.

\subsection{Consistency with negative-result experiments}

For the models of Figure \ref{FiGDAMACoGeNTResults} to be acceptable, we have to ensure that they satisfy the constraints set by the experiments
that do not observe any signal, as XENON100, LUX, CDMS-II/Ge and CRESST-II. Since these are able to discriminate between nuclear and electron recoils and,
as seen in Chapter \ref{DirectSearches}, bound-state-formation events producing electron recoils in such detectors will be considered as backgrounds. Therefore, if some events remain, they should still have a smaller rate than the observed background. 

XENON100 and LUX have similar detectors, but LUX puts the strongest constraint with expected and observed electron-recoil backgrounds respectively of $(2.6\pm0.2_{stat}\pm0.4_{syst})\times 10^{-3}$ and $(3.1\pm 0.2_{stat})\times 10^{-3}$ cpd/kg/keV in the $(0.9-5.3)$ keV range. This leaves the possibility of an additional contribution to the expected background of at most $5.72\times 10^{-3}$ cpd/kg in that energy interval. Computing the constant part $\Gamma^{0}$ of the rate from \eqref{gamma0} for $^{132}$Xe and at the operating temperature $T_{op}=175$ K, and rejecting the models leading to higher rates, does not change the ranges of parameters previously found from the reproduction of the experiments with positive results.

This model predicts strongly suppressed event rates in cryogenic detectors, such as CDMS and CRESST, where temperatures $T_{op}\simeq 1$ mK give rise to much too low thermal energies for the dark atoms to tunnel through the Coulomb barrier and be captured. As an example, the rates computed with $^{74}$Ge at $T_{op}=1$ mK are effectively consistent with zero and are therefore in agreement with the negative results from CDMS-II/Ge and superCDMS. The same result can be extrapolated to CRESST.

However, the three nuclear recoils observed by CDMS-II/Si cannot be reproduced in the framework of this model based on electron recoils. For these, we should see if it is possible that the observed nuclear recoils may be misinterpreted background events occurring near the edge of the detector. Another possibility could be that they are due to a fluctuation of the background, since there is a probability of $5\%$ that three background events leak into the WIMP-search region.

\subsection{Constraints on milli-charges}\label{subsecConstraints}

Ref. \cite{Essig:2013lka} summarizes the constraints on milli-charged particles from direct laboratory tests and from cosmological and astrophysical observations. Very strong limits on $\epsilon$ can be obtained from the cooling of red giants and white dwarfs, from the alteration of the baryon-to-photon ratio during Big Bang Nucleosynthesis or from the invisible decay mode of orthopositronium into a pair of milli-charged particles, but they hold all together for masses smaller than 1 MeV, while we are interested here by masses $\mpp$ of \pplus~larger than 10 GeV. The mass $\mem$ of the light species \eminus~is not directly constrained by the direct searches but only the product of $\mem$ and $\dot{e}^2$ through the Bohr radius $\ad$ of the dark atoms. However, we can make the reasonable assumption $\dot{e}\simeq e$, in which case the value $\ad=1$ \AA~leads to $\mem\simeq 1$ MeV. In this case, accelerators put upper limits on $\epsilon$ which lie just in the interval deduced in Section \ref{subsecResultsMilli}. If it is so, we could therefore be close to a discovery of milli-charges in accelerators via the component \eminus. The same constraints are not a problem for the heavy species \pplus~since they let a large allowed window with $\epsilon<0.1$ for masses larger than 1 GeV. An additional constraint comes from the disruptions of the acoustic peaks of the Cosmic Microwave Background in presence of milli-charged particles \cite{Dolgov:2013una}. In view of the Planck data, it sets an upper limit on the cosmological density $\Omega_{mc}$ of milli-charges: $\Omega_{mc}h^2<0.001$ ($95\%$ C.L.), but is assumes that the milli-charged dark matter is fully ionized. This should be somewhat weakened here, as the oppositely charged particles form neutral atomic structures while only an ionized fraction remains. The fact that we are dealing with a subdominant dark matter also facilitates the satisfaction of this constraint.

On can derive a constraint on $\tilde{\eta}=\frac{\eta g_Y}{\dot{g}_Y}$ from unseen vector meson disintegrations. A direct consequence of the mass mixing term in \eqref{Lmix2} is that a certain fraction of $\sigma$ mesons can convert into \hd~scalars and then evade in the dark sector. This can be seen in the disintegrations of quarkonium states such as the $J/\psi$ meson and the $1s$ and $3s$ resonances of the $\Upsilon$ meson. The studied and unseen processes are generically represented by:

\begin{equation}
\begin{array}{ccccc}
Q\bar{Q} & \rightarrow & \sigma \bar{\sigma} & \rightarrow & \dot{h}^0 \bar{\dot{h}}^0, \\
Q\bar{Q} & \rightarrow & \gamma \sigma & \rightarrow & \gamma \dot{h}^0,
\end{array}\label{disintegrations}
\end{equation}
where $Q\bar{Q}=\Upsilon(1s)$, $\Upsilon(3s)$ or $J/\psi(1s)$. Because of the parity $-1$ of these states, the disintegration in two particles of parity $+1$ is forbidden, and one hence avoids the constraints from the first process. From \cite{Balest:1994ch}, \cite{Aubert:2008as} and \cite{Insler:2010jw}, the 90$\%$ C.L. upper limits on the branching ratios of the second process are respectively:

\begin{equation}
\begin{array}{lcc}
B\left(\Upsilon(1s)\rightarrow \gamma \dot{h}^0\right)& < & 5.6\times 10^{-5}, \\
B\left(\Upsilon(3s)\rightarrow \gamma \dot{h}^0\right)& < & 15.9\times 10^{-6}, \\
B\left(J/\psi(1s)\rightarrow \gamma \dot{h}^0\right)& < & 4.3\times 10^{-6}.
\end{array}\label{branchingConstraint}
\end{equation}
In the limit where the momenta of the constituent quarks are zero ($p=(m_{Q\bar{Q}}/2,\vec{0})$, where $m_{Q\bar{Q}}$ is the mass of the $Q\bar{Q}$ meson), we get:

\begin{equation}
\frac{B\left(Q\bar{Q}\rightarrow \gamma \dot{h}^0 \right)}{B\left(Q\bar{Q}\rightarrow e^{+}e^{-}\right)}=\frac{2\beta}{\alpha}\frac{m_{Q\bar{Q}}(m_{Q\bar{Q}}^2-\mhd^2)}{(m_{Q\bar{Q}}^2+2m_e^2)\sqrt{m_{Q\bar{Q}}^2-4m_e^2}}\frac{\tilde{\eta}^2(m_{\sigma}^2+\mhd^2)^2}{(\mhd^2-m_{\sigma}^2)^2},\label{branchingEquation}
\end{equation}
where $B\left(Q\bar{Q}\rightarrow e^{+}e^{-}\right)$ is the branching ratio of the disintegration of $Q\bar{Q}$ into an electron-positron pair and $m_e$ is the mass of the electron. $B\left(Q\bar{Q}\rightarrow e^{+}e^{-}\right)=(2.38\pm 0.11)\%$, $(2.03\pm 0.20)\%$ and $(5.94\pm 0.06)\%$ \cite{Amsler:2008zzb}, respectively for $Q\bar{Q}=\Upsilon(1s)$, $\Upsilon(3s)$ and $J/\psi(1s)$. Putting together \eqref{branchingConstraint} and \eqref{branchingEquation}, one gets allowed regions for parameters $\tilde{\eta}$  and $\mhd$ from processes \eqref{disintegrations}. But for the rather small values of $\mhd$ considered here, expression \eqref{branchingEquation} turns out to be independent of the mass of the scalar particle and the most stringent constraint comes from the disintegration of $J/\psi(1s)$:

\begin{equation}
\tilde{\eta}<1.2\times 10^{-4}.\label{etacons}
\end{equation}
In the same way as for $\mem$, \eqref{etacons} is not directly applicable to $\eta$ since it is not constrained by the direct searches, but a reasonable choice would consist in posing $\dot{g}_Y=g_Y$. In this case, the constraint translates directly to $\eta$ and we see that all the previous models satisfy it easily, by two or three orders of magnitude.

\section{On the realization of a $\sigma$-\hd~mass mixing}

The $\sigma$-\hd~mass mixing in \eqref{Lmix2} is clearly written in the framework of an effective theory, and one should find a fundamental mechanism that reduces to it in the low-energy limit. The exchange of a Higgs boson, mixed with a new scalar particle, gives a coupling that is several orders of magnitude too small to reproduce the results of the direct-search experiments. One could also imagine a mechanism similar to the one proposed in \cite{Holdom:1985ag} for the $\gamma$-\gammad~kinetic mixing , where the $\sigma$ and the \hd~states (\hd~being therefore seen as a composite object) emit respectively two $SU(3)$ gluons and two $SU(3)_d$ dark gluons, which are connected to a loop of heavy quarks charged under $SU(3)\times SU(3)_d$. The quark loop is actually a box which is proportional to $\frac{1}{M^4}$, where $M$ is the mass of the heavy quark. Therefore, if we assume a quark-gluon coupling of the same order as the standard one, we find that values of $M$ of the order of 1 GeV are needed, which is not allowed experimentally.

While looking for an acceptable mechanism, another possibility is to simply remove the mass mixing and formulate a new model that can do without it. This will be the purpose of Chapter \ref{AntiH}.

\chapter{Dark anti-atoms}\label{AntiH}

In this Chapter, relating paper \cite{Wallemacq:2014sta}, I show that the existence of a subdominant form of dark matter, made of dark ``anti-atoms'' of mass $m_{\dot{p}^{-}}\simeq 1$ TeV and size $\dot{a}_0\simeq 30~\mathrm{fm}$, can explain the results of direct detection experiments, with a positive signal in DAMA and no signal in other experiments. The signal comes from the binding of the dark anti-atoms to thallium, a dopant in DAMA, and is not present for the constituent atoms of other experiments. The dark anti-atoms are made of two particles oppositely charged under a dark $U(1)_d$ symmetry and can bind to terrestrial atoms because of a kinetic mixing between the photon and the massless dark photon, such that the dark particles acquire an electric milli-charge of value $\simeq 5\times 10^{-4}e$. This milli-charge enables them to bind to  high-$Z$ atoms via radiative capture, after they thermalize in terrestrial matter through elastic collisions.

\section{Modifying the milli-interacting model}\label{Modifying}

In Chapter \ref{Millicharges} about milli-interacting dark matter, we saw that a mass mixing between the standard $\sigma$ meson and a dark scalar could give a good framework to address the results of the direct-search experiments. While looking for a theory to interpret the effective limit represented by the second term of the mixing Lagrangian \eqref{Lmix2}, we can try to avoid the problem by removing the mass mixing, which considerably simplifies the model since the parameters $\eta$ and $\mhd$ are not part of the model anymore.

As we want to keep the same ideas as in the previous model and in particular explain the direct searches by the formation of bound states between ordinary and standard matter, we have to find another way to produce an attractive interaction, which was due to the $\sigma$-\hd~mass mixing. A possibility is to change the signs of the charges in the dark atom, in which case the attraction is mainly due to the interaction between the positively charged standard nucleus and the negatively charged dark one. To do so, we could change the sign of the kinetic mixing parameter $\epsilon$, but we prefer to keep it positive and rather change the signs of the dark charges of \pplus~and \eminus: \pplus~acquires a charge $-\dot{e}$ and \eminus ~gets a charge $+\dot{e}$. The electric milli-charges become therefore ``anti''-milli-charges and \pplus~and \eminus~are respectively renamed \newcommand{\pminus}{$\dot{p}^{-}$}\pminus~and \newcommand{\eplus}{$\dot{e}^{+}$}\eplus. The bound states they form can hence be seen as dark anti-hydrogen atoms.

A consequence is that now the repulsion is due to the interaction between the nucleus of an atom and the electron distribution of the other, which produces a much smaller barrier in the atom-dark atom potential, leading to a greatly enhanced capture rate. In order to reach the rate of DAMA, we will have to use rather low values of the Bohr radius $\ad$ of the dark anti-atoms, since atomic capture cross sections depend on the typical size of the atoms. But the size of the attractive potential well between the nucleus and \pminus~is determined by $\ad$, since the charge of \pminus~is screened by the distribution of \eplus, which is of the order of the Bohr radius in the $1s$-state of a hydrogen-like configuration. It is then not guaranteed that opposite charges will have bound states in such a narrow well. For the bound states to exist, the potential must be deep enough, i.e. the atomic nucleus needs to have a high enough charge $Z$. Although DAMA is mostly made of sodium and iodine, it also contains traces of thallium ($Z=81$), at the $10^{-3}$ level, as we saw in Section \ref{secDAMA} (NaI(Tl) crystal). This is much heavier than the elements constituting the other detectors, and it will turn out that dark anti-atoms bind to thallium and not to the others, so that only DAMA will observe a signal.

As in the milli-interacting model, the signal will come from the emission of photons produced by the radiative captures of dark anti-atoms by atoms of the active medium, causing electron recoils. No bound states will exist with xenon and germanium, making the model consistent with the negative-result experiments (XENON100, LUX, CDMS-II/Ge, superCDMS and CRESST-II), but it is therefore clear that we will not be able anymore to explain the excess of events seen by CoGeNT, for which we will have to make a personal bias. The same will hold true for CDMS-II/Si, the three nuclear-recoil events of which are still incompatible with the phenomenology of this model.

Note that when we built the model, we paid a particular attention to the barrier due to the nucleus-\eplus~and electron-\pminus repulsions, in case it would have produced the same kind of effect as in the previous model despite its reduced height. As it is much smaller here, it is more challenging to estimate its effect exactly, so we needed to be as precise as possible in our description of the atomic structures. While the approximation with uniformly charged spheres of finite size of Chapter \ref{Millicharges} was sufficient, we rather adopted in the following a description based on electromagnetic form factors fitted on data in order to take the complex electron distributions of standard atoms into account.

\section{Dark sector}

In view of the dark sector presented in Section \ref{secDarkSect1} and of the discussion of Section \ref{Modifying}, we take a dark sector in which two fermions \pminus~and \eplus, of masses \newcommand{\mpm}{m_{\dot{p}^{-}}}$\mpm$~and \newcommand{\mep}{m_{\dot{e}^{+}}}$\mep$, have respective charges $-\dot{e}$ and $+\dot{e}$ ($\dot{e}>0$) under a dark $U(1)_d$. This gives the dark interaction Lagrangian:

\begin{equation}
\mathcal{L}_{int}^{dark}=\dot{e}\bar{\psi}_{\dot{p}}\gamma^{\mu}\dot{A}_{\mu}\psi_{\dot{p}}+\dot{e}\bar{\psi}_{\dot{e}}\gamma^{\mu}
\dot{A}_{\mu}\psi_{\dot{e}},\label{Lintdark2}
\end{equation}
where $\psi_{\dot{p}}$ and $\psi_{\dot{e}}$ are the fermionic fields that contain \pminus~and \eplus~and $\dot{A}$ is the vectorial field of the dark massless photon \gammad.

The dark anti-proton \pminus~and the dark positron~\eplus~bind to each other and form dark anti-hydrogen atoms. In that system, \pminus~plays the role of the nucleus, i.e. we assume $\mpm\gg \mep$. The mass $m_{\dot{p}^{-}\dot{e}^{+}}$ of the atom is $m_{\dot{p}^{-}\dot{e}^{+}}\simeq \mpm$, and its Bohr radius $\ad$ is given by $\ad=1/(\mep\dot{\alpha})$, where $\dot{\alpha}=\dot{e}^{2}/4\pi$.

In order to link non-gravitationally the dark and the visible sectors, the dark photon $\dot{\gamma}$ is kinetically mixed with the standard photon $\gamma$ through:

\begin{equation}
\mathcal{L}_{mix}=\frac{\tilde{\epsilon}}{2}F^{\mu\nu}\dot{F}_{\mu\nu},\label{Lmix3}
\end{equation}
where $F$ and $\dot{F}$ are the electromagnetic-field-strength tensors of $\gamma$ and \gammad~respectively and $\tilde{\epsilon}$ is the dimensionless parameter of the kinetic $\gamma$-\gammad~mixing and is assumed to be small compared with unity. The mixing Lagrangian \eqref{Lmix3} makes \pminus~and \eplus~behave like electric milli-charges of values $-\tilde{\epsilon}\dot{e}\equiv -\epsilon e$ and $+\tilde{\epsilon}\dot{e}\equiv +\epsilon e$, where $\epsilon$ is the mixing parameter redefined in terms of the electric charge of the proton $e$.

The three relevant parameters of the model are therefore $\mpm$, $\epsilon$ and $\ad$. Here also, the mass $\mep$ of the lighter species \eplus~is not constrained by the direct searches, but the value $\ad\simeq 30$ fm that will be of interest leads, if one assumes $\dot{e}\simeq e$, to $\mep\simeq 1$ GeV. Given this and the fact that $\mpm\simeq 1$ TeV and $\epsilon\simeq 5\times 10^{-4}$, it is clear that all the constraints on electric milli-charges discussed in Section \ref{subsecConstraints} are satisfied.

Similarly to the milli-interacting model, these dark anti-atoms constitute a kind of self-interacting dark matter with a cooling mechanism, which has therefore to be subdominant and is expected to form a dark disk. Here we allow the local dark matter density not to be realized at 100$\%$ by the dark disk and we denote by $f\in]0,1]$ the fraction of $\rho_{\odot}=0.3$ GeV/cm$^3$ that actually corresponds to dark anti-atoms. The rest of the dark matter in the galactic halo is made of collisionless particles that do not produce any nuclear recoil in underground detectors nor any interaction of any type.

Finally, the velocity distribution in the dark disk is neglected here just as in Chapter~\ref{Millicharges}, with the same argument that all the particles thermalize in terrestrial matter and end up with the same thermal distribution.

\section{Binding to very heavy elements}\label{Binding}

\begin{figure}
\begin{center}
\includegraphics[scale=0.7]{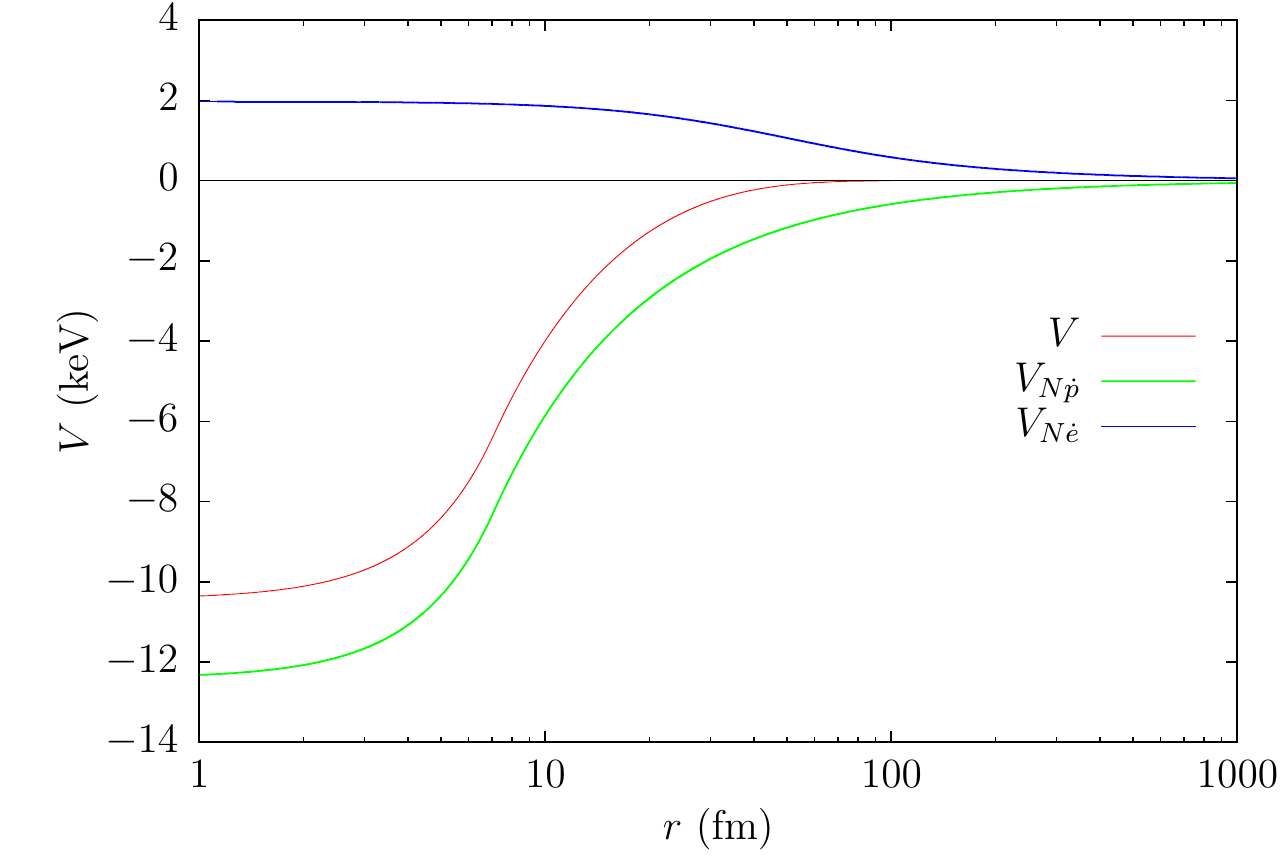}
\end{center}
\caption{Total thallium-dark atom potential $V$ (red) as a function of the radial distance $r$ for typical parameters, $\epsilon=5\times 10^{-4}$ and $\dot{a}_0=30$ fm. The two dominant terms, corresponding to the nucleus-$\dot{p}^{-}$ attraction $V_{N\dot{p}^{-}}$  (green) and to the nucleus-$\dot{e}^{+}$ repulsion $V_{N\dot{e}^{+}}$ (blue), are also shown.}
\label{atatpotential}
\end{figure}

The dark anti-atoms, after they have lost most of their energy in terrestrial matter, bind to the atoms of the active medium of a detector. At 
long distances, the atoms and dark anti-atoms are neutral, and the potential is zero. As the anti-atom approaches, and neglecting van der Waals 
forces which are extremely small, the electron cloud and the dark positron cloud start to overlap, without any Fermi repulsion as the electrons and dark positrons are different particles. This causes first a rather weak repulsive force, but as the two systems get closer, an attractive  interaction of the nucleus with the negatively charged $\dot{p}^{-}$ develops at distances close to the radius of the dark anti-atom, which will turn out to be of the order of $30$ fm to reproduce the event rate of DAMA. The attractive force reaches a maximum at distances of the order of the radius of the nucleus. Hence we have a rather narrow attractive potential, shown in Figure~\ref{atatpotential}, which will be the source of the capture cross section.

The first condition is that bound states must exist, and it is well known that the narrower the potential, the deeper it has to be. So in this case not only is the effective charge reduced by $\epsilon$, but also the narrowness of the potential implies that it must be deep, i.e. that the charge $Z$ of the nucleus must be large. The second condition will be that we capture the anti-atom on a $p$-bound state, starting from an $s$-state of the continuum, so that it can emit a photon during an electric dipole transition (E1), exactly as in the milli-interacting model (see Section \ref{subsecBstateFormation}). This in turn will require larger values of $Z$.

To determine precisely the  interaction potential between the atom and the dark anti-atom, we consider the four interacting charges. The milli-charges are easy to model: we take the dark anti-proton as a point charge $-\epsilon e$, and assume that the dark positron, of charge $\epsilon e$, is in a $1s$ hydrogen-like orbital. For the visible atom of mass number $A$, we take its nucleus as a uniform charge distribution of radius $R_N$(fm)$=1.2\times A^{1/3}$ and charge $+Ze$, and use explicit atomic form factors \cite{Formfact:2006} to treat the electron cloud. As the interaction between atom and dark anti-atom is rather weak, we assume that the charge structure is not modified during the interaction. The total atom-dark atom interaction potential is then the sum of four terms: the nucleus-$\dot{p}^{-}$ potential $V_{N\dot{p}^{-}}$, the nucleus-$\dot{e}^{+}$ potential $V_{N\dot{e}^{+}}$, the electron-$\dot{p}^{-}$ potential $V_{e\dot{p}^{-}}$ and the electron-$\dot{e}^{+}$ potential $V_{e\dot{e}^{+}}$:

\begin{equation}
 V(r)=V_{N\dot{p}^{-}}(r)+V_{N\dot{e}^{+}}(r)+V_{e\dot{p}^{-}}(r)+V_{e\dot{e}^{+}}(r),\label{eq:1}
 \end{equation}
where $r$ is the distance between the center of the nucleus and $\dot{p}^{-}$. This treatment of the atomic structures can be seen as an improvement of what had been made in Section~\ref{secInteractionPotentials}.

The first term in \eqref{eq:1} corresponds to the potential between a point charge and a uniform sphere, and is given by:

\begin{align}
V_{N\dot{p}^{-}}(r)&=-\frac{Z\epsilon\alpha}{2R_N}\left(3-\frac{r^{2}}{R_N^{2}}\right),~\mathrm{for}~r<R_N\nonumber \\
&=-\frac{Z\epsilon\alpha}{r},~\mathrm{for}~r\geq R_N\label{eq:2}
\end{align}
The other terms involve diffuse distributions and can be calculated through the use of form factors. It is shown in Appendix \ref{AppendixB} that 
the Fourier transform $\tilde V(\vec q)$ of the electrostatic potential $V(\vec D)$ between two charge distributions $\rho_1(\vec x)$ and $
\rho_2(\vec y)$ is related to their Fourier transforms $F_1(\vec q)$ and $F_2(\vec q)$ , i.e. to their form factors, through:

\begin{equation}
\tilde V(\vec q)=\frac{F_1(\vec q)F_2(-\vec q)}{q^2}\label{eq:3},
\end{equation}
where $\vec D$ is the position vector between two arbitrary points taken in each distribution and where $\vec x$ and $\vec y$ locate 
respectively the charges in the distributions $1$ and $2$ with respect to these arbitrary points. The Fourier variable $\vec q$ is the transferred 
momentum, i.e. the momentum of the exchanged photon. When the distributions are spherical, their form factors and the potential depend  
respectively only on $q=|\vec q|$ and $D=|\vec D|$, if the latter is taken as the distance between the centers of the two distributions.
One then gets, using \eqref{eq:3}:

\begin{align}
V(D) & = \int{\frac{\dd \vec q}{(2\pi)^3}\tilde V(q)}e^{-i\vec q \cdot \vec D}\nonumber \\
& = \frac{1}{2\pi^2}\int_{0}^{\infty}{\dd qF_1(q)F_2(q)\frac{\sin(qD)}{qD} }\nonumber \\
& = \frac{1}{i(2\pi)^2}\int_{-\infty}^{\infty}{\frac{\dd q}{qD}F_1(q)F_2(q)e^{iqD}},
%& = & \frac{1}{2\pi^2D}\int{\frac{dq}{q}F_1(q)F_2(q)\sin(qD)}
\label{eq:4}
\end{align}
by extending $q$ to the negative values in the last step.

The form factor of a point-like particle is simply the value of its charge, so that $$F_{\dot{p}^{-}}(q)=-\epsilon e.$$ That of a hydrogen-like distribution in the ground state, of Bohr radius $\dot{a}_0$ and charge $+\epsilon e$, is given by $$F_{\dot{e}^{+}}(q)=+\epsilon e \frac{16}{\left(4+\dot{a}_0^2q^2\right)^2}.$$ For the electronic distribution of the  atoms, Ref. \cite{Formfact:2006} fits the form factors to a sum of Gaussians. These are not directly useful, as their functional form  prevents the analytic calculation of equation (\ref{eq:4}). We thus refitted these form factors (to better than 2 \%) to three hydrogen-like form factors:

\begin{equation}F_{e}(q)=-Ze\sum\limits_{i=1}^3 c_i \frac{16}{\left(4+a_i^2q^2\right)^2},\label{eq:feq}
\end{equation}
with $c_3=1-c_1-c_2$. The best-fit parameters for several useful elements are given in Table \ref{Table1}. 

This enables us to perform analytically the integral \eqref{eq:4} by residues. We then get for the three last terms of \eqref{eq:1}:

\begin{eqnarray}
V_{N\dot{e}^{+}}(r) & = & +\frac{Z\epsilon \alpha}{r}\left(1-\left(1+\frac{r}{\dot{a}_0}\right)e^{-2r/\dot{a}_0}\right)\label{eq:5}, \\
V_{e\dot{p}^{-}}(r) & = & +\frac{Z\epsilon \alpha}{r}\sum\limits_{i=1}^3c_i\left(1-\left(1+\frac{r}{a_i}\right)e^{-2r/a_i}\right)\label{eq:6}, 
\\
V_{e\dot{e}^{+}}(r) & = & -\frac{Z\epsilon \alpha}{r}\sum\limits_{i=1}^3c_i\left(1+\frac{1}{2}\frac{e^{-2r/a_i}S_i(r)+e^{-2r/\dot{a}_0}
T_i(r)}{U_i} \right)\label{eq:7},
\end{eqnarray}
where $S_i(r)$ and $T_i(r)$ are order-1 polynomials in $r$ and $U_i$ is a constant depending on $\dot{a}_0$ and $a_i$, as defined in 
Appendix \ref{AppendixC} where the development for \eqref{eq:7} is shown as an example. Note that in \eqref{eq:5}, we have assumed that the   
nucleus is point-like, so that we can analytically integrate. This is a good approximation: as the N-$\dot{e}^{+}$ potential is subdominant, the 
approximation changes the total potential by less than 3 \% as long as $\dot{a}_0\geq 20 ~{\rm fm}$.

\begin{table}
\begin{center}
\begin{tabular}{l|c|c|c|c|c}
\hline
\hline
 \multicolumn{1}{c|}{$Z$} & $a_1$ (\AA) & $a_2$ (\AA) & $a_3$ (\AA) & $c_1$ & $c_2$ \\
\hline
\hline
 11 (sodium $1+$) & 0.04052 & 0.1961 & 0.2268 & 0.1639 & -1.1537 \\
 14 (silicon) & 0.1777 & 0.9300 & 0.001500 & 0.6295 & 0.2994 \\
 18 (argon) & 0.1082 & 0.0001260 & 0.5452 & 0.4567 & 0.02783 \\
 26 (iron) & 0.3125 & 0.05739 & 1.4872 & 0.6531 & 0.2878 \\
 32 (germanium) & 0.2157 & 0.03320 & 0.9148 & 0.6876 & 0.1834 \\
 53 (iodine 1-) & 0.9232 & 0.2728 & 0.06153 & 0.1385 & 0.4776 \\
 54 (xenon) & 0.05898 & 0.2543 & 0.7585 & 0.3755 & 0.4690 \\
 81 (thallium) & 0.01509 & 0.1199 & 0.4176 & 0.1454 & 0.5336 \\
 82 (lead) & 0.01880 & 0.1193 & 0.4171 & 0.1505 & 0.5264 \\
 \hline
 \hline
\end{tabular}
\end{center}
\caption{Best-fit parameters for the electronic form factors of several relevant atoms or ions. We consider Na$^{+}$ (sodium 1+) and I$^{-}$ 
(iodine 1-) as they are present under their ionized forms in the NaI(Tl) crystals of DAMA.}
\label{Table1}
\end{table}

The atom-dark anti-atom potential $V$ is therefore the sum of two attractive terms, $V_{N\dot{p}^{-}}$ and $V_{e\dot{e}^{+}}$, and two repulsive ones, $V_{N\dot{e}^{+}}$ and $V_{e\dot{p}^{-}}$. We show the total potential and the two dominant terms in Figure \ref{atatpotential} for a thallium atom and for the typical parameters $\epsilon=5\times 10^{-4}$ and $\dot{a}_0=30$ fm. We see that the potential is zero at large distances, as the two neutral atomic structures are well separated. As the electron cloud starts to merge with the dark positron one, the repulsion between the nucleus of the atom and the dark positron cloud that is located between the nucleus and the dark anti-proton induces a very small potential barrier, within the thickness of Figure \ref{atatpotential}. After the nucleus enters the dark positron cloud, the potential becomes attractive. For $r\leq \dot{a}_0$, the interaction between the nucleus and the dark anti-proton $V_{N\dot{p}^{-}}$ dominates. 

We numerically solve the radial Schr\"odinger equation for the wave function of the atom-dark anti-atom relative motion with the potential 
$V(r)$ of equation \eqref{eq:1} to find the bound states. As usual, we take the wave function to be $\psi_{n,l,m}(\vec r)=\frac{u_{n,l} (r)}{r}Y_l^m(\theta,\phi)$. For a state of a given $l$, one knows that for $r\rightarrow 0$, $u_{n,l} (r)\rightarrow N_0 r^{l+1}$. Asymptotically, for $r\rightarrow\infty$, one knows that $u_{n,l} (r)\rightarrow N_\infty \exp(-\sqrt{-2E\mu} r)$, with $\mu$ the reduced mass of the atom-dark anti-atom system and $E<0$ the energy. The energy equals the potential at the turning point, which we shall call $r=a_t$, and it defines two regions. To fix the constants $N_0$ and $N_\infty$, we calculate the wave function numerically, using a Runge-Kutta-Dormand-Prince (RKDP \cite{RKDP}) method, starting forward from 0 to $a_t$ for the small-$r$ region, and backwards from $r_{max}=5 a_t$ to $a_t$ for the large-$r$ region. At $a_t$, we impose that $u_{n,l}(r)$ is continuous, and we scan the value of the binding energy until the derivatives at $a_t$ are also symmetric, which gives us the energy that is solution. We finally normalize the wave function to unity. Note that the method has been checked to give results consistent with those obtained through the WKB approximation.

\begin{figure}
\begin{center}
\includegraphics[scale=0.7]{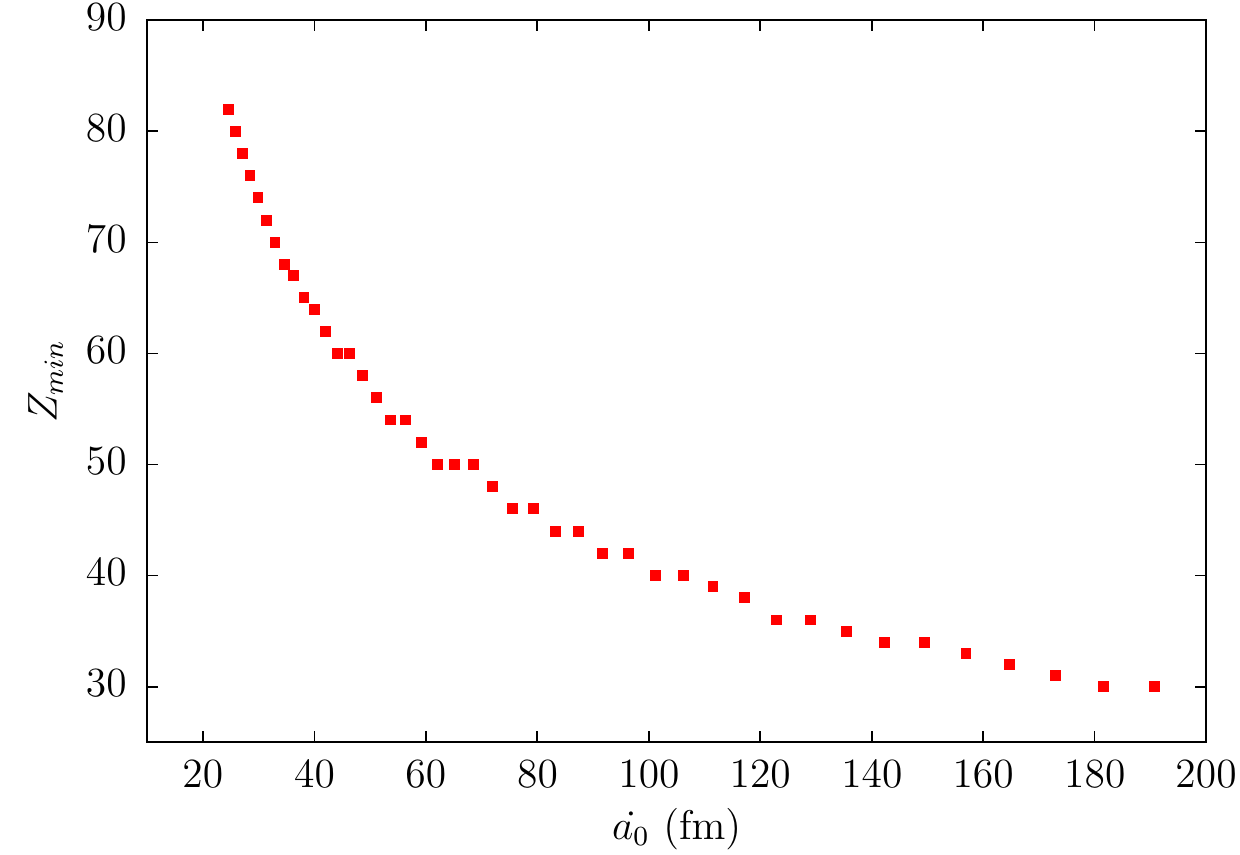}
\end{center}
\caption{Atomic number $Z_{min}$ from which at least one $p$-bound state exists, and hence radiative capture is possible, as a function of the 
Bohr radius $\dot{a}_0$ of the dark anti-atom, for $\mpm=1000$ GeV and $\epsilon=5\times 10^{-4}$.}
\label{Zmina0p}
\end{figure}

As we saw in Section \ref{subsecBstateFormation}, the radiative capture of the dark atoms by visible ones requires the existence of at least one $p$-state in the potential $V(r)$. For $\dot{a}_0$ between $20$ and $200$ fm  and for $\mpm=1000$ GeV and $\epsilon=5\times 10^{-4}$, we sought the first stable element $Z_{min}$ for which a $p$-bound state appears, and we show the result in Figure \ref{Zmina0p}. We see that $Z_{min}=74$ 
(tungsten) for $\dot{a}_0=30$ fm, showing that for compact dark anti-atoms, binding is possible only with very heavy elements.

\section{Thermalization and DAMA event rate}\label{secDAMArate}

Because of the orbital motion of the earth around the sun and of the sun around the galactic center, dark anti-atoms hit the surface of the earth 
continuously. Just as in the O-helium and milli-interacting models, when they penetrate the crust, dark anti-atoms undergo collisions with terrestrial atoms. As bound states can form only with very heavy elements and not with the rather light ones constituting the crust, these collisions are purely elastic. Only lead is sufficiently heavy ($Z=82$) and abundant ($10$ ppm) to be considered and its effect, which is in fact negligible, will be discussed in Section \ref{AnElem}. Due to the repeated collisions, the dark atoms deposit their energy in terrestrial matter until they thermalize completely. This happens before the dark atoms have reached the underground detectors, located at about $1$ km deep, and results in a descending cloud of dark anti-atoms driven by gravity towards the center of the earth. When that cloud enters a detector, the thermal energies are too low to give rise to nuclear recoils, but the emission of photons caused by the radiative capture of the dark anti-atoms by atoms of the active medium (thallium for DAMA) produces the signal. The constraints from the other experiments are avoided by the absence of bound states with their constituent atoms.

To simplify the problem, we use here also the result of Ref. \cite{Sigurdson:2004zp} saying that it is enough to approximate the crust of the earth as being made of pure silicon, with atomic and mass numbers $Z_{Si}=14$ and $A_{Si}=28$. In the Born approximation, we saw in equation \eqref{dsigmaFourier} that the differential elastic cross section $\frac{d\sigma}{d\Omega}$ is related to the Fourier transform of the interaction potential \eqref{eq:1} and so to the form factors of the silicon and dark anti-atoms as seen in \eqref{eq:3}:

\begin{eqnarray}
\frac{d\sigma}{d\Omega} & = & \frac{\mu^2_{Si}}{4\pi^2}\left|\int d\vec{r}~V(r) e^{-i\vec{q}\cdot \vec{r}}\right|^2\nonumber 
\\
& = & \frac{\mu^2_{Si}}{4\pi^2}\frac{(F_{N}(q)+F_e(q))^2(F_{\dot{p}^{-}}(q)+F_{\dot{e}^{+}}(q))^2}{q^4},\label{dsigma}
\end{eqnarray}
where $\mu_{Si}=m_{Si}\mpm/(m_{Si}+\mpm)$ is the reduced mass of the silicon-dark anti-atom system, $m_{Si}$ is the mass of a silicon 
nucleus and $\vec{q}$ is the transferred momentum. Here we take a point-like silicon nucleus, i.e. $F_N(q)=14e$, which is a very good approximation since $qR_N\ll 1$ for the whole thermalization process. The parameters for the electronic form factor $F_e(q)$ of a silicon atom are given in Table \ref{Table1}. In the center-of-mass frame, $q=2p\sin\theta/2$ as usual, where $p$ is the initial momentum and $\theta$ is the deflection angle with respect to the collisional axis.

The discussion about the thermalization and the drift in the terrestrial crust in the same as in Section \ref{secFromSpaceTo}, where equation \eqref{penLength}, together with equation \eqref{dEdx} for the energy loss $\frac{dE}{dx}$, allows us to calculate the penetration length $x$ with the condition that it is smaller than 1 km. After which the thermalized dark anti-atoms drift down towards the center of the earth at the drift velocity\footnote{The same statement as in the OHe and milli-interacting models holds true, i.e. there is a correction factor $\frac{\mpm}{m_{Si}}\simeq 40$ in the drift velocity $V_d$ that was used. Although larger capture cross sections are easily obtained in this model, there is a way to correct the results without changing the regions of Figure \ref{regions} and Figure \ref{regions_gold} that will be obtained in Section \ref{subsecResultsAntiH} and Section \ref{super}: as we allow dark-anti atoms to represent only a fraction $f<1$ of the local dark matter density, we can compensate the factor $1/40$ in the underground density of dark anti-atoms by a factor 40 in the factor $f$ contained in the incident flux at the surface.} $V_d$ similar to \eqref{Vdrift3} with the cross section \eqref{dsigma}. Typically, $V_d\simeq 1$ cm/s, so that a change in the incident flux at the surface is felt at most one day after for a detector located at a depth of $1$ km. The number density $n$ of dark anti-atoms in the detector is still obtained by balancing the incident flux at the surface of the earth with the down-drifting thermalized flux but, as we try here to reproduce the results of DAMA only (the other experiments being seen as with negative results, CoGeNT included), we evaluate the average flux at the precise location of DAMA, which brings a factor $\xi$ to be determined instead of the factor $\frac{1}{4}$ resulting from the average over all latitudes:

\begin{equation}
n_{\odot}\xi|\vec{V}_{\odot} + \vec{V}_{\oplus}|=nV_d,\label{xi}
\end{equation}
where $n_{\odot}$ (cm$^{-3}$)$=0.3f/\mpm$ (GeV) is the local number density of dark anti-atoms and $\vec{V}_{\odot}$, $\vec{V}_{\oplus}$ have been defined in Section \ref{OHeunderground}. The average suppression factor $\xi$ comes from the occultation of the dark matter stream by the surface of the earth and from the averaging of the cosine of the angle between the dark matter stream and the normal to the earth. Using the parameters of Ref. \cite{DAMAday}, we obtain an average value $\xi=0.5$ at the latitude of DAMA. Due to the periodic orbital motion of the earth around the sun, the norm $|\vec{V}_{\odot} + \vec{V}_{\oplus}|$ is annually modulated (equation \eqref{norm}) and we get a modulation of $n$, with a period of one year and a phase $t_0\simeq$ June 2 just as measured by DAMA, similar to \eqref{neq}-\eqref{nm} except that the factors $\frac{1}{4}$ in \eqref{n0} and \eqref{nm} have to be replaced by $\xi$.

Since there are no bound states with the sodium ($Z=11$) and iodine ($Z=53$) components of the DAMA detector, the signal is entirely due to the thallium dopant, present at the $10^{-3}$ level, i.e. with a number density $n_{Tl}=10^{-3}n_{NaI}$, where $n_{NaI}$ is the number density of the sodium iodide crystal. Thermal collisions within the detector between dark atoms and thallium give rise to a rate per unit volume for bound-state-formation:

\begin{equation}
\Gamma(t)=n_{Tl}~n(t)\left<\sigma_{capt}v\right>=\Gamma^0+\Gamma^m\cos(\omega (t-t_0)),
\label{eq:14}
\end{equation}
where $\sigma_{capt}$ is the radiative capture cross section and $v$ the relative velocity. This rate is also modulated just as $n$. The product $\sigma_{capt}v$ is thermally averaged over two Maxwell-Boltzmann velocity distributions, for thallium and the dark atoms that have entered the detector. Both distributions are for temperature $T_{op}=300$ K, as DAMA operates at room temperature.

The capture cross section in \eqref{eq:14} is obtained exactly in the same manner as had been presented in Section \ref{subsecBstateFormation}, i.e. it is given by \eqref{sigmacapt} particularized to thallium:

\begin{equation}
\sigma_{capt}=\frac{32\pi^{2}Z^{2}\alpha}{3\sqrt{2}}\left(\frac{\mpm}{m_{Tl}+\mpm}\right)^{2}\frac{1}{\sqrt{\mu}}\frac{(E-E_{p})^{3}}{E^{3/2}}\left|\mathcal{M}\right|^{2},\label{sigmacapt2}
\end{equation}
where $Z=81$ for thallium, $m_{Tl}$ is the mass thallium and $\mu=m_{Tl}\mpm/(m_{Tl}+\mpm)$ is the reduced mass of the thallium-dark anti atom system. We solve the radial Schr\"odinger equation for the incident energy $E$ in the center-of-mass frame by the RKDP method \cite{RKDP} to obtain the radial part $R(r)$ of the initial diffusion eigenfunction and with the method discussed in Section~\ref{Binding} to get $E_p$ and the radial part $R_p(r)$ of the wave function of the final $p$-bound state. The capture cross section is then obtained by computing the matrix element $\mathcal{M}=\int_{0}^{\infty} rR_p(r)R(r)r^2\dd r$. Note that here we improve the computation of $\sigma_{capt}$ with respect to the milli-interacting model as we sum over all the final $p$-states that are eventually present in the potential \eqref{eq:1}. After this first E1 transition, releasing a photon of maximum energy $|E-E_1|\simeq |E_1|$, where $E_1$ is the energy of the lowest $p$-bound state, a second E1 transition produces the emission of a photon of energy $E_1-E_0$, where $E_0$ is the energy of the ground state. We then use the conditions \eqref{transitionDAMA} and \eqref{threshDAMA} to obtain a signal in the energy window observed by DAMA without producing events that could be seen as double-hit events.

Finally, the expressions for the constant and the modulated parts $\Gamma^0$ and $\Gamma^m$ of $\Gamma$, in cpd/kg, are similar to \eqref{gamma0} and \eqref{gammam} except that the constant $C$ has now an additional factor $10^{-3}$ coming from the number density of thallium in the NaI crystal:

\begin{eqnarray}
\Gamma^{0} & = & Cn^{0}\int_{0}^{\infty}\sigma_{capt}(E)Ee^{-E/T{op}}\dd E,\label{gamma02} \\
\Gamma^{m} & = & Cn^{m}\int_{0}^{\infty}\sigma_{capt}(E)Ee^{-E/T_{op}}\dd E,\label{gammam2} \\
C & = & \frac{9.71\times 10^{11}}{M_{NaI}\sqrt{\mu}(T_{op})^{3/2}}\nonumber,
\end{eqnarray}
where  $M_{NaI}=150$ g/mol is the molar mass of NaI.

Note that in this model, we do not expect the capture rate to be suppressed in cryogenic detectors since there is no more Coulomb barrier, or at least the remaining barrier is very low. In colder detectors where $p\ad\ll 1$, the capture cross section is actually expected to rise as $\sigma_{capt}\propto 1/v$, so that the rate, which is in $\sigma_{capt}v$ from \eqref{eq:14}, is constant at low relative velocity. The average in \eqref{eq:14} does not change this result since from \eqref{gamma02} or \eqref{gammam2}:

\begin{align*}
\Gamma \propto & \frac{1}{T_{op}^{3/2}}\int_0^{\infty} \sqrt{E} e^{-E/T{op}}\dd E \\
&  \frac{1}{T_{op}^{3/2}}\times \frac{1}{2} \sqrt{\pi} T_{op}^{3/2},
\end{align*}
which is independent on $T_{op}$. We have used the fact that $E=\frac{1}{2}\mu v^2$, so that $\sigma_{capt}\propto 1/\sqrt{E}$. We checked numerically that, for some detector composition, the capture rate reaches a constant lower value when $T_{op}\rightarrow 0$.

\subsection{Results}\label{subsecResultsAntiH}

\begin{figure}
\begin{minipage}{0.48\linewidth}
\centerline{\includegraphics[width=1\linewidth]{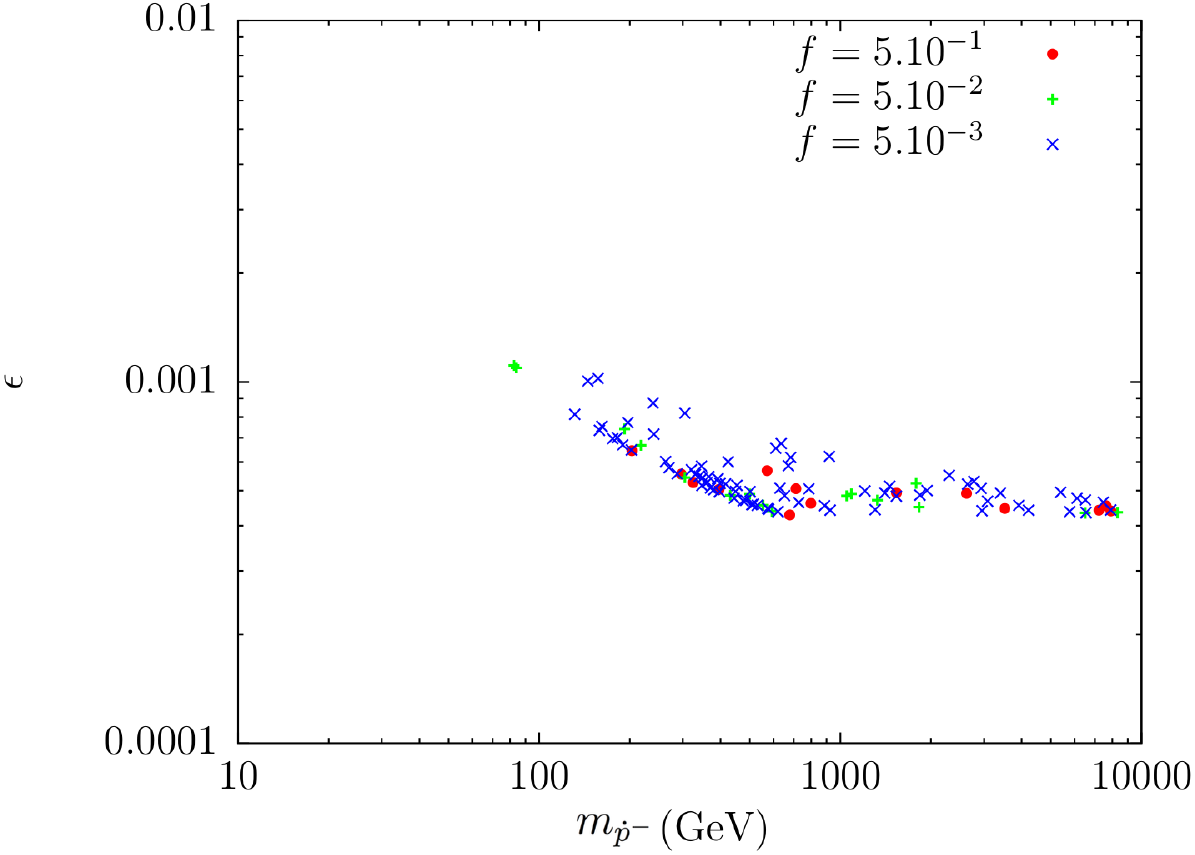}}
\end{minipage}
\begin{minipage}{0.48\linewidth}
\centerline{\includegraphics[width=1\linewidth]{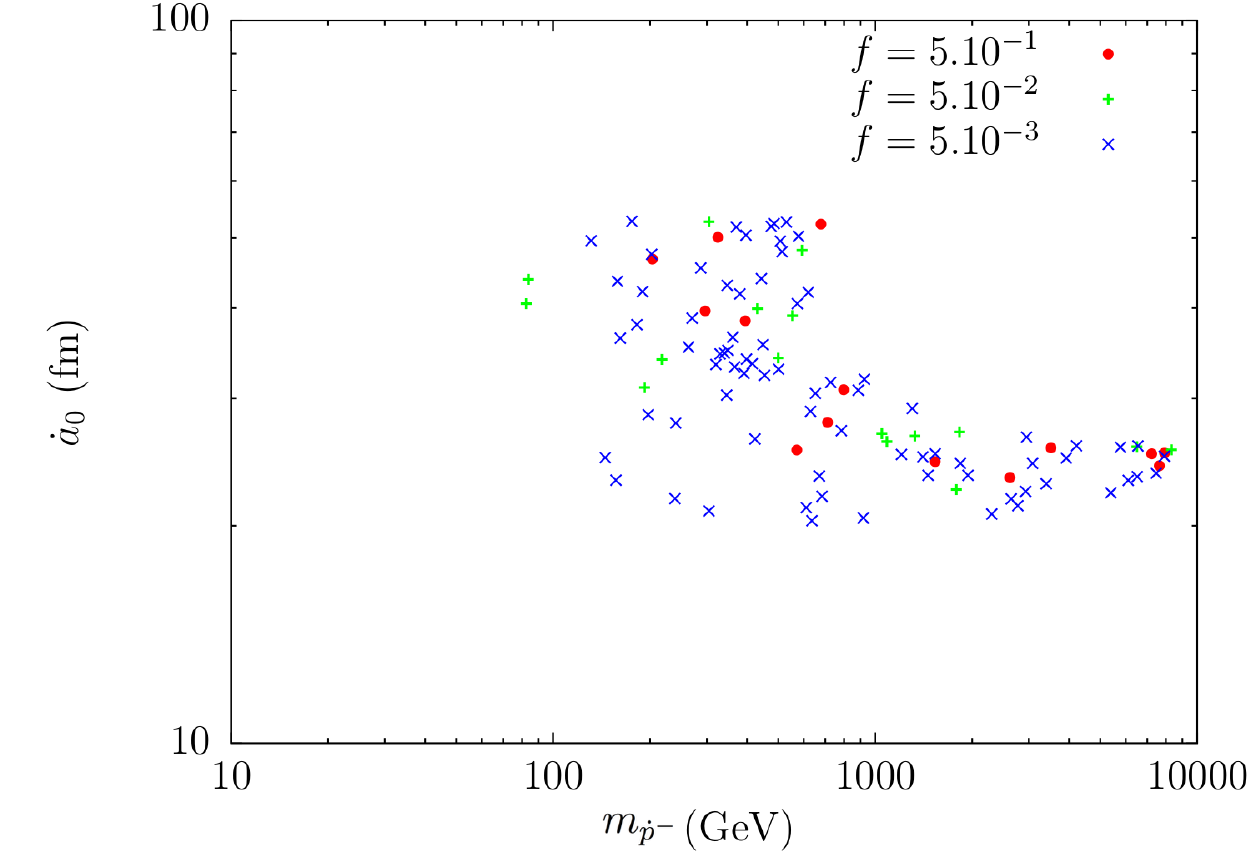}}
\end{minipage}
\begin{center}
\includegraphics[scale=0.6]{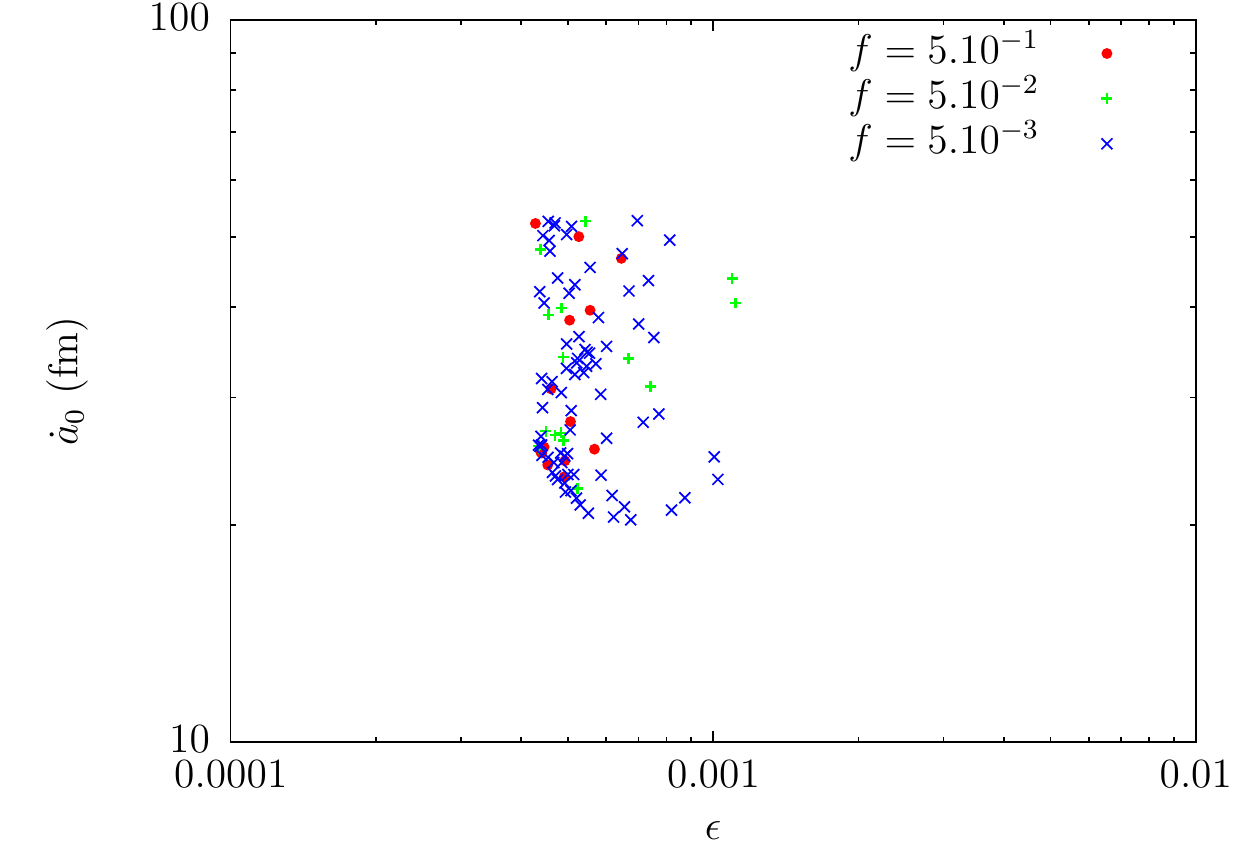}
\end{center}
\caption{Regions of the three-dimensional parameter space that reproduce the DAMA results with thallium at the $2\sigma$ level, projected on the $(\mpm,\epsilon )$ plane (top left), on the $(\mpm,\dot{a}_0 )$ plane (top right) and on the $(\epsilon,\dot{a}_0)$ plane (bottom center). The models corresponding to $f=5\times 10^{-1},5\times 10^{-2}$ and $5\times 10^{-3}$ are shown respectively with red dots, green crosses and blue crosses.}
\label{regions}
\end{figure}

To reproduce the rate observed by DAMA at the $2\sigma$ level, we randomly sampled the $(\mpm,\epsilon,\dot{a}_0)$ space for $1$ GeV $\leq \mpm\leq 50$ TeV, $10^{-5}\leq \epsilon \leq 10^{-2}$ and $10$ fm $\leq \dot{a}_0\leq 1$~\AA~and generated several millions of models. For each of them, we required the existence of at least one $p$-state with thallium, an energy for the first E1 transition below the DAMA threshold, a second E1 transition in the observed energy range, with the correct rate, thermalization before $1$ km and the absence of bound states with sodium and iodine. The parameters of the successful models are shown in Figure \ref{regions} for different fractions of dark anti-atoms, $f=5\times 10^{-1},5\times 10^{-2}$ and $5\times 10^{-3}$. 

The models that fulfill all the conditions are characterized by $100$ GeV $\leq \mpm\leq 10~\mathrm{TeV}$, $4\times 10^{-4}\leq \epsilon \leq 10^{-3}$ and $20$ fm $\leq \dot{a}_0\leq 50$ fm. The cutoff on $\mpm$ around $10~\mathrm{TeV}$ is due to the requirement of thermalization before $1$ km while the range of values of $\epsilon$ is a direct consequence of the existence of energy levels in the keV region, such that the transition lies between $2$ and $6$ keV. For $\dot{a}_0$, rather small values are needed to decrease the usually large atomic capture cross sections and reach the event rate of DAMA. As larger capture cross sections are easily obtained, we expect that, in the case of models where this is compensated by a 
smaller number of incoming particles ($f<1$), the regions in the parameter space are more extended and denser, which we verify in Figure  \ref{regions} by going from $f=5\times 10^{-1}$ to $f=5\times 10^{-2}$ and $f=5\times 10^{-3}$.

We also checked that there are no bound states with xenon, and hence with the elements of atomic numbers $Z\leq 54$, so that the negative results of CDMS-II/Ge and superCDMS (both detectors made of germanium) and XENON100 and LUX (both detectors made of xenon) can be naturally explained. In such a situation, the excess of events seen by CoGeNT cannot be reproduced, but an independent analysis of the data \cite{Davis:2014bla} has shown that it is possible that surface events at near-threshold energies have been misinterpreted as bulk events, so that the signal that has been reported by the collaboration could actually be due to background.

For each of the models of Figure \ref{regions}, there are no bound states with the oxygen and calcium components of the CRESST-II detector and some of them do not make it possible to bind to tungsten either. Those are therefore consistent with the null results of CRESST-II while the others bring a contribution to the electron-recoil background that should be studied in detail. For the latter case, the typical model $\mpm=1000$ GeV, $\epsilon=5\times 10^{-4}$, $\ad=30$ fm is an example, since the lightest element that admits a bound state in just tungsten ($Z_{min}=74$), as seen in Section \ref{Binding}. In order to clearly identify the regions such that $Z_{min}>74$, one should improve the statistics by increasing the number of trial models by a factor 10, which we postpone to further work.

\subsection{Absorption by lead and anomalous elements}\label{AnElem}
Models involving capture will always lead to anomalous heavy isotopes, to a corresponding decrease in the flux of dark matter, and to an eventual signal originating from the shielding of the detector. We shall now examine these effects, and show that they are not a problem, but may instead offer a way to test this model. In the following discussion, we concentrate on typical values of the parameters as in Figure \ref{Zmina0p}: $\mpm=1000$ GeV, $\epsilon=5\times 10^{-4}$ and $\dot{a}_0=30$ fm.

\subsubsection{Absorption in the terrestrial crust}

We have seen in Section \ref{Binding} that the dark anti-atoms bind only to very heavy elements. This can be seen in Figure \ref{Zmina0p} for the interesting values of $\dot{a}_0$. For the parameters that we chose, the first element that has bound states is tungsten, with $Z=74$. 
In the terrestrial crust, the only element beyond tungsten that is sufficiently abundant to be considered is lead, with $Z=82$, which is present at the level of about $10$ ppm. We shall first check that the absorption by this element on the way down through terrestrial matter does not 
significantly reduce the available dark matter flux.

Let us consider the case of a dark matter column above the DAMA detector with the same section. As the capture cross section is maximal at low energy, an upper bound on absorption can be obtained by taking all the dark atoms at the thermal energy, as if they thermalized as soon as they touched the surface of the earth. 

To estimate the reduction of the initial flux, we first need the flux of dark anti-atoms in the column. The available flux at the surface of the earth, for a fraction $f$ and dark atoms of mass $\mpm=1000$ GeV, is given by $\phi_i=n_{\odot}\xi V_{\odot}$ which must be equal to the thermalized down drifting flux $\phi_d(0)=n(0)V_d$, so that $$n(0)=3300f {\rm cm}^{-3},$$ if we take $V_d=1$ cm/s. $n(0)$ is here the 
available number density of thermalized dark atoms just below the surface.

The capture cross section can also be extracted directly from the data. DAMA observes a modulation amplitude of $0.0448$ cpd/kg, which corresponds to a constant part of the rate $\Gamma^0_{DAMA}$ of $0.672$ cpd/kg since $\Gamma^0/\Gamma^m\simeq15$, or to $2.85\times 10^{-8}$ counts/s/cm$^3$ using the density of sodium iodide $\rho_{NaI}=3.67$ g/cm$^3$. Thallium is present at the $10^{-3}$ level in the detector, i.e. $n_{Tl}=10^{-3}n_{NaI}=1.5\times 10^{19}$ cm$^{-3}$. The capture rate can be roughly expressed as $\Gamma^0_{DAMA}=n_{Tl}~n(0)\sigma_{capt}v_t$, where $v_t=\sqrt{\frac{8T_{op}}{\pi \mu_{Tl\dot{p}^{-}}}}= 1.7\times 10^4$ cm/s is the mean relative velocity between dark anti-atoms and thallium in the detector at temperature $T_{op}=300$ K. We can therefore access the capture cross section of the dark anti-atoms by thallium at thermal energy:

\begin{equation}
\sigma_{capt}=3.45\times 10^{-35}/f {\rm cm}^2.
\end{equation}

Since thallium and lead are very close elements ($Z=81$ and $Z=82$ and similar masses), we will assume that their capture cross sections are the same. $\sigma_{capt}$ will therefore be used along the whole thermalized column from the surface to the detector to estimate the absorption by lead. The variation of the down drifting dark matter flux $\phi_d=n(x)V_d$ at depth $x$ is given by:

\begin{equation}
 V_\dd \frac{\dd n(x)}{\dd x}=-n_{Pb}n(x)v_t \sigma_{capt},
\end{equation} 
from which we obtain $n(x)=n(0)\exp(-n_{Pb}\left(v_t/V_d\right)\sigma_{capt}x)$, where $n_{Pb}$ is the number density of lead in the terrestrial crust. Since its abundance is $10$ ppm, $n_{Pb}\sim 10^{17}$ cm$^{-3}$. As the DAMA detector is located at a depth $L=1$ km, we finally get the following estimate for the relative variation of the dark matter flux:

\begin{align}
\left|\frac{\phi_d(L)-\phi_d(0)}{\phi_d(0)}\right|&=\left|\frac{n(L)-n(0)}{n(0)}\right|\nonumber \\
&=1-e^{-n_{Pb}\left(v_t/V_d\right)\sigma_{capt}L}\simeq n_{Pb}\left(v_t/V_d\right)\sigma_{capt}L=5.87\times 10^{-9}/f.
\label{eq:19}
\end{align}
Clearly, the absorption by the earth is negligible: for the values of $f$ that we considered, it does not exceed $1.17\times 10^{-6}$, and this is why we considered elastic collisions only in the thermalization process.

\subsubsection{Absorption in the shield of the detector}

\begin{figure}
\begin{center}
\includegraphics[scale=0.7]{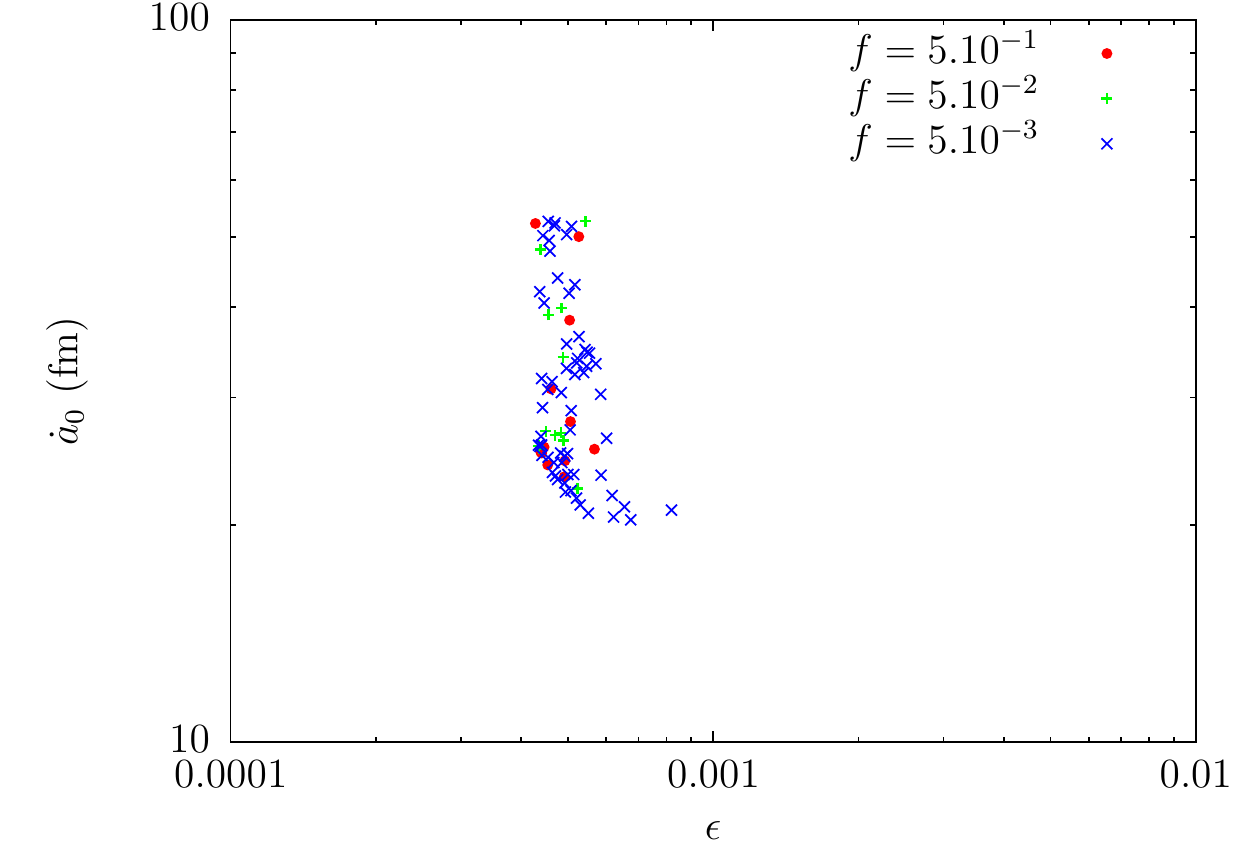}
\end{center}
\caption{Region of the three-dimensional parameter space that reproduces the DAMA results with thallium at the $2\sigma$ level, projected on the $(\epsilon,\dot{a}_0)$ plane, with the account for the constraint on $\mpm$ from anomalously heavy isotopes of gold: $\mpm>300$ GeV. The models corresponding to $f=5\times 10^{-1},5\times 10^{-2}$ and $5\times 10^{-3}$ are shown respectively with red dots, green crosses and blue crosses.}
\label{regions_gold}
\end{figure} 

Another source of lead, which could be more problematic, is the shield of the detector. Indeed, underground dark matter detectors are usually 
shielded with several layers of lead in order to isolate them from cosmic rays or environmental radioactivity, as we saw in Chapter \ref{DirectSearches}.

A column of rock of height $1$ km has a lead density of $3.5$ g/cm$^2$ while the layer of $15$ cm of lead on top of the DAMA apparatus has a density of $170.25$ g/cm$^2$. From \eqref{eq:19}, this corresponds to a relative correction of the event rate of $\frac{170.25}{3.5}\times5.87\times 10^{-9}/f=2.85\times 10^{-7}/f$, which is still negligible with respect to the experimental uncertainty, even for the smallest considered value of $f$. As the experimental error on the rate of DAMA is of the order of $10\%$ of the central measured value, absorption by lead in the shield could become important for $f<10^{-5}$.

Note that there are other materials constituting the shield of the detector, such as copper or cadmium, but these are lighter and therefore do not 
contribute to the reduction of the flux since they do not have any bound states with the dark anti-atoms.

\subsubsection{Limits from superheavy elements}\label{super}

Since the birth of the earth, dark anti-atoms have been binding to the heavy stable nuclei that constitute it ($Z\geq 74$ for our typical model), resulting in the accumulation of anomalous superheavy isotopes of known elements. To estimate their abundance today, we can use the formation rate of superheavy isotopes of thallium from DAMA: dark atoms bind to thallium at a rate of $0.672$ counts per day and per kilogram of NaI, i.e. $0.672\times \frac{150}{1000}=0.1$ counts per day and per mole of NaI. As in $1$ mole of components of the detector, $0.001$ is made of thallium, this gives $0.1\times1000=100$ superheavy isotopes of thallium formed per day and per mole of thallium. Therefore, over the whole history of the earth, which is $4.5\times 10^{9}$ years old, $100\times365\times4.5\times 10^{9}=1.64\times 10^{14}$ anomalous isotopes have been produced per mole, which gives an abundance of $\frac{1.64\times 10^{14}}{6.02\times 10^{23}}=2.7\times 10^{-10}$.

In Ref. \cite{PhysRevD.65.072003}, limits on the abundance of superheavy gold have been obtained by analyzing several terrestrial samples in an accelerator mass spectrometer. These limits depend on the mass $M$ of the superheavy isotope and get weaker as the mass increases. Since thallium and gold are close elements ($Z=81$ and $Z=79$ respectively), it is reasonable to use the abundance of superheavy thallium as an approximation for gold. For that abundance, $M>500$ GeV is required. As $M=m_{Au}+\mpm$, where $m_{Au}\simeq 200$ GeV is the mass of a gold atom, this gives $\mpm>300$ GeV from the search for anomalously heavy isotopes of gold.

Therefore, the two upper subfigures of Figure \ref{regions} should be viewed with a lower limit on $\mpm$ at $300$ GeV, while the bottom  subfigure of the $(\epsilon,\dot{a}_0)$ plane, taking into account this new constraint, becomes Figure \ref{regions_gold}.

\section{Some predictions}\label{secPredictions}

The model is easily falsifiable, as it lies on the edge of the existing limits for superheavy anomalous elements \cite{PhysRevD.65.072003}: the 
lower bound on $\mpm$ coming from the limits on the terrestrial abundances of superheavy isotopes of known elements. Furthermore, the addition of a heavy isotope in the active part of any direct detection experiment would dramatically increase the signal or the electronic noise. 

As the occultation of the flux of dark matter by the earth suppresses the flux, and as the angle of incidence also changes it, our model and the two previous ones have a diurnal effect, but the diffusion process through the earth to the detector will largely wash it out, and the event rate is compatible with that reported in \cite{DAMAday}. It could be however that the diurnal variation is observable if one averages not on a whole year, but on a fraction of a year.

Note that this model also implies some X-ray emission from the center of galaxies. The dark anti-atoms considered here have small sizes $\dot{a}_0$, about $1000$ times smaller than that of standard hydrogen. If we assume that $\dot{\alpha}\simeq \alpha =1/137$, then we expect their binding energy $\dot{E}_I=-\frac{1}{2}\frac{\dot{\alpha}}{\dot{a}_0}$ to be shifted from the eV region to the keV region. Collisions between dark anti-atoms in the central regions of dark halos, where the velocities and densities of dark matter are expected to be higher, will lead to the excitation of the dark atoms, and a fraction of them will de-excite through the emission of photons, as the dark current has a milli-charge $\epsilon e$. This will result in the emission of a weak thin line in the X-ray region. This is reminiscent of the recently observed 3.5 keV line \cite{Bulbul:2014sua,Boyarsky:2014jta}. However, a rough estimate of the emitted flux indicates that it is at least four times smaller than the reported one.

%\cleardoublepage

\chapter*{Conclusion}\label{Conclusion}
\markboth{\MakeUppercase{Conclusion}}{}
\addcontentsline{toc}{chapter}{Conclusion}
%\counterwithout{table}{chapter}

The direct searches for dark matter have provided challenging results. When interpreted in terms of Weakly Interacting Massive Particles (WIMPs) scattering off nuclei producing nuclear recoils in underground detectors, many of them excluded any observation and continue to strengthen their limits on WIMP parameters over a broad range of masses going from a few GeV up to 1 TeV in new runs or upgraded experiments, while some others keep on increasing the statistical significance of their signal.

In response to these tensions, alternative models have been developed in order to reinterpret the data, but none of them is, to our knowledge, able to reconcile all the experiments together in a single picture. In this thesis, we considered the possibility of composite dark matter, where dark matter particles form neutral bound states, or dark atoms, that interact with standard matter in non-trivial ways. The first model that has been considered was the O-helium scenario in Chapter \ref{OHe}, where stable particles O, of charge $-2$, bind to primordial helium just after Big Bang Nucleosynthesis to form hydrogen-like O-helium (OHe) dark atoms. This model features novel characteristics since the OHe atoms thermalize in the terrestrial crust by elastic collisions with atomic nuclei before they reach underground detectors. Then, they start drifting towards the center of the earth by gravity and arrive in detectors with thermal energies so that they cannot produce nuclear recoils. Instead, they form bound states with nuclei of the active medium by radiative capture, which causes the emission of photons that produce the signals by interacting with the atomic electrons, hence reinterpreting the direct-search experiments in terms of electron recoils.

The model could give a reconciliation framework through the existence of weakly bound states between OHe and some specific nuclei present in only some experiments and not in the others, as well as by the suppression of the capture rate in cryogenic detectors. It could also explain, as we have shown in this thesis, several indirect observations such as the excess in the electron-positron annihilation line from the galactic bulge observed by INTEGRAL or the unidentified 3.5-keV line seen in the XMM-newton data by OHe de-excitations after its excitations by OHe-OHe inelastic collisions in the central parts of galaxies. But all this would only be possible if we did not show that there is no stabilization mechanism in the interaction between OHe and ordinary nuclei. In this case, the O and/or He components can fall into the deep nuclear well of the nucleus where they can deposit large amounts of energy (tens of MeV), which is dramatic for direct searches. Moreover, we show that in that case, inelastic processes during the early universe lead to O-beryllium ions, which then capture free electrons and behave like heavy isotopes of helium, forming a kind of atomic dark matter with standard size and cross sections, which is totally excluded.

In view of the interesting features of the OHe scenario and the fact that it is not far from successful on many levels, we developed two new models, based on OHe and on other alternative models presented in the Introduction, in which milli-charged particles, oppositely charged under a dark $U(1)$ symmetry, bind and form hydrogen-like dark atoms. In the milli-interacting model, presented in Chapter \ref{Millicharges}, the milli-charges are obtained via two different mixings: a kinetic mixing between the standard photon and the dark massless photon and a mass mixing between the $\sigma$ meson and a dark scalar. In the dark anti-atom scenario, discussed in Chapter \ref{AntiH}, only the kinetic mixing between photons is present. Both scenarios are similar to the OHe one for the thermalization in the crust and the formation of bound states between thermal dark atoms and atoms in underground detectors, but the difference is that while OHe realized the full dark matter density, these dark sectors are subdominant and have to be considered together with a dominant fraction of collisionless particles that do not produce any nuclear recoil, mainly due to constraints on self-interacting dark matter.

The milli-interacting model can reproduce the annual modulations of the event rates of DAMA and CoGeNT in the keV range without any contradiction with the negative results of XENON100, LUX, CDMS-II/Ge, superCDMS and CRESST-II. This is principally done by playing on the detector composition for the existence of bound states and on the operating temperature to adjust the efficiency of tunneling through the Coulomb barrier of the atom-dark atom potential, allowing to reduce the capture rate in experiments such as XENON100 or LUX, which are colder, with respect to DAMA, which operates at room temperature, while they are made of almost the same nuclei.

While the attractive interaction allowing for the formation of bound states is due to the mass mixing in the milli-interacting model, it is only caused by the electrostatic interaction between the standard and the dark nuclei in the dark anti-atom scenario, but the signs of the charges inside the dark atoms have thus to be reversed, hence giving dark ``anti''-atoms. The mass mixing is therefore removed here, which gives a simpler model with fewer parameters. We could reproduce the positive results of DAMA and the negative results from XENON100, LUX, CDMS-II/Ge, superCDMS and CRESST-II. No signal is possible in CoGeNT, but this case is not clear since background events could have been misinterpreted as a signal. In case the signal is actually not present, the consequences on the milli-interacting model would be that it only relaxes the constraints that we derived on the corresponding parameters. The success of this simplified dark anti-atom model in reconciling DAMA with the negative-search experiments is due to the fact that bound states exist only with very heavy elements. Indeed, the rate of DAMA is due to the formation of bound states with its thallium dopant, present as an impurity in the NaI crystal, while no bound states exist with the elements that constitute the other detectors.

The table summarizes the orders of magnitude of several quantities common to the three models. All of them feature hydrogen-like dark atoms, made of a heavy and a light species of masses $m_{heavy}$ and $m_{light}$ and of typical size given by the Bohr radius. The components are directly coupled to the standard photon with a multiple of the charge $e$ of the proton (OHe) or via a mixing and hence with a fraction of it (milli-interacting dark matter and dark anti-atoms). The dark atoms have self-interactions with an elastic cross section denoted here by $\sigma_{dark-dark}$ and interactions with atoms of the terrestrial crust (silicon) with the elastic cross section $\sigma_{standard-dark}$, given at thermal energy. This makes them lose their energy until they thermalize at a depth $x$ below the surface. In underground detectors, they get captured by nuclei with a capture cross section $\sigma_{capt}$ given here for iodine (milli-interacting) or for thallium (dark anti-atoms) from DAMA and at room temperature (300 K). While the quantities above the intermediate horizontal line are deduced from direct-search experiments, the ones below that line, such as the mass of the light species and the binding energy of the dark atom, can only be derived by assuming a value for the dark charge (coupling of the dark particle to the dark massless photon), for which we took the standard value as an example (this does not apply to the OHe model which is entirely determined by known physics as soon as the mass of O is given).

\begin{table}
\begin{center}
\begin{tabular}{c|ccc}
& O-helium & Milli-interacting & Dark anti-atoms \\
\hline
\hline
$m_{heavy}$ & 1 TeV & 10 GeV - 2 TeV & 1 TeV \\
Size & 2 fm & 1 \AA & 30 fm \\
Charge & $2e$ & $3\times 10^{-5}e$ - $2\times 10^{-4}e$ & $5\times 10^{-4}e$ \\
$\sigma_{dark-dark}$ & $10^{-25}$ cm$^2$ & $10^{-15}$ cm$^2$ & $10^{-22}$ cm$^2$ \\
$\sigma_{standard-dark}$ (thermal) & $10^{-25}$ cm$^2$ & $10^{-20}$ cm$^2$ - $10^{-15}$ cm$^2$ & $10^{-24}$ cm$^2$ \\
$x$ & 10 m & 1 cm - 1 km & 100 m \\
%$V_d$ & 10 cm/s & & 1 cm/s \\
$\sigma_{capt}$ (thermal) & - & $10^{-38}$ cm$^2$ & $10^{-35}$ cm$^2$ \\
Fraction of DM density & 100$\%$ & $<5\%$ & $<5\%$ \\
\hline
Dark charge & - & $e$ & $e$ \\
$m_{light}$ & 3.7 GeV & 1 MeV & 1 GeV \\
Binding energy & -1.6 MeV & -10 eV & -10 keV \\ 
\hline
\hline
\end{tabular}
\end{center}
%\caption{Ranges and orders of magnitude of several relevant quantities common to the O-helium (second column), the milli-interacting (third column) and the dark anti-atoms (fourth column) models.\label{TablOrders}}
Ranges and orders of magnitude of several relevant quantities common to the O-helium (second column), the milli-interacting (third column) and the dark anti-atoms (fourth column) models.
\end{table}

But more than giving constraints on the parameters of a specific model, it has been shown in this thesis that it is possible, in the framework of dark matter models containing a sector with a richness and a complexity similar to ours, to reconcile experiments such as DAMA/LIBRA and XENON100 or LUX, that seem contradictory when interpreted in terms of WIMPs. This gives composite dark matter a particular importance and is encouraging for further studies, especially as these kinds of models are characterized by a number of typical signatures.

Focusing on milli-interacting dark matter and dark anti-atoms as the OHe scenario has been ruled out, we saw in Section \ref{super} that limits on the terrestrial abundances of anomalously heavy isotopes of known elements could lead, if they are improved, to a falsification of the models or to a discovery of milli-charges through the existence of superheavy isotopes of heavy elements. Note that these limits are weaker in the case of milli-interacting dark matter since the predicted formation rate of superheavy isotopes is at least 1000 times smaller than in the dark anti-atoms scenario because the dark atoms bind to iodine in DAMA and not to the thallium component present at the $10^{-3}$ level, meaning that the capture needs to be less efficient to reproduce the correct event rate. Therefore, taking the abundance derived in Section \ref{super} divided by 1000, we get $2.7\times 10^{-13}$, which in fact does not lead to any lower limit on the mass of the dark atoms from Ref. \cite{PhysRevD.65.072003} since for that abundance all the masses are allowed. An additional suppression factor actually comes in when transposing the formation rate of heavy iodine to higher-$Z$ elements such as gold, due to the higher Coulomb barrier that reduces the efficiency of tunneling.

As we already mentioned in Section \ref{secPredictions}, in the case of dark anti-atoms, the simple addition of a high-$Z$ element in the active part of any direct detection experiment would dramatically increase the signal or the electronic noise and this would be an indication of their existence. Indeed, as the full rate of DAMA is reproduced with a subdominant component only, the addition of substantial amounts of a high-$Z$ material would easily be visible provided that the increased background (because of additional radioactivity) is still well controlled. Also, going to a colder detector does not reduce the event rate, as shown in Section \ref{secDAMArate}. This would not necessarily be the case for milli-interacting dark matter, since the signals do not originate here from subdominant components but also because elements with a $Z$ higher than iodine get less captured because of an increased Coulomb barrier and bound-state-formation rates are suppressed in colder detectors.

Let us recall that, for simplicity, we assumed that the electric dipole (E1) transition that was responsible for the signals was the dominant one, i.e. from the lowest $p$-state to the ground state in potentials \eqref{VN} and \eqref{eq:1} for the milli-interacting and dark anti-atoms models respectively. This is clearly a simplification since, if there are several $p$- and $s$-states, all the allowed E1 transitions of different energies should in principle give rise to a line spectrum. Lowering the resolution of the detectors would allow to resolve these lines, which would be a typical signature of the models directly testable in earth-based experiments. Another particularity of these models is that at each bound-state-formation event, two photons are emitted, the first one being required to be below threshold in order to produce single-hit events. By lowering the thresholds typically below 1 keV (in electron equivalent units), one could seek this soft photon and the correlation with the second one in the keV range would betray the presence of dark (anti-) atoms. In the particular case of the milli-interacting model, there is a natural contribution to the expected electron background of the XENON100 and LUX experiments, even if the model is build to reduce it in order to be consistent with the observed one. Therefore, a precise temporal study of the electron background could highlight a modulation at energies close to the ones observed by DAMA, since DAMA is explained with its iodine component which is very similar to the xenon constituent of XENON100 and LUX.

We mentioned in Section \ref{secPredictions} that collisions between dark anti-atoms in the central regions of dark halos, where the velocities and densities of dark matter are expected to be higher, will lead to their excitation, and a fraction of them will de-excite through the emission of photons. The same statement holds for the dark atoms of the milli-interacting model and, in view of the value of the binding energies in the table, we expect the emission of forests of weak lines respectively in the keV and in the eV ranges. These could be looked for in X-ray and visible astronomy, even if the intensities of the lines are expected to be reduced at least by a factor $10^{-6}$ with respect to the background from baryonic matter, due to the $\epsilon^2$ factor present in the emission rate coming from the conversion of a dark photon into a standard one.

A last unavoidable effect of these models is the presence of a diurnal variation of the event rate in an underground detector. Indeed, as the occultation of the incident flux of dark matter by the earth suppresses the flux, and as the angle of incidence may also change it, we expect a diurnal modulation of the underground dark matter density due to the rotation of the earth around its own axis. But we already mentioned in Section~\ref{secPredictions} that diffusion processes through the earth to the detector largely wash this effect out at the latitude of DAMA, and  the event rate is compatible with that reported in \cite{DAMAday}. However, for some periods of the year, the dark matter flux passes a fraction of a day below the horizon, accentuating the modulation effect. By averaging not on a whole year, but on a fraction of a year, it could therefore be that the diurnal variation becomes observable.

In fact, the DAMA detector, as well as all the current detectors, are particularly well placed for the direct detection of this kind of thermalizing dark matter since they are all located in the northern hemisphere at latitudes between 42$^{\circ}$N and 47$^{\circ}$N, while the dark matter flux has a mean inclination of about 42$^{\circ}$ with the north pole. This means that during a rotation of the earth on itself, the dark matter flux is most of the time above the horizon and is not occulted, being therefore observable. This is why we predict, in upcoming experiments such as ANAIS \cite{Cuesta:2014vna,Amare:2013lca}, which will be the sister experiment of DAMA, using 250 kg of NaI(Tl) scintillators at the Canfranc Underground Laboratory in Spain (42$^{\circ}$N) to look for an annual modulation, similar results as for DAMA. However, things get more complicated as one goes towards the southern hemisphere, where other experiments are planned. To smaller latitudes, the incident flux is more and more occulted by the earth and the coefficient $\xi$ in equation \eqref{xi} that was used to quantify this effect decreases progressively before vanishing at a latitude of 50$^{\circ}$S. We therefore predict no signal in the DM-Ice experiment \cite{Cherwinka:2014xta}, i.e. the reproduction of the DAMA experiment at the south pole, due to a total occultation of the dark matter stream by terrestrial matter.

\appendix

\chapter[$e^{+} e^{-}$ production in the galactic bulge]{$e^{+} e^{-}$ production in the galactic bulge by OHe collisions}\label{AppendixA}

The studied reaction is the inelastic collision between two incident OHe atoms in their ground states $1s$ giving rise to an OHe in an excited $s$-state $ns$, while the other one remains in its ground state:
\begin{equation}
\mathrm{OHe}(1s)+\mathrm{OHe}(1s)\rightarrow \mathrm{OHe}(1s)+\mathrm{OHe}(ns).\label{OHecoll}
\end{equation}
If we work in the rest frame of the OHe that gets excited, and if we neglect its recoil after the collision, the differential cross section of the process is given by \cite{Landau}:

\begin{equation}
\mathrm{d}\sigma(1s\rightarrow ns)=2\pi\left|\left\langle ns,\vec{p'}|U|1s,\vec{p}\right\rangle \right|^{2}\delta\left(\frac{p'^{2}}{2m_O}+E_{ns}-\frac{p^{2}}{2m_O}-E_{1s}\right)\frac{\mathrm{d}\vec{p'}}{(2\pi)^{3}},\label{dsigma1sns}
\end{equation}
where $m_O$ is the mass of OHe, $\vec{p}$, $\vec{p'}$ are the momenta of the incident OHe before and after the collision, $E_{1s}$, $E_{ns}$ are the ground-state and excited-state energies of the target OHe and $U$ is the interaction potential between the incident and the target OHe's. As the incident OHe does not get excited, we shall neglect its internal structure, so that its wavefunctions are plane waves.  $\psi_{\vec{p}}$ is normalized to obtain a unit incident current density and the normalization of $\psi_{\vec{p'}}$ is chosen for it to be pointlike, i.e. the Fourier transform of $\delta^{(3)}(\vec {r})$~\cite{Landau}:

\begin{equation}
\begin{array}{cll}
\psi_{\vec{p}} & = & \sqrt{\frac{m_O}{p}}e^{i\vec{p}.\vec{r}},\\
\\\psi_{\vec{p'}} & = & e^{i\vec{p'}.\vec{r}},
\end{array}\label{normalization}
\end{equation}
where $\vec{r}$ is the position vector of the incident OHe and $p=\left|\vec{p}\right|$. As O is much more massive than the helium component, the origin of the rest frame of the target OHe coincides with the position of O.

For the purpose of this calculation, we describe the OHe that gets excited as a hydrogenoid atom, so that the energy levels $E_{ns}\equiv E^H_n=-\frac{1}{2}m_{He}\left(Z_{He}Z_{O}\alpha\right)^{2}/n^{2}$ as in Section \ref{secSpectrum} and the initial and final bound state wavefunctions are the ones of a hydrogenoid atom, denoted here by $\psi_{1s}$ and $\psi_{ns}$, with a Bohr radius $r_o$ given in \eqref{ro}. Note that $E_{1s}$ is the binding energy of the hydrogenoid OHe atom $E_o$ given in \eqref{EbindingOHe}.

The incident OHe interacts with the O and He components in the target OHe respectively with the interaction potentials $U_{O}$ and $U_{He}$, so that the total interaction potential $U$ is the sum of two contributions:

\begin{equation}
U(\vec{r},\vec{r}_{He})=U_{O}(\vec{r})+U_{He}(\vec{r}-\vec{r}_{He}),\label{Utot}
\end{equation}
where $\vec{r}_{He}$ is the position vector of the He component. The first term $U_{O}$ gives a zero contribution to the integral of expression
\eqref{dsigma1sns} since the states $\psi_{1s}$ and $\psi_{ns}$ are orthogonal. For the second term, we treat the incident OHe as a heavy neutron colliding on a helium nucleus through short-range nuclear forces. The interaction potential can then be written in the form of a contact term:

\begin{equation}
U_{He}(\vec{r}-\vec{r}_{He})=-\frac{2\pi}{m_{He}}r_o\delta(\vec{r}-\vec{r}_{He}),\label{UHe}
\end{equation}
where we have normalized the delta function to obtain an OHe-He elastic cross section equal to $4\pi r_o^{2}$ when using the Born approximation to link the elastic cross section to the potential (see equation \eqref{dsigmaFourier}).

%\begin{equation}
%\frac{\mathrm{d}\sigma_{He}}{\mathrm{d}\Omega}=\frac{m_{He}^2}{4\pi^2}\left|\int U_{He}(\vec{r}-\vec{r}_{He})e^{-i\vec{k}\cdot(\vec{r}-\vec{r}_{He})}\mathrm{d}(\vec{r}-\vec{r}_{He})\right|^2=r_o^2=\frac{\sigma_{He}}{4\pi},
%\end{equation}
%where $\vec{k}$ is the transferred momentum in the OHe-He collision.
 
Going to spherical coordinates for $\vec{p'}$ and integrating over $p'=|\vec{p'}|$ in the differential cross section \eqref{dsigma1sns}, together with the previous expressions \eqref{normalization}, \eqref{Utot} and \eqref{UHe}, we get:
\begin{equation}
\mathrm{d}\sigma(1s\rightarrow ns)=\left(\frac{m_O}{m_{He}}\right)^{2}r_{o}^{2}\left(\frac{p'}{p}\right)\left|\int e^{-i\vec{q}.\vec{r}_{He}}\psi_{ns}^{*}\psi_{1s}\mathrm{d}\vec{r}_{He}\right|^{2}\mathrm{d}\Omega ,\label{dsigma1sns2}
\end{equation}
where $\vec{q}=\vec{p'}-\vec{p}$ is the transferred momentum and $\mathrm{d}\Omega$ is the solid angle. From the integration over the delta
function in \eqref{dsigma1sns}, we have obtained the conservation of energy during the process:

\begin{equation}
p'^{2}=p^{2}+2m_O\left(E_{1s}-E_{ns}\right).\label{pp2}
\end{equation}
It leads to the threshold incident energy corresponding to $p'^{2}=0$ and hence to a minimum incident velocity $v_{min}=\sqrt{{2\left(E_{ns}-E_{1s}\right)/m_O}}$.
The previous expression for $p'$ allows us to express the squared
modulus of $\vec{q}$ as:
\begin{equation}q^{2}=2\left(p^{2}+m_O\left(E_{1s}-E_{ns}\right)-p\sqrt{p^{2}+2m_O\left(E_{1s}-E_{ns}\right)}\cos\theta\right),\label{q2}
\end{equation}
where $\theta$ is the deviation angle of the incident OHe with respect to the collision axis in the rest frame of the target OHe.

$e^{+} e^{-}$ pairs will be dominantly produced if OHe is excited to a $2s$-state, since the only de-excitation channel is in this case from $2s$ to $1s$. As $e^{+} e^{-}$-pair production is the dominant channel, the differential pair-production cross section $\mathrm{d}\sigma_{ee}$
is approximately equal to the differential collisional excitation cross section. By particularizing expression \eqref{dsigma1sns2} to the case $n=2$, one gets:

\begin{equation}
\frac{\mathrm{d}\sigma_{ee}}{\mathrm{d}\cos\theta}=512^{2}\left(\frac{2\pi m_{O}^{2}}{m_{He}^{2}}\right)r_{o}^{6}\left(\frac{p'}{p}\right)\frac{q^{4}}{2\left(4r_{o}^{2}q^{2}+9\right)^{6}}.\label{dsigmaee}
\end{equation}

The total $e^{+} e^{-}$-pair-production rate in the galactic bulge is therefore given by:

\begin{equation}
\Gamma_{ee}=\int_{V_{b}}\frac{\rho_{DM}^{2}\left(\vec{R}\right)}{m_{O}^{2}}\left\langle \sigma_{ee}v\right\rangle \left(\vec{R}\right)\mathrm{d}\vec{R},
\label{rateee}
\end{equation}
where $V_{b}$ is the volume of the galactic bulge, which is a sphere of radius $R_{b}=1.5$ kpc, $\rho_{DM}$ is the energy density distribution of dark matter in the galactic halo and $\left\langle \sigma_{ee}v\right\rangle $ is the pair production cross section $\sigma_{ee}$ times the relative velocity $v$ averaged over the velocity distribution of dark matter particles. The total pair-production cross section $\sigma_{ee}$
is obtained by integrating \eqref{dsigmaee} over the diffusion angle. Its dependence on the relative velocity $v$ is contained in $p$, $p'$
and $q$ through $p=m_Ov$ and the expressions \eqref{pp2} and \eqref{q2} of $p'$ and $q$ in terms of $p$.

We use a Burkert \cite{Burkert:1995yz} flat, cored, dark matter density profile known to reproduce well the kinematics of disk systems in massive spiral galaxies and supported by recent simulations including supernova feedback and radiation pressure of massive stars \cite{Maccio2012} in response to the core/cusp halo problem \cite{Flores:1994gz,Moore:1994yx}:

\begin{equation}
\rho_{DM}\left(R\right)=\rho_{0}\frac{R_{0}^{3}}{\left(R+R_{0}\right)\left(R^{2}+R_{0}^{2}\right)},\label{rhoBurkert}
\end{equation}
where $R$ is the distance from the galactic center. The central dark matter density $\rho_{0}$ is left as a free parameter and $R_{0}$ is determined by requiring that the local dark matter density at $R=R_{\odot}=8$ kpc is $\rho_{\odot}=0.3$ GeV/cm$^{3}$. The dark matter mass enclosed in a sphere of radius $R$ is therefore given by:

\begin{equation}
M_{DM}\left(R\right)=\rho_{0}\pi R_{0}^{3}\left\{ \log\left(\frac{R^{2}+R_{0}^{2}}{R_0^{2}}\right)+2\log\left(\frac{R+R_{0}}{R_{0}}\right)-2\arctan\left(\frac{R}{R_{0}}\right)\right\}.\label{MDM}
\end{equation}

For the baryons in the bulge, we use an exponential profile \cite{Gnedin:2004cx} of the form:

\begin{equation}
\rho_{B}\left(R\right)=\frac{M_{b}}{8\pi R_{b}^{3}}e^{-R/R_{b}},\label{rhoBaryons}
\end{equation}
where $M_{b}=10^{10}$ M$_{\odot}$ \cite{vandenBosch} is the mass of the bulge. This gives the baryonic mass distribution in the galactic
bulge:

\begin{equation}
M_{B}\left(R\right)=M_{b}\left\{ 1-e^{-R/R_{b}}\left(1+\frac{R}{R_{b}}+\frac{R^{2}}{R_{b}^{2}}\right)\right\}.\label{MB}
\end{equation}

We assume a Maxwell-Boltzmann velocity distribution for the dark matter particles of the galactic halo, with a velocity dispersion $u\left(R\right)$ and a cutoff at the galactic escape velocity $v_{esc}\left(R\right)$:

\begin{equation}
f\left(R,\vec{v}_{h}\right)=\frac{1}{C\left(R\right)}e^{-v_{h}^{2}/u^{2}(R)},\label{fRv}
\end{equation}
where $\vec{v}_{h}$ is the velocity of the dark matter particles in the frame of the halo and
\newcommand{\erf}{\mathrm{erf}} $C(R)=\pi u^{2}\left(\sqrt{\pi}u\, \erf(v_{esc}/u)-2v_{esc}e^{-v_{esc}^{2}/u^{2}}\right)$
is a normalization constant such that $\int_{0}^{v_{esc}(R)}f\left(R,\vec{v}_{h}\right)d\vec{v}_{h}=1$. The radial dependence of the velocity dispersion is obtained via the virial theorem:

\begin{equation}
u\left(R\right)=\sqrt{\frac{GM_{tot}\left(R\right)}{R}},\label{uR}
\end{equation}
where $M_{tot}=M_{DM}+M_{B}$, while $v_{esc}=\sqrt{2}u$.

Using the velocity distribution \eqref{fRv}, going to center-of-mass and relative velocities $\vec{v}_{CM}$ and $\vec{v}$ and performing
the integrals over $\vec{v}_{CM}$, we obtain for the mean pair-production cross section times relative velocity:

\begin{equation}
\left\langle \sigma_{ee}v\right\rangle =\frac{1}{u^{2}}\frac{\sqrt{2\pi}u\, \erf\left(\sqrt{2}v_{esc}/u\right)-4v_{esc}e^{-2v_{esc}^{2}/u^{2}}}{\left(\sqrt{\pi}u\, \erf\left(v_{esc}/u\right)-2v_{esc}e^{-v_{esc}^{2}/u^{2}}\right)^{2}}\int_{0}^{2v_{esc}}\sigma_{ee}\left(v\right)v^{3}e^{-v^{2}/2u^{2}}\dd v,\label{sigmaeev}
\end{equation}
which is also a function of $R$ through $u$ and $v_{esc}$. Putting \eqref{dsigmaee}, \eqref{rhoBurkert}, \eqref{MDM}, \eqref{MB}, \eqref{uR} and \eqref{sigmaeev} together allows us to compute the pair-production rate in the galactic bulge defined in \eqref{rateee} as a function of $\rho_{0}$ and $m_O$.

\chapter{Electromagnetic form factors}\label{AppendixB}

Let $\rho_1(\vec x)$ and $\rho_2(\vec y)$ be two charge distributions localized by $\vec x$ and $\vec y$
from two arbitrary reference points. We can parameterize the potential energy of the electrostatic interaction 
between the two distributions, or the electrostatic interaction potential $V$, via the vector $\vec D$ that 
joins the two reference points, although it is independent on the choice of the latter. We have:

\begin{equation}
V(\vec D)=\frac{1}{4\pi}\int \dd \vec x \dd \vec y \frac{\rho_1(\vec x)\rho_2(\vec y)}{|\vec D + \vec y - \vec x|}.
\end{equation}
We can then write $\rho_1(\vec x)$ and $\rho_2(\vec y)$ in terms of their Fourier transforms $\tilde{\rho}_1(\vec k)$ and $\tilde{\rho}_2(\vec l)$ and  use  the properties of the $\delta$-distribution, to get:

\begin{eqnarray}
V(\vec D) & = & \frac{1}{4\pi}\int\frac{\dd \vec k \dd\vec l}{(2\pi)^6}\int \dd\vec x \dd\vec y \dd\vec z \frac{\tilde{\rho}_1(\vec k)\tilde{\rho}_2(\vec l)}{|\vec z|}e^{-i\vec k \cdot \vec x}e^{-i\vec l \cdot \vec y}\delta(\vec z - \vec D - \vec y + \vec x)\nonumber \\
& = & \int\frac{\dd \vec k \dd\vec l \dd\vec q}{(2\pi)^9}\int \dd\vec x \dd\vec y \dd\vec z \frac{\tilde{\rho}_1(\vec k)\tilde{\rho}_2(\vec l)}{q^2}e^{-i\vec k \cdot \vec x}e^{-i\vec l \cdot \vec y}e^{-i\vec q \cdot \vec z}\delta(\vec z - \vec D - \vec y + \vec x)\nonumber,
\end{eqnarray}
where we have used the fact that the Fourier transform of $\frac{1}{|\vec z|}$ is $\frac{4\pi}{q^2}$ at the second line. Performing the integral over $\vec z$ and rearranging the exponentials leads to:

\begin{eqnarray}
V(\vec D) & = & \int\frac{\dd \vec k \dd\vec l \dd\vec q}{(2\pi)^9}\int \dd\vec x \dd\vec y \frac{\tilde{\rho}_1(\vec k)\tilde{\rho}_2(\vec l)}{q^2}e^{-i\vec k \cdot \vec x}e^{-i\vec l \cdot \vec y}e^{i\vec q \cdot \vec x}e^{-i\vec q \cdot \vec y}e^{-i\vec q \cdot \vec D}\nonumber \\
& = & \int\frac{\dd \vec k \dd\vec l \dd\vec q}{(2\pi)^9}\int \dd\vec x \dd\vec y \frac{\tilde{\rho}_1(\vec k)\tilde{\rho}_2(\vec l)}{q^2}e^{-i(\vec k - \vec q)\cdot \vec x}e^{-i(\vec l + \vec q)\cdot \vec y}e^{-i\vec q \cdot \vec D}\nonumber, 
\end{eqnarray}
where we recognize the inverse Fourier transforms of the $\delta$-functions:

\begin{eqnarray}
V(\vec D) & = & \int\frac{\dd \vec k \dd\vec l \dd\vec q}{(2\pi)^3}\frac{\tilde{\rho}_1(\vec k)\tilde{\rho}_2(\vec l)}{q^2}\delta(\vec k - \vec q)\delta(\vec l + \vec q)e^{-i\vec q \cdot \vec D}\nonumber \\
& = & \int\frac{\dd\vec q}{(2\pi)^3}\frac{\tilde{\rho}_1(\vec q)\tilde{\rho}_2(-\vec q)}{q^2}e^{-i\vec q \cdot \vec D}.\nonumber
\end{eqnarray}
As the form factors $F_1$ and $F_2$ of the distributions $\rho_1$ and $\rho_2$ are by definition their Fourier transforms, this proves that the Fourier transform $\tilde V$ of the potential $V$ is given by:

\begin{equation}
\tilde V(\vec q)=\frac{F_1(\vec q)F_2(-\vec q)}{q^2}.
\end{equation}

\chapter{Calculation of $V_{e\dot{e}^{+}}$}\label{AppendixC}

Using \eqref{eq:4} and the expressions for the form factors $F_e$ and $F_{\dot{e}^{+}}$ of the visible and dark electronic clouds, we have:

\begin{equation}
V_{e\dot{e}^{+}}(r)=-\frac{Z\epsilon \alpha}{i\pi r}\sum\limits_{k=1}^3 c_k\int_{-\infty}^{\infty}\frac{\dd q}{q}\frac{256}{\left(4+\dot{a}_0^2q^2\right)^2\left(4+a_k^2q^2\right)^2}e^{iqr}.\label{eq:28}
\end{equation}

This can be calculated by analytic continuation of the integrand in the complex plane and by residues. The function $f_k(q)=\frac{e^{iqr}}{q}\frac{256}{\left(4+\dot{a}_0^2q^2\right)^2\left(4+a_k^2q^2\right)^2}$ has poles in $q=0$ (order $1$), $q=\pm \frac{2i}{\dot{a}_0}$ (order $2$) and $q=\pm \frac{2i}{a_k}$ (order $2$) and can be rewritten as:

\begin{equation}
f_k=\frac{e^{iqr}}{q}\frac{256}{\dot{a}_{0}^{4}a_{k}^{4}\left(q-i\frac{2}{\dot{a}_0}\right)^2\left(q+i\frac{2}{\dot{a}_0}\right)^2\left(q-i\frac{2}{a_k}\right)^2\left(q+i\frac{2}{a_k}\right)^2}.
\end{equation}

\begin{figure}
\begin{center}
\includegraphics[scale=0.25]{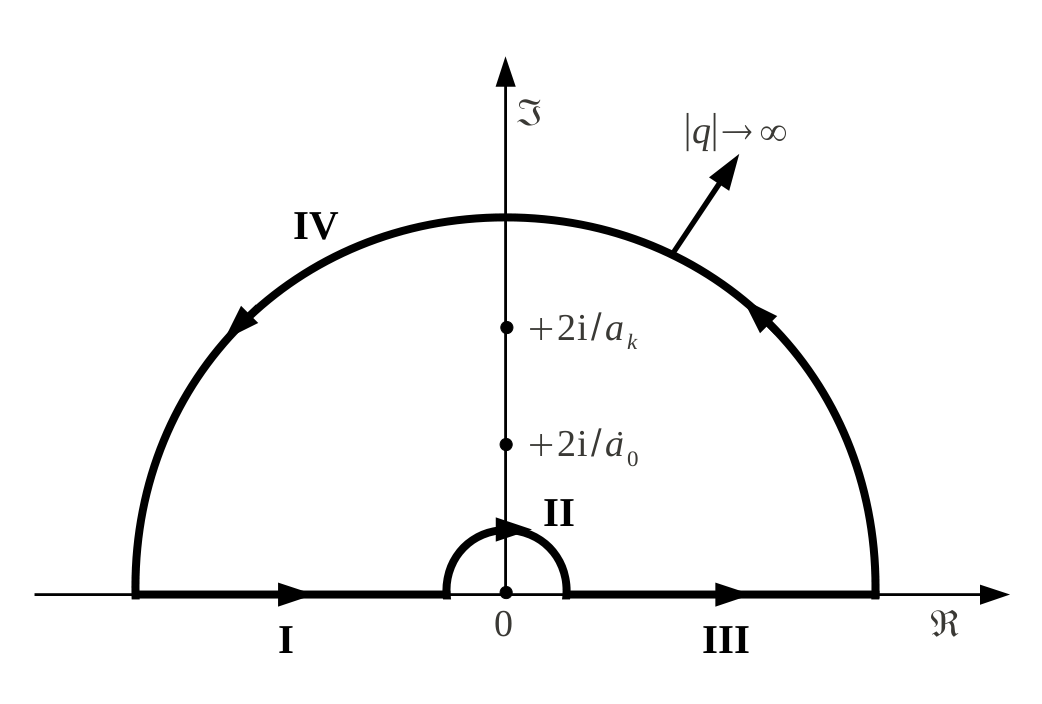}
\end{center}
\caption{Contour in the complex plane that encloses the poles of order $2$ in $q=+\frac{2i}{\dot{a}_0}$ and $q=+\frac{2i}{a_k}$ of the function $f_k(q)$. The limit of this contour for $|q| \rightarrow \infty$ on part IV is taken to calculate the integral \eqref{eq:28}.}
\label{contour}
\end{figure}

Taking the limit $|q| \rightarrow \infty$ of the contour represented in Figure \ref{contour}, we have:

\begin{equation}
\int_{\mathrm{I}}f_k \dd q+\int_{\mathrm{II}}f_k \dd q+\int_{\mathrm{III}}f_k \dd q+\int_{\mathrm{IV}}f_k \dd q=2i\pi \left(\mathrm{Res}_{f_k}\left(q=+\frac{2i}{\dot{a}_0}\right)+\mathrm{Res}_{f_k}\left(q=+\frac{2i}{a_k}\right)\right)\label{eq:30},
\end{equation}
with $\int_{\mathrm{IV}}f_k \dd q \rightarrow 0$ as the part IV of the contour is sent to regions where $\Im(q)>0$. Here, $\mathrm{Res}_{f_k}(q=z)$ denotes the residue of $f_k$ in $q=z$. Part II is realized clockwise and on half a turn around the pole in $q=0$, and thus:

\begin{equation}
\int_{\mathrm{II}}f_k \dd q=\left(-\frac{1}{2}\right)2i\pi \mathrm{Res}_{f_k}\left(q=0\right).\label{eq:31}
\end{equation}

The residues in $q=0$, $+\frac{2i}{\dot{a}_0}$ and $+\frac{2i}{a_k}$ are respectively given by:

\begin{eqnarray}
\mathrm{Res}_{f_k}\left(q=0\right) & = & 1\label{eq:32}, \\
\mathrm{Res}_{f_k}\left(q=+\frac{2i}{\dot{a}_0}\right) & = & e^{-2r/\dot{a}_0}\frac{T_k(r)}{4U_k}\label{eq:33}, \\
\mathrm{Res}_{f_k}\left(q=+\frac{2i}{a_k}\right) & = & e^{-2r/a_k}\frac{S_k(r)}{4U_k}\label{eq:34},
\end{eqnarray}
where $S_k(r)=(-2\dot{a}_0^4a_k^3+4\dot{a}_0^2a_k^5-2a_k^7)r-6\dot{a}_0^4a_k^4+8\dot{a}_0^2a_k^6-2a_k^8$ and $T_k(r)=(-2\dot{a}_0^3a_k^4+4\dot{a}_0^5a_k^2-2\dot{a}_0^7)r-6\dot{a}_0^4a_k^4+8\dot{a}_0^6a_k^2-2\dot{a}_0^8$ are polynomials of order $1$ in $r$ and $U_k=\dot{a}_0^8-4\dot{a}_0^6a_k^2+6\dot{a}_0^4a_k^4-4\dot{a}_0^2a_k^6+a_k^8$. Therefore, using \eqref{eq:30} to \eqref{eq:34}, we get for the integral of $f_k$ along the real axis:

\begin{eqnarray}
\int_{-\infty}^{\infty}f_k \dd q & = & \int_{\mathrm{I}}f_k \dd q+\int_{\mathrm{III}}f_k \dd q\nonumber \\
 & = & -\int_{\mathrm{II}}f_k \dd q+2i\pi \left(\mathrm{Res}_{f_k}\left(q=+\frac{2i}{\dot{a}_0}\right)+\mathrm{Res}_{f_k}\left(q=+\frac{2i}{a_k}\right)\right)\nonumber \\
  & = & i\pi \left(1+\frac{1}{2}\frac{e^{-2r/a_k}S_k(r)+e^{-2r/\dot{a}_0}T_k(r)}{U_k}\right),\label{eq:37}
\end{eqnarray}
and finally, inserting \eqref{eq:37} into \eqref{eq:28}, we get the desired expression for $V_{e\dot{e}^{+}}$:

\begin{equation}
V_{e\dot{e}^{+}}(r)=-\frac{Z\epsilon \alpha}{r}\sum\limits_{k=1}^3 c_k\left(1+\frac{1}{2}\frac{e^{-2r/a_k}S_k(r)+e^{-2r/\dot{a}_0}T_k(r)}{U_k}\right).
\end{equation}

\newpage
\bibliographystyle{hope}
\addcontentsline{toc}{chapter}{Bibliography}
\bibliography{references}

\end{document}